\documentclass[a4paper,final,12pt]{book}%
\pdfoutput=1
\usepackage{amsfonts}
\usepackage{amsmath}
\usepackage{amssymb}
\usepackage{graphicx}
\usepackage{fancyheadings}
\usepackage{hyperref}%
\providecommand{\U}[1]{\protect\rule{.1in}{.1in}}
\providecommand{\U}[1]{\protect\rule{.1in}{.1in}}
\providecommand{\U}[1]{\protect\rule{.1in}{.1in}}
\providecommand{\U}[1]{\protect\rule{.1in}{.1in}}
\providecommand{\U}[1]{\protect\rule{.1in}{.1in}}
\providecommand{\U}[1]{\protect\rule{.1in}{.1in}}
\providecommand{\U}[1]{\protect\rule{.1in}{.1in}}
\renewcommand{\chaptermark}[1]{\markboth{#1}{}}

\renewcommand{\sin}{\mathop{\rm sen}\nolimits}
\makeindex
\pdfoutput=1 
\ifx\pdfoutput\relax\let\pdfoutput=\undefined\fi
\newcount\msipdfoutput
\ifx\pdfoutput\undefined\else
\ifcase\pdfoutput\else
\ifx\paperwidth\undefined\else
\ifdim\paperheight=0pt\relax\else\pdfpageheight\paperheight\fi
\ifdim\paperwidth=0pt\relax\else\pdfpagewidth\paperwidth\fi
\fi\fi\fi
\begin{document}

\frontmatter
\title{\vspace{-3cm}Decoherence of many-spin systems in NMR:\\From molecular characterization to an environmentally induced quantum dynamical phase transition \bigskip\ \bigskip\\{\small por}\bigskip\ \bigskip\\\textbf{Gonzalo Agust\'{\i}n \'{A}lvarez}\\{\large \bigskip\ \bigskip}\\{\large Presentado ante la Facultad de Matem\'{a}tica, Astronom\'{\i}a y
F\'{\i}sica}\\{\large como parte de los requerimientos para acceder al grado de }\\{\Large Doctor en F\'{\i}sica}\\{\large de la}\\\vspace{2.4cm}{\LARGE Universidad Nacional de C\'{o}rdoba}}
\date{Marzo de 2007\\
\copyright \ FaMAF - UNC 2007\bigskip\ \bigskip\\
{\large Directora: Dra. Patricia Rebeca Levstein}}
\maketitle
\tableofcontents
\listoffigures
\frontmatter

\chapter{Abstract}

\begin{center}
{\Large Decoherence of many-spin systems in NMR:}

{\Large From molecular characterization to an environmentally induced quantum
dynamical phase transition}
\end{center}

The control of open quantum systems has a fundamental relevance for fields
ranging from quantum information processing to nanotechnology. Typically, the
system whose coherent dynamics one wants to manipulate, interacts with an
environment that smoothly degrades its quantum dynamics. Thus, a precise
understanding of the inner mechanisms of this process, called
\textquotedblleft decoherence\textquotedblright, is critical to develop
strategies to control the quantum dynamics.

In this thesis we solved the generalized Liouville-von Neumann quantum master
equation to obtain the dynamics of many-spin systems interacting with a spin
bath. We also solve the spin dynamics within the Keldysh formalism. Both
methods lead to identical solutions and together gave us the possibility to
obtain numerous physical predictions that contrast well with Nuclear Magnetic
Resonance experiments. We applied these tools for molecular characterizations,
development of new numerical methodologies and the control of quantum dynamics
in experimental implementations. But, more important, these results
contributed to fundamental physical interpretations of how quantum
dynamics\ behaves in open systems. In particular, we found a manifestation of
an environmentally induced quantum dynamical phase transition.

\chapter{Resumen}

\begin{center}
{\Large Decoherencia en sistemas de espines interactuantes en RMN:}

{\Large De la caracterizaci\'{o}n molecular a una transici\'{o}n de fase en la
din\'{a}mica cu\'{a}ntica inducida por el ambiente}
\end{center}

El control de sistemas cu\'{a}nticos abiertos tiene una relevancia fundamental
en campos que van desde el procesamiento de la informaci\'{o}n cu\'{a}ntica
hasta la nanotecnolog\'{\i}a. T\'{\i}picamente, el sistema cuya din\'{a}mica
coherente se desea manipular, interact\'{u}a con un ambiente que suavemente
degrada su din\'{a}mica cu\'{a}ntica. Es as\'{\i} que el entendimiento preciso
de los mecanismos internos de este proceso, llamado decoherencia, es
cr\'{\i}tico para el desarrollo de estrategias para el control de la
din\'{a}mica cu\'{a}ntica.

En esta tesis usamos la ecuaci\'{o}n maestra cu\'{a}ntica generalizada de
Liouville-von Neumann para resolver la din\'{a}mica de sistemas de muchos
espines interactuando con un ba\~{n}o de espines. Tambi\'{e}n obtuvimos la
din\'{a}mica de espines dentro del formalismo de Keldysh. Ambos m\'{e}todos
nos llevaron a id\'{e}nticas soluciones y juntos nos dieron la posibilidad de
realizar numerosas predicciones que concuerdan con las observaciones de
experimentos de Resonancia Magn\'{e}tica Nuclear. Estos resultados son usados
para la caracterizaci\'{o}n molecular, el desarrollo de nuevas
metodolog\'{\i}as num\'{e}ricas y el control de la din\'{a}mica cu\'{a}ntica
en implementaciones experimentales. Pero a\'{u}n m\'{a}s importante es el
surgimiento de interpretaciones f\'{\i}sicas fundamentales de la din\'{a}mica
cu\'{a}ntica de sistemas cu\'{a}nticos abiertos, tales coma la
manifestaci\'{o}n de una transici\'{o}n de fase en la din\'{a}mica
cu\'{a}ntica inducida por el ambiente.

\chapter{Acknowledgments}

I wish to express my gratitude to many people, who in different ways, have contributed to the realization of this work. From the beginning of my thesis, one of my main motivations was to train myself as a physicist; in this aspect, from my point of view, a strong complementation between theoretical and experimental tools is essential to attack the diverse problems of nature. For that reason, I am specially grateful to my director, Patricia Levstein, and my co-director, Horacio Pastawski, who offered me their knowledge and the ways to see and do Physics. Patricia has contributed from an experimental point of view while Horacio has done so from the theoretical one, thus, helping me to generate a theoretical and experimental background to face Physics. In addition, I am indebted to Patricia for having helped me in the polishing of the English version of this thesis.

\bigskip

	I am also very thankful to the examining committee that evaluated my thesis: Prof. Dr. Carlos Balseiro, Prof. Dr. Guido Raggio, Prof. Dr. Juan Pablo Paz and Prof. Dr. Pablo Serra, who read my work and contributed with very interesting comments.

	I wish to thank J\'{e}sus Raya, with whom it was very pleasing and enriching to work during my stay in France, and who gave me a complementary view with respect to the experimental measurements. Also, I would like to thank J\'{e}r\^{o}me Hirschinger for his hospitality and comments.

	I offer my grateful thanks to Lucio Frydman for his hospitality during the time I worked in his laboratory but, most important of all, for having contributed in my training and having shared his style of working with me.

\bigskip

	I am also deeply grateful
\begin{itemize} 
\item To my group partners: especially the oldest ones, Fernando Cucchietti, Luis Foa Torres, Ernesto Danieli and Elena Rufeil Fiori and the newest ones, Claudia
S\'{a}nchez, Bel\'{e}n Franzoni, Hern\'{a}n Calvo, Yamila Garro Linck, Axel
Dente and Guillermo Ludue\~{n}a, who not only contributed to my training by sharing together our knowledge, but also have contributed to a warm environment of work.
\item To the staff at Lanais: Gustavo Monti, Mariano Zuriaga, N\'{e}stor Veglio, Karina Chattah, Rodolfo Acosta and Fernando Zuriaga who numerous times helped me with my rebellious computer.
\item To the administration people who always, with their better attitude, helped me a lot.
\item To my office mates: Fernando Bonetto, Ana Majtey, Alejandro Ferr\'{o}n, Santiago Pighin, Santiago G\'{o}mez, Marianela Carubelli and Josefina Perlo who have collaborated to create a pleasant atmosphere at work.
\end{itemize}
\bigskip

	Very special thanks
\begin{itemize} 
\item To my family, who have unconditionally supported me in everything and have always given me their kindest support.
\item To all my friends for their love and moments of amusement. In special to Lucas, Eduardo, Andr\'{e}s and Sandra.
\item But the ones I am most grateful to are Valeria, who was close to me most of my life and while I was doing this thesis (thanks for your support); Sol, who stood next to me at a very critical moment, helping me to re-focus my effort; and Any who supported me and helped me keep my critical state at the culmination of this work.
\end{itemize}
\bigskip

	I am thankful to CONICET for the financial support, offered through a doctoral fellowship, to do this work possible. Also I wish to thank CONICET, ANPCyT, SECyT and Fundaci\'{o}n Antorchas for their financial support for my education in my country and abroad.

\bigskip

	Finally, I wish to thank all of those who, in one way or another, have supported and encouraged me to make this thesis come true. To everybody:

\bigskip

THANK YOU VERY MUCH....

\chapter{Agradecimientos}

Deseo expresar mi agradecimiento a muchas personas, que en diferentes \textquotedblleft formas
y medidas\textquotedblright, fueron contribuyendo a la finalizaci\'{o}n de este trabajo. Desde
el comienzo del mismo, una de mis principales motivaciones fue formarme como
f\'{\i}sico; en este aspecto, desde mi punto de vista es esencial una fuerte
complementaci\'{o}n entre herramientas te\'{o}ricas y experimentales para
atacar los diversos problemas de la naturaleza. Es por ello, que estoy en
especial muy agradecido con mi directora, Patricia Levstein, y mi co-director,
Horacio Pastawski; quienes me brindaron su conocimiento y las formas de ver y
hacer f\'{\i}sica. Patricia contribuyendo desde su punto de vista experimental
y Horacio desde el te\'{o}rico, ayud\'{a}ndome as\'{\i} a generar una
formaci\'{o}n te\'{o}rica-experimental de c\'{o}mo encarar la f\'{\i}sica. Le
agradezco mucho a Patricia, adem\'{a}s, por haberme ayudado en el pulido de la
escritura de esta tesis, en el idioma ingl\'{e}s.

\bigskip

Estoy muy agradecido tambi\'{e}n con el jurado, que evalu\'{o} mi tesis, el
Dr. Carlos Balseiro, Dr. Guido Raggio, Dr. Juan Pablo Paz y Dr. Pablo Serra,
quienes leyeron mi trabajo y me aportaron comentarios muy interesantes.

Tambi\'{e}n le agradezco a J\'{e}sus Raya, con quien fue muy grato e
enriquecedor trabajar en mi estad\'{\i}a en Francia, quien me dio una
visi\'{o}n complementaria a la de Patricia con respecto a las mediciones
experimentales. A J\'{e}r\^{o}me Hirschinger por su hospitalidad y comentarios.

Le agradezco a Lucio Frydman, por su hospitalidad en mi pasant\'{\i}a en su
laboratorio; pero mucho m\'{a}s importante por su contribuci\'{o}n en mi
formaci\'{o}n y por haber compartido conmigo su forma de trabajo.

\bigskip

Agradezco tambi\'{e}n a mis compa\~{n}eros de grupo, empezando por los m\'{a}s
antiguos: Fernando Cucchietti, Luis Foa Torres, Ernesto Danieli y Elena Rufeil
Fiori, quienes no s\'{o}lo contribuyeron en mi formaci\'{o}n compartiendo
entre todos nuestro conocimiento, sino tambi\'{e}n por haber aportado calidez
al ambiente de trabajo. Lo mismo agradezco a los m\'{a}s nuevos: Claudia
S\'{a}nchez, Bel\'{e}n Franzoni, Hern\'{a}n Calvo, Yamila Garro Linck, Axel
Dente y Guillermo Ludue\~{n}a.

A la gente del Lanais: Gustavo Monti, Mariano Zuriaga, N\'{e}stor Veglio,
Karina Chattah, Rodolfo Acosta y a Fernando Zuriaga, quien numerosas veces me
ayud\'{o} con mi rebelde computadora.

A la gente de administraci\'{o}n, que con su mejor onda me ayudaron siempre.

A mis compa\~{n}eros de oficina: Fernando Bonetto, Ana Majtey, Alejandro
Ferr\'{o}n, Santiago Pighin, Santiago G\'{o}mez, Marianela Carubelli, Josefina
Perlo por haber colaborado para generar un espacio grato de trabajo.

\bigskip

Un muy especial agradecimiento a mi familia, por haberme bancado y apoyado en
todo incondicionalmente y por su apoyo afectivo.

A todos mis amigos por su afecto y momentos de descuelgue. En especial a
Lucas, Eduardo, Andr\'{e}s y Sandra.

A quienes m\'{a}s tengo que agradecerles es: a Valeria, quien estuvo a mi lado
gran parte de mi vida y de este trabajo, gracias por tu sost\'{e}n; a Sol, que
estuvo, en un momento muy cr\'{\i}tico ayud\'{a}ndome a reenfocar mi esfuerzo
y a Any que aguant\'{o} y sostuvo mi estado cr\'{\i}tico durante la
culminaci\'{o}n de este trabajo.

\bigskip

Agradezco a CONICET por el apoyo econ\'{o}mico, brindado a trav\'{e}s de una
beca doctoral para realizar este trabajo. A la instituciones, CONICET, ANPCyT,
SECyT y Fundaci\'{o}n Antorchas por el soporte econ\'{o}mico para mi
formaci\'{o}n, tanto aqu\'{\i} como en el exterior.

\bigskip

Y a todos aquellos, que de una manera u otra me fueron apoyando y alentando
para concretar este trabajo. A todos MUCHAS GRACIAS....

\mainmatter

\renewcommand{\chaptermark}[1]{\markboth{\chaptername\ \thechapter. #1}{}}

\chapter{Introduction\label{Mark_introduction}}

Quantum Mechanics was developed to describe the behavior of matter at very
small scales, around the size of single atoms. Today, it is applied to almost
every device that improves our quality of life, from medical to communication
technology. Since it involves laws and concepts that challenge our intuition,
it keeps having a revolutionary impact on the formulation of new philosophical
and scientific concepts not totally solved today \cite{Omnes92,Schlosshauer04}%
. While the foundations of quantum mechanics were established in the early
20$^{\mathrm{th}}$ century, many fundamental aspects of the theory are still
actively studied and this thesis intends to contribute to this knowledge.

\section{What is quantum physics?}

One of the main characteristics of quantum mechanics is that it involves many
counterintuitive concepts such as the superposition states. They were
illustrated by the Austrian physicist Erwin Schr\"{o}dinger in 1935 by his
famous Schr\"{o}dinger's cat thought experiment. In his words
\cite{Schrodinger35}:

\begin{quote}
\emph{\textquotedblleft One can even set up quite ridiculous cases. A cat is
penned up in a steel chamber, along with the following device (which must be
secured against direct interference by the cat): in a Geiger counter there is
a tiny bit of radioactive substance, so small, that perhaps in the course of
the hour one of the atoms decays, but also, with equal probability, perhaps
none; if it happens, the counter tube discharges and through a relay releases
a hammer which shatters a small flask of hydrocyanic acid. If one has left
this entire system to itself for an hour, one would say that the cat still
lives if meanwhile no atom has decayed. The psi-function of the entire system
would express this by having in it the living and dead cat (pardon the
expression) mixed or smeared out in equal parts.}

\emph{It is typical of these cases that an indeterminacy originally restricted
to the atomic domain becomes transformed into macroscopic indeterminacy, which
can then be resolved by direct observation. That prevents us from so naively
accepting as valid a "blurred model" for representing reality. In itself it
would not embody anything unclear or contradictory. There is a difference
between a shaky or out-of-focus photograph and a snapshot of clouds and fog
banks.\textquotedblright}

\qquad\qquad\qquad\qquad\qquad\qquad\qquad\qquad\qquad\qquad\qquad Erwin Schr\"{o}dinger
\end{quote}

%

\begin{figure}
[tbh]
\begin{center}
\includegraphics[
height=2.5278in,
width=5.028in
]%
{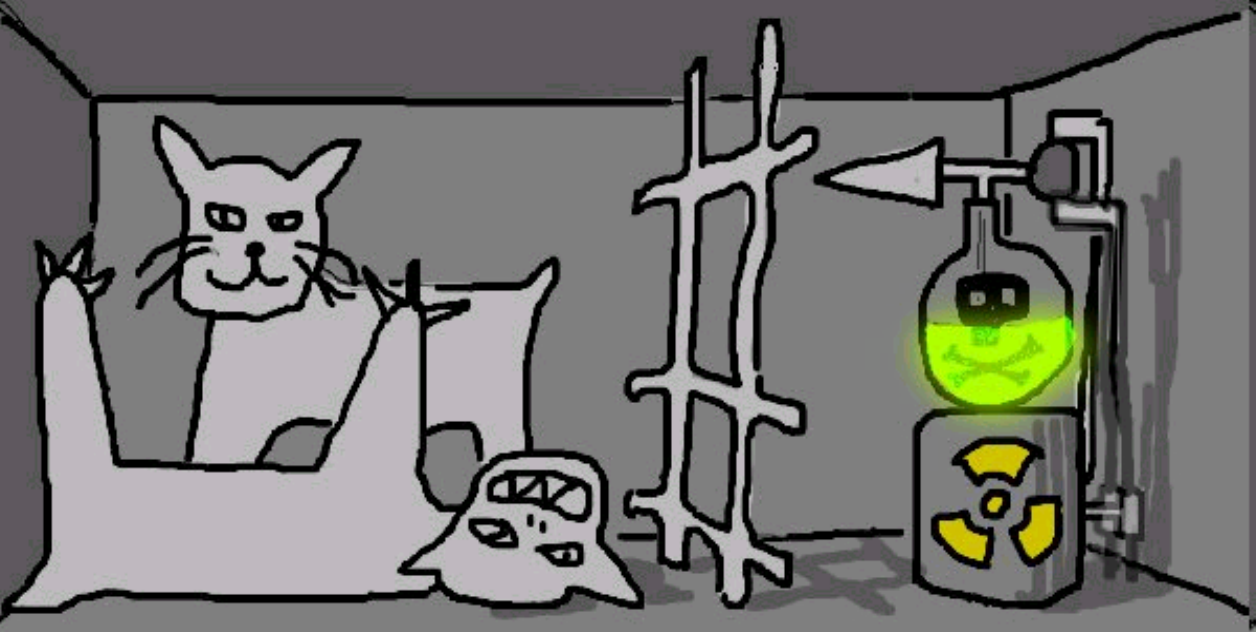}%
\caption[Cartoon description of the Schr\"{o}dinger's cat paradox.]{Cartoon
description of the Schr\"{o}dinger's cat paradox. After an hour the cat is in
a quantum superposition of coexisting alive and dead states. Only after
opening the box we found the cat in a defined state. Figure extracted from http://en.wikipedia.org/wiki/Image:Katze.jpg.}%
\label{Fig_schrodinger_cat}%
\end{center}
\end{figure}

Essentially, he states that if we put an alive cat in a box where, isolated
from external interference, is in a situation where death has an appreciable
probability, the cat's state can only be described as a superposition of the
possible state results (dead and alive), i.e. the two states at the same time.
This situation is sometimes called quantum indeterminacy or the observer's
paradox: the observation or measurement itself affects an outcome, so that it
can never be known what the outcome would have been, if it were not observed.
The Schr\"{o}dinger paper \cite{Schrodinger35} was part of a discussion of the
Einstein, Podolsky and Rosen's paradox \cite{Einstein35} that attempted to
demonstrate the incompleteness of quantum mechanics. They said that quantum
mechanics has a non-local effect on the physical reality. However, recent
experiments refuted the principle of locality, invalidating the EPR's paradox.
The property that disturbed the authors was called entanglement (a
superposition phenomenon) that could be described briefly as a
\textquotedblleft spooky action at a distance\textquotedblright\ as expressed
in ref. \cite{Einstein35}. This was a very famous counterintuitive effect of
quantum mechanics which leads very important physicists to mistrust of quantum
theory. The entanglement property could be schematized by adding some
condiments to the Schr\"{o}dinger's cat thought experiment. First of all, we
may consider that the indeterminacy on the cat's state is correlated with the
state of the flask of hydrocyanic acid, i.e. if the cat is alive the flask is
intact but if the cat is dead the flask is broken. We have here two elements
or systems (the cat and the flask) in a superposition state $\left\vert
\text{cat alive,flask intact}\right\rangle $ and $\left\vert \text{cat
dead,flask broken}\right\rangle $ existing at the same time. Assuming that
after an hour we can divide the box with a slide as shown in figure
\ref{Fig_schrodinger_cat} and deactivate the trigger, we can separate as we
want the two boxes. Then, if someone opens the cat's box and sees the cat's
state, the state of the flask will be determined instantaneously without
concerning the distance between them. This is only a cartoon description of
what quantum entanglement is about, but for a further description we refer to
Nielsen and Chuang (2000) \cite{Nielsen00} or
chapter\ \ref{Mark_entanglement_vs_ensemble}.

\begin{figure}
[pth]
\begin{center}
\includegraphics[
height=4.1883in,
width=5.9075in
]%
{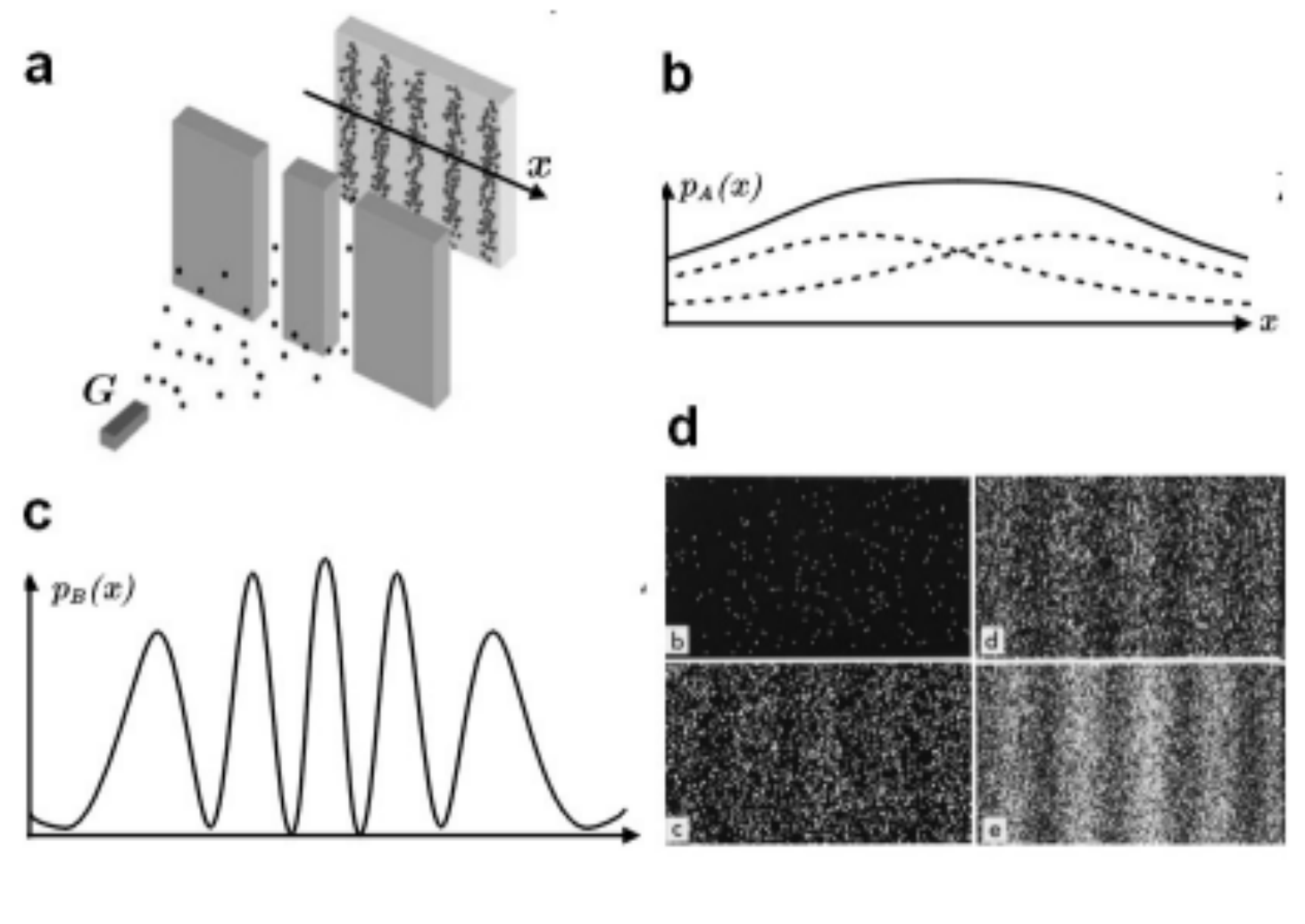}%
\caption[The double slit experiment.]{The double slit experiment. a) Schematic
representation of the double slit device. b) The solid line is the classical
probability prediction which is the sum of the individual one-slit
probabilities. c) Interference pattern predicted by quantum probabilities
accounting superposition. Panels b) and c) describe mathematical
probabilities, in panel d) the physical reality is shown. The experiments were
performed by A. Tonomura, \emph{et al.} \cite{Tonomura89} where they showed
that single electron events build up to form an interference pattern in the
double-slit experiments. The number of electrons detected are 100, 3000, 20000
and 70000 respectively. Reprinted with permission from A. Tonomura, \emph{et al.} Amer. J. Phys. Vol. 57, Issue 2, Page 117, 1989. Copyright 1989, American Association of Physics Teachers.}%
\label{Fig_two_slit_experiment}%
\end{center}
\end{figure}

One of the most interesting effects of quantum superposition is the
interference phenomenon consequence of the information indeterminacy of the
quantum state (dead or alive). The famous double slit ideal experiment, as
Richard Feynman said, contains everything you need to know about quantum
mechanics. As shown in fig. \ref{Fig_two_slit_experiment} a), the experiment
consists of a double slit where a particle (photon, electron, etc.) can pass
and a screen where it is detected.
Behind it, there is a screen where we can register where the particle arrives.
If only one of the slits is open, we have certainty that the particle only can
pass through this slit. The probability to arrive to different places of the
screen is shown in figure \ref{Fig_two_slit_experiment} b). There, we see that
the most probable place for the particle arrival is obtained projecting the
center of the slit to the register screen. Moving away from it, the
probability decreases monotonically. The reciprocal situation occurs if only
the other slit is open. However, if we leave the two slits open an
interference pattern appears as in figure \ref{Fig_two_slit_experiment} c).
Figures \ref{Fig_two_slit_experiment} b) and c) represent mathematical
probabilities (mathematical reality) describing the physical reality shown in
figure \ref{Fig_two_slit_experiment} d) \cite{Tonomura89}.

Paul Kwiat, Harald Weintfurter and Anton Zeilinger making reference to quantum
interference, in ref. \cite{Zeilinger96}, express:

\begin{quote}
\textquotedblleft\emph{According to the rules of quantum mechanics,
interference occurs whenever there is more than one possible way for a given
outcome to happen, and the ways are not distinguishable by any means (this is
a more general definition of interference than is often given in textbooks).
In the double-slit experiment, light can reach the screen in two possible ways
(from the upper or the lower slit), and no effort is made to determine which
photons pass through which slit. If we somehow could determine which slit a
photon passed through, there would be no interference, and the photon could
end up anywhere on the screen. As a result, no fringe pattern would emerge.
Simply put, without two indistinguishable paths, interference cannot
occur.\textquotedblright}

\qquad\qquad\qquad\qquad\qquad Paul Kwiat, Harald Weinfurter and Anton Zeilinger
\end{quote}

Thus, the quantum mechanics is the physics of potentialities. When we have
determinacy of some event, the classical physics appears.
Nowadays, this appearance of the classical physics and state determinacy is
considered a consequence of a phenomenon called decoherence \cite{Zurek03}
which is the central topic of this thesis.

\section{Decoherence: the degradation of quantum superpositions}

The gedanken experiments introduced above must involve a perfect shielding
from external influences allowing the existence of quantum superposition.
Realistic quantum systems are never isolated, because they are immersed in an
environment that continuously interacts with them. A typical environment
consists of a system with many degrees of freedom that are hardly fully
controlled or are not relevant for the observation. The system-environment
(SE) interaction degrades the quantum superposition leading to the phenomenon
called decoherence \cite{Zurek03,Schlosshauer04}. Actually, the measurement
process to observe if the cat is dead or alive involves an interaction between
the system (cat, acid, box, hammer, etc.) and the environment (observer,
apparatus to observe, etc.). When the observation is performed, the cat is
found either dead or alive, but not in the state dead and alive. The last one
is a pure-state and the first one is a mixed-state. The decoherence process
leads the system from a pure-state to a mixed-state. It is important to
emphasize that, although quantum mechanics is open to many interpretations,
decoherence by itself is neither an interpretation nor a modification of the
theory. Thus, their existence can be taken as a well-confirmed fact. However,
the implications that derive from decoherence could need some interpretations
and this is one of the reasons why nowadays many researchers are devoted to
its study \cite{Zurek03,Schlosshauer04}.

Decoherence does not exist if we consider the entire system. It arises when we
are interested in a particular part of the system leading to the consideration
of a system plus an environment which is called an open system. Looking at the
properties of the system, the environment modifies them leading to
decoherence. It is at this point when the concept of the reduced density
operator appears as a tool to mathematically describe the quantum world. A
system is described by an entity called density operator, but the density
operator of the Universe is impossible to obtain, thus one decides to reduce
it to describe a relevant subsystem. The concept of the reduced density
operator appeared together with quantum mechanics introduced by Lev Landau
1927 \cite{Landau27} and further developed by John von Neumann 1932
\cite{Neumann32} and W.H. Furry 1936 \cite{Furry36}. To illustrate the idea of
how the reduced density matrix works, and why by observing at a subsystem we
can not distinguish between a pure and a mixed-state, we consider a system
with two entangled elements in a pure-state\footnote{This entanglement is
consequence of a previous interaction between the two elements.}:
\begin{equation}
\left\vert \Psi\right\rangle =\frac{1}{\sqrt{2}}\left(  \left\vert
+\right\rangle _{1}\left\vert -\right\rangle _{2}-\left\vert -\right\rangle
_{1}\left\vert +\right\rangle _{2}\right)  .
\end{equation}
For an observable $\hat{O}$ that belongs only to the system $1$, i.e. $\hat
{O}=\hat{O}_{1}\otimes\hat{1}_{2},$ the expectation value is given by%
\begin{equation}
\left\langle \hat{O}\right\rangle _{\Psi}=\mathrm{Tr}\left\{  \hat{\rho}%
\hat{O}\right\}  ,
\end{equation}
where the density operator of the pure-state is defined by%
\begin{equation}
\hat{\rho}=\left\vert \Psi\right\rangle \left\langle \Psi\right\vert .
\end{equation}
This statistical expectation value is defined as the sum of the values of the
possible outcomes, multiplied by the probability of that outcome. The same
statistics is applied to the reduced density operator that is obtained by
tracing over the degrees of freedom of the system $2.$ Thus, we obtain
\begin{equation}
\left\langle \hat{O}\right\rangle _{\Psi}=\mathrm{Tr}\left\{  \hat{\rho}%
\hat{O}\right\}  =\mathrm{Tr}_{1}\left\{  \hat{\sigma}_{1}\hat{O}_{1}\right\}
,
\end{equation}
where the reduced density operator is%
\begin{equation}
\hat{\sigma}_{1}=\mathrm{Tr}_{2}\left\{  \left\vert \Psi\right\rangle
\left\langle \Psi\right\vert \right\}  =~_{2}\left\langle +|\Psi\right\rangle
\left\langle \Psi|+\right\rangle _{2}+~_{2}\left\langle -|\Psi\right\rangle
\left\langle \Psi|-\right\rangle _{2}.
\end{equation}
Therefore, when the observer has access to a particular part of the system
(system $1$), all the information obtainable through the subsystem is
contained in the reduced density matrix (this assumes a statistical
expectation value).

Noting that the states of the system $2$ are orthogonal, $_{2}\left\langle
+|-\right\rangle _{2}=0,$ the reduced density matrix becomes diagonal
\begin{equation}
\hat{\sigma}_{1}=\mathrm{Tr}_{2}\left\{  \left\vert \Psi\right\rangle
\left\langle \Psi\right\vert \right\}  =\frac{1}{2}\left(  \left\vert
+\right\rangle \left\langle +\right\vert \right)  _{1}+\frac{1}{2}\left(
\left\vert -\right\rangle \left\langle -\right\vert \right)  _{1}.
\label{mixed-state-intro}%
\end{equation}
This result corresponds to the density matrix of a mixed-state of the system
$1$, i.e. in either one of the two states $\left\vert +\right\rangle _{1}$ and
$\left\vert -\right\rangle _{1}$ with equal probabilities as opposed to the
superposition state $\left\vert \Psi\right\rangle .$ A suitable interference
experiment could confirm if it is a pure or a mixed-state, but if the
observable belongs only to system $1$, the previous calculation demonstrates
that it is impossible to distinguish between a pure or a mixed-state. We
should not forget that this would not happen if the two elements, the system
($1$) and the environment ($2$) were not entangled. This demonstration could
be extended to an arbitrary system of $N$ elements as discussed in ref.
\cite{Schlosshauer04}. While eq. (\ref{mixed-state-intro}) could be
misinterpreted as it means that the state of the system is in both states at
the same time, it is important to remark that the density matrix is a
mathematical tool to calculate the probability distribution of a set of
outcomes of a measurement of the physical reality but it\ does not represent a
specific state of the system.
\begin{figure}
[htbf]
\begin{center}
\includegraphics[
height=3.557in,
width=3.4947in
]%
{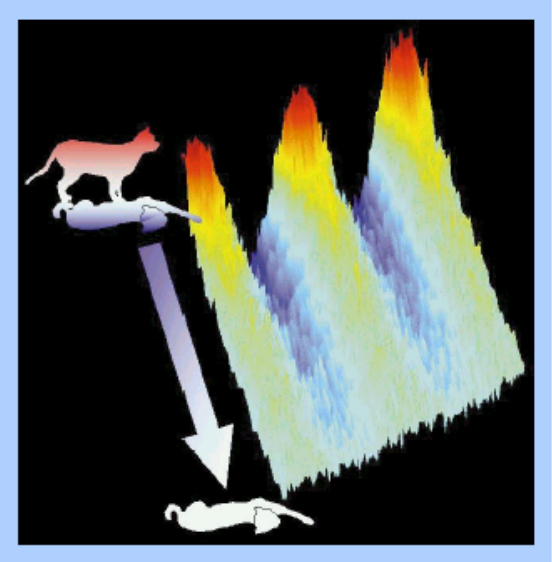}%
\caption[Schematic representation of decoherence.]{Schematic representation of
decoherence showed in ref. \cite{Schleich00}. Decoherence slides the quantum
world into the classical world. The oscillations from a quantum superposition
of a single atom, being in two places at once, gradually die out to lead the
atom to one definite place (either position). Reprinted by permission from Macmillan Publishers Ltd: Nature (W. P. Schleich, Nature {\bf 403}, (2000) 256), copyright (2000).}%
\label{Fig_schrodinger_cat_decoherence}%
\end{center}
\end{figure}

Thus, the interaction of a quantum system with an environment destroys the
quantum superposition leading the system to a statistical mixture of states.
This process called decoherence has many implications in the foundations of
quantum mechanics like the problem of quantum measurements, the quantum to
classical transition and irreversibility \cite{Zurek03,Schlosshauer04}.
But questions arise not only at a basic level. As a real quantum system can
never be isolated, when the technology gives us the possibility to work with
systems where quantum phenomena appear, the understanding of decoherence
becomes relevant to exploit the potentialities of quantum superpositions
.

In all the examples treated up to this point, the states constituting the
quantum superposition have the same probability to exist. However, what
happens when the probabilities are different? Moreover, what happens if the
probabilities are time dependent? This leads to temporal interferences that
appear in numerous experiments. For example, if we open each of the cat's
boxes in an ensemble, at one minute or after one hour the probability
distribution of the cat's state found is different. The same happens in the
double slit experiment if there is an obstacle that blocks the slit
oscillating between them. The interference pattern will be different depending
on the time of observation. What happens now if we consider the environment
effects? Including the SE interaction the quantum evolution is more
complicated. There is no simple explanation for the appearance of decoherence
because as we said previously one deals with an environment that has many
degrees of freedom. More importantly, the decoherence affects the phases of
the quantum superposition states, whose consequences are difficult to observe
and understand. The first quantitative evaluation was given by Feynman and
Veron (1963) \cite{Feynman63} where they calculated dissipation through an
environment of harmonic oscillators. Then, there were contributions from other
people like K. Hepp and E.H. Lieb (1974) \cite{Hepp74} and Wojciech Zurek
(1981,1982) \cite{Zurek81,Zurek82} who, while using less realistic models,
suggest the universality of the effect and the relation with the measurement
theory. However, the most complete work, in my opinion, was done by Caldeira
and Legget (1983) \cite{Caldeira83a,Caldeira83b,Caldeira83c}.

One of the first techniques, if not the first, in allowing the experimental
control of the temporal evolution of quantum states was the nuclear magnetic
resonance (NMR). In this thesis, we consider NMR experiments in connection
with the physical reality of the theoretical interpretations.

\section{NMR: The workhorse of quantum mechanics}

\begin{figure}
[tbh]
\begin{center}
\includegraphics[
height=4.6544in,
width=4.2194in
]%
{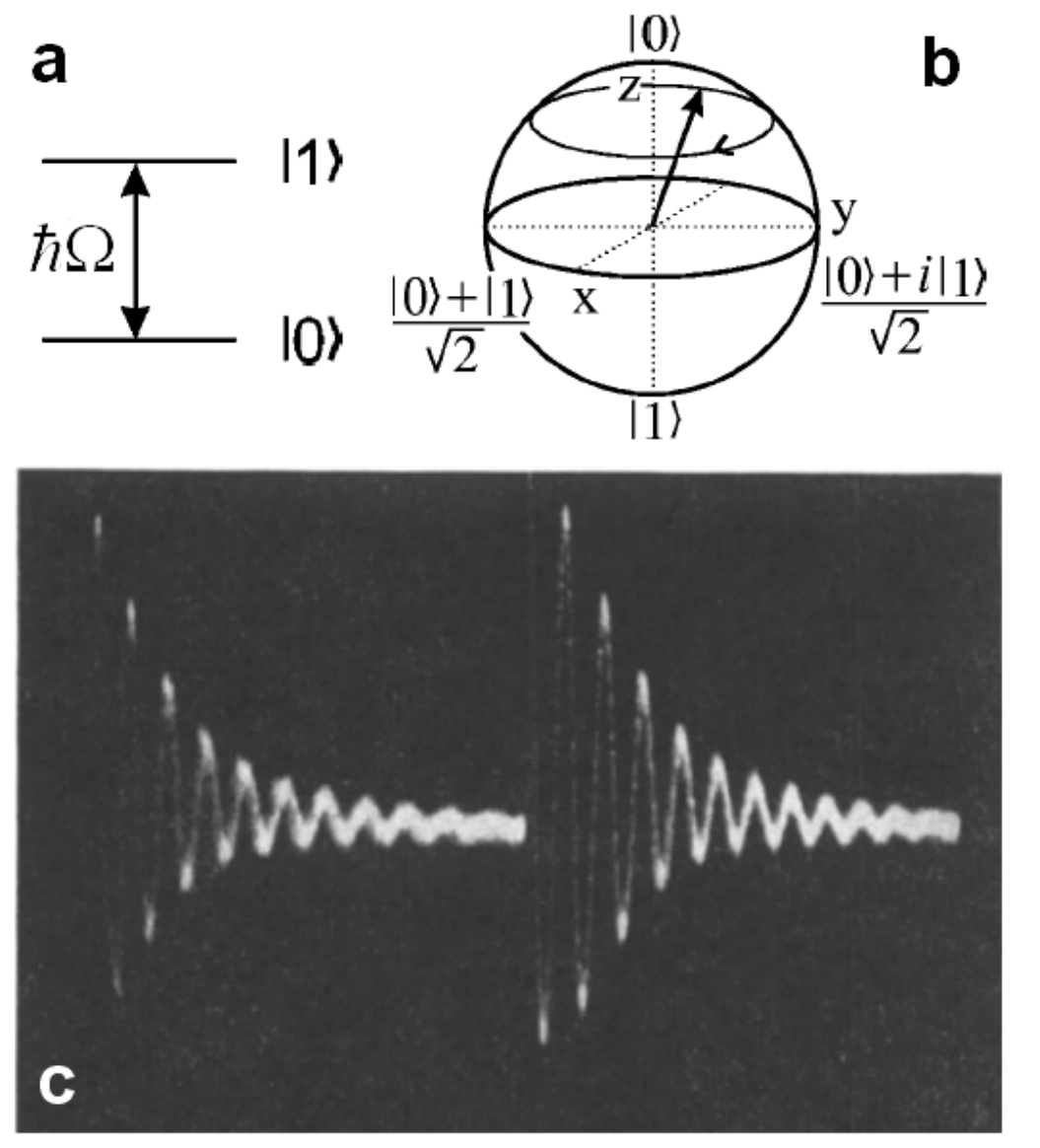}%
\caption[Oscillation between two states and the free induction decay.]%
{Oscillation between two states and the free induction decay. a) Energy
splitting, $\hbar\Omega$, of the states of a spin $1/2.$b) Scheme of the spin
precession around the static field. c) Experimental free induction decay
obtained by E. Hahn (1950) \cite{Hahn50a}. Reprinted figure with permission from E. L. Hahn, Phys. Rev. {\bf 77}, 297 (1950). Copyright (1950) by the American Physical Society.}%
\label{Fig_Hahn_experiment_FID_intro}%
\end{center}
\end{figure}

The origins of Nuclear Magnetic Resonance dates from 1930 when Isidor Isaac
Rabi discovered a technique for measuring the magnetic characteristics of
atomic nuclei. Rabi's technique was based on the resonance principle first
described by Irish physicist Joseph Larmor, and it enabled more precise
measurements of nuclear magnetic moments than had ever been previously
possible. Rabi's method was later independently improved upon by physicists
Edward Purcell and Felix Bloch in 1945
\cite{Bloch46a,Bloch46b,Bloch46c,Purcell46a,Purcell46b}. Later on, the
technique was improved by the advent of fast computers and the development of
pulse techniques that, through the Fourier transform, used the temporal
evolution of the signal to notably optimize the acquisition time. The first
experimental observations of the temporal evolution of a two-state system were
done by H.C. Torrey (1949) \cite{Torrey49} and Erwin Hahn (1950)
\cite{Hahn50a} where essentially a $1/2$-spin system (two-state system) is
under the presence of a static field $H_{0},$ which splits the energy levels
of the states $\left\vert +\right\rangle $ and $\left\vert -\right\rangle $ of
each spin [see fig. \ref{Fig_Hahn_experiment_FID_intro} a)].%
Then, through a transversal field with a radio-frequency (RF) pulse, one can
build a superposition state $a\left\vert +\right\rangle +b\left\vert
-\right\rangle $ whose dynamics can be interpreted as a classical precession
around the static field direction with the Larmor frequency $\Omega$ [see fig.
\ref{Fig_Hahn_experiment_FID_intro} b)]. Fig.
\ref{Fig_Hahn_experiment_FID_intro} c) shows the original experimental data
taken by Hahn \cite{Hahn50a}, where one can observe, after detection, a
manifestation of the oscillation between the two states in an ensemble of
spins. The attenuation of the oscillations is a consequence of the interaction
with the environment, the other degrees of freedom that are not controlled and
not observed. The simplest description of the experiment is to consider one
spin and the other spins representing a spin-bath (the environment) whose
interaction with the system (the selected spin) leads to decohere at a
characteristic time $T_{2}$ called the spin-spin relaxation time.

From its fundamental beginnings, the NMR technique turned out soon into a
precise spectroscopy of complex molecules which triggered impressive
instrumental developments.
However, nuclear spins and NMR keep providing wonderful models and continued
inspiration for the advance of coherent control over other coupled quantum
systems. It has gained the role of the workhorse of quantum dynamics. NMR was
involved in the beginning of the experimental quantum information processing
(QIP) applications, although nowadays, it is not considered feasible because
its difficult scalability \cite{QCRoadmap04}. However, in Vandersypen and
Chuang words \cite{Chuang04}, NMR

\begin{quote}
\emph{\textquotedblleft being one of the oldest areas of quantum physics}[,
give us the possibility to play with quantum mechanics because it]\emph{ made
possible the application of a menagerie of new and previously existing control
techniques, such as simultaneous and shaped pulses, composite pulses,
refocusing schemes, and effective Hamiltonians. These techniques allow control
and compensation for a variety of imperfections and experimental artifacts
invariably present in real physical systems, such as pulse imperfections,
Bloch-Siegert shifts, undesired multiple-spin couplings, field
inhomogeneities, and imprecise system Hamiltonians.}

\emph{The problem of control of multiple coupled quantum systems is a
signature topic for NMR and can be summarized as follows: given a system with
Hamiltonian }$\widehat{\mathcal{H}}=\widehat{\mathcal{H}}_{\mathrm{sys}%
}+\widehat{\mathcal{H}}_{\mathrm{control}}$\emph{, where }$\widehat
{\mathcal{H}}_{\mathrm{sys}}$\emph{ is the Hamiltonian in the absence of any
active control, and }$\widehat{\mathcal{H}}_{\mathrm{control}}$\emph{
describes terms that are under external control, how can a desired unitary
transformation }$\widehat{U}$\emph{ be implemented, in the presence of
imperfections, and using minimal resources? Similar to other scenarios in
which quantum control is a welldeveloped idea, such as in laser excitation of
chemical reactions [Walmsley and Rabitz, 2003], }$\widehat{\mathcal{H}%
}_{\mathrm{control}}$\emph{ arises from precisely timed sequences of multiple
pulses of electromagnetic radiation, applied phase-coherently, with different
pulse widths, frequencies, phases, and amplitudes. However, importantly, in
contrast to other areas of quantum control, in NMR }$\widehat{\mathcal{H}%
}_{\mathrm{sys}}$\emph{ is composed from multiple distinct physical pieces,
i.e., the individual nuclear spins, providing the tensor product Hilbert space
structure vital to quantum computation. Furthermore, the NMR systems employed
in quantum computation are better approximated as being closed, as opposed to
open quantum systems.\textquotedblright}

\qquad\qquad\qquad\qquad\qquad\qquad\qquad\qquad Vandersypen and Chuang.
\end{quote}

\begin{figure}
[tbh]
\begin{center}
\includegraphics[
height=4.6363in,
width=3.1185in
]%
{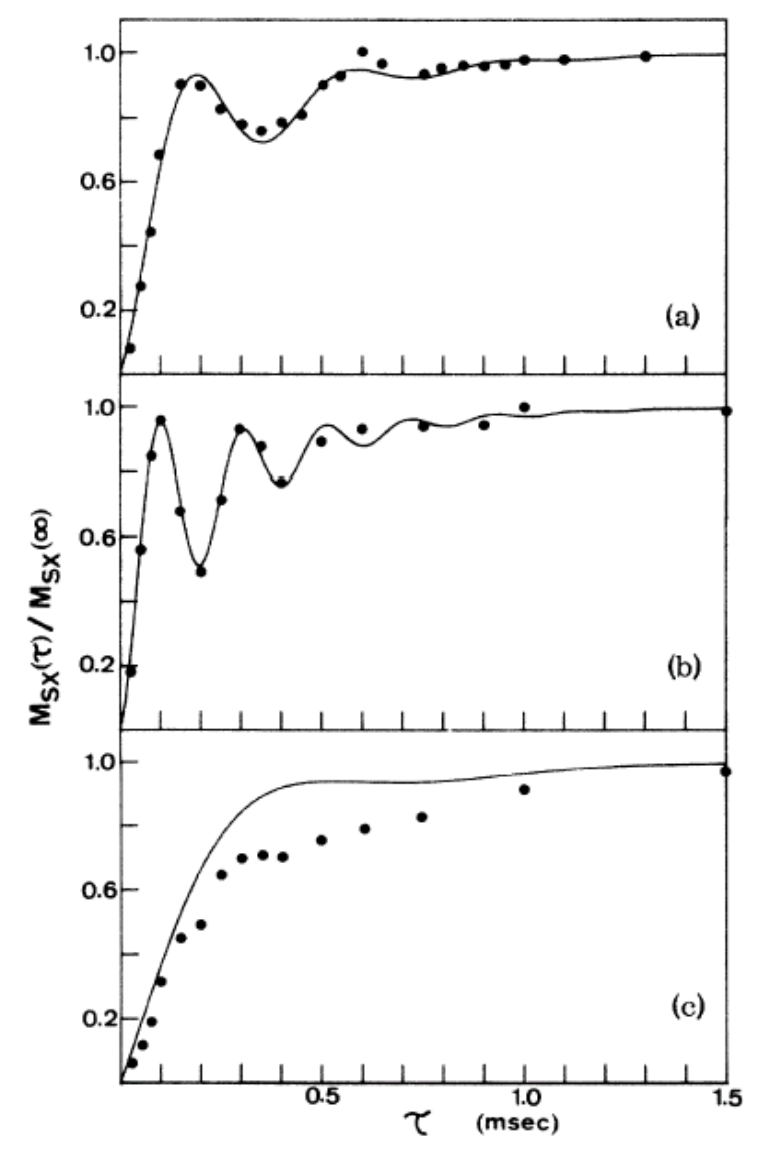}%
\caption[Transient oscillations in a cross-polarization experiment by
M\"{u}ller, Kumar, Baumann and Ernst (1974)]{Transient oscillations in a
cross-polarization experiment by M\"{u}ller, Kumar, Baumann and Ernst (1974)
\cite{MKBE74}. The two-spin dynamics coupled to a spin-bath is shown for three
different internal couplings. Reprinted figure with permission from L. M\"{u}ller, A. Kumar, T. Baumann and R. R. Ernst, Phys. Rev. Lett. {\bf 32}, 1402 (1974). Copyright (1974) by the American Physical Society.}%
\label{Fig_MKBE_CP_intro}%
\end{center}
\end{figure}

\begin{figure}
[tbh]
\begin{center}
\includegraphics[
height=4.3656in,
width=3.448in
]%
{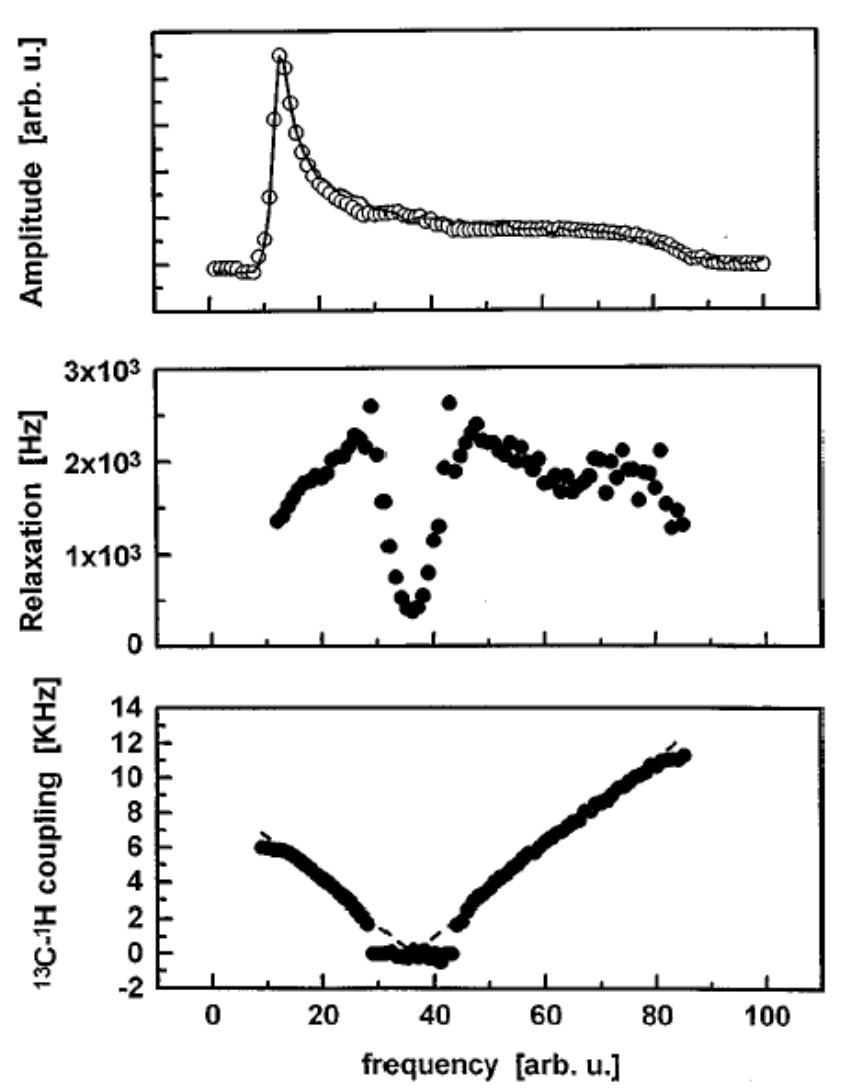}%
\caption[Fitting parameters of a two-spin cross-polarization experiment
performed by P. R. Levstein, G. Usaj and H. M. Pastawski (1998)]{Fitting
parameters of a two-spin cross-polarization experiment performed by P. R.
Levstein, G. Usaj and H. M. Pastawski (1998) \cite{JCP98}. Striking changes in
the relaxation and in the oscillation frequency behaviour are observed. These
effect are described in chapter \ref{Sec_QDPT}. Reprinted with permission from P. Levstein, G. Usaj and H. M. Pastawski, J. Chem. Phys. Vol. 108, Issue 7, Page 2718, 1998. Copyright 1998, American Institute of Physics.}%
\label{Fig_JCP98_original}%
\end{center}
\end{figure}

Thus NMR inspired other techniques in the methodology of quantum control
\cite{Petta05}.
In fact, the first realization of a SWAP operation in solids, an essential
building block for QIP, could be traced back to a pioneer NMR experiment by
M\"{u}ller, Kumar, Baumann and Ernst (1974)\footnote{A similar work where
transient oscillation where observed was presented the next year by D. E.
Denco, J. Tegenfeldt and J. S. Waugh \cite{Waugh75}.} \cite{MKBE74}. While
they did not intended it as a QIP operation, they described theoretically and
experimentally the swapping dynamics (cross polarization) of two strong
interacting spin systems and had to deal with the coupling to a spin-bath.
Until that moment, all the experiments considering two interacting spins were
treated through hydrodynamical equations \cite{Forster90} using the
spin-temperature hypothesis that leads to a simple exponential dynamics.
M\"{u}ller, \emph{et al}. (MKBE) showed that, in a case where the coupling
between two spins is stronger than with the rest, one has to consider quantum
coherences in the quantum calculations. They modeled the experiment treating
quantum mechanically the two-spin system and considering the coupling with the
spin-bath in a phenomenological way as a relaxation process. The original
figure published in the paper is shown in fig. \ref{Fig_MKBE_CP_intro},
where typical cross-polarization (swapping) dynamics for three different
internal interactions (coupling between the two-spins) in ferrocene are
displayed. One can clearly observe the frequency change of the quantum
oscillation. More recent experiments, spanning the internal interaction
strength were done by P. R. Levstein, G. Usaj and H. M. Pastawski
\cite{JCP98}. By using the model of MKBE \cite{MKBE74}, they obtained the
oscillation frequency and the relaxation for different interaction strengths.
These results are shown in fig. \ref{Fig_JCP98_original}%
where one can observe striking changes in the relaxation time and frequency as
a function of the control parameter. Since this discontinuous change is not
predicted by the standard model of MKBE, it remained unexplained. The
description and interpretation of this striking behavior are among the main
results of this thesis.

Thus, in view of possible applications to fields like quantum information
processing\ \cite{Kane98,BD2000}, the experimental manifestation of these
dynamical interference phenomena in qubit clusters of intermediate size has a
great interest.
However, experimental realizations and control of a pure-state dynamics is
still one of the challenges in nowadays physics \cite{QCRoadmap04}. Therefore,
one generally has to deal with an ensemble evolution, which is the case of the
states involved in NMR, i.e. the dynamics of an initial mixed-state.
One can generate mixed-states that are called pseudo-pure because they are
constituted by a pure-state plus a mixed-state density operator. Numerous spin
dynamics NMR experiments have shown surprising quantum phenomena
\cite{Pastawski95,Madi97,Ramanathan05}. The difficulty to produce pure-states
in a high temperature sample leads to the development of the ensemble quantum
computation \cite{Cirac04,Suter05}. However, as we mention previously if the
system is too complex, it is hard to mathematically describe its temporal
evolution. This is a consequence of the exponential growing of the Hilbert
space dimension
as a function of the number of elements in the system. In order to overcome
this limitation, we take profit of the quantum parallelism
\cite{Loss02parallelism} and the fragility of the quantum superpositions to
develop a method that describes ensemble dynamics.

\begin{figure}
[tbh]
\begin{center}
\includegraphics[
height=4.6579in,
width=3.5206in
]%
{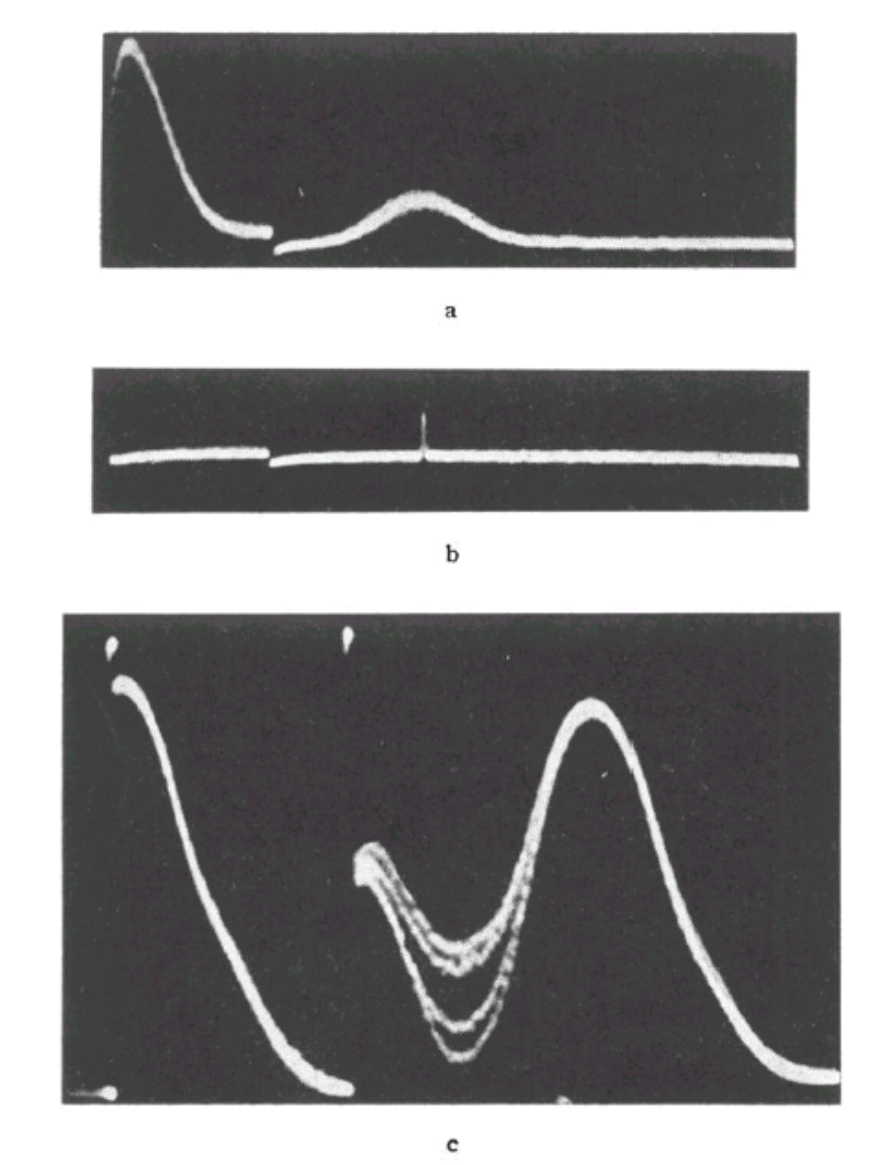}%
\caption[First experimental time reversion in NMR: The Hahn's echo.]{First
experimental time reversion in NMR: The Hahn's echo \cite{Hahn50b}. Reprinted figure with permission from E. L. Hahn, Phys. Rev. {\bf 80}, 580 (1950). Copyright (1950) by the American Physical Society.}%
\label{Fig_Hahn_echo_experiment}%
\end{center}
\end{figure}

As the dimension of the system increases, the sensitivity of the quantum
superposition might lead to the inference that quantum phenomena will not
manifest at macroscopic scales \cite{Myatt00,Schleich00}. In contrast, an
experimental demonstration of macroscopic quantum states done by Y. Nakamura,
\emph{et al.} \cite{Nakamura99,Averin99} shows the opposite.
Indeed, there is no doubt about the high sensitivity of the quantum
superposition states in large systems which paves the way for an effective
decoherence when there are interactions with the environment. As any
environment usually has many degrees of freedom, it is very difficult to
reverse the SE interaction constituting the dominant source of irreversibility
in nature \cite{Zurek03,Schlosshauer04}. Numerous works are related to this
topic, but we should begin discussing the pioneer work that made a temporal
reversion of a quantum dynamics: the Hahn's echo experiment. It is based on
the reversion of the dephasing among rotating spins due to inhomogenities of
the static field \cite{Hahn50b}. He observed an echo in the NMR polarization
signal (see fig. \ref{Fig_Hahn_echo_experiment})
manifesting the deterministic nature of quantum mechanics, but with an
attenuation rate proportional to the spin-spin coupling. The forward dynamics
is a consequence of the interaction of the spins with the static field and the
spin-spin interactions, but only the interactions with the static field are
reverted. Thus, the dipolar interaction remains working. Within the NMR field,
there were many experiments using the deterministic nature of quantum
mechanics to take out some interactions that disturb the relevant system
evolution.
But, the first work that emphasizes the deterministic nature of quantum
mechanics, invalidating the spin temperature hypothesis
(thermodynamical approaches), was done by W. -K. Rhim and A. Pines and J. S.
Waugh \cite{Pines70}. They called a \textquotedblleft Loschmidt
daemon\textquotedblright\ to the process of reversion of the dipolar
interaction in the \textquotedblleft magic echoes\textquotedblright%
\ experiment. There, they observed an echo signal after undoing (reversion
control) the evolution under spin-spin interactions that remain untouched in
the Hahn's echo experiment. The previous experiments evolve from multi-spin
initial excitations. The local initial excitation version of the
\textquotedblleft magic echoes\textquotedblright\ was done by S. Zhang, B. H.
Meier and R. R. Ernst (1992) \cite{ZME92}. They called this experiment as
\textquotedblleft the polarization echo\textquotedblright\ where they used a
very ingenious idea to observe a local magnetization \cite{ZME92,Zhang92}.
They used a rare nucleus, $^{13}$C, bonded to a $^{1}$H nucleus (abundant) as
a local probe to create and observe the local polarization. However, we have
to remark that while one increases the quantum control on the Hamiltonians, a
minimal decay of the echoes can not be avoided. Experiments performed in
C\'{o}rdoba suggest that the quantum states are so sensitive to perturbations
that even a very small uncontrolled perturbation generates an intrinsic
irreversibility characterized by the own system dynamics
\cite{JCP98,Usaj-Mol98,PhysicaA}. By considering an analogy with the behavior
of a simpler one body chaotic system, this was interpreted
\cite{JalPas01,Beenakker01,Cucchietti04} as the onset of a Lyapunov phase,
where $1/\tau_{\phi}=\min\left[  1/\tau_{\mathrm{SE}},\lambda\right]  $ is
controlled by the system's own complexity $\lambda$. However, a theoretical
answer for many-body systems that do not have a classical analogue
characterized by Lyapunov exponent remains open. This is also a topic that
enters in this thesis' motivation: the improvement of our comprehension and
control of decoherence processes and irreversibility. The complexity of
many-body systems leads us to study the forward dynamics of open systems to
characterize the decoherence process before studying the time reversal.

\section{Our contribution\label{Mark_our_contribution}}

In this thesis, we solve the dynamics of many-spin systems interacting with a
spin-bath through the generalized Liouville-von Neumann quantum master
equation beyond the standard approximation. Further consideration of the
explicit dynamics of the bath helps us to solve the spin dynamics within the
Keldysh formalism, where the interaction with the bath is taken into account
in a precisely perturbative method based on Feynman diagrams. Both methods
lead to identical solutions and together gave us the possibility to obtain
numerous physical interpretations contrasting with NMR experiments. We used
these solutions in conjunction with experimental data\ to design new protocols
for molecular characterization, develop new numerical methodologies and
control the quantum dynamics for experimental implementations. But, most
important, these developments contributed to improve the fundamental physical
interpretations of the dynamics in a quantum open system under the presence of
an environment. In particular, we show a manifestation of an environmentally
induced quantum dynamical phase transition.

\subsection{Organization of this thesis}

In \textbf{Chapter 2 }we use the standard formalism of density matrix to solve
the spin dynamics using the generalized Liouville-von Neumann quantum master
equation. In the first part of the chapter, the spin dynamics of a two-spin
system coupled with a fast fluctuating spin-bath is solved. This system
describes the cross-polarization experiment of MKBE \cite{MKBE74}. We start
using the standard approximations and then we extend the solution without
these restrictions. We compare the solutions and remark the main differences.
We analyze the spin dynamics for different anisotropies of the SE interactions
given by the different contributions of the Ising and the XY interaction. We
show how the rates of decoherence and dissipation change depending on the
anisotropy ratio between the Ising and XY coupling. In the second part of the
chapter, we extend the solution to a three-spin system coupled with a
spin-bath. The solutions obtained are applied to experimental data to get more
detailed information for molecular characterization. In particular, we use the
three-spin solution to characterize the liquid crystal $8$CB and incorporating
some memory effects, we conclude that the spin-bath has a slow dynamics.

In \textbf{Chapter 3 }we solve the spin dynamics within the Keldysh formalism
\cite{Keldysh64}. The Keldysh formalism is well established in the electron
transport description. Through the Jordan-Wigner transformation
\cite{Jordan28}, we map the two-spin system of chapter $2$ into a fermion
system. We find how to describe the SE interaction within the wide band
approximation (fast fluctuation inside the bath) and we obtain a solution for
the spin dynamics that improves the standard solution of the generalized
Liouville-von Neumann quantum master equation. Here, we use a microscopic
model to obtain the spin dynamics that avoids using a phenomenological
description of the SE interaction. However, we obtain the same solution going
beyond the standard approximation within the density matrix formalism. Then,
we solve the spin dynamics of a linear chain including all the degrees of
freedom of the environment in the calculations and we show how the memory
effects induce a time dependence in the oscillation frequency as is observed
experimentally. We develop a stroboscopic model to describe decoherence which
is optimized for numerical applications. This model converges to the
continuous expression.

In \textbf{Chapter 4 }based on the solutions obtained in previous chapters we
describe a manifestation of an environmentally induced quantum dynamical phase
transition. We show the experimental evidence and interpret the phenomenon in
detail. In particular, we show how the anisotropy of the SE interaction has an
important role in the critical point of the phase transition. An extension of
this phenomenon to a three-spin system shows how to vary the control parameter
to \textquotedblleft isolate\textquotedblright\ two of them from the environment.

In \textbf{Chapter 5}, inspired in the stroboscopic model developed in chapter
3, we propose a new NMR pulse sequence to improve the transfer of polarization
through a specific pathway in a system of many interacting spins. The sequence
effectively prunes branches of spins, where no polarization is required,
during the polarization transfer procedure. Simulations of the spin dynamics
in the $^{13}$C backbone of leucine are performed. Possible applications and
potential fundamental contributions to engineered decoherence are discussed.

In \textbf{Chapter 6 }we develop a novel numerical method to obtain the spin
dynamics of an ensemble. It overcomes the limitations of standard numerical
calculations for large number of spins because it does not involve ensemble
averaging.\textbf{ }We exploit quantum parallelism \cite{Loss02parallelism}
and the fragility of a randomly correlated entangled state to reproduce an
ensemble dynamics.

In the \textbf{final part of each chapter} a brief summary of the main
original contributions including references to publications is included.

In \textbf{Chapter 7} we summarize the whole work emphasizing the main
conclusions and perspectives.

\chapter{Many-spin quantum dynamics within the density matrix
formalism\label{Mark_spin_dynamics_Density_matrix}}

The exact quantum dynamics of small quantum systems has regained interest
during the last years \cite{Altshuler},
due to the technological advances that give us the opportunity to observe
quantum phenomena. Spin systems are good candidates in this respect and
provide beautiful playgrounds for fundamental studies. Besides, several
challenging applications require a very fine knowledge of the spin
interactions, such as molecular characterization,
spin control in nanodevices \cite{Awschalom01,Loss02}
and quantum computation \cite{Chuang97,Cory98,BD2000}.
In the introduction became evident the limitations of simple thermodynamical
arguments \cite{Forster90} based on the spin temperature\emph{ }hypothesis.
The experiment of MKBE \cite{MKBE74} showed the need to consider the system
quantum mechanically keeping the quantum coherences to describe the transient
oscillations. However, the first work that showed the weakness of the
\textquotedblleft spin temperature\textquotedblright\ hypothesis was done in
1970 \cite{Pines70}. In it, a time reversal of the spin-spin interactions was
performed. It was followed by numerous nuclear magnetic resonance (NMR)
experiments that have demonstrated the time reversibility of the dipolar
(many-spin) evolution \cite{ZME92,MErnst98,MErnst98-2,JCP98,Usaj-Mol98}
leading to revise the concept of \textquotedblleft spin
diffusion\emph{\textquotedblright\ }%
\cite{Pastawski95,Pastawski96,Madi97,Waugh98}. More importantly, by selecting
appropriate systems and pulse sequences, one can investigate the sources of
quantum decoherence \cite{Zurek03,Schlosshauer04}, ergodicity
\cite{Pastawski95,Pastawski96,Waugh98}, and quasi-equilibrium
\cite{Sakellariou98}.

From a practical point of view, spin dynamics observed by NMR has proved very
powerful in order to characterize molecular structures and dynamics
\cite{Spiess}. Experimental observations together with simple analytical
solutions for few-spin dynamics can provide detailed information on the intra
and intermolecular interactions \cite{MKBE74,JCP98,Usaj-Mol98}. This is
particularly important for the characterization of complex fluids in their
native state, where one uses cross-polarization (CP) dynamics
\cite{Hartmann62,Slichter} to evaluate order parameters \cite{Pratima96}.
However, the reliability of these and other structural and dynamical
parameters depends on the accuracy of the spin dynamics description to which
the experimental data are fitted.

In this chapter, we use the standard formalism of density matrix to solve the
spin dynamics using the generalized Liouville-von Neumann quantum master
equation \cite{Abragam,Ernst}. In the first part of the chapter, we solve the
spin dynamics of a two-spin system coupled to a fast fluctuating spin-bath.
This system describes the cross-polarization experiment of MKBE \cite{MKBE74}.
As a first step, we use the standard approximations and then we extend the
solution releasing these restrictions. We compare the solutions and remark the
main differences. We analyze the spin dynamics for different SE interactions
consisting of different Ising and XY contributions. We show how the
decoherence and dissipation rates change depending on the anisotropy ratio
between the Ising and XY couplings. In the second part of the chapter, we
extend the solutions to a three-spin system coupled to a spin-bath. The
solutions are applied to get more detailed information from our NMR
experimental data. This leads to new methodologies for molecular
characterization. In particular, we use the three-spin solution to
characterize the liquid crystal $8$CB. The slow dynamics of the smectic phase,
experimentally observed, lead us to include some spin-bath memory effects.

\section{Quantum dynamics of a two-spin system}

For didactical reasons, we start solving the spin dynamics of an isolated
two-spin system. Then, we will include the interactions with the spin-bath.

\subsection{Quantum evolution of an isolated two-spin system
\label{Mark_2spin_isolated}}

We solve the evolution of an isolated two-spin system during
cross-polarization (CP).

In this procedure, two different species of spins, $S$-$I,$ which here will
correspond to a $^{13}$C-$^{1}$H system are coupled in such a way that they
\ \textquotedblleft believe\textquotedblright\ that they are of the same
species \cite{Abragam,Slichter,Ernst}. In that situation, the most efficient
polarization transfer can occur.
The system Hamiltonian, in presence of a static field $H_{0}$ and the radio
frequency fields of amplitudes $H_{1,S}$ and $H_{1,I}$ with frequencies
$\omega_{\mathrm{rf},S}$ and $\omega_{\mathrm{rf},I}$ respectively, is given
by \cite{Abragam,Slichter}
\begin{multline}
\widehat{\mathcal{H}}_{\mathrm{S}}\mathcal{=-}\hbar\Omega_{0,S}\hat{S}%
^{z}-\hbar\Omega_{0,I}\hat{I}^{z}\\
-\hbar\Omega_{1,S}\left\{  \hat{S}^{x}\cos\left(  \omega_{\mathrm{rf}%
,S}~t\right)  +\hat{S}^{y}\sin\left(  \omega_{\mathrm{rf},S}~t\right)
\right\}  -\hbar\Omega_{1,I}\left\{  \hat{I}^{x}\cos\left(  \omega
_{\mathrm{rf},S}~t\right)  +\hat{I}^{y}\sin\left(  \omega_{\mathrm{rf}%
,S}~t\right)  \right\} \\
+2b\hat{I}^{z}\hat{S}^{z},
\end{multline}
where
\begin{equation}
\Omega_{0,i}=\gamma_{i}H_{0,i},i=S,I
\end{equation}
are the precession Larmor frequencies in the static field and
\begin{equation}
\Omega_{1,i}=\gamma_{i}H_{1,i},i=S,I
\end{equation}
are the Zeeman (nutation) frequencies of the RF fields. The last term is the
truncated dipolar interaction assuming that
\begin{equation}
\left\vert \hbar\Omega_{0,I}-\hbar\Omega_{0,S}\right\vert \gg\left\vert
b\right\vert .
\end{equation}
The amplitude of the interaction is \cite{Slichter}
\begin{equation}
b=-\frac{1}{2}\left(  \frac{\mu_{0}\gamma_{I}\gamma_{S}\hbar^{2}}{4\pi r^{3}%
}\right)  \left(  3\cos^{2}\theta-1\right)  , \label{dipolar_coupling}%
\end{equation}
where $r$ is the internuclear distance and $\theta$ the angle between the
static field and the internuclear vector. In the double rotating frame
\cite{Slichter}, at the frequencies of the RF fields, the system Hamiltonian
becomes
\begin{equation}
\widehat{\mathcal{H}}_{\mathrm{S}}\mathcal{=}\hbar\Delta\Omega_{S}\hat{S}%
^{z}+\hbar\Delta\Omega_{I}\hat{I}^{z}-\hbar\Omega_{1,S}\hat{S}^{x}-\hbar
\Omega_{1,I}\hat{I}^{x}+2b\hat{S}^{z}\hat{I}^{z}%
\end{equation}
where
\begin{equation}
\Delta\Omega_{i}=\Omega_{0,i}-\omega_{\mathrm{rf},i} \label{offresonance}%
\end{equation}
are the respective off-resonance shifts with $i=I,S$ . We assume the
conditions
\begin{equation}
\Delta\Omega_{I}=\Delta\Omega_{S}=0
\end{equation}
and%
\begin{equation}
\hbar\left\vert \Omega_{1,S}+\Omega_{1,I}\right\vert \gg\left\vert
b\right\vert
\end{equation}
which are obtained when the RF fields are applied on-resonance and when the RF
power is much bigger than the dipolar interaction.
Thus, the doubly truncated Hamiltonian becomes%
\begin{align}
\widehat{\mathcal{H}}_{\mathrm{S}}  &  \mathcal{=}\tfrac{1}{2}\left(
\Sigma-\Delta\right)  \hat{S}^{x}+\tfrac{1}{2}\left(  \Sigma+\Delta\right)
\hat{I}^{x}+b\left(  \hat{S}^{z}\hat{I}^{z}+\hat{S}^{y}\hat{I}^{y}\right)
\label{H_CH_truncado}\\
&  =\tfrac{1}{2}\Sigma\left(  \hat{S}^{x}+\hat{I}^{x}\right)  +\tfrac{1}%
{2}\Delta\left(  \hat{I}^{x}-\hat{S}^{x}\right)  +b\left(  \hat{S}^{z}\hat
{I}^{z}+\hat{S}^{y}\hat{I}^{y}\right)  ,
\end{align}
where the non-secular elements of the dipolar interaction with respect to the
$\Sigma\left(  S^{x}+I^{x}\right)  $ term have been neglected. Here, $\Sigma$
and $\Delta$ are defined by%
\begin{equation}
\Sigma=-\hbar\left(  \Omega_{1,S}+\Omega_{1,I}\right)  ~~~~~~~\text{and}%
~~~~~~~\Delta=\hbar\left(  \Omega_{1,S}-\Omega_{1,I}\right)  .
\end{equation}
Within the Hartmann-Hahn condition \cite{Hartmann62,Slichter}, $\Delta=0,$ the
two spins act as they belong to the same species improving the polarization
transfer between them. To obtain the quantum evolution, we solve the
Liouville-von Neumann equation \cite{Abragam,Ernst}
\begin{equation}
\frac{\mathrm{d}}{\mathrm{d}t}\hat{\rho}(t)=-\frac{\mathrm{i}}{\hbar}\left[
\widehat{\mathcal{H}}_{\mathrm{S}},\hat{\rho}(t)\right]  ,
\label{ec_maestra_sin_difusion}%
\end{equation}
where $\hat{\rho}$ is the density matrix operator of the two-spin system. Its
solution is given by%
\begin{equation}
\hat{\rho}(t)=\hat{U}(t)\hat{\rho}(0)\hat{U}^{-1}(t),
\end{equation}
with
\begin{equation}
\hat{U}(t)=\exp\left(  -\frac{\mathrm{i}}{^{\hbar}}\widehat{\mathcal{H}%
}_{\mathrm{S}}t\right)
\end{equation}
the evolution operator and $\hat{\rho}\left(  0\right)  $ the initial
condition. For the last, we consider the $^{1}$H totally polarized and the
$^{13}$C depolarized. This can be experimentally achieved by rotating, with a
$\pi/2$ pulse, the equilibrium polarization of the $^{1}$H with the Zeeman
field $H_{0}$ in the $z$ direction to the XY plane assisted by a cycling pulse
sequence \cite{JCP98}.
The initial condition (immediately after the $\pi/2$ pulse) is expressed as
\begin{equation}
\hat{\rho}\left(  0\right)  =\frac{\hat{1}}{\mathcal{Z}}\exp\left(
-\frac{\hbar\Omega_{0,I}\hat{I}^{x}}{k_{\mathrm{B}}T}\right)  ,
\label{sigma_ini_T}%
\end{equation}
where
\begin{equation}
\mathcal{Z}=\mathrm{tr}\left\{  \exp\left(  -\frac{\hbar\Omega_{0,I}\hat
{I}^{x}}{k_{\mathrm{B}}T}\right)  \right\}
\label{partition_function_sigma_ini_T}%
\end{equation}
is the partition function.

In the high temperature approximation%
\begin{equation}
\hat{\rho}\left(  0\right)  =\frac{\hat{1}+\beta_{\mathrm{B}}\hbar\Omega
_{0,I}\hat{I}^{x}}{\mathrm{Tr}\left\{  \hat{1}\right\}  }, \label{sigma_ini}%
\end{equation}
where $\beta_{\mathrm{B}}=1/k_{\mathrm{B}}T$\emph{.}

\begin{figure}
[tbh]
\begin{center}
\includegraphics[
height=4.4944in,
width=5.1361in
]%
{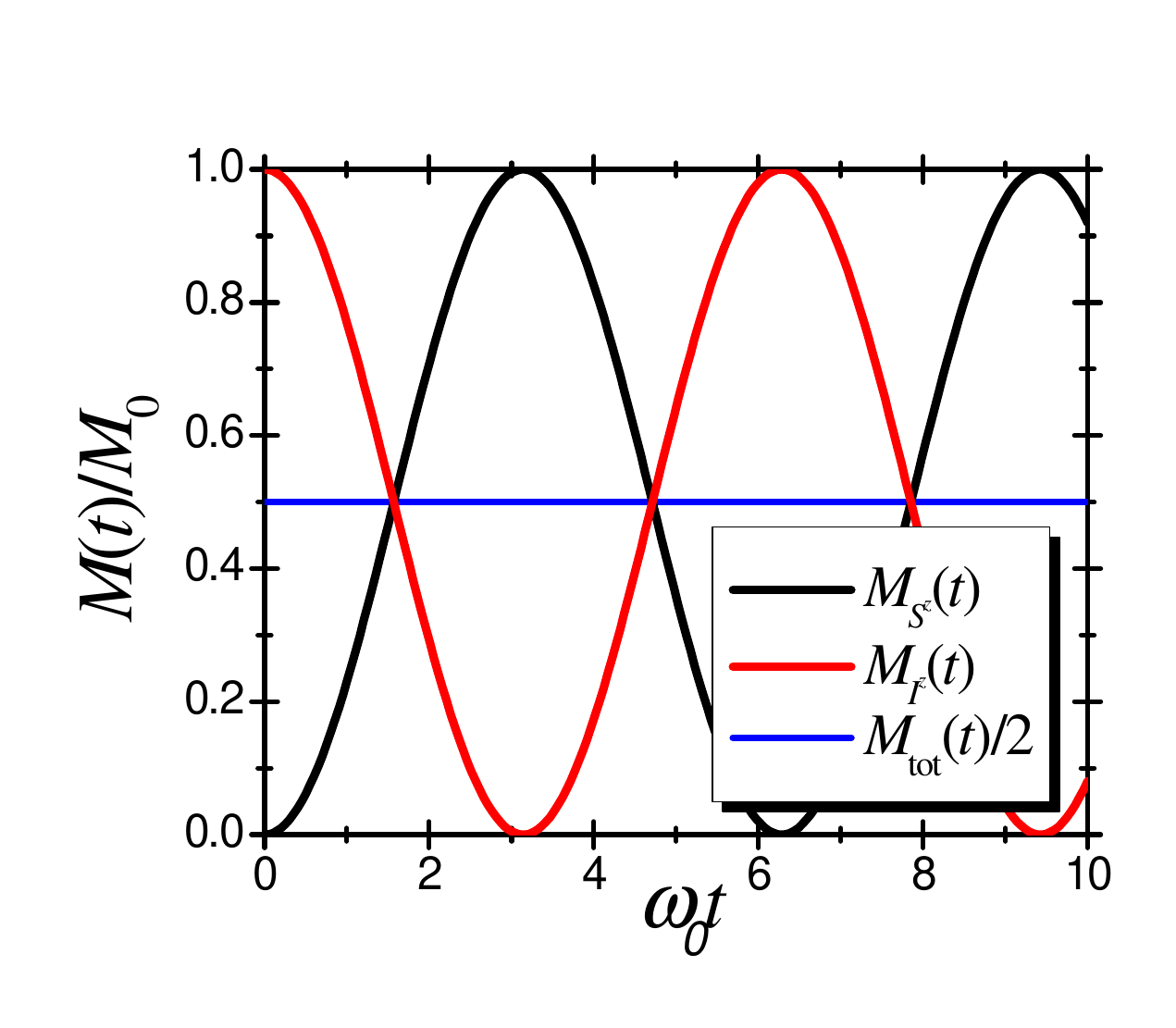}%
\caption[Polarization evolution of an isolated two-spin system$.$%
]{Polarization evolution of the $S$ spin (black line) and the $I$ spin (red
line). The oscillation frequency is $\omega_{0}=b/\hbar$ and the maximum
amplitude is given by the initial polarization of the spin $I.$ }%
\label{fig_CH_isolated_DM}%
\end{center}
\end{figure}

Thus, calculating the evolution operator, we obtain the magnetization of the
$^{13}$C as a function of the contact time $t$ of the cross-polarization
\begin{equation}
M_{S^{x}}\left(  t\right)  =\mathrm{Tr}\left\{  \hat{\rho}\left(  t\right)
\hat{S}^{x}\right\}  =M_{0}\frac{\left[  1-\cos\left(  \omega_{0}t\right)
\right]  }{2}, \label{Msx_CH_isolated}%
\end{equation}
and for the $^{1}$H we obtain%
\begin{equation}
M_{I^{x}}\left(  t\right)  =\mathrm{Tr}\left\{  \hat{\rho}\left(  t\right)
\hat{I}^{x}\right\}  =M_{0}\frac{\left[  1+\cos\left(  \omega_{0}t\right)
\right]  }{2}. \label{Mix_CH_isolated}%
\end{equation}
Here,
\begin{equation}
\omega_{0}=b/\hbar
\end{equation}
is the Rabi frequency and
\begin{equation}
M_{0}=\frac{1}{4}\beta_{\mathrm{B}}\hbar\Omega_{0,I}%
\end{equation}
is the initial magnetization at the $^{1}$H. Figure \ref{fig_CH_isolated_DM}
shows these two curves as a function of $t.$%
\ The Hamiltonian (\ref{H_CH_truncado}) has only Zeeman fields along the $x$
direction. Thus, by changing the axis names: $x\rightarrow z,$ $y\rightarrow
x$ and $z\rightarrow y$ the Hamiltonian becomes%
\begin{align}
\widehat{\mathcal{H}}_{\mathrm{S}}  &  \mathcal{=}\tfrac{1}{2}\left(
\Sigma-\Delta\right)  \hat{S}^{z}+\tfrac{1}{2}\left(  \Sigma+\Delta\right)
\hat{I}^{z}+b\left(  \hat{S}^{x}\hat{I}^{x}+\hat{S}^{y}\hat{I}^{y}\right)
\label{Hs_isolated_two_spin}\\
&  =\tfrac{1}{2}\Sigma\left(  \hat{S}^{z}+\hat{I}^{z}\right)  +\tfrac{1}%
{2}\Delta\left(  \hat{I}^{z}-S^{z}\right)  +\tfrac{1}{2}b\left(  \hat{S}%
^{+}\hat{I}^{-}+\hat{S}^{-}\hat{I}^{+}\right)  .
\end{align}
Now, it is evident that the dipolar interaction is an XY (flip-flop) term that
splits the energy level of the states $\left\vert \uparrow,\downarrow
\right\rangle $ and $\left\vert \downarrow,\uparrow\right\rangle $ and induces
an oscillation between them. This is manifested in the oscillation of figure
\ref{fig_CH_isolated_DM} where the magnetization is totally transferred forth
and back from the $^{1}$H to the $^{13}$C with the Rabi (swapping) frequency
$\omega_{0}$. Within this representation\footnote{Remember that now the $z$
direction is the originally $x$ direction.}, in the new basis of $\hat{I}%
^{z},\hat{S}^{z}\,$given by $\{|+,+\rangle,$ $|+,-\rangle,$ $|-,+\rangle,$
$|-,-\rangle\}$, where $|\pm,\pm\rangle=|\pm\rangle\otimes|\pm\rangle
=|M_{I,z}\rangle\otimes|M_{S,z}\rangle$ with $\hat{I}^{z}|M_{I,z}%
\rangle=M_{I,z}|M_{I,z}\rangle,\,M_{I,z}=\pm\frac{1}{2},$ the magnetization of
the $^{13}$C spin can be expressed as%

\begin{equation}
M_{S^{z}}\left(  t\right)  =\mathrm{Tr}\left\{  \hat{\rho}\left(  t\right)
\hat{S}^{z}\right\}  =\frac{1}{2}\left(  \rho_{++,++}\left(  t\right)
-\rho_{--,--}\left(  t\right)  +\rho_{-+,-+}\left(  t\right)  \right)
-\rho_{+-,+-}\left(  t\right)  ,
\end{equation}
with $\rho_{\pm\pm,\pm\pm}\left(  t\right)  =\langle\pm,\pm|\rho(t)|\pm
,\pm\rangle$. Within this basis, the diagonal terms (populations) of the
density matrix contribute positively to the magnetization when the carbon is
in the state up and negatively when it is down. This means that the carbon
magnetization is given by the difference between the populations of the state
up and down. We can see that the elements of the density matrix $\rho
_{++,++}\left(  t\right)  $ and $\rho_{--,--}\left(  t\right)  $ are constants
of motion and give the first term of eq. (\ref{Msx_CH_isolated}). The
difference between $\rho_{+-,+-}\left(  t\right)  \ $and $\rho_{-+,-+}\left(
t\right)  $ (coherences) gives the oscillatory term, which describes the
transition between $\left\vert \uparrow,\downarrow\right\rangle $ and
$\left\vert \downarrow,\uparrow\right\rangle .$ The $\mathrm{Tr}\left\{
\hat{\rho}\left(  t\right)  \hat{S}^{z}\right\}  $ is given by
\begin{equation}
M_{S^{z}}\left(  t\right)  =\frac{1}{2}\left[  \rho_{11}\left(  t\right)
-\rho_{44}\left(  t\right)  +\rho_{32}\left(  t\right)  +\rho_{23}\left(
t\right)  \right]  =\frac{1}{2}\left[  \rho_{11}\left(  t\right)  -\rho
_{44}\left(  t\right)  +2\operatorname{Re}\left\{  \rho_{32}\left(  t\right)
\right\}  \right]  \label{MSx_base_H_expression}%
\end{equation}
in the basis of the eigenstates of the Hamiltonian,
\begin{equation}
\left\{  |1\rangle=|+,+\rangle,|2\rangle=\frac{1}{\sqrt{2}}\left(
|+,-\rangle+|-,+\rangle\right)  ,|3\rangle=\frac{1}{\sqrt{2}}\left(
|+,-\rangle-|-,+\rangle\right)  ,|4\rangle=|-,-\rangle\right\}  .
\label{eigenstates_H_CH}%
\end{equation}
The states $|1\rangle$ and $|4\rangle,$ previously defined as $|+,+\rangle$
and $|-,-\rangle,$ give again the first term of eq. (\ref{Msx_CH_isolated}).
However, the oscillatory term is given by the real part of the $\rho_{32}$
coherence. The magnetization on the $I$ spin has the sign of the oscillatory
term changed. Thus, the total magnetization of the system is a constant of
motion described by%
\begin{equation}
M_{\mathrm{tot.}}\left(  t\right)  =\mathrm{Tr}\left\{  \hat{\rho}\left(
t\right)  \left(  \hat{S}^{z}+\hat{I}^{z}\right)  \right\}  =\rho
_{++,++}\left(  t\right)  -\rho_{--,--}\left(  t\right)  =\rho_{11}\left(
t\right)  -\rho_{44}\left(  t\right)  =M_{0},
\end{equation}
where $M_{0}$ is the initial magnetization. The blue line in fig.
\ref{fig_CH_isolated_DM} represents $M_{0}/2$ which constitutes the mean
magnetization of each species.

\subsection{A two-spin system interacting with a
spin-bath\label{M_2-spin_spin_bath}}

We use the model proposed by M\"{u}ller \emph{et al. }\cite{MKBE74} to
describe the experimental spin dynamics of a molecule of ferrocene
and to characterize the quantum dynamics of a two-spin system interacting with
a spin-bath. The model assumes that only one spin, $I_{1},$ interacts with the
spin-bath which is described in a phenomenological way. The modeled
Hamiltonian is%

\begin{align}
\widehat{\mathcal{H}}  &  =\widehat{\mathcal{H}}_{\mathrm{S}}+\widehat
{\mathcal{H}}_{\mathrm{SE}}+\widehat{\mathcal{H}}_{\mathrm{E}},\\
\widehat{\mathcal{H}}_{\mathrm{S}}  &  =\hbar\Omega_{\mathrm{Z}}\left(
\hat{S}_{{}}^{z}+\hat{I}_{1}^{z}\right)  +\tfrac{1}{2}b\left(  \hat{S}_{{}%
}^{+}\hat{I}_{1}^{-}+\hat{S}_{{}}^{-}\hat{I}_{1}^{+}\right)  ,\label{H_system}%
\\
\widehat{\mathcal{H}}_{\mathrm{SE}}  &  =\alpha\hat{I}_{1}^{z}\hat{F}_{{}}%
^{z}+\tfrac{1}{2}\beta\left(  \hat{I}_{1}^{+}\hat{F}_{{}}^{-}+\hat{I}_{1}%
^{-}\hat{F}_{{}}^{+}\right)  ,\label{H_SE_2_spin_spin_bath_DM}\\
\widehat{\mathcal{H}}_{\mathrm{E}}  &  =\hbar\Omega_{\mathrm{Z}}%
{\textstyle\sum_{i>1}}
\hat{I}_{i}^{z}+%
{\textstyle\sum_{\genfrac{}{}{0pt}{}{i>1}{j>i}}}
b_{ij}\left[  2\hat{I}_{i}^{z}\hat{I}_{j}^{z}-\tfrac{1}{2}\left(  \hat{I}%
_{i}^{+}\hat{I}_{j}^{-}+\hat{I}_{i}^{-}\hat{I}_{j}^{+}\right)  \right]  ,
\label{H_E_2_spin_spin_bath_DM}%
\end{align}
where $\hbar\Omega_{\mathrm{Z}}$ is the Zeeman interaction and we are assuming
the Hartmann-Hahn condition \cite{Hartmann62,Slichter}%
\begin{equation}
\Omega_{1,S}=\Omega_{1,I}=\Omega_{\mathrm{Z}}.
\end{equation}
$\widehat{\mathcal{H}}_{\mathrm{S}}$ is the system Hamiltonian of the two
coupled spins, $\widehat{\mathcal{H}}_{\mathrm{E}}$ is the spin-bath
Hamiltonian with a truncated dipolar interaction and $\widehat{\mathcal{H}%
}_{\mathrm{SE}}$ is the system-environment (SE) interaction with
\begin{equation}
\hat{F}^{u}=\sum_{i}b_{1i}^{{}}\hat{I}_{i}^{u},~u=x,y,z
\end{equation}
and
\begin{equation}
\hat{F}^{\pm}=\left(  \hat{F}^{x}\pm\mathrm{i}\hat{F}^{y}\right)  .
\end{equation}
$\widehat{\mathcal{H}}_{\mathrm{SE}}$ is an Ising interaction if $\beta
/\alpha=0$ and an XY, isotropic (Heisenberg) or the truncated dipolar
interaction if $\alpha/\beta=0,1,-2$ respectively. This last case is typical
in solid-state NMR experiments \cite{Abragam,Slichter,Ernst}. In a quantum
mechanical relaxation theory the terms $\hat{F}^{u}$ are bath operators while
in the semi-classical theory \cite{Abragam,Ernst} the $F^{u}\left(  t\right)
$ represent classical stochastic forces. The experimental conditions justify a
high temperature approximation, and hence the semiclassical theory coincides
with a quantum treatment \cite{Abragam}.
By tracing on the bath variables, the random SE\ interaction Hamiltonian is
written as
\begin{equation}
\widehat{\mathcal{H}}_{\mathrm{SE}}\left(  t\right)  =\alpha F_{{}}^{z}\left(
t\right)  \hat{I}_{1}^{z}+\tfrac{1}{2}\beta\left[  F_{{}}^{-}\left(  t\right)
\hat{I}_{1}^{+}+F_{{}}^{+}\left(  t\right)  \hat{I}_{1}^{-}\right]  .
\end{equation}
The time average of these random processes satisfies
\begin{equation}
\overline{F^{u}\left(  t\right)  }=0,
\end{equation}
where their correlation functions are
\begin{equation}
g^{\left(  u,v\right)  }\left(  \tau\right)  =\overline{F^{u}\left(  t\right)
F^{v\ast}\left(  t+\tau\right)  }.
\end{equation}
\emph{\ }Following the usual treatment to second order approximation, the
dynamics of the reduced density operator is given by the generalized
Liouville-von Neumann differential equation \cite{Abragam,Ernst}
\begin{equation}
\frac{\mathrm{d}}{\mathrm{d}t}\hat{\sigma}\left(  t\right)  =-\frac
{\mathrm{i}}{\hbar}[\widehat{\mathcal{H}}_{\mathrm{S}},\hat{\sigma}\left(
t\right)  ]-\frac{1}{\hbar}\widehat{\widehat{\Gamma}}\left\{  \hat{\sigma
}\left(  t\right)  -\hat{\sigma}_{\infty}\right\}  , \label{master}%
\end{equation}
where $\hat{\sigma}\left(  t\right)  $ is the reduced density operator%
\begin{equation}
\hat{\sigma}\left(  t\right)  =\mathrm{Tr}_{\mathrm{E}}\left\{  \hat{\rho
}\left(  t\right)  \right\}
\end{equation}
with $\mathrm{Tr}_{\mathrm{E}}$ denoting a partial trace over the environment
variables. The relaxation superoperator $\widehat{\widehat{\Gamma}}$ is given
by the SE interaction. It accounts for the dissipative interactions between
the reduced spin system and the spin-bath and it imposes the relaxation of the
density operator towards its equilibrium value $\hat{\sigma}_{\infty}$.

We assume that the correlation times of the fluctuations are extremely short
compared with all the relevant transition rates between eigenstates of the
Hamiltonian, i.e. frequencies of the order of $\Omega_{\mathrm{Z}}$ and
$\omega_{0}=b/\hbar$. In this extreme narrowing regime or fast fluctuation
approximation we obtain
\begin{equation}
\widehat{\widehat{\Gamma}}\left\{  \hat{\sigma}\right\}  =\frac{1}{2}%
\sum_{u,v}\xi_{u,v}\mathcal{J}^{\left(  u,v\right)  }\left(  0\right)  \left[
\hat{I}_{1}^{u},\left[  \hat{I}_{1}^{v},\hat{\sigma}\right]  \right]
,\nonumber
\end{equation}
where
\begin{equation}
\mathcal{J}^{\left(  u,v\right)  }\left(  \omega\right)  =\int_{-\infty
}^{\infty}\frac{d\tau}{\hbar}g^{\left(  u,v\right)  }\left(  \tau\right)
\exp\left\{  -\mathrm{i}\omega\tau\right\}
\end{equation}
is the spectral density and
\begin{equation}
\xi_{u,v}=\left[  \alpha\delta_{u,z}+\beta\left(  \delta_{u,x}+\delta
_{u,y}\right)  \right]  \left[  \alpha\delta_{v,z}+\beta\left(  \delta
_{v,x}+\delta_{v,y}\right)  \right]  .
\end{equation}
Assuming that the spatial directions are statistically independent, i.e.
$g^{\left(  u,v\right)  }\left(  \tau\right)  =0$ if $u\neq v$, the
superoperator $\widehat{\widehat{\Gamma}}$\emph{\ }can be written as
\begin{equation}
\widehat{\widehat{\Gamma}}\left\{  \hat{\sigma}\right\}  =\alpha
^{2}\mathcal{J}^{z}\left[  \hat{I}_{1}^{z},\left[  \hat{I}_{1}^{z},\hat
{\sigma}\right]  \right]  +\beta^{2}\mathcal{J}^{x}\left[  \hat{I}_{1}%
^{x},\left[  \hat{I}_{1}^{x},\hat{\sigma}\right]  \right]  +\beta
^{2}\mathcal{J}^{y}\left[  \hat{I}_{1}^{y},\left[  \hat{I}_{1}^{y},\hat
{\sigma}\right]  \right]  ,
\end{equation}
where
\begin{equation}
\mathcal{J}^{u}=\dfrac{1}{2}\mathcal{J}^{\left(  u,u\right)  }\left(
0\right)  .
\end{equation}
Notice that the axial symmetry of $\widehat{\mathcal{H}}_{\mathrm{S}}$ around
the $z$ axis leads to the impossibility to evaluate separately $\mathcal{J}%
^{x}$ and $\mathcal{J}^{y}$, so they will appear only as the averaged value
\begin{equation}
\mathcal{J}^{xy}=\frac{1}{2}\left(  \mathcal{J}^{x}+\mathcal{J}^{y}\right)  .
\end{equation}
Thus, we obtain
\begin{equation}
\widehat{\widehat{\Gamma}}\left\{  \hat{\sigma}\right\}  =\Gamma_{\mathrm{ZZ}%
}\left[  \hat{I}_{1}^{z},\left[  \hat{I}_{1}^{z},\hat{\sigma}\right]  \right]
+\Gamma_{\mathrm{XY}}\left(  \left[  \hat{I}_{1}^{x},\left[  \hat{I}_{1}%
^{x},\hat{\sigma}\right]  \right]  +\left[  \hat{I}_{1}^{y},\left[  \hat
{I}_{1}^{y},\hat{\sigma}\right]  \right]  \right)  ,
\end{equation}
where
\begin{equation}
\Gamma_{\mathrm{ZZ}}=\alpha^{2}\mathcal{J}^{z}\;\;\;\;\;\text{and}%
\;\;\;\;\;\Gamma_{\mathrm{XY}}=\beta^{2}\mathcal{J}^{xy}.
\label{Gammas_spectral_density_2spin}%
\end{equation}
Note that $\Gamma_{\mathrm{ZZ}}$ and $\Gamma_{\mathrm{XY}}$ contain the
different sources of anisotropy. The usual approximation considers
$\mathcal{J}^{x}=\mathcal{J}^{y}=\mathcal{J}^{z}$ (identical correlations in
all the spatial directions) and $\alpha=\beta=1$ (isotropic interaction
Hamiltonian) \cite{MKBE74}. A better approximation considers a dipolar
interaction Hamiltonian, i.e. $\alpha=-2\beta=-2$
\cite{JCP03,JCP06,Alvarez07a}.
This is in excellent agreement with previous experiments in polycrystalline
samples where fittings to phenomenological equations have been performed
\cite{JCP98,Hirschinger97}. In particular, in the case of isotactic
polypropylene \cite{Hirschinger97}, a fitting where $R_{\mathrm{dp}}$
corresponds to $\alpha^{2}\mathcal{J}^{z}/\hbar$ and $R_{\mathrm{df}}%
=\beta^{2}\mathcal{J}^{xy}/\hbar$, gives $R_{\mathrm{dp}}/R_{\mathrm{df}}%
\sim4.$

We consider the experimental initial local polarization (\ref{sigma_ini}) on
the spin $I_{1}$,
\begin{equation}
\hat{\sigma}\left(  0\right)  =\frac{\hat{1}+\beta_{\text{\textrm{B}}}%
\hbar\Omega_{0,I}\hat{I}_{1}^{z}}{\mathrm{Tr}\left\{  \hat{1}\right\}  },
\label{2spin_DM_initial_condition}%
\end{equation}
and the spin-bath polarized, where $\beta_{\mathrm{B}}=1/\left(
k_{\mathrm{B}}T\right)  $. By neglecting other relaxation processes
($T_{1},T_{1\rho},$ etc.), the final state reaches the temperature of the
spin-bath yielding
\begin{equation}
\hat{\sigma}_{\infty}=\frac{\hat{1}+\beta_{\mathrm{B}}\hbar\Omega_{0,I}\left(
\hat{S}^{z}+\hat{I}_{1}^{z}\right)  }{\mathrm{Tr}\left\{  \hat{1}\right\}  }.
\label{2spin_DM_final_condition}%
\end{equation}
Here, $\hat{\sigma}_{\infty}$ commutes with $\widehat{\mathcal{H}}%
_{\mathrm{S}}$, not containing coherences with $\Delta M\geq1$.

\subsubsection{Neglecting non-secular terms in the relaxation
superoperator\label{Mark_2spin_secular}}

\begin{figure}
[pth]
\begin{center}
\includegraphics[
height=4.4918in,
width=5.431in
]%
{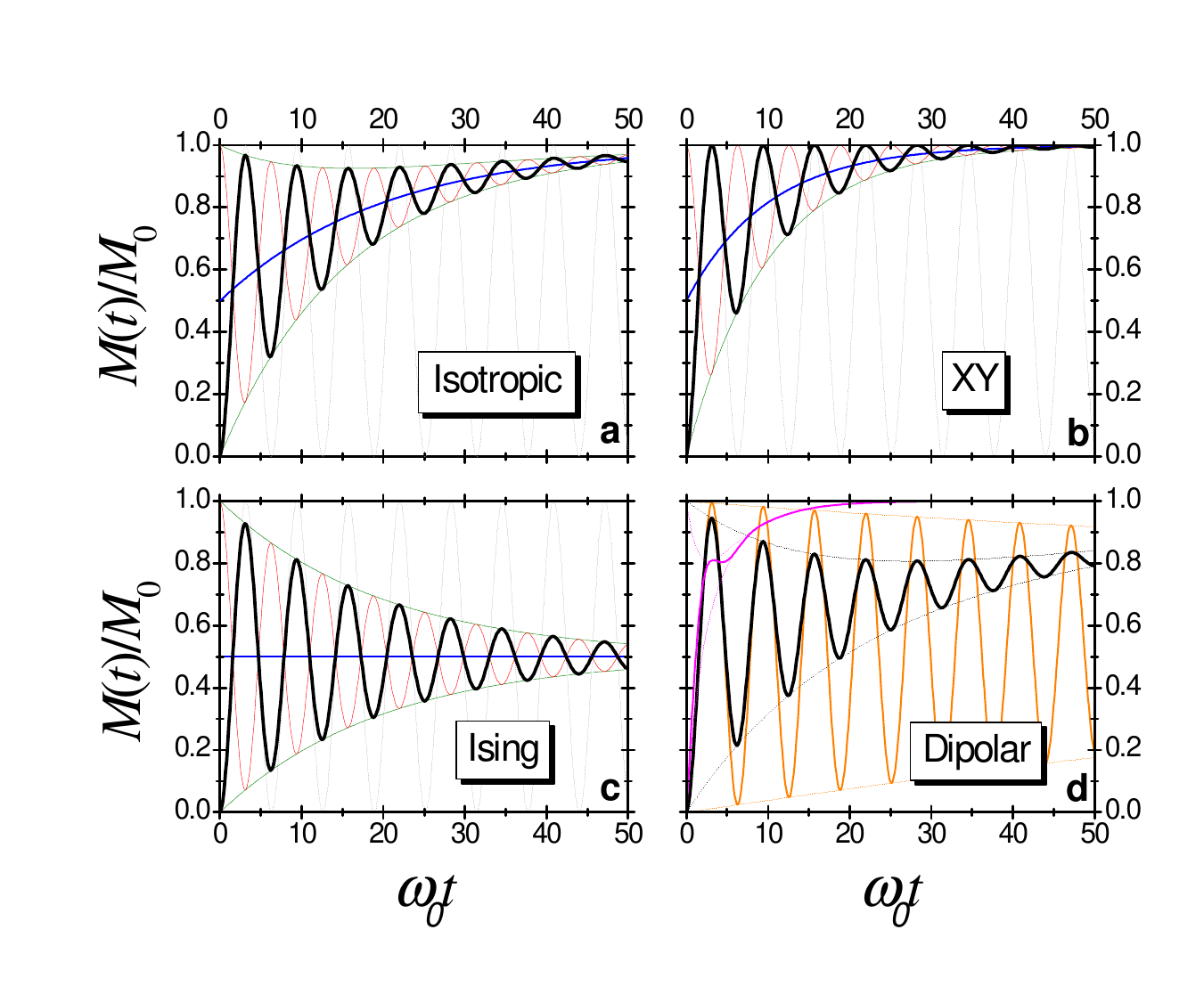}%
\caption[Temporal evolution of the polarization in the two-spin system
interacting with a spin-bath for different system-environment interactions.]%
{Temporal evolution of the polarization in the two-spin system interacting
with a spin-bath for different system-environment interactions. Panels a), b)
and c) show the polarization curves for the isotropic, XY and Ising SE
Interaction, where $p_{\mathrm{XY}}$ takes the values $1/2$, $1$ and $0$
respectively. The black line is the magnetization of the spin $S$ and the red
line is the magnetization of $I_{1}$ for a ratio $\Gamma_{\mathrm{SE}}/\left(
\omega_{0}\hbar\right)  =0.1.$ The blue line is the temporal evolution of the
mean magnetization at each site. The green lines show the coherence decays
with respect to the mean magnetization evolution. Panel c) shows the evolution
of the $S$ magnetization with a dipolar SE interaction for different values of
the ratio $\Gamma_{\mathrm{SE}}/\left(  \omega_{0}\hbar\right)  $. The orange,
black and magenta lines show the temporal evolution for the ratios $1$, $0.1$
and $0.01$ respectively.}%
\label{Fig_2spin_bath}%
\end{center}
\end{figure}

Following the standard formalism \cite{Abragam,Ernst}, we write the
superoperator $\widehat{\widehat{\Gamma}}$ using the basis of eigenstates of
the system Hamiltonian (\ref{H_system}). As the SE Hamiltonian is secular with
respect to the RF fields a block structure results. If $M$ is the total spin
projection in the $z$ direction, the first block couples the populations and
off-diagonal elements with $\Delta M=0$, zero quantum transitions (ZQT), of
the density matrix. Each of the following blocks couples one order $\left\vert
\Delta M\right\vert \geq1$ of the off-diagonal elements of the density matrix
among themselves. As the initial and final conditions do not contain
coherences with $\left\vert \Delta M\right\vert \geq1,$
the evolution of the density operator is reduced to a Liouville space
restricted to populations and ZQT. Thus, in the Hamiltonian eigenstate basis
(\ref{eigenstates_H_CH}), the generalized Liouville-von Neumann quantum master
equation (\ref{master}) restricted to this subspace becomes
\begin{equation}
\frac{\mathrm{d}}{\mathrm{d}t}\mathbf{\sigma}=\left\{  -\frac{\mathrm{i}%
}{\hbar}\mathcal{H}_{\mathrm{S}}\mathcal{-}\frac{1}{\hbar}\mathbf{\Gamma
}\right\}  \left(  \mathbf{\sigma-\sigma}_{\infty}\right)  ,
\end{equation}

\begin{multline}
\frac{\mathrm{d}}{\mathrm{d}t}\left[
\begin{tabular}
[c]{c}%
$\sigma_{11}$\\
$\sigma_{22}$\\
$\sigma_{33}$\\
$\sigma_{44}$\\\hline
$\sigma_{23}$\\
$\sigma_{32}$%
\end{tabular}
\right]  =\label{gQME_non_sec}\\
\left[
\begin{tabular}
[c]{cccc|cc}%
$-\frac{\Gamma_{\mathrm{XY}}}{\hbar}$ & $\frac{\Gamma_{\mathrm{XY}}}{2\hbar}$
& $\frac{\Gamma_{\mathrm{XY}}}{2\hbar}$ & $0$ & $\frac{\Gamma_{\mathrm{XY}}%
}{2\hbar}$ & $\frac{\Gamma_{\mathrm{XY}}}{2\hbar}$\\
$\frac{\Gamma_{\mathrm{XY}}}{2\hbar}$ & $-\frac{\left(  \Gamma_{\mathrm{ZZ}%
}+2\Gamma_{\mathrm{XY}}\right)  }{2\hbar}$ & $\frac{\Gamma_{\mathrm{XY}}%
}{2\hbar}$ & $\frac{\Gamma_{\mathrm{XY}}}{2\hbar}$ & $0$ & $0$\\
$\frac{\Gamma_{\mathrm{XY}}}{2\hbar}$ & $\frac{\Gamma_{\mathrm{XY}}}{2\hbar}$
& $-\frac{\left(  \Gamma_{\mathrm{ZZ}}+2\Gamma_{\mathrm{XY}}\right)  }{2\hbar
}$ & $\frac{\Gamma_{\mathrm{XY}}}{2\hbar}$ & $0$ & $0$\\
$0$ & $\frac{\Gamma_{\mathrm{XY}}}{2\hbar}$ & $\frac{\Gamma_{\mathrm{XY}}%
}{2\hbar}$ & $-\frac{\Gamma_{\mathrm{XY}}}{\hbar}$ & $-\frac{\Gamma
_{\mathrm{XY}}}{2\hbar}$ & $-\frac{\Gamma_{\mathrm{XY}}}{2\hbar}$\\\hline
$\frac{\Gamma_{\mathrm{XY}}}{2\hbar}$ & $0$ & $0$ & $-\frac{\Gamma
_{\mathrm{XY}}}{2\hbar}$ & $-\mathrm{i}\omega_{0}-\frac{\left(  \Gamma
_{\mathrm{ZZ}}+2\Gamma_{\mathrm{XY}}\right)  }{2\hbar}$ & $0$\\
$\frac{\Gamma_{\mathrm{XY}}}{2\hbar}$ & $0$ & $0$ & $-\frac{\Gamma
_{\mathrm{XY}}}{2\hbar}$ & $0$ & $\mathrm{i}\omega_{0}-\frac{\left(
\Gamma_{\mathrm{ZZ}}+2\Gamma_{\mathrm{XY}}\right)  }{2\hbar}$%
\end{tabular}
\right] \\
\cdot\left[
\begin{tabular}
[c]{c}%
$\sigma_{11}-\sigma_{\infty_{11}}$\\
$\sigma_{22}-\sigma_{\infty_{22}}$\\
$\sigma_{33}-\sigma_{\infty_{33}}$\\
$\sigma_{44}-\sigma_{\infty_{44}}$\\\hline
$\sigma_{23}-\sigma_{\infty_{23}}$\\
$\sigma_{32}-\sigma_{\infty_{32}}$%
\end{tabular}
\right]  .
\end{multline}
Here, the superoperator $\widehat{\widehat{\Gamma}}$ is given by
\begin{equation}
\mathbf{\Gamma}=\left[
\begin{tabular}
[c]{cccc|cc}%
$\Gamma_{\mathrm{XY}}$ & $-\frac{\Gamma_{\mathrm{XY}}}{2}$ & $-\frac
{\Gamma_{\mathrm{XY}}}{2}$ & $0$ & $-\frac{\Gamma_{\mathrm{XY}}}{2}$ &
$-\frac{\Gamma_{\mathrm{XY}}}{2}$\\
$-\frac{\Gamma_{\mathrm{XY}}}{2},$ & $\frac{\left(  \Gamma_{\mathrm{ZZ}%
}+2\Gamma_{\mathrm{XY}}\right)  }{2}$ & $-\frac{\Gamma_{\mathrm{XY}}}{2}$ &
$-\frac{\Gamma_{\mathrm{XY}}}{2}$ & $0$ & $0$\\
$-\frac{\Gamma_{\mathrm{XY}}}{2}$ & $-\frac{\Gamma_{\mathrm{XY}}}{2}$ &
$\frac{\left(  \Gamma_{\mathrm{ZZ}}+2\Gamma_{\mathrm{XY}}\right)  }{2}$ &
$-\frac{\Gamma_{\mathrm{XY}}}{2}$ & $0$ & $0$\\
$0$ & $-\frac{\Gamma_{\mathrm{XY}}}{2}$ & $-\frac{\Gamma_{\mathrm{XY}}}{2}$ &
$\Gamma_{\mathrm{XY}}$ & $\frac{\Gamma_{\mathrm{XY}}}{2}$ & $\frac
{\Gamma_{\mathrm{XY}}}{2}$\\\hline
$-\frac{\Gamma_{\mathrm{XY}}}{2}$ & $0$ & $0$ & $\frac{\Gamma_{\mathrm{XY}}%
}{2}$ & $\frac{\left(  \Gamma_{\mathrm{ZZ}}+2\Gamma_{\mathrm{XY}}\right)  }%
{2}$ & $0$\\
$-\frac{\Gamma_{\mathrm{XY}}}{2}$ & $0$ & $0$ & $\frac{\Gamma_{\mathrm{XY}}%
}{2}$ & $0$ & $\frac{\left(  \Gamma_{\mathrm{ZZ}}+2\Gamma_{\mathrm{XY}%
}\right)  }{2}$%
\end{tabular}
\right]
\end{equation}
and the superoperator of the system Hamiltonian, that is diagonal in its
basis, results%
\begin{equation}
\mathcal{H}_{\mathrm{S}}=\left[
\begin{tabular}
[c]{cccc|cc}%
$0$ & $0$ & $0$ & $0$ & $0$ & $0$\\
$0$ & $0$ & $0$ & $0$ & $0$ & $0$\\
$0$ & $0$ & $0$ & $0$ & $0$ & $0$\\
$0$ & $0$ & $0$ & $0$ & $0$ & $0$\\\hline
$0$ & $0$ & $0$ & $0$ & $b$ & $0$\\
$0$ & $0$ & $0$ & $0$ & $0$ & $-b$%
\end{tabular}
\right]  ,
\end{equation}
where its elements are transition frequencies (energy differences). After
neglecting the rapidly oscillating non-secular terms with respect to the
Hamiltonian, i.e.
\begin{equation}
\Gamma_{\mathrm{ZZ}},\Gamma_{\mathrm{XY}}\ll b,
\end{equation}
a kite structure results \cite{Ernst}.
All the non-diagonal terms coupling the population block with the ZQT block
are non-secular and can be neglected because the Hamiltonian (\ref{H_system})
does not have degenerate eigenenergies. Although in this case the ZQT block is
diagonal, in a general case only the diagonal terms of this block contribute
to the evolution if there are not degenerate transitions. The differential eq.
(\ref{gQME_non_sec}) is now%
\begin{multline}
\frac{\mathrm{d}}{\mathrm{d}t}\left[
\begin{tabular}
[c]{c}%
$\sigma_{11}$\\
$\sigma_{22}$\\
$\sigma_{33}$\\
$\sigma_{44}$\\\hline
$\sigma_{23}$\\
$\sigma_{32}$%
\end{tabular}
\right]  =\\
\left[
\begin{tabular}
[c]{cccccc}%
$-\frac{\Gamma_{\mathrm{XY}}}{\hbar}$ & $\frac{\Gamma_{\mathrm{XY}}}{2\hbar}$
& $\frac{\Gamma_{\mathrm{XY}}}{2\hbar}$ & $0$ & \multicolumn{1}{|c}{$0$} &
$0$\\
$\frac{\Gamma_{\mathrm{XY}}}{2\hbar}$ & $-\frac{\left(  \Gamma_{\mathrm{ZZ}%
}+2\Gamma_{\mathrm{XY}}\right)  }{2\hbar}$ & $\frac{\Gamma_{\mathrm{XY}}%
}{2\hbar}$ & $\frac{\Gamma_{\mathrm{XY}}}{2\hbar}$ & \multicolumn{1}{|c}{$0$}
& $0$\\
$\frac{\Gamma_{\mathrm{XY}}}{2\hbar}$ & $\frac{\Gamma_{\mathrm{XY}}}{2\hbar}$
& $-\frac{\left(  \Gamma_{\mathrm{ZZ}}+2\Gamma_{\mathrm{XY}}\right)  }{2\hbar
}$ & $\frac{\Gamma_{\mathrm{XY}}}{2\hbar}$ & \multicolumn{1}{|c}{$0$} & $0$\\
$0$ & $\frac{\Gamma_{\mathrm{XY}}}{2\hbar}$ & $\frac{\Gamma_{\mathrm{XY}}%
}{2\hbar}$ & $-\frac{\Gamma_{\mathrm{XY}}}{\hbar}$ & \multicolumn{1}{|c}{$0$}
& $0$\\\cline{1-5}\cline{5-5}%
$0$ & $0$ & $0$ & $0$ & \multicolumn{1}{|c}{$-\mathrm{i}\omega_{0}%
-\frac{\left(  \Gamma_{\mathrm{ZZ}}+2\Gamma_{\mathrm{XY}}\right)  }{2\hbar}$}
& \multicolumn{1}{|c}{$0$}\\\cline{5-6}%
$0$ & $0$ & $0$ & $0$ & $0$ & \multicolumn{1}{|c}{$\mathrm{i}\omega_{0}%
-\frac{\left(  \Gamma_{\mathrm{ZZ}}+2\Gamma_{\mathrm{XY}}\right)  }{2\hbar}$}%
\end{tabular}
\right] \\
\cdot\left[
\begin{tabular}
[c]{c}%
$\sigma_{11}-\sigma_{\infty_{11}}$\\
$\sigma_{22}-\sigma_{\infty_{22}}$\\
$\sigma_{33}-\sigma_{\infty_{33}}$\\
$\sigma_{44}-\sigma_{\infty_{44}}$\\\hline
$\sigma_{23}-\sigma_{\infty_{23}}$\\
$\sigma_{32}-\sigma_{\infty_{32}}$%
\end{tabular}
\right]  ,
\end{multline}
where its solution, in the Hamiltonian eigenstate basis
(\ref{eigenstates_H_CH}), is
\begin{multline}
\mathbf{\sigma}(t)=\\
\left[
\begin{array}
[c]{cccc}%
\frac{1}{4}+M_{0}\left(  1-\frac{e^{-\Gamma_{\mathrm{XY}}t/\hbar}}{2}\right)
& 0 & 0 & 0\\
0 & \frac{1}{4} & \frac{-M_{0}e^{-\mathrm{i}\omega_{0}t}e^{-\frac{\left(
\Gamma_{\mathrm{ZZ}}+2\Gamma_{\mathrm{XY}}\right)  }{2\hbar}t}}{2} & 0\\
0 & \frac{-M_{0}e^{\mathrm{i}\omega_{0}t}e^{-\frac{\left(  \Gamma
_{\mathrm{ZZ}}+2\Gamma_{\mathrm{XY}}\right)  }{2\hbar}t}}{2} & \frac{1}{4} &
0\\
0 & 0 & 0 & \frac{1}{4}-M_{0}\left(  1-\frac{e^{-\Gamma_{\mathrm{XY}}t/\hbar}%
}{2}\right)
\end{array}
\right]  .
\end{multline}
At time $t=0,$ we obtain the initial state of the reduced density matrix
(\ref{2spin_DM_initial_condition}) and when $t\longrightarrow\infty,$ it goes
to the final state $\hat{\sigma}_{\infty}$ of eq.
(\ref{2spin_DM_final_condition}). The elements $\sigma_{11}\left(  t\right)  $
and $\sigma_{44}\left(  t\right)  ,$ which are the populations of the states
$|+,+\rangle$ and $|-,-\rangle$ respectively, go to the equilibrium state
$\hat{\sigma}_{\infty}$ with a rate $\Gamma_{\mathrm{XY}}/\hbar.$ This
accounts for the net transfer of magnetization from the spin-bath to the
system and it contains the information of the net magnetization inside the
system.
The coherences $\sigma_{23}$ and $\sigma_{32},$ that take into account the
swapping between the states $\left\vert \downarrow,\uparrow\right\rangle $ and
$\left\vert \uparrow,\downarrow\right\rangle $ with the natural frequency
$\omega_{0},$ decay to zero with a decoherence rate
\begin{equation}
1/\tau_{\phi}=\frac{\left(  \Gamma_{\mathrm{ZZ}}+2\Gamma_{\mathrm{XY}}\right)
}{2\hbar}.
\end{equation}
Note that the coherences decay faster than the time that the system takes to
arrive to the equilibrium state. We calculate the magnetization of the spin
$S$ obtaining an extension \cite{Alvarez07a} of the result given in ref.
\cite{MKBE74}
\begin{equation}
M_{S^{z}}\left(  t\right)  =\mathrm{Tr}\left\{  \hat{\sigma}\left(  t\right)
\hat{S}^{z}\right\}  =M_{0}\left[  1-\frac{1}{2}e^{-\Gamma_{\mathrm{XY}%
}t/\hbar}-\frac{1}{2}e^{-\frac{1}{2}\left(  \Gamma_{\mathrm{ZZ}}%
+2\Gamma_{\mathrm{XY}}\right)  t/\hbar}\cos\left(  \omega_{0}t\right)
\right]  , \label{Msz_2spin_MKBEg}%
\end{equation}
where our essential contribution is that we specifically account for the
anisotropy arising from the nature of the SE interaction reflected in
$\Gamma_{\mathrm{ZZ}}=\alpha^{2}\mathcal{J}^{z}$ and $\Gamma_{\mathrm{XY}%
}=\beta^{2}\mathcal{J}^{xy}$.
The first two terms of eq. (\ref{Msz_2spin_MKBEg}) are given by $\left[
\sigma_{11}\left(  t\right)  -\sigma_{44}\left(  t\right)  \right]  /2$ and
the oscillatory one by $\operatorname{Re}\left\{  \sigma_{32}\left(  t\right)
\right\}  $ [see eq. (\ref{MSx_base_H_expression})]. The sum of the first two
terms are the mean magnetization at each site or, multiplied by two, it
represents the total magnetization of the $2$-spin system which is given by%
\begin{equation}
M_{\mathrm{tot.}}\left(  t\right)  =\mathrm{Tr}\left\{  \hat{\sigma}\left(
t\right)  \left(  \hat{S}_{{}}^{z}+\hat{I}_{1}^{z}\right)  \right\}
=\sigma_{11}\left(  t\right)  -\sigma_{44}\left(  t\right)  =2M_{0}\left(
1-\frac{1}{2}e^{-\Gamma_{\mathrm{XY}}t/\hbar}\right)  .
\end{equation}
The time dependence of this quantity is due to a \textquotedblleft diffusion
process\textquotedblright\ from the spin-bath that injects magnetization at a
rate $\Gamma_{\mathrm{XY}}/\hbar$ through the XY SE interaction term. We
define the SE interaction rate as
\begin{equation}
\Gamma_{\mathrm{SE}}=\Gamma_{\mathrm{ZZ}}+\Gamma_{\mathrm{XY}}%
\end{equation}
and the weight of the XY interaction as
\begin{equation}
p_{\mathrm{XY}}=\Gamma_{\mathrm{XY}}/\Gamma_{\mathrm{SE}}.
\end{equation}
An Ising, dipolar, isotropic, and XY SE interactions are obtained when
$p_{\mathrm{XY}}=0,\frac{1}{5},\frac{1}{2},1$ respectively. Equation
(\ref{Msz_2spin_MKBEg}) becomes
\begin{equation}
M_{S^{z}}\left(  t\right)  =M_{0}\left[  1-\frac{1}{2}e^{-p_{\mathrm{XY}%
}\Gamma_{\mathrm{SE}}t/\hbar}-\frac{1}{2}e^{-\frac{1}{2}\left(
1+p_{\mathrm{XY}}\right)  \Gamma_{\mathrm{SE}}t/\hbar}\cos\left(  \omega
_{0}t\right)  \right]  .
\end{equation}
Figure \ref{Fig_2spin_bath}
shows typical curves for different SE interactions (black lines). The blue
line is the temporal evolution of the total magnetization divided by two or
equivalently the mean magnetization at each site. We see that the curves go to
$M_{0}$ for long times manifesting that the system arrives to the equilibrium
state $\hat{\sigma}_{\infty}$ with the exception of the Ising SE interaction.
In this case, the system has no injection from the spin-bath and goes to the
quasi-equilibrium state of the $2$-spin system, i.e. the initial magnetization
is spread over both sites. This quasi-equilibrium is described by%
\begin{equation}
\hat{\sigma}^{\mathrm{qe}}=\frac{\exp\{-\widehat{\mathcal{H7}}_{\mathrm{S}%
}/(k_{\mathrm{B}}T_{\mathrm{qe}})\}}{\mathrm{Tr}\{\exp\{-\widehat{\mathcal{H}%
}_{\mathrm{S}}/(k_{\mathrm{B}}T_{\mathrm{qe}})\}\}}
\label{quasi_equilibrium_2spin}%
\end{equation}
with
\begin{equation}
T_{\mathrm{qe}}=2(\Omega_{1,I}/\Omega_{0,I})T.
\end{equation}
In general, there is a competition between the Ising and the XY SE interaction
terms that leads the system to a $2$-spin quasi-equilibrium state or to the
total system equilibrium state respectively \cite{JCP03,Alvarez07a}.
This quasi-equilibrium is time dependent and is given by the mean
magnetization, $M_{\mathrm{tot}}\left(  t\right)  /2$, represented by the blue
line in fig. \ref{Fig_2spin_bath}. The green lines in the figure show the
coherence decay relative to the mean magnetization at each site. We see that
the XY interaction is the most coherent because its decoherence rate is equal
to the magnetization transfer rate, while in the other cases, decoherence is
faster than magnetization transfer. The red lines show the magnetization on
the spin $I_{1}$ described by the expression%
\begin{equation}
M_{I^{z}}\left(  t\right)  =M_{0}\left[  1-\frac{1}{2}e^{-p_{\mathrm{XY}%
}\Gamma_{\mathrm{SE}}t/\hbar}+\frac{1}{2}e^{-\frac{1}{2}\left(
1+p_{\mathrm{XY}}\right)  \Gamma_{\mathrm{SE}}t/\hbar}\cos\left(  \omega
_{0}t\right)  \right]  ,
\end{equation}
where only the sign of the oscillatory term $\operatorname{Re}\left\{
\sigma_{32}\left(  t\right)  \right\}  $ changes. Figure \ref{Fig_2spin_bath}
c) shows curves with dipolar SE interaction for different values of the ratio
$\Gamma_{\mathrm{SE}}/\left(  \omega_{0}\hbar\right)  .$ It shows how the
decoherence and magnetization transfer are stronger as $\Gamma_{\mathrm{SE}%
}/\left(  \omega_{0}\hbar\right)  $ becomes higher. Here, we observe the
decoherence's role described in the Introduction. The temporal interference
pattern is described by the oscillatory term which contains the entangled
two-spin superposition. A strong SE interaction leads to an efficient
degradation of the two-spin quantum entanglement. This drives the system to a
mixed-state, described by the diagonal elements of the density matrix, which
constitutes the quasi-equilibrium state represented by the blue line. When
decoherence is not too strong, we observe that it is not necessary to wait
long times to obtain the maximum magnetization at the spin $S$ (totally
polarized). It is enough to wait for a maximum of the oscillation at time
$\pi/\omega_{0}$ where the magnetization reaches a value close to the maximum
obtainable ($M_{0}$). But a more important result is that for an XY SE
interaction, one can achieve the biggest gain of polarization at the first
maximum of the oscillation. This is a consequence of the different behavior of
the decoherent processes arising on the Ising or XY interactions.
Moreover, for an XY SE interaction, expression (\ref{Msz_2spin_MKBEg}) yields
all the maxima of the oscillation equal to $M_{0},$regardless of the magnitude
of the SE interaction. However, we should not forget that this expression is
valid only for $\Gamma_{\mathrm{ZZ}},\Gamma_{\mathrm{XY}}\ll b$.

\subsubsection{Non-secular solution\label{Mark_non_secular_2spin}}

If we release the condition $\Gamma_{\mathrm{ZZ}},~\Gamma_{\mathrm{XY}}\ll b,$
i.e. we do not neglect the non-secular terms for the superoperator
$\widehat{\widehat{\Gamma}}$, the dynamics still occurs in the Liouville space
of the populations and ZQT.
The solution of the generalized quantum master equation is now,%
\begin{equation}
M_{S^{z}}\left(  t\right)  =M_{0}\left(  1-a_{0}e^{-R_{0}t}-a_{1}\cos\left[
\left(  \omega+\mathrm{i}\eta\right)  t+\phi_{0}\right]  e^{-R_{1}t}\right)  ,
\end{equation}
where the real functions $\omega,$ $R_{0},R_{1}$ and $\eta$ as well as $a_{0}%
$, $a_{1}$ and $\phi_{0}$ depend exclusively on $b,$ $1/\tau_{\mathrm{SE}%
}=2\left(  \Gamma_{\mathrm{ZZ}}+\Gamma_{\mathrm{XY}}\right)  /\hbar$ and
$p_{\mathrm{XY}}=\Gamma_{\mathrm{XY}}/\left(  \Gamma_{\mathrm{ZZ}}%
+\Gamma_{\mathrm{XY}}\right)  $. This expression will be discussed in chapter
\ref{Marker_Spin_within_keldysh}, where it is obtained from a microscopic
derivation. However, it is important to remark that the short time evolution,
$t\ll\tau_{\mathrm{SE}},$ of the secular expression does not satisfy the
correct quadratic quantum behavior while the non-secular expression does. The
relevance of this inertial property reflected in the quadratic short time
evolution will become evident in chapter \ref{Sec_QDPT}. We will see how it
leads to the manifestation of something that we called an environmentally
induced quantum dynamical phase transition \cite{JCP06,SSC07}.

\section{Three-spin quantum dynamics\label{M_Q_dyn_in_8CB}}

In this section, we analyze theoretically and experimentally the quantum
dynamics of a three-spin system coupled to a spin-bath during
cross-polarization (CP) \cite{Hartmann62,MKBE74}. Our analysis takes into
account a pure Hamiltonian behavior for a carbon $^{13}$C coupled to two
protons $^{1}$H, while the coupling to a spin-bath is treated in the fast
fluctuation approximation. This model is inspired and then applied to the
methylene and biphenyl groups of the smectic and nematic phases of the liquid
crystal $4$-n-octyl-$4$'-cyanobiphenyl ($8$CB). We make use of the
Hartmann-Hahn CP technique as a function of contact time
to measure $^{1}$H-$^{13}$C and $^{1}$H-$^{1}$H effective dipolar
interactions. This technique has proved very useful in order to evaluate order
parameters in liquid crystals \cite{Pratima96}. Most of the previous works
where transient oscillations were observed during CP were analyzed in terms of
a single $^{1}$H-$^{13}$C interaction incorporating the interaction with other
protons as a thermal bath or reservoir in a phenomenological way. However,
many liquid crystals have alkyl chains and aromatic groups in their
structures, where the carbon is coupled to more than one proton and the
carbon-proton and proton-proton dipolar interactions are of the same order of
magnitude. This led us to consider a set of three strongly dipolar coupled
spins $1/2$ as the main system, which in turns interacts with the protons of
the bath. Combining detailed calculations of the three-spin dynamics with
structural information which provide the relative sign of the $^{1}$H-$^{13}$C
couplings, we are able to obtain separately the $^{1}$H-$^{13}$C and $^{1}%
$H-$^{1}$H effective interactions. In order to test the suitability of the
formula obtained, we compare the values of the $^{1}$H-$^{13}$C couplings
obtained by two procedures. One involves fitting of the data from a standard
CP experiment to the calculated dynamics while in the other the $^{1}$%
H-$^{13}$C couplings are obtained directly from a CP under Lee-Goldburg
conditions, i.e. when the dipolar proton-proton interactions have been
cancelled out. The advantages and disadvantages of each procedure are discussed.

An interesting aspect we could observe during the CP dynamics in $8$CB is that
the rate of attenuation of the oscillations (representing the coherences) is
much faster than that of the polarization transfer from the bath in a factor
several times larger
than the one calculated assuming isotropic interaction with the bath
\cite{MKBE74}. We analyze here, the origin of this highly anisotropic
behavior, not observed in solid molecular crystals \cite{JCP98,Hirschinger94}.
A well differentiated relaxation behavior among the two phases seems to
indicate that while the extreme narrowing approximation is appropriate for the
nematic phase, the description of the smectic phase requires the consideration
of the slow motion limit.

\subsection{An isolated three-spin system}

We will consider the quantum evolution of a system of three spins $1/2$
coupled through the magnetic dipolar interaction during the contact time in a
cross-polarization experiment \cite{Hartmann62,MKBE74}. The system is
constituted by one spin $S$ and two spins $I$ representing a carbon-$13$ and
two protons, respectively, under the presence of a static magnetic field
$H_{0}$ in the $z$ direction and radio-frequency (RF) magnetic
fields\textrm{\ }$H_{1,I}$ and $H_{1,S}$ in the $x$ direction.
The Hamiltonian including the dipolar interactions truncated with respect to
the Zeeman field $H_{0}$ and in a double rotating frame \cite{Slichter} can be
written as
\begin{equation}
\widehat{\mathcal{H}}_{\mathrm{S}}=\hbar\Delta\Omega_{I}\hat{I}_{{}}^{z}%
+\hbar\Delta\Omega_{S}\hat{S}_{{}}^{z}-\hbar\Omega_{1,I}\hat{I}_{{}}^{x}%
-\hbar\Omega_{1,S}\hat{S}_{{}}^{x}+2\sum_{k=1,2}b_{k}\hat{S}_{{}}^{z}\hat
{I}_{k}^{z}+d(2\hat{I}_{1}^{z}\hat{I}_{2}^{z}-\hat{I}_{1}^{x}\hat{I}_{2}%
^{x}-\hat{I}_{1}^{y}\hat{I}_{2}^{y}),
\end{equation}
where, as in the previous section,
\begin{equation}
\Delta\Omega_{I}=\Omega_{0,I}-\omega_{\mathrm{rf},I}\;\;\;\;\;\text{and}%
\;\;\;\;\;\Delta\Omega_{S}=\Omega_{0,S}-\omega_{\mathrm{rf},S}%
\end{equation}
are the resonance offsets,
\begin{equation}
\hat{I}^{u}=\hat{I}_{1}^{u}+\hat{I}_{2}^{u}%
\end{equation}
with $\,u=x,y,z$,
\begin{equation}
\Omega_{1,I}=\gamma_{I}H_{1,I}\;\;\;\;\;\text{and}\;\;\;\;\;\Omega
_{1,S}=\gamma_{S}H_{1,S}%
\end{equation}
where $\gamma_{I},\gamma_{S}$ are the gyromagnetic factors of the $I$ and $S$
spins. The constant, as defined in eq. (\ref{dipolar_coupling}),
\begin{equation}
b_{k}=-\frac{1}{2}\frac{\mu_{0}\gamma_{I}\gamma_{S}\hbar^{2}}{4\pi
}\left\langle \frac{(3\cos^{2}(\theta_{Sk})-1)}{r_{Sk}^{3}}\right\rangle
\qquad k=1,2
\end{equation}
and
\begin{equation}
d=-\frac{1}{2}\frac{\mu_{0}\gamma_{I}^{2}\hbar^{2}}{4\pi}\left\langle
\frac{(3\cos^{2}(\theta_{12})-1)}{r_{12}^{3}}\right\rangle
\end{equation}
are the heteronuclear and homonuclear effective dipolar couplings
respectively. However, here the angular brackets in the equations indicate
that the dipolar couplings in liquid crystals are averaged over both molecular
tumbling and any internal bond rotations. Thus, the molecular variation of the
spin-spin distance, $r_{ij},$ and the angle between the internuclear vector
and the external field, $\theta_{ij},$ are taken into account. Because of a
special geometry of the oriented $n$CB liquid crystals, we will consider two
different cases where the dipolar constants are related by $b_{1}=b_{2}=b$ and
$b_{1}=-b_{2}=b.$

As in the previous section, for a standard CP\ experiment, one can neglect the
resonance offsets,
and considering that
\begin{equation}
|\Omega_{1,I}+\Omega_{1,S}|\gg|b_{k}|,|d|,
\end{equation}
the truncated Hamiltonian can be written as
\begin{equation}
\widehat{\mathcal{H}}_{\mathrm{S}}=\tfrac{1}{2}\Sigma\left(  \hat{S}_{{}}%
^{x}+\hat{I}_{{}}^{x}\right)  +\tfrac{1}{2}\Delta\left(  \hat{I}_{{}}^{x}%
-\hat{S}_{{}}^{x}\right)  +\sum_{k=1,2}b_{k}\left(  \hat{S}_{{}}^{z}\hat
{I}_{k}^{z}+\hat{S}_{{}}^{y}\hat{I}_{k}^{y}\right)  -\tfrac{1}{2}d(2\hat
{I}_{1}^{x}\hat{I}_{2}^{x}-\hat{I}_{1}^{z}\hat{I}_{2}^{z}-\hat{I}_{1}^{y}%
\hat{I}_{2}^{y}), \label{Hsecular}%
\end{equation}
with
\begin{equation}
\Sigma=-\hbar\left(  \Omega_{1,S}+\Omega_{1,I}\right)  \;\;\;\;\;\text{and}%
\;\;\;\;\;\Delta=\hbar\left(  \Omega_{1,S}-\Omega_{1,I}\right)  .
\end{equation}
As the Hamiltonian (\ref{Hsecular}) has only Zeeman fields along the $x$
direction, we change the names of the axis as we did in section
\S \ \ref{Mark_2spin_isolated}: $x\rightarrow z,$ $y\rightarrow x$ and
$z\rightarrow y.$ Hence, the Hamiltonian becomes%
\begin{equation}
\widehat{\mathcal{H}}_{\mathrm{S}}=\tfrac{1}{2}\Sigma\left(  \hat{S}_{{}}%
^{z}+\hat{I}_{{}}^{z}\right)  +\tfrac{1}{2}\Delta\left(  \hat{I}_{{}}^{z}%
-\hat{S}_{{}}^{z}\right)  +\sum_{k=1,2}b_{k}\left(  \hat{S}_{{}}^{x}\hat
{I}_{k}^{x}+\hat{S}_{{}}^{y}\hat{I}_{k}^{y}\right)  -\tfrac{1}{2}d(2\hat
{I}_{1}^{z}\hat{I}_{2}^{z}-\hat{I}_{1}^{x}\hat{I}_{2}^{x}-\hat{I}_{1}^{y}%
\hat{I}_{2}^{y}). \label{Hsecular_3spin_isolated_axis_changed}%
\end{equation}
In eq. (\ref{Hsecular_3spin_isolated_axis_changed}) the non-secular elements
of the dipolar interaction with respect to the $\Sigma\left(  \hat{I}^{z}%
+\hat{S}^{z}\right)  $ term have been neglected. Similar as in the previous
section, this allows us to write the matrix representation of the Hamiltonian
in a simple block structure using the basis $\left\{  |M_{I},M_{S}%
\rangle\right\}  $, with $M_{I}=M_{1}+M_{2}$ and $M_{S}$ denoting the spin
projections of the $I$ and $S$ systems in the direction of their respective RF
fields. Now, each block is characterized by the total spin projection
$M=M_{I}+M_{S},$ i.e. nonzero matrix elements exist only between states with
the same magnetic quantum numbers $M$. Thus, the heteronuclear dipolar
Hamiltonian has non-diagonal terms different from zero generating transitions
between spin states $\left\{  |M_{I},M_{S}\rangle\right\}  $ and $\left\{
|M_{I}\pm1,M_{S}\mp1\rangle\right\}  .$ The eigenstates of this Hamiltonian
can be denoted in the form $|M,n_{M}\rangle$, with $n_{M}=1,..,g_{M}$, where
$g_{M}$ is the degeneracy of $M$ ($n_{\pm3/2}=1$\emph{\ }and \emph{\ }%
$n_{\pm1/2}=1,2,3$). It is very interesting to note that in each space of
$M=\pm1/2$ there are only two of the three eigenstates that are involved in
the dipolar transitions that give rise to the oscillations. This is a
consequence of the symmetry of the system, i.e. the flip-flop can occur only
between the carbon and one (the symmetric or the antisymmetric) combination of
the proton states depending on the relative signs of the heteronuclear
couplings ($b_{1}=b_{2}$ or $b_{1}=-b_{2}$). The symmetric and antisymmetric
combination of the proton are
\begin{align}
\left\vert S\right\rangle  &  =1/\sqrt{2}\left(  \left\vert +,-\right\rangle
+\left\vert -,+\right\rangle \right)  ,\\
\left\vert A\right\rangle  &  =1/\sqrt{2}\left(  \left\vert +,-\right\rangle
-\left\vert -,+\right\rangle \right)  ,
\end{align}
where the vectors are denoted by $\left\vert M_{1},M_{2}\right\rangle .$
Hence, in the ordered basis
\begin{equation}
\left\{  \left\vert S\right\rangle \otimes\left\vert +\right\rangle
;\left\vert +,+\right\rangle \otimes\left\vert -\right\rangle ;\left\vert
A\right\rangle \otimes\left\vert +\right\rangle \right\}
\end{equation}
with $\left\vert M_{1},M_{2}\right\rangle \otimes\left\vert M_{S}\right\rangle
$, the $M=1/2$ block of the system Hamiltonian is given by%
\begin{equation}
\mathcal{H}_{\mathrm{S},M=\frac{1}{2}}=\left[
\begin{array}
[c]{ccc}%
\frac{1}{4}\left(  \Sigma-\Delta\right)  +\frac{1}{2}d & \frac{\sqrt{2}}%
{8}\left(  b_{1}+b_{2}\right)  & 0\\
\frac{\sqrt{2}}{8}\left(  b_{1}+b_{2}\right)  & \left(  \frac{1}{4}%
\Sigma+\frac{3}{4}\Delta\right)  -\frac{1}{4}d & \frac{\sqrt{2}}{8}\left(
b_{2}-b_{1}\right) \\
0 & \frac{\sqrt{2}}{8}\left(  b_{2}-b_{1}\right)  & \frac{1}{4}\left(
\Sigma-\Delta\right)
\end{array}
\right]  ,
\end{equation}
and similarly for the $M=-1/2$ block. It is evident from the previous equation
that under the conditions $b_{1}=b_{2}$ or $b_{1}=-b_{2}$ one of the states,
the symmetric or antisymmetric, is involved in the dipolar transition.

The Liouville-von Neumann equation \cite{Abragam,Ernst} for the density matrix
of the system is (\ref{ec_maestra_sin_difusion})%
\begin{equation}
\frac{\mathrm{d}}{\mathrm{d}t}\hat{\rho}\left(  t\right)  =-\frac{\mathrm{i}%
}{\hbar}[\widehat{\mathcal{H}}_{\mathrm{S}},\hat{\rho}\left(  t\right)  ],
\label{Liouville}%
\end{equation}
where, similarly as in the $2$-spin case (\ref{sigma_ini}), the initial
density operator $\hat{\rho}(0)$ considering the situation after the $\pi/2$
pulse in the $I$ system is given by\footnote{Remember that this initial
condition is given in the high temperature approximation.}
\begin{equation}
\hat{\rho}\left(  0\right)  =\frac{\hat{1}+\beta_{\mathrm{B}}\hbar\Omega
_{0,I}\hat{I}^{z}}{\mathrm{Tr}\left\{  \hat{1}\right\}  }.
\label{initial_condition}%
\end{equation}
The solution of this equation is given by%
\begin{equation}
\rho(t)=\hat{U}(t)\rho(0)\hat{U}^{-1}(t),
\end{equation}
where $\hat{U}(t)=\exp\left(  -\frac{\mathrm{i}}{^{\hbar}}\widehat
{\mathcal{H}}_{\mathrm{S}}t\right)  $.

\begin{figure}
[tbh]
\begin{center}
\includegraphics[
height=4.4987in,
width=5.7847in
]%
{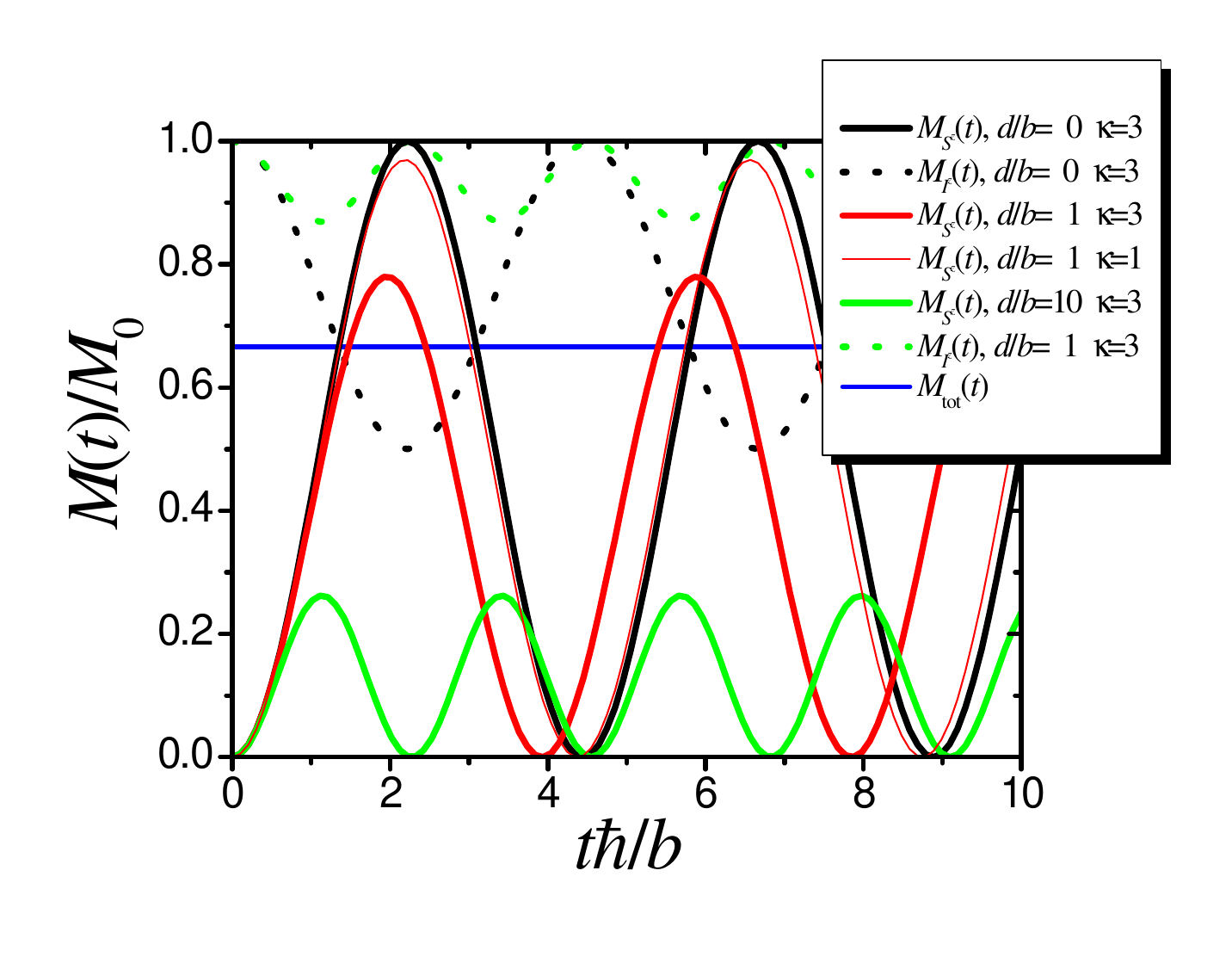}%
\caption[Temporal evolution of the polarization in a $3$-spin system.]%
{Temporal evolution of the polarization in a $3$-spin system. The black, red
and blue lines are the $S$ (solid) and $I$ (dotted) magnetization with
$b_{1}=b_{2}=b$ ($\kappa=3$) under the ratio $d/b=0,1$ and $10$ respectively.
The thin red line represents the $S$ magnetization with $d/b=1$ and
$b_{1}=-b_{2}=b$ $\left(  \kappa=1\right)  .$ The blue line is the mean
magnetization at each site $M_{\mathrm{tot}}\left(  t\right)  /3=2/3M_{0}.$}%
\label{Fig_3spin_isolated}%
\end{center}
\end{figure}

In the simplest case, where the Hartmann-Hahn condition is exactly fulfilled,
$\Delta=0$, the exact solution for the evolution of the observed magnetization
$M_{S^{z}}(t)$ is
\begin{equation}
M_{S^{z}}(t)=\mathrm{Tr}\left\{  \hat{S}^{z}\hat{\sigma}\left(  t\right)
\right\}  =M_{0}~f~\frac{\left[  1-\cos(\omega_{0}t)\right]  }{2},
\label{MSx_isolated}%
\end{equation}
where
\begin{equation}
\omega_{0}=\sqrt{\left(  \frac{\kappa}{4}\right)  ^{2}\left(  \frac{d}{\hbar
}\right)  ^{2}+2\left(  \frac{b}{\hbar}\right)  ^{2}}\text{\qquad and}\qquad
f=2\left(  \frac{b}{\hbar}\right)  ^{2}/\omega_{0}^{2} \label{freqCP}%
\end{equation}
with
\begin{equation}
\kappa=\left\{
\begin{array}
[c]{l}%
1\quad\text{if }b_{1}=-b_{2}=b\\
3\quad\text{if }b_{1}=b_{2}=b
\end{array}
\right.  . \label{kapa}%
\end{equation}
The natural frequency $\omega_{0}$ of the polarization transfer corresponds to
the transitions between the eigenstates mentioned above. Now, it is clear that
the symmetry of the system manifests directly in the frequency, where the
difference between the two situations is represented through the $\kappa$
parameter. So, the relative signs of the heteronuclear couplings lead to a
characteristic contribution of the homonuclear coupling being three times
bigger when $b_{1}=b_{2}$ than when $b_{1}=-b_{2}$.\emph{\ }The constant
\begin{equation}
M_{0}=\beta_{\mathrm{B}}\hbar\Omega_{0,I}/4
\end{equation}
corresponds to the initial magnetization of one $I$ spin. Eq.
(\ref{MSx_isolated}) shows that the magnetization of $S$ is attenuated by the
factor $f$, and it takes its maximum value $M_{0}$ when $d=0$, i.e. when there
is no $I_{1}-I_{2}$ interaction. The fact that the homonuclear interaction
decreases the transferred magnetization was already noticed in ref.
\cite{Pastawski96}. We can see that the constant term in eq.
(\ref{MSx_isolated}) is proportional to the differences in populations between
the relevant eigenstates of the system, while the time dependent term
corresponds to the coherences representing the transitions from $\left\{
|M_{I},M_{S}\rangle\right\}  $ to $\left\{  |M_{I}\pm1,M_{S}\mp1\rangle
\right\}  $\emph{.}
The magnetization in the spins $I_{1}$ and $I_{2}$ is given by%
\begin{equation}
\left.
\begin{array}
[c]{c}%
M_{I_{1}^{z}}\left(  t\right)  =\mathrm{Tr}\left\{  \hat{I}_{1}^{z}\hat
{\sigma}\left(  t\right)  \right\} \\
M_{I_{2}^{z}}\left(  t\right)  =\mathrm{Tr}\left\{  \hat{I}_{2}^{z}\hat
{\sigma}\left(  t\right)  \right\}
\end{array}
\right\}  =M_{I^{z}}\left(  t\right)  =M_{0}\left\{  1-\frac{1}{2}%
~f~\frac{\left[  1-\cos\left(  \omega_{0}t\right)  \right]  }{2}\right\}
=M_{0}-\frac{1}{2}M_{S^{z}}\left(  t\right)  \label{MIz_CH2_isolated}%
\end{equation}
and the total magnetization by%
\begin{equation}
M_{\mathrm{tot}}\left(  t\right)  =\mathrm{Tr}\left\{  \left(  \hat{S}%
^{z}+\hat{I}_{1}^{z}+\hat{I}_{2}^{z}\right)  \hat{\sigma}\left(  t\right)
\right\}  =2M_{0}.
\end{equation}
Thus, the total magnetization is given by the initial state and the mean
magnetization in each site is given by $M_{\mathrm{tot}}\left(  t\right)
/3=\frac{2}{3}M_{0}.$ Because of the symmetry of the system, each of the
proton transfers forth and back the same polarization to the carbon-$13$ that
is half magnitude of the magnetization observed at site $S.$ Figure
\ref{Fig_3spin_isolated}%
shows typical curves of the $M_{S^{z}}\left(  t\right)  $ and $M_{I^{z}%
}\left(  t\right)  $ magnetization. There, we can see curves for different
values of the ratio $d/b,$ i.e. different factors $f.$ (the higher the ratio
$d/b$, the lower the value of $f$). The red lines show the difference, with
the same values of $b$ and $d,$ between the evolution with $b_{1}=b_{2}$
(solid line) and $b_{1}=-b_{2}$ (dashed line). The mean magnetization at each
site is show by the blue line.

\subsection{A three-spin system coupled to a spin-bath}

In this section, we add to the three-spin system some interaction with other
spins using an extension of the model proposed by M\"{u}ller \emph{et al}.
\cite{MKBE74}, see section \S \ \ref{M_2-spin_spin_bath}. The model assumes
that the dipolar interactions of the $S$ spin with the $I$ spins are neglected
except for the coupling to $I_{1}$ and $I_{2}$.
The interaction of these particular spins with the bath or the infinite
reservoir of $I$ spins is considered in a phenomenological way. All kind of
spin-lattice relaxations are neglected. The system-environment
(SE)\ interaction Hamiltonian can be represented by
\begin{align}
\widehat{\mathcal{H}}_{\mathrm{SE}}  &  =\sum_{k=1,2}\left[  \alpha\hat{I}%
_{k}^{z}\hat{F}_{k}^{z}+\beta\left(  \hat{I}_{k}^{x}\hat{F}_{k}^{x}+\hat
{I}_{k}^{y}\hat{F}_{k}^{y}\right)  \right] \nonumber\\
&  =\sum_{k=1,2}\left[  \alpha\hat{I}_{k}^{0}\hat{F}_{k}^{0}+\tfrac{1}{2}%
\beta\left(  \hat{I}_{k}^{-}\hat{F}_{k}^{+}+\hat{I}_{k}^{+}\hat{F}_{k}%
^{-}\right)  \right]  \label{Hint}%
\end{align}
with
\begin{equation}
\hat{F}_{k}^{u}=\sum_{l>2}d_{kl}^{{}}I_{l}^{u},~~u=x,y,z
\end{equation}
and
\begin{equation}
\hat{F}_{k}^{\pm}=\left(  \hat{F}_{k}^{x}\pm\mathrm{i}\hat{F}_{k}^{y}\right)
\text{ \ \ \ \ \ \ \ \ \ \ \ \ }\hat{F}_{k}^{0}=\hat{F}_{k}^{z}%
\end{equation}
where the subscript $l$ corresponds to the spins within the bath. As in
section \S \ \ref{M_2-spin_spin_bath}, $\widehat{\mathcal{H}}_{\mathrm{SE}}$
is an Ising interaction if $\beta/\alpha=0$ and a XY, isotropic (Heisenberg)
or the truncated dipolar interaction\footnote{It corresponds to the fact that
we have neglected the non-secular terms with respect to the RF field.} if
$\alpha/\beta=0,1,-2$ respectively. Following the procedure of section
\S \ \ref{M_2-spin_spin_bath}, in the semi-classical theory by tracing on the
bath variables, $\hat{F}_{k}^{u}$ are treated as temporal functions $F_{k}%
^{u}\left(  t\right)  $ representing classical random processes. However, the
NMR experimental conditions justify a high temperature approximation, and
hence the semiclassical theory coincides with a quantum treatment
\cite{Abragam}.
Then, the random SE interaction Hamiltonian is written as
\begin{equation}
\widehat{\mathcal{H}}_{\mathrm{SE}}\left(  t\right)  =\sum_{k=1,2}\left[
\alpha F_{k}^{z}\left(  t\right)  \hat{I}_{k}^{z}+\tfrac{1}{2}\beta\left(
F_{k}^{x}\left(  t\right)  \hat{I}_{k}^{x}+F_{k}^{y}\left(  t\right)  \hat
{I}_{k}^{y}\right)  \right]  .
\end{equation}
\emph{\ }Here, the interaction of the system with the spins of the bath has
been taken into account. Any influence of the bath coming from others degrees
of freedom (rotations, translations, etc.) will manifest through this
interaction. These random processes satisfy
\begin{equation}
\overline{F_{k}^{u}\left(  t\right)  }=0
\end{equation}
where the bar denotes time average, and their correlation functions are
\begin{equation}
g_{k}^{\left(  u,v\right)  }\left(  \tau\right)  =\overline{F_{k}^{u}\left(
t\right)  F_{k}^{\nu\ast}\left(  t+\tau\right)  }.
\end{equation}
\emph{\ }The dynamics of the reduced density operator $\hat{\sigma}\left(
t\right)  $, following the usual treatment to second order approximation, is
\cite{Abragam,Blum,Ernst}
\begin{equation}
\frac{\mathrm{d}}{\mathrm{d}t}\hat{\sigma}\left(  t\right)  =-\frac
{\mathrm{i}}{\hbar}[\widehat{\mathcal{H}}_{\mathrm{S}},\hat{\sigma}\left(
t\right)  ]-\frac{1}{\hbar}\widehat{\widehat{\Gamma}}\left\{  \hat{\sigma
}\left(  t\right)  -\hat{\sigma}_{\infty}\right\}  . \label{master_3-spins}%
\end{equation}
The relaxation superoperator $\widehat{\widehat{\Gamma}}$ generated by
$\widehat{\mathcal{H}}_{\mathrm{SE}}\left(  t\right)  $, that accounts for the
dissipative interactions between the reduced spin system and the bath, drives
the density operator towards its equilibrium value $\hat{\sigma}_{\infty}$.

In the following, we assume\emph{\ }that the correlation times of the
fluctuations are extremely short compared with all the relevant transitions
rates between eigenstates of the Hamiltonian, i.e. frequencies of the order of
$\Sigma/2$ and $\omega_{0}$. In this extreme narrowing regime we obtain
\begin{equation}
\widehat{\widehat{\Gamma}}\left\{  \hat{\sigma}\right\}  =\frac{1}{2}\sum
_{k}\sum_{u,v}\xi_{u,v}\mathcal{J}_{k}^{\left(  u,v\right)  }\left(  0\right)
\left[  \hat{I}_{k}^{u},\left[  \hat{I}_{k}^{\nu},\hat{\sigma}\right]
\right]  ,\nonumber
\end{equation}
where
\begin{equation}
\mathcal{J}_{k}^{\left(  u,v\right)  }\left(  \omega\right)  =\int_{-\infty
}^{\infty}\frac{d\tau}{\hbar}g_{k}^{\left(  u,v\right)  }\left(  \tau\right)
\exp\left\{  -\mathrm{i}\omega\tau\right\}
\end{equation}
is the spectral density and
\begin{equation}
\xi_{u,v}=\left(  \alpha\delta_{u,z}+\beta\left(  \delta_{u,x}+\delta
_{u,y}\right)  \right)  \left(  \alpha\delta_{v,z}+\beta\left(  \delta
_{v,x}+\delta_{v,y}\right)  \right)  .
\end{equation}
If we suppose the spatial directions statistically independent, i.e.
\begin{equation}
g_{k}^{\left(  u,v\right)  }\left(  \tau\right)  =0\;\;\;\;\;\text{if}%
\;\;u\neq v,
\end{equation}
the superoperator $\widehat{\widehat{\Gamma}}$\emph{\ }can be written as
\begin{equation}
\widehat{\widehat{\Gamma}}\left\{  \hat{\sigma}\right\}  =\sum_{k=1,2}%
\alpha^{2}\mathcal{J}_{k}^{z}\left[  \hat{I}_{k}^{z},\left[  \hat{I}_{k}%
^{z},\hat{\sigma}\right]  \right]  +\beta^{2}\mathcal{J}_{k}^{x}\left[
\hat{I}_{k}^{x},\left[  \hat{I}_{k}^{x},\hat{\sigma}\right]  \right]
+\beta^{2}\mathcal{J}_{k}^{y}\left[  \hat{I}_{k}^{y},\left[  \hat{I}_{k}%
^{y},\hat{\sigma}\right]  \right]  ,
\end{equation}
where
\begin{equation}
\mathcal{J}_{k}^{u}=\frac{1}{2}\mathcal{J}_{k}^{\left(  u,u\right)  }\left(
0\right)  .
\end{equation}
Here, as in the two-spin case, the axial symmetry of $\widehat{\mathcal{H}%
}_{\mathrm{S}}$ around the $z$ axis leads to the impossibility to evaluate
separately $\mathcal{J}_{k}^{x}$ and $\mathcal{J}_{k}^{y}$, so they will
appear only as the averaged value
\begin{equation}
\mathcal{J}_{k}^{xy}=\left(  \mathcal{J}_{k}^{x}+\mathcal{J}_{k}^{y}\right)
/2.
\end{equation}
Taking into account the symmetry of our system $b_{1}=\pm b_{2}$, an extra
simplification can be done by
\begin{equation}
\mathcal{J}^{u}=\left(  \mathcal{J}_{1}^{u}+\mathcal{J}_{2}^{u}\right)  /2.
\end{equation}
Thus, we obtain
\begin{equation}
\widehat{\widehat{\Gamma}}\left\{  \hat{\sigma}\right\}  =\sum_{k=1,2}%
\alpha^{2}\mathcal{J}^{z}\left[  \hat{I}_{k}^{z},\left[  \hat{I}_{k}^{z}%
,\hat{\sigma}\right]  \right]  +\beta^{2}\mathcal{J}^{xy}\left(  \left[
\hat{I}_{k}^{x},\left[  \hat{I}_{k}^{x},\hat{\sigma}\right]  \right]  +\left[
\hat{I}_{k}^{y},\left[  \hat{I}_{k}^{y},\hat{\sigma}\right]  \right]  \right)
. \label{gamma}%
\end{equation}
Although we could absorb the constant $\alpha^{2}$ and $\beta^{2}$ in
$\mathcal{J}^{z}$ and $\mathcal{J}^{xy}$ respectively, we will keep it to
emphasize the different sources of the anisotropy in eq. (\ref{gamma}). As we
discuss in section \S \ \ref{M_2-spin_spin_bath}, the most usual approximation
is to consider $\mathcal{J}^{x}=\mathcal{J}^{y}=\mathcal{J}^{z}$ (identical
correlations in all the spatial directions) and $\alpha=\beta=1$ (isotropic
interaction Hamiltonian) \cite{MKBE74}, however, a better approximation
considers a dipolar interaction Hamiltonian, i.e. $\alpha=-2\beta=-2.$ As in
eq. (\ref{Gammas_spectral_density_2spin}), we define
\begin{equation}
\Gamma_{\mathrm{ZZ}}=\alpha^{2}\mathcal{J}^{z}\;\;\;\;\;\text{and}%
\;\;\;\;\;\Gamma_{\mathrm{XY}}=\beta^{2}\mathcal{J}^{xy}.
\end{equation}

\subsubsection{Neglecting non-secular terms\label{M_SI2_bath_secular}}

Following the formalism in Abragam and Ernst \emph{et al}. books
\cite{Abragam,Ernst} that was used in section \S \ \ref{Mark_2spin_secular},
we write the superoperator $\widehat{\widehat{\Gamma}}$ using the basis of
eigenstates of the Hamiltonian (\ref{Hsecular}). After neglecting the rapidly
oscillating non-secular terms with respect to the Hamiltonian, i.e.,
$\Gamma_{\mathrm{XY}},\Gamma_{\mathrm{ZZ}}\ll\left\vert b\right\vert
,\left\vert d\right\vert $, a block structure results. The first block couples
the populations and off-diagonal elements with $\Delta M=0$, Zero Quantum
Transitions (ZQT), of the density matrix. Each of the following blocks couples
one order $\Delta M\geq1$ of off-diagonal elements of the density matrix among
themselves. Because the Hamiltonian (\ref{Hsecular}) does not have degenerate
eigenenergies,\emph{\ }all the non-diagonal terms coupling the population
block with the ZQT block are non-secular and can be neglected. As the initial
condition (\ref{initial_condition}) does not contain coherences with $\Delta
M\geq1,$ we only need to study the evolution of the density operator into a
Liouville space restricted to populations and ZQT. When there are no
degenerate transitions, the secular ZQT block is diagonal. However, in our
case there are degenerate transitions between eigenstates within the sets with
$M=\pm1/2.$ Thus, some non-diagonal terms in the ZQT block cannot be
neglected.

In the final condition, the $SI_{2}$ system reaches the temperature of the $I$
spins reservoir as was described in the section \S \ \ref{M_2-spin_spin_bath}%
:
\begin{equation}
\hat{\sigma}_{\infty}=\frac{\hat{1}+\beta_{\mathrm{B}}\hbar\Omega_{0,I}\left(
\hat{S}^{z}+\hat{I}^{z}\right)  }{\mathrm{Tr}\left\{  \hat{1}\right\}  }.
\label{final}%
\end{equation}
It is easily seen that $\hat{\sigma}_{\infty}$ commutes with $\widehat
{\mathcal{H}}_{\mathrm{S}}$, not containing coherences with $\Delta M\geq1$.

By using the present formalism under the considered approximations, we will
solve eq. (\ref{master}) for the cases relevant to our liquid crystal study.

\paragraph{Isotropic system-environment interaction rate.}

\begin{figure}
[tbh]
\begin{center}
\includegraphics[
height=4.1883in,
width=5.6844in
]%
{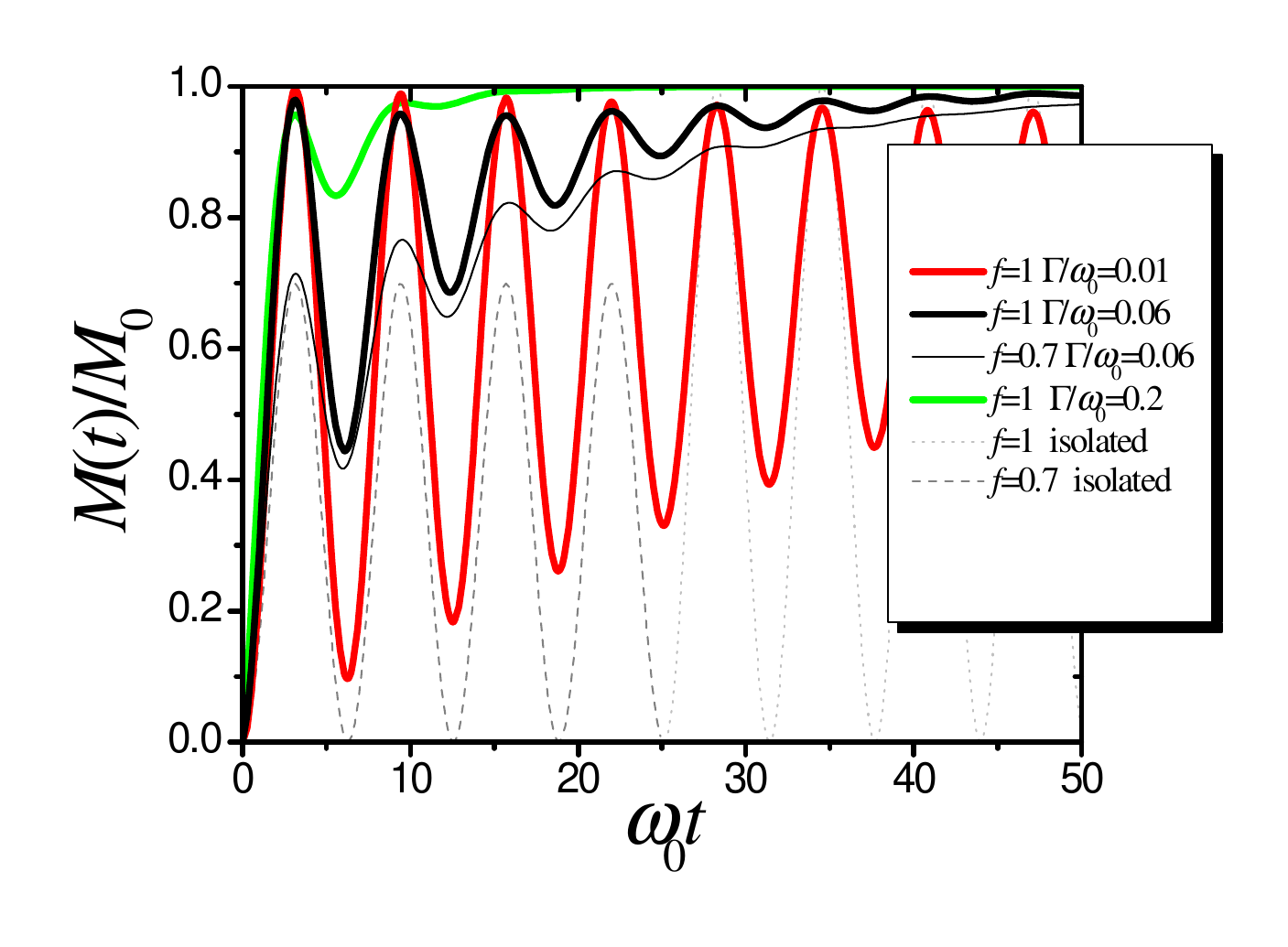}%
\caption[Typical curves of the magnetization of the $S$ spin in a three-spin
system coupled to a spin-bath under an isotropic system-environment
interaction.]{Typical curves of the magnetization of the $S$ spin in a
three-spin system coupled to a spin-bath under an isotropic SE interaction.
The thick red, black and green lines are the polarization evolution with $f=1$
and the ratios $\Gamma/\left(  \omega_{0}\hbar\right)  =0.01,$ $0.06$ and
$0.2$ respectively. The black thin line represents the magnetization for
$f=0.7$ and $\Gamma/\left(  \omega_{0}\hbar\right)  =0.06$ and the light and
dark gray dotted lines are the isolated dynamics for $f=1$ and $0.7$
respectively.}%
\label{Fig_3spin_isotropic}%
\end{center}
\end{figure}

\begin{figure}
[tbh]
\begin{center}
\includegraphics[
height=3.998in,
width=5.7968in
]%
{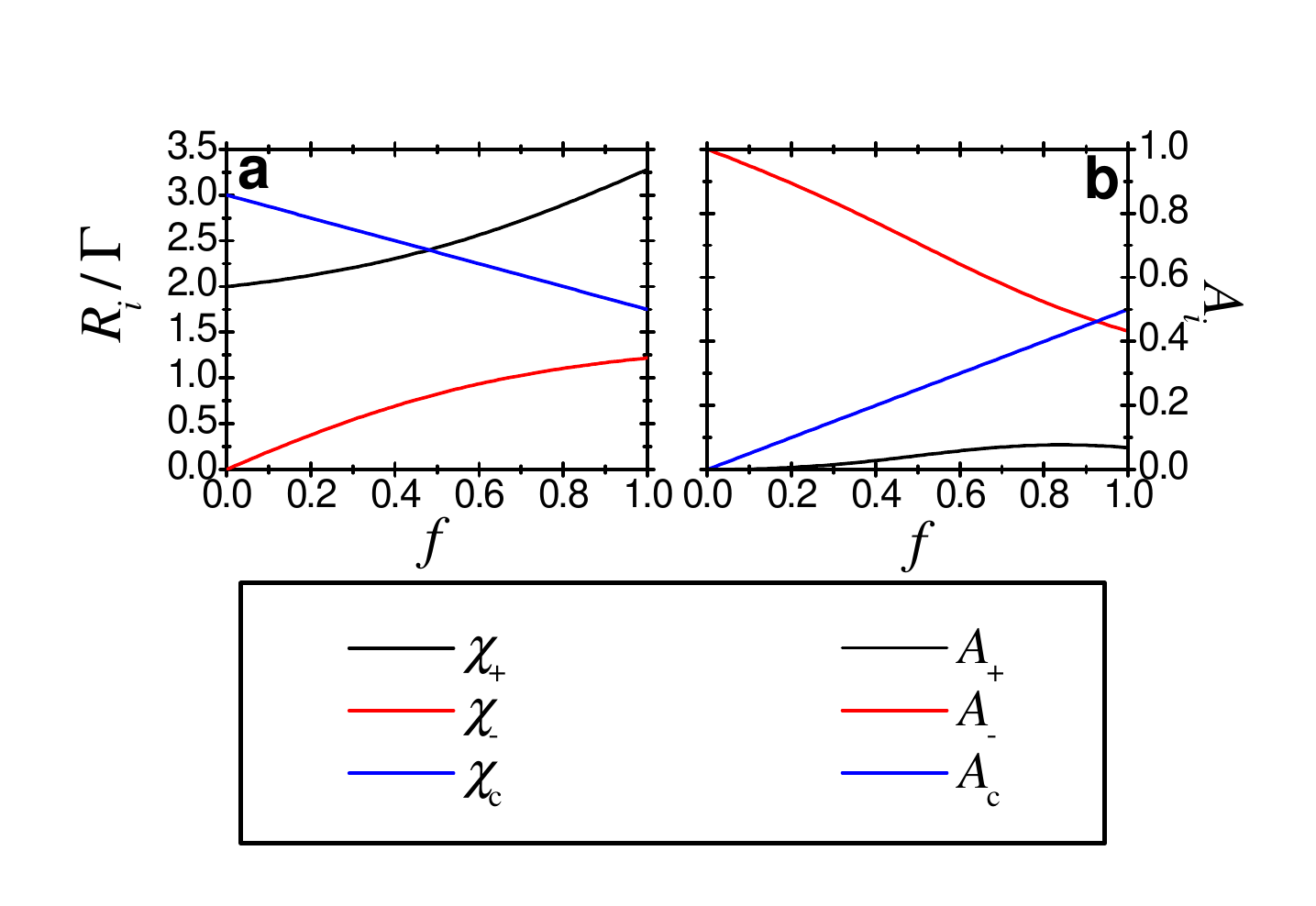}%
\caption[Coefficients $A_{i}$ and relaxation rates $R_{i}$ of the $S$
polarization expression for an isotropic system-environment interaction as a
function of the relation between homo and heteronuclear spin couplings $f$%
.]{Coefficients $A_{i}$ [panel a)] and relaxation rates $R_{i}$ [panel b)] of
the $S$ polarization expression (\ref{M_isotropic}) for an isotropic
system-environment interaction as a function of the relation between homo and
heteronuclear spin couplings $f$.}%
\label{Fig_SI2_iso_coef}%
\end{center}
\end{figure}

\begin{figure}
[tbh]
\begin{center}
\includegraphics[
height=2.6662in,
width=6.1211in
]%
{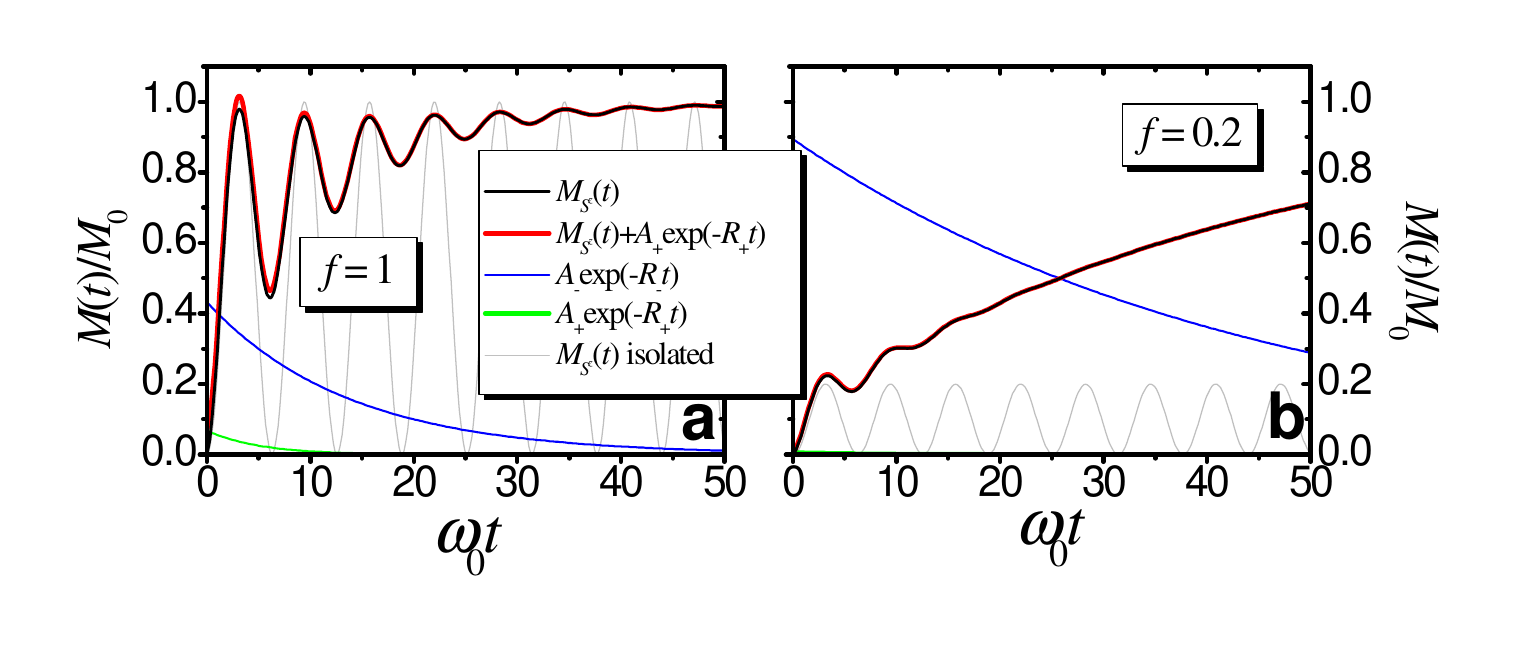}%
\caption[Polarization evolution of the $S$ magnetization of a $3$-spin system
coupled to a spin-bath.]{Polarization evolution of the $S$ magnetization
(black lines) of a $3$-spin system coupled to a spin-bath. The figure shows
comparisons between $M_{S^{z}}\left(  t\right)  $ and the aproximate solution
(red line) where was neglected one of the pure exponetial term. The pure
exponetial terms that are contained in the $M_{S^{z}}\left(  t\right)  $
expression are showed by the blue and green line. It is evident why the term
described by the green line could be neglected. The light gray line shows the
isolated evolution of the $S$ magnetization to observe that the first maximun
is essentially $fM_{0}$. Panel a) shows the evolution for $f=1$ and panel b)
for $f=0.2$ where in the last one the aproximate solution is more better.}%
\label{Fig_3spin_iso_comparacio}%
\end{center}
\end{figure}

Considering
\begin{equation}
\Gamma_{\mathrm{ZZ}}=\Gamma_{\mathrm{XY}}=\Gamma
\;\;\;~\ \;\;\;\;\ \;\;\;\ \ \ (\mathcal{J}^{z}=\mathcal{J}^{xy}%
\;\;\;\text{and\ \ \ }\alpha=\beta=1),
\end{equation}
the time evolution of the $S$ magnetization results
\begin{equation}
M_{S^{z}}\left(  t\right)  =M_{0}\left[  1-A_{+}e^{-R_{+}t}-A_{-}e^{-R_{-}%
t}-A_{\mathrm{c}}\cos\left(  \omega t\right)  \,e^{-R_{\mathrm{c}}t}\right]  ,
\label{M_isotropic}%
\end{equation}
$\allowbreak$where
\begin{align}
\omega &  =\omega_{0},\\
R_{\pm}  &  =\chi_{\pm}^{{}}\Gamma/\hbar,\\
R_{\mathrm{c}}  &  =\chi_{\mathrm{c}}^{{}}\Gamma/\hbar,
\end{align}
with
\begin{align}
\chi_{\pm}^{{}}  &  =1+\frac{5}{4}f\pm\sqrt{\left(  \frac{5}{4}f-1\right)
^{2}+f}\\
\chi_{\mathrm{c}}^{{}}  &  =3-\frac{5}{4}f
\end{align}
and
\begin{align}
A_{\pm}  &  =\frac{1}{2}\left\{  \left(  1-\frac{f}{2}\right)  \pm
\frac{\left[  \frac{5}{4}f\left(  1-\frac{f}{2}\right)  -1\right]  }%
{\sqrt{\left(  \frac{5}{4}f-1\right)  ^{2}+f}}\right\} \\
A_{\mathrm{c}}  &  =\frac{1}{2}f.
\end{align}
The figure \ref{Fig_3spin_isotropic}
shows typical curves of eq. (\ref{M_isotropic}) for different values of the
ratio $\Gamma/\left(  \omega_{0}\hbar\right)  .$ As we observed in the
$2$-spin case, we see that the oscillations are attenuated when the ratio is
bigger and the net transfer of polarization is faster. The black lines compare
two different curves with $f=1$ (thick line) and $f=0.7$ (thin line) for a
fixed value of $\Gamma/\left(  \omega_{0}\hbar\right)  .$ We observe that the
maximum of the oscillation is smaller as $f$ decreases but the final
magnetization is the same for both curves. The figure \ref{Fig_SI2_iso_coef}
shows the dependence of the coefficients $A_{i}$ and the relaxation rates
$R_{i}$ as a function of $f.$ Notice that $\chi_{\pm}^{{}},\chi_{\mathrm{c}%
}^{{}}\geq0.$ Using the initial condition $M_{S^{z}}\left(  0\right)  =0,$ it
is easy to see that the positive constants $A_{+},A_{-},$ $A_{\mathrm{c}}$
satisfy $1-A_{+}-A_{-}-A_{\mathrm{c}}=0$.\emph{\ }In general $A_{+}\ll A_{-}$
and $R_{+}>R_{-}$, so the first exponential term can be neglected as can be
observed in fig. \ref{Fig_3spin_iso_comparacio}.
This approximate solution is excellent for $f\ll1$, but even in the worst case
($f\sim1$), it differs about $7\%$ from the exact solution (see fig.
\ref{Fig_3spin_iso_comparacio}).

The first maximum in the magnetization $M_{S^{z}}\left(  t\right)  $ is
approximately $f\,M_{0}$ and the oscillation has frequency $\omega_{0}$ as can
be observed comparing with the isolated evolution in figs.
\ref{Fig_3spin_isotropic} and \ref{Fig_3spin_iso_comparacio}. The oscillations
have an amplitude $\frac{f}{2}e^{-R_{\mathrm{c}}t}$ that represents the
attenuation of the coherences of the $SI_{2}$ system mounted over
non-oscillatory terms. These terms take into account the effect of the bath,
not only by transferring magnetization but also breaking coherences and
leading to a quasi-equilibrium. This quasi-equilibrium state is given by
\begin{equation}
\hat{\sigma}^{\mathrm{qe}}=\exp\{-\widehat{\mathcal{H7}}_{\mathrm{S}%
}/(k_{\mathrm{B}}T_{\mathrm{qe}})\}/\mathrm{Tr}\{\exp\{-\widehat{\mathcal{H}%
}_{\mathrm{S}}/(k_{\mathrm{B}}T_{\mathrm{qe}})\}\}
\label{quasi-equilibrium_3spin_DM}%
\end{equation}
with
\begin{equation}
T_{\mathrm{qe}}=3/2(\Omega_{1,I}/\Omega_{0,I})T
\end{equation}
the temperature of the three-spin system.

In the particular case when $f=1,$ i.e. no $I_{1}$-$I_{2}$ interaction,
$A_{\pm}=\frac{1}{4}\left(  1\mp\frac{3}{\sqrt{17}}\right)  $, $A_{\mathrm{c}%
}=f/2,$ $R_{\pm}=\frac{1}{4}\left(  9\pm\sqrt{17}\right)  \Gamma/\hbar$ ,
$R_{\mathrm{c}}=7/4\Gamma/\hbar$ and $\omega_{0}=\sqrt{2}b/\hbar$, showing
that only under this condition the frequency given in ref. \cite{Pratima96} is
valid.
But even under this condition, our results show that the equation obtained by
M\"{u}ller \emph{et al.} for the $SI$ case cannot be directly applied to the
$SI_{2}$ system. In this last case the attenuation of the oscillations and the
transfer of polarization to the system is slightly faster than in the $SI$ case.

\paragraph{Anisotropic system-environment interaction rate.}

Considering
\begin{equation}
\Gamma_{\mathrm{ZZ}}\neq\Gamma_{\mathrm{XY}},
\end{equation}
we obtain
\begin{equation}
M_{S^{z}}\left(  t\right)  =M_{0}\left[  1-A_{1}e^{-R_{1}t}-A_{2}e^{-R_{2}%
t}-A_{3}e^{-R_{3}t}-A_{\mathrm{c}}\cos\left(  \omega t\right)
\,e^{-R_{\mathrm{c}}t}\right]  , \label{MSaniso}%
\end{equation}
where the $A_{i},$ $i=1,2,3$ are functions of $f$ and $\Gamma_{\mathrm{ZZ}%
}/\Gamma_{\mathrm{XY}}$ and
\begin{align}
\omega &  =\omega_{0},\\
R_{\mathrm{c}}  &  =\left(  2-f\right)  \Gamma_{\mathrm{XY}}/\hbar+\left(
1-\frac{1}{4}f\right)  \Gamma_{\mathrm{ZZ}}/\hbar,\\
A_{\mathrm{c}}  &  =\frac{1}{2}f.
\end{align}
The expressions for $A_{i}=A_{i}(f,\Gamma_{\mathrm{ZZ}}/\Gamma_{\mathrm{XY}})$
and $R_{i}=R_{i}(f,\Gamma_{\mathrm{ZZ}}/\hbar,\Gamma_{\mathrm{XY}}/\hbar)$ are
too long to be included here but they are available as supplementary
material.

The transfer of polarization from the bath to the system depends on the
non-oscillatory terms of eq. (\ref{MSaniso}). In the $\Gamma_{\mathrm{ZZ}%
}/\Gamma_{\mathrm{XY}}\geq1$ case, at long times ($R_{\mathrm{c}}t>>1$), only
one of the three exponential terms contributes.
In this regime, the transfer is essentially given by $\Gamma_{\mathrm{XY}}$,
although there is a slight dependence on $\Gamma_{\mathrm{ZZ}}$. This differs
from the $SI$ behavior where the polarization transfer from the bath depends
exclusively on $\Gamma_{\mathrm{XY}}$ (see section
\S \ \ref{M_2-spin_spin_bath}). This is a consequence of the fact that in the
$SI$ system the quasi-equilibrium $S^{z}$ polarization, $(1/2)M_{0}$ (the mean
magnetization at each site), coincides with the time averaged value of the
isolated system. As $\Gamma_{\mathrm{XY}}$ is associated to the flip-flop term
in the SE interaction Hamiltonian (\ref{Hint}), its role transferring
polarization can be easily interpreted. The effect of $\Gamma_{\mathrm{ZZ}}$
is more subtle, it can be associated to a process where the environment
\emph{observes}
the system breaking its coherences. This process that involves the operator
$\hat{F}^{z}\hat{I}^{z}$ in $\widehat{\mathcal{H}}_{\mathrm{SE}},$ which is a
like number operator, is discussed in chapter \ref{Marker_Spin_within_keldysh}%
.

\begin{figure}
[tbh]
\begin{center}
\includegraphics[
height=4.3898in,
width=6.0269in
]%
{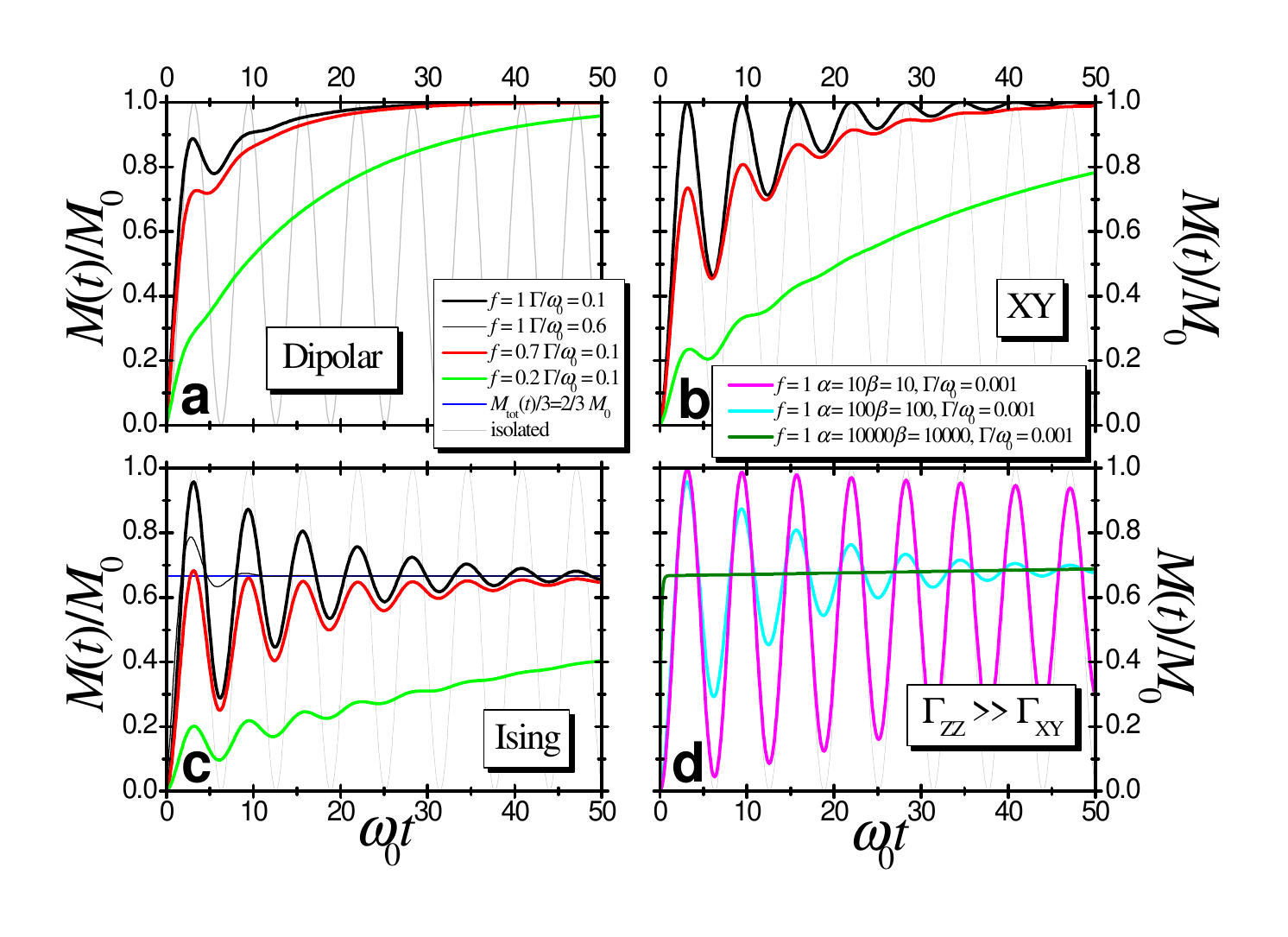}%
\caption[Typical curves of the $S$ polarization of the $SI_{2}$ system for
different SE interactions and values of $f$ (the relation between homo and
heteronuclear spin couplings).]{Typical curves of the $S$ polarization
(\ref{MSaniso}) for different SE interactions and values of $f$ (the relation
between $b$ and $d$). The dipolar, XY and isotropic SE are showed in panel a),
b) and c) respectively. In the limit $\Gamma_{\mathrm{ZZ}}\gg\Gamma
_{\mathrm{XY}}$ [highly anisotropic case, panel d)] it is possible to
distinguish two time regimes: one in which the system decoheres and reaches a
quasi-equilibrium, $[\widehat{\mathcal{H}}_{\mathrm{S}},\hat{\sigma
}^{\mathrm{qe}}]=0$, and other in which polarization transfer from the bath is
completed with a rate $\Gamma_{\mathrm{XY}}/\hbar.$}%
\label{Fig_3spin_ani_polarizati}%
\end{center}
\end{figure}

Figure \ref{Fig_3spin_ani_polarizati}
shows typical curves of the $S$ polarization (\ref{MSaniso}) for different SE
interactions and values of $f$ (the relation between $b$ and $d$). In the
limit $\Gamma_{\mathrm{ZZ}}\gg\Gamma_{\mathrm{XY}}$ [highly anisotropic case,
panel d)] it is possible to distinguish two time regimes: one in which the
system decoheres and reaches a quasi-equilibrium, $[\widehat{\mathcal{H}%
}_{\mathrm{S}},\hat{\sigma}^{\mathrm{qe}}]=0$ \cite{Sakellariou98},
characterized by $\Gamma_{\mathrm{ZZ}}$ and other in which polarization
transfer from the bath is completed with a rate $\Gamma_{\mathrm{XY}}/\hbar.$
In this situation, it is possible to see that $R_{\mathrm{c}}$ is proportional
to $\Gamma_{\mathrm{ZZ}}$, as it occurs with the $R_{i}$ corresponding to the
fastest exponential terms. The dependence of the non-oscillatory terms on
$\Gamma_{\mathrm{ZZ}}$ observed in the $SI_{2}$ system can be assigned to the
fact that the quasi-equilibrium carbon polarization $(2/3)M_{0}$ (the mean
magnetization at each site) does not coincide with its time averaged value
$(f/2)M_{0}$ in the isolated three-spin system [eq. (\ref{MSx_isolated})].

Equations (\ref{M_isotropic}) and (\ref{MSaniso}) will be used to fit the
experimental data in order to extract the relevant parameters of our system.

\subsection{Many-spin quantum dynamics during Cross-Polarization in
8CB\label{Mark_3spin_experimental_8CB}}

\begin{figure}
[th]
\begin{center}
\includegraphics[
height=1.6354in,
width=5.1655in
]%
{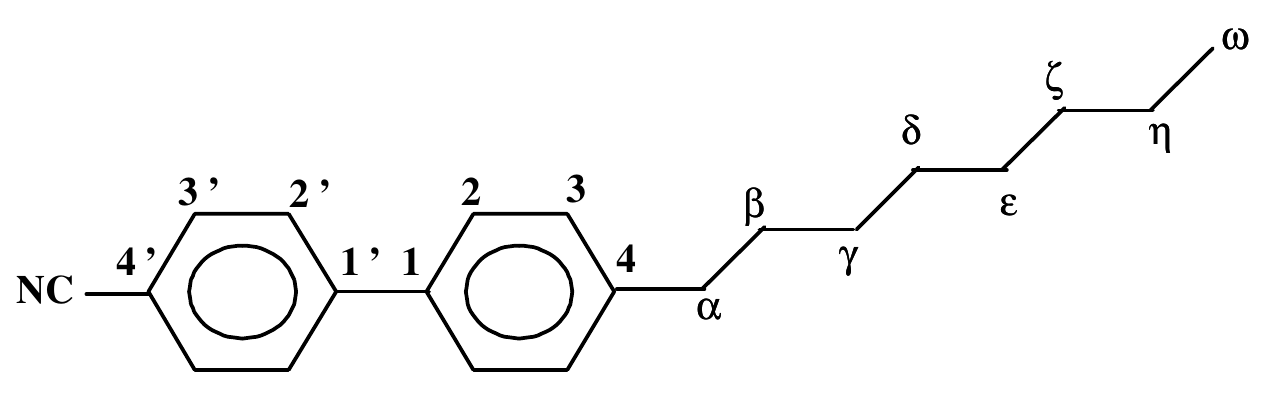}%
\caption[Chemical structure of $4$-n-octyl-$4$'-cyanobiphenyl ($8$%
CB).]{Chemical structure of $4$-n-octyl-$4$'-cyanobiphenyl ($8$CB) showing the
labels used in the $^{13}$C-NMR spectra.}%
\label{Fig_8CB}%
\end{center}
\end{figure}

Nuclear Magnetic Resonance experiments were carried out by people of our group
in $4$-n-octyl-$4$'-cyanobiphenyl, also called $8$CB (see fig. \ref{Fig_8CB})
obtained from Sigma (Chemical, Co) and used without further purification. This
system presents the mesophases smectic A (SA) and nematic (N),
with transition temperatures at $294.5%
\operatorname{K}%
$ (K-SA), $306.5%
\operatorname{K}%
$ (SA-N) and $313.5%
\operatorname{K}%
$ (N-I).

$^{1}$H-$^{13}$C cross-polarization measurements as a function of contact time
$t_{\mathrm{c}}$ were performed in the smectic and nematic mesophases. In all
the cases, the CP sequence was performed in static conditions and combined
with the sequence SPINAL-$64$
to perform an efficient proton decoupling during acquisition without
appreciable heating of the sample \cite{Fung00}.

In the smectic phase, standard CP experiments were performed at $300%
\operatorname{K}%
$ in a Bruker MSL-$300$. The acquisition time was $92%
\operatorname{ms}%
$, with $60%
\operatorname{ms}%
$ for decoupling and a recycling time of $15%
\operatorname{s}%
$. The Hartmann-Hahn (H-H) condition \cite{Hartmann62,Slichter} was set with
an RF field amplitude for carbons corresponding to $\Omega_{1,S}/2\pi=67.7%
\operatorname{kHz}%
$. During the experiment the contact time was varied in the range $2%
\operatorname{\mu s}%
$ $<t_{\mathrm{c}}<5%
\operatorname{ms}%
$.

\begin{figure}
[tbh]
\begin{center}
\includegraphics[
height=4.6812in,
width=4.3457in
]%
{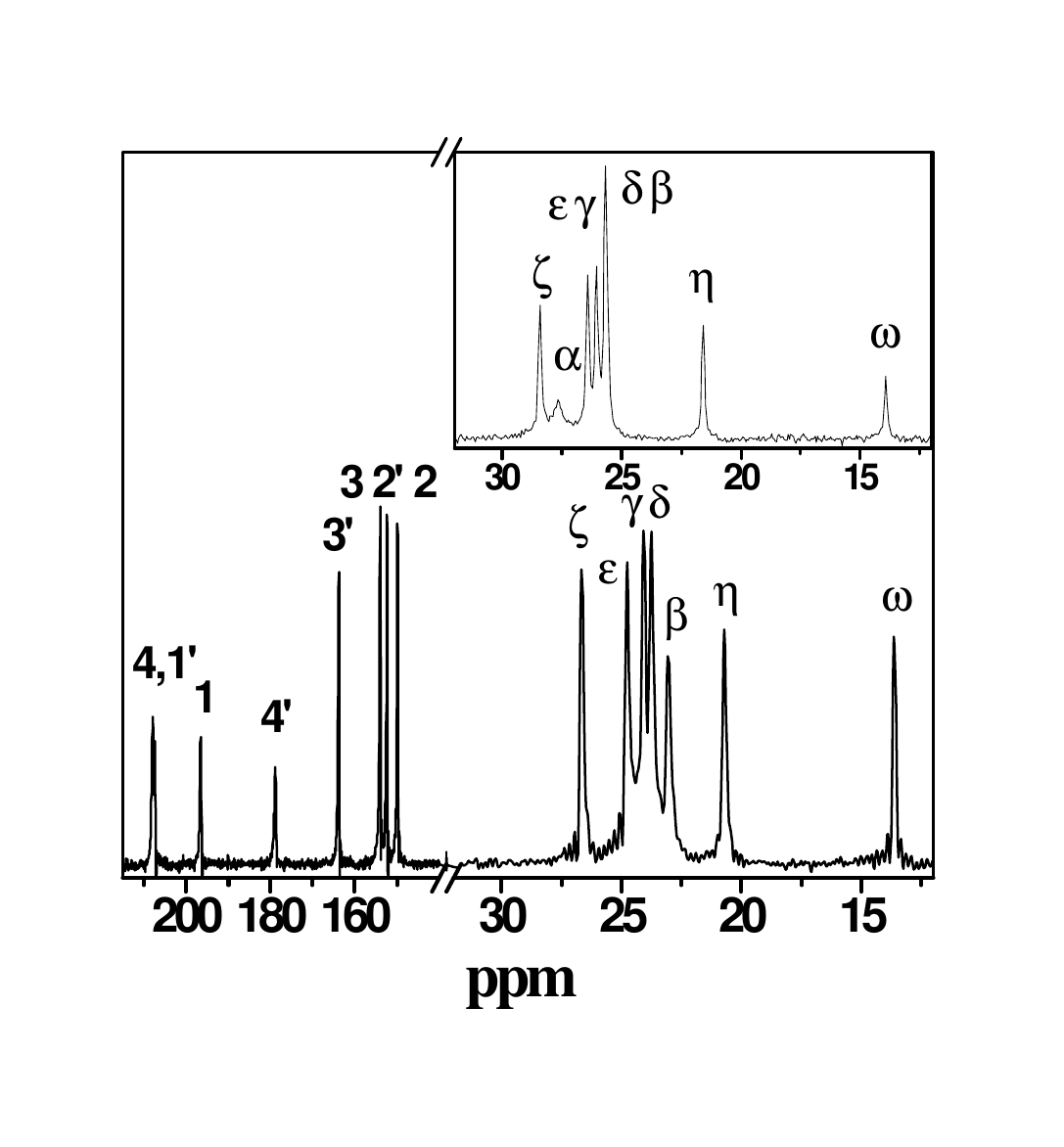}%
\caption[$^{13}$C-NMR spectra of $8$CB in the smectic and nematic
mesophases.]{$^{13}$C-NMR spectra of $8$CB in the smectic mesophase at
$300\operatorname{K}$. In the inset we see the aliphatic part of the spectrum
corresponding to $311.5{}\operatorname{K}$ (nematics), which is the only part
that changes at the working temperatures. The assignments for the aliphatic
part of the spectrum have been done in previous works. Labels refer to fig.
\ref{Fig_8CB}.}%
\label{Fof-8CB2}%
\end{center}
\end{figure}

In the nematic phase at $311.5%
\operatorname{K}%
$ two types of CP experiments were performed in a Bruker AVANCE DSX-$500$. The
first was a standard CP with protons on-resonance. The second was a CP
experiment with irradiation for protons in the Lee-Goldburg (LG) condition,
i.e. the off-resonance for protons was set to have an effective field at the
magic-angle with the static field $H_{0}$. The acquisition and decoupling
times were $74%
\operatorname{ms}%
$. In the standard CP, the H-H condition was set with $\Omega_{1,S}/2\pi=60.3%
\operatorname{kHz}%
$ while in the Lee-Goldburg $\Omega_{1,S}/2\pi=74%
\operatorname{kHz}%
$. In both sets of experiments the H-H condition was optimized for $C(\gamma)$
(see figs. \ref{Fig_8CB} and \ref{Fof-8CB2}), and the contact time
$t_{\mathrm{c}}$ varied up to $2%
\operatorname{ms}%
$.%

In all these experiments the temperatures were calibrated using the N-I
temperature transition. Also a $^{13}$C-NMR spectrum of $8$CB in the isotropic
phase (at $320%
\operatorname{K}%
$) was taken as reference using a single pulse sequence with $^{1}$H
decoupling (direct $^{13}$C polarization).

\subsubsection{Comparison between experiments and theoretical results}

The $^{13}$C-NMR spectrum of $8$CB in smectic phase can be seen in fig.
\ref{Fof-8CB2}. In the inset of this figure, the alkyl part of the spectrum at
a temperature corresponding to the nematic mesophase is displayed. The
aromatic part of the spectrum keeps the same features up to the
nematic-isotropic transition temperature. The position of C($\omega$), methyl
group, which in the isotropic phase is at $14.1$ ppm has been taken as
reference because, as experimentally observed, it does not vary at the working
temperatures due to its high mobility. The aromatic part of the carbon
spectrum for $8$CB is similar to those of other members of the $n$CB series,
so we consider the up-dated assignments reported for $5$CB in previous works
\cite{Fung00}. A detailed temperature dependence of the $^{13}$C chemical
shifts in $8$CB as well as their complete assignments have been reported
previously \cite{Chattah02}. The assignments for the alkyl carbons are
supported by the segmental order parameters obtained from experimental CP
frequencies (see below), and deuterium NMR experiments \cite{Dong,mol85}.

In order to analyze the experimental data, it is necessary to correlate the
geometry/symmetry of the molecular interactions with the cases presented in
the theoretical section associated to different values of $\kappa,$ eq.
(\ref{kapa}). For each methylene carbon in the aliphatic chain, the geometry
of the molecule and the rapid rotations around the C-C bonds, which lead to
the trans-gauche isomerizations, allows us to take a single averaged value for
both heteronuclear couplings i.e. $b_{1}\approx$ $b_{2}$. Then, for carbons
C($\alpha$) to C($\eta$) in the aliphatic chain the relation of signs of the
heteronuclear couplings corresponds to the case where $\kappa=3$. It is also
possible to see from simulations done with a similar molecule ($4$%
-n-pentyl-$4$'-cyanobiphenyl or $5$CB) and from geometrical considerations
that the homonuclear dipolar interaction $d$ between protons belonging to the
same methylene in general, cannot be neglected \cite{Stevensson01}.
For each non-quaternary carbon in the phenyls rings C($3^{\prime}$),
C($2^{\prime}$), C($2$) and C($3$), we have one directly bonded proton
interacting with a dipolar coupling $b_{1}$; however, a careful analysis
indicates that neither the $^{1}$H-$^{1}$H interaction $d$ nor the coupling
between the $^{13}$C and the nearest non-bonded $^{1}$H, $b_{2},$ can be
neglected. Considering the rigidity of the phenyl ring and the orientation of
each internuclear vector with respect to the external field, we see that
$b_{1}\approx-b_{2}.$ Then, each non-quaternary aromatic carbon is part of a
three-spin system, where both heteronuclear couplings can be considered having
an \emph{averaged} magnitude $b=\sqrt{\left(  b_{1}^{2}+b_{2}^{2}\right)  /2}$
and different relative signs, leading to $\kappa=1$.

\begin{figure}
[tbh]
\begin{center}
\includegraphics[
height=4.5602in,
width=6.0208in
]%
{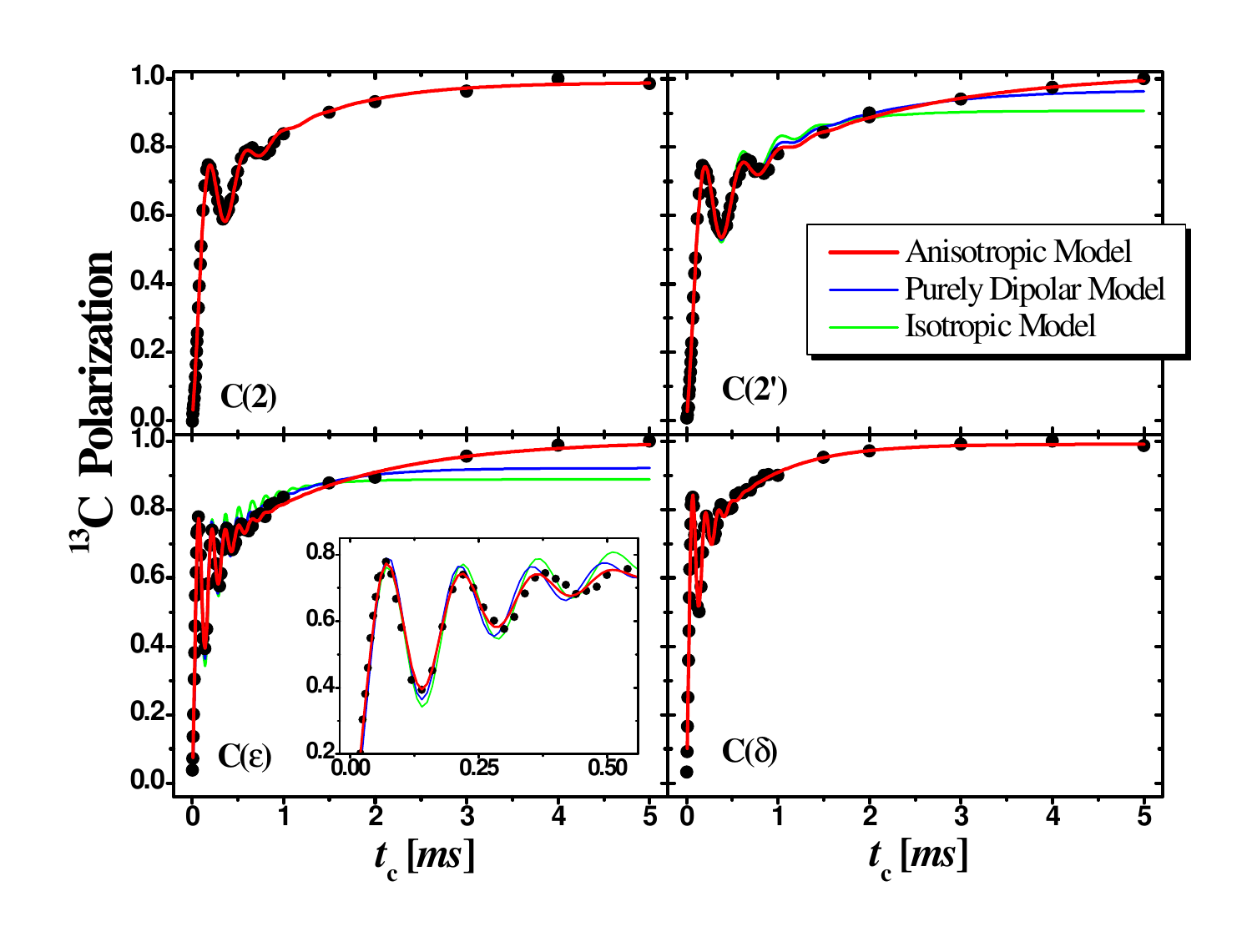}%
\caption[$^{13}$C polarization as a function of contact time $t_{\mathrm{c}}$
for aromatic and aliphatic carbons of $8$CB in a standard CP experiment at
$300{}\operatorname{K}$ (smectic phase).]{$^{13}$C polarization as a function
of contact time $t_{\mathrm{c}}$ for aromatic and aliphatic carbons in a
standard CP experiment at $300{}\operatorname{K}$ (smectic phase). Fittings of
the experimental data to the anisotropic, isotropic and purely dipolar models
described in the text, are displayed.}%
\label{Fog-CP8CB}%
\end{center}
\end{figure}

Firstly, let's consider the experimental results corresponding to the smectic
phase. In this case, typical oscillations and relaxation of the $^{13}$C
polarization vs. contact time $t_{\mathrm{c}}$ are shown in fig.
\ref{Fog-CP8CB}. We remark that neither the quaternary carbons nor C($\omega$)
show oscillations in the CP experiment. In the last case, the reason is the
high mobility of the methyl group that averages to zero the effective
carbon-proton interaction. In all these experiments, the $^{13}$C polarization
has essentially reached its asymptotic value at $5%
\operatorname{ms}%
$. The absence of a decay in these curves allows us to neglect spin-lattice
relaxation in the rotating frame ($T_{1\rho}$) in the time regime analyzed. We
also note that the CP frequencies corresponding to methylene groups are higher
than those of the aromatic cores. This is due to a particularly unfavorable
angle ($\sim60^{\circ}$) between the internuclear carbon-proton vector and the
external magnetic field in the case of the phenyl rings. Besides, the
contribution of the homonuclear coupling to the frequency is much smaller when
$\kappa=1$. In each dynamical curve shown in fig. \ref{Fog-CP8CB}, the $^{13}%
$C polarization at the first maximum is lower than its asymptotic value. As we
have seen in the theoretical section, this fact indicates that the homonuclear
coupling is not negligible and it will allow us to evaluate it.%

We have fitted the experimental CP data to the equations derived for the
isotropic (eq. (\ref{M_isotropic})) and anisotropic models (eq. (\ref{MSaniso}%
)) presented in section \S \ \ref{M_SI2_bath_secular}. In the last case, we
have distinguished a purely dipolar anisotropy ($\alpha=-2\beta=-2$ and
$\mathcal{J}^{xy}=\mathcal{J}^{z},$ i.e. $\Gamma_{\mathrm{ZZ}}/\Gamma
_{\mathrm{XY}}=4$) from the most general case. For the smectic phase, it is
seen in fig. \ref{Fog-CP8CB} that the isotropic model separates from the
experimental points after approximately the time of the second maximum,
fitting very poorly the asymptotic behavior. The dipolar model constitutes an
improvement over the isotropic one, without adding extra free parameters.
However, a much better fitting is obtained using the anisotropic model which
follows very closely the first oscillations of the magnetization and it is the
best suited in the asymptotic regime.

In the case of the standard CP\ experiments performed in the nematic phase up
to $2%
\operatorname{ms}%
$ (not shown in the figure), the behavior is already well fitted with the
purely dipolar model, not showing any appreciable improvement by the use of
the completely anisotropic model.

Taking into account the fitting parameters corresponding to each experiment,
we can obtain the effective dipolar couplings and the relaxation constants for
each carbon showing oscillations during CP. In tables
\ref{Tab_8CB_dip_coupling_smectic} and \ref{Tab_dip_coupling_nematic}, we show
the heteronuclear and homonuclear dipolar couplings at different temperatures.
The values for the heteronuclear couplings are comparables with those obtained
for $8$CB in ref. \cite{Fung86}, where a different experimental technique was
applied.
\begin{table}[tbp] \centering
$%
\begin{tabular}
[c]{|c|c|c|}\hline
Carbons & $%
\begin{array}
[c]{c}%
\frac{1}{2\pi}b/\hbar\\%
\operatorname{kHz}%
\end{array}
$ & $%
\begin{array}
[c]{c}%
\frac{1}{2\pi}d/\hbar\\%
\operatorname{kHz}%
\end{array}
$\\\hline
C($3^{\prime}$) & $1.30\pm0.03$ & $4.9\pm0.3$\\\hline
C($2^{\prime}$) & $1.49\pm0.04$ & $4.8\pm0.3$\\\hline
C($2$) & $1.58\pm0.06$ & $4.9\pm0.3$\\\hline
C($3$) & $1.45\pm0.04$ & $5.3\pm0.4$\\\hline
C($\beta$) & $4.6\pm0.1$ & $4.5\pm0.6$\\\hline
C($\gamma$) & $4.8\pm0.1$ & $5.9\pm0.7$\\\hline
C($\delta$) & $4.7\pm0.3$ & $2.9\pm0.3$\\\hline
C($\varepsilon$) & $4.3\pm0.2$ & $4.0\pm0.5$\\\hline
C($\varsigma$) & $3.1\pm0.1$ & $3.7\pm0.4$\\\hline
C($\eta$) & $3.2\pm0.1$ & $2.2\pm0.3$\\\hline
\end{tabular}
\ $%
\caption{Effective heteronuclear and homonuclear dipolar couplings $b$ and $d$ obtained by fitting the data of the standard CP experiment performed at 300K  (smectic phase) to the anisotropic model. The signal corresponding to C($\alpha$) does not appear in the smectic spectra. The errors have been assigned taking into account eqs. (\ref{despeje}) for $b$, $d$ in terms of the fitted parameters.\label{Tab_8CB_dip_coupling_smectic}}%
\end{table}%
\begin{table}[tbp] \centering
$%
\begin{tabular}
[c]{|c|c|c|c|c|}\hline
Carbons & $%
\begin{array}
[c]{c}%
\text{Standard CP}\\
\frac{1}{2\pi}b/\hbar\\%
\operatorname{kHz}%
\end{array}
$ & $%
\begin{array}
[c]{c}%
\text{LG-CP}\\
\frac{1}{2\pi}b/\hbar\\%
\operatorname{kHz}%
\end{array}
$ & $%
\begin{array}
[c]{c}%
\text{Standard CP}\\
\frac{1}{2\pi}d/\hbar\\%
\operatorname{kHz}%
\end{array}
$ & $%
\begin{array}
[c]{c}%
\text{LG-CP}\\
\frac{1}{2\pi}d/\hbar\\%
\operatorname{kHz}%
\end{array}
$\\\hline
C($\alpha$) & $4.0\pm0.2$ & $4.18\pm0.04$ & $2.7\pm0.5$ & $2.9\pm0.6$\\\hline
C($\beta,\delta$) & $2.7\pm0.1$ & $2.95\pm0.03$ & $3.5\pm0.6$ & $2.8\pm
0.6$\\\hline
C($\gamma$) & $3.0\pm0.2$ & $3.34\pm0.03$ & $3.5\pm0.7$ & $2.5\pm0.5$\\\hline
C($\varepsilon$) & $2.8\pm0.1$ & $2.95\pm0.03$ & $3.1\pm0.5$ & $2.8\pm
0.6$\\\hline
C($\varsigma$) & $2.5\pm0.1$ & $2.29\pm0.03$ & $2.4\pm0.5$ & $2.8\pm
0.5$\\\hline
C($\eta$) & $2.4\pm0.1$ & $2.17\pm0.03$ & $1.5\pm0.3$ & $2.3\pm0.4$\\\hline
\end{tabular}
\ $%
\caption{Effective heteronuclear and homonuclear dipolar couplings $b$ and $d$ for the aliphatic carbons obtained from the experiments performed at 311.5K  (nematic phase). In the standard CP experiment the values are obtained by fitting the data to the purely dipolar model (see text). In LG-CP the values of $b$ are obtained directly from eq. (\ref{LG}) while the values of $d$ requires a combination of the standard CP with the LG-CP frequencies (eq. (\ref{resta})). In this mesophase the signal of C($\beta$) merges to the signal of C($\delta$), so the same value of coupling has been assigned for both carbons.\label{Tab_dip_coupling_nematic}}%
\end{table}%

It is important to emphasize here, that the frequency of the oscillations
$\omega$ is a quite independent and robust parameter, leading to values which
vary less than $4\%$ using different models. This allows one to determine the
heteronuclear coupling $b$ with very small error. The homonuclear coupling,
however, is much more sensitive to the relation between the oscillatory and
the asymptotic regimes and so, more dependent on the model and the extent of
the experimental data. To see clearly this fact, we notice that the frequency
and the attenuation factor $f$ given in eq. (\ref{freqCP}) are the fitting
parameters that yield$\ b$ and $d$ regardless of the model. The dependence
between these parameters and the couplings are
\begin{equation}
\frac{1}{2\pi}\frac{b}{\hbar}=\frac{1}{\sqrt{2}}\frac{\omega}{2\pi}\sqrt
{f}\text{ \qquad and \qquad}\frac{1}{2\pi}\frac{d}{\hbar}=\frac{4}{\kappa
}\frac{\omega}{2\pi}\sqrt{\left(  1-f\right)  }. \label{despeje}%
\end{equation}
As in our cases, $f$ is always bigger than $1/2$, the error in $f$ (approx.
$2\%$ for the experiment in smectic and $5\%$ for the experiment in nematics)
affects less to $b$ than $d$. Now, taking into account that the CP frequency
$\omega$ is the best parameter, fitted with an error lower than $2\%$, we can
assign the error of the other parameters of interest. Then, the coupling $b$
is obtained with a relative error between $3\%$ and $5\%$ while the relative
error of $d$ reaches in some cases a value of about $20\%$.

\begin{figure}
[tbh]
\begin{center}
\includegraphics[
height=4.3241in,
width=3.6383in
]%
{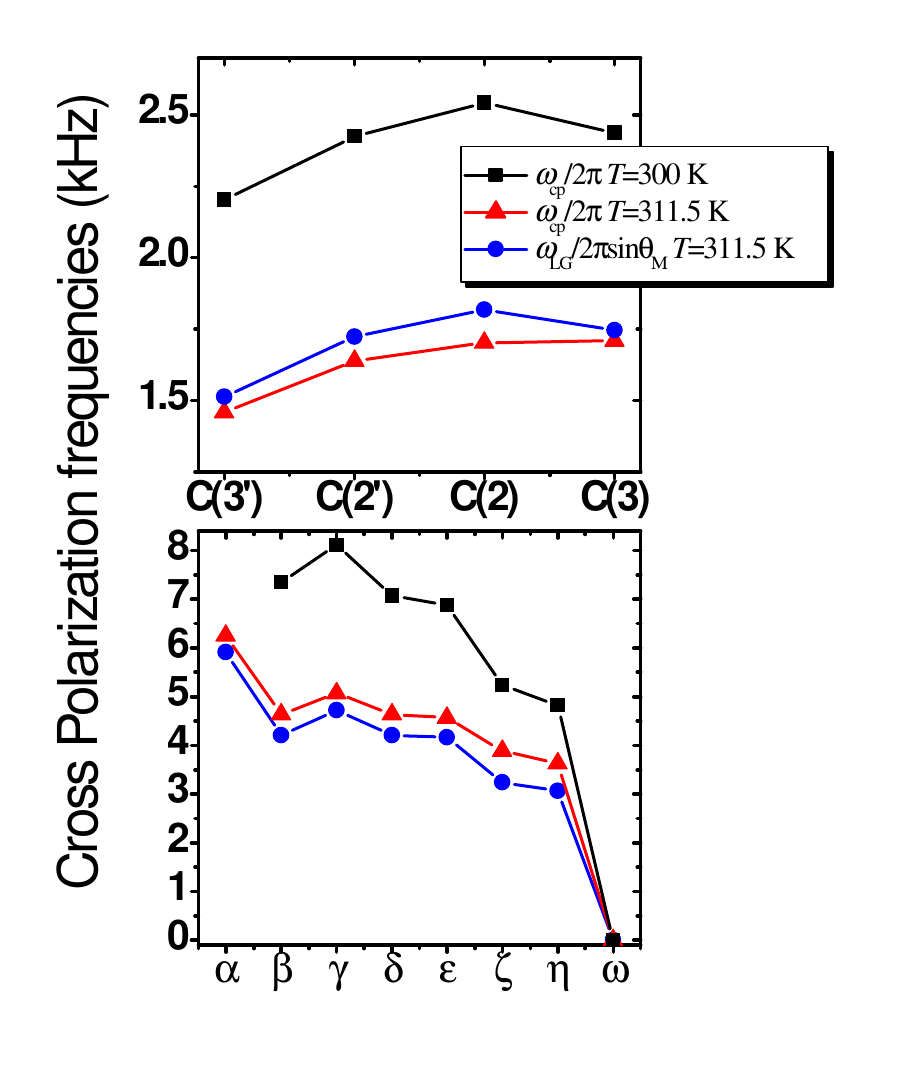}%
\caption[Cross-Polarization (CP) frequencies obtained from the Lee-Goldburg
and standard CP experiments for all the non-quaternary carbons in the $8$CB
molecule.]{Cross-Polarization frequencies obtained from the LG and standard CP
experiments for all the non-quaternary carbons in the molecule. Note that the
signal corresponding to C($\alpha$) does not appear at $300{}\operatorname{K}%
$; also note that the signal of C($\beta$) merges with that of C($\delta$) at
$311.5\operatorname{K}$ so the same value of frequency has been assigned to
both carbons.}%
\label{Fig_w_8CB_std_and_LG}%
\end{center}
\end{figure}

Fig. \ref{Fig_w_8CB_std_and_LG} displays the tendency of the CP frequency
obtained for the different experiments. In particular, for the aliphatic
carbons, the expected zig-zag pattern is observed \cite{Dong,mol85}. For the
aromatic part of the molecule we have $\omega(3^{\prime})\lessapprox
\omega(2^{\prime})\approx\omega(2)\approx\omega(3)$, this is expected because
C($3\prime$), being the closest to the cyano group forms a bigger C-C-H angle,
giving rise to a smaller dipolar interaction \cite{Caldarelli96}. On the other
hand, when increasing temperature, i.e. going from smectic to nematic, a
further averaging of the dipolar interactions occurs, manifested in the
decrease of the CP frequencies. For comparison, fig.
\ref{Fig_w_8CB_std_and_LG} shows the results of the CP experiment at the
Lee-Goldburg\ condition, performed in the nematic phase at the same
temperature. In the LG experiment the CP frequency ($\omega_{\mathrm{LG}}$) is
only related to the heteronuclear coupling \cite{Slichter}
because the homonuclear contribution has been quenched, then
\begin{equation}
\omega_{\mathrm{LG}}=\sqrt{2\left(  b/\hbar\right)  ^{2}(\sin\theta
_{\mathrm{m}})^{2}}, \label{LG}%
\end{equation}
where $\theta_{\mathrm{m}}=54.7^{0}$ is the magic angle \cite{Slichter}.
This angular factor comes from the projection of the RF field into the
direction of the effective field for protons irradiated off-resonance at the
LG condition \cite{Slichter}. Then, considering eq. (\ref{freqCP}), we expect
the relation
\begin{equation}
\omega>\omega_{\mathrm{LG}}/\sin\theta_{\mathrm{m}}. \label{ineq}%
\end{equation}

As shown in fig. \ref{Fig_w_8CB_std_and_LG},
this relation is not satisfied for aromatic carbons. Considering that
C($\gamma$) is exactly on-resonance and that the H-H condition was optimized
for that carbon, the disagreement can be attributed to a non negligible
mismatch of the Hartmann-Hahn condition for carbons in the aromatic part of
the spectrum. Taking into account a mismatch $\Delta$ for the aromatic
carbons, the CP frequency for the LG experiment becomes,
\[
\omega_{\mathrm{LG}}=\sqrt{2\left(  b/\hbar\right)  ^{2}(\sin\theta
_{\mathrm{m}})^{2}+\Delta^{2}}.
\]
In the standard CP experiment, two frequencies appear, where the observable
one is
\begin{equation}
\omega=\frac{\omega^{+}+\omega^{-}}{2}%
\end{equation}
with
\[
\omega^{\pm}=\sqrt{\frac{\left[  \left(  d/\hbar\right)  \pm4\Delta\right]
^{2}}{16}+2\left(  b/\hbar\right)  ^{2}}.
\]
The other modulating frequency is too low to be observed in the presence of
relaxation. Although it is difficult to quantify exactly the magnitude of
$\Delta$, we can see that the effect of the mismatch is greater for
$\omega_{\mathrm{LG}}$ than for $\omega$ where there is a partial compensation
between the two contributing frequencies $\omega^{\pm}$. This explains why
$\omega_{\mathrm{LG}}\gtrapprox\omega$ for the carbons in the aromatic part,
in contrast with eq. (\ref{ineq}).

For carbons irradiated on-resonance (aliphatic part), the values of the fitted
parameters $b$ and $d$ obtained from standard CP experiments can be compared
with the parameters obtained from the LG-CP. In the later case, the parameter
of interest is $\omega_{\mathrm{LG}}.$ Then, $b$ is obtained in a direct way
from expression (\ref{LG}) and $d$ can be obtained comparing $\omega
_{\mathrm{LG}}$ with the corresponding value of $\omega$ at the same
temperature. Using both experiments, $d$ is calculated from
\begin{equation}
\frac{d}{\hbar}=\frac{4}{\kappa}\sqrt{\omega^{2}-\frac{\omega_{\mathrm{LG}%
}^{2}}{(\sin\theta_{\mathrm{m}})^{2}}}. \label{resta}%
\end{equation}
The values of $d$ calculated in this way with $\kappa=1$ (aliphatic carbons)
can be compared with those obtained by fitting eq. (\ref{MSaniso}) to the
standard CP data, i.e. coming from a single experiment.

\begin{figure}
[tbh]
\begin{center}
\includegraphics[
height=4.2056in,
width=2.7233in
]%
{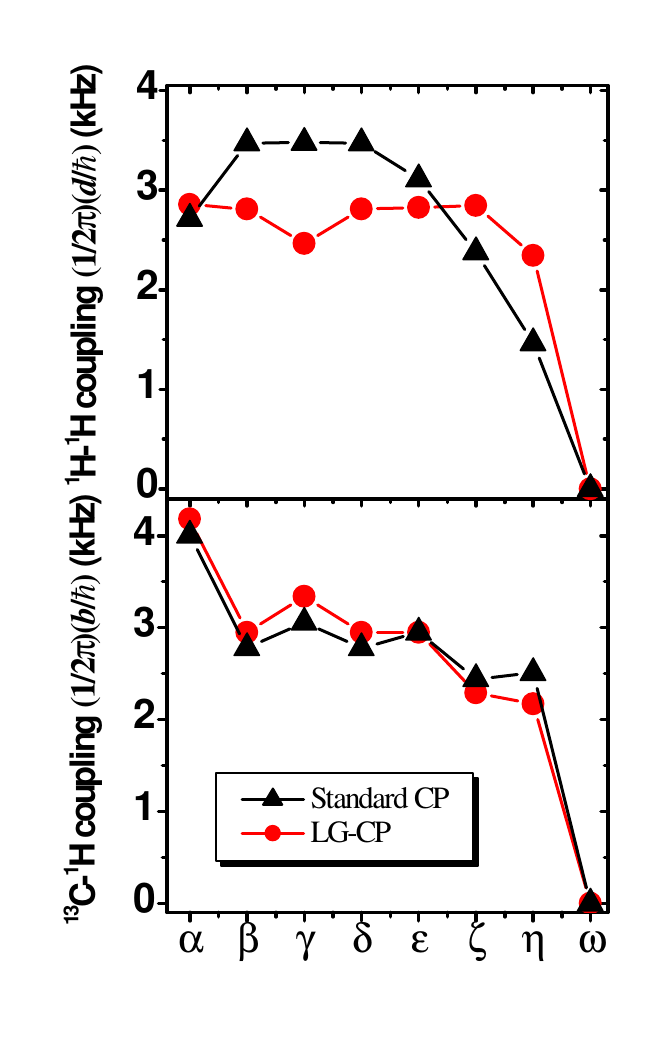}%
\caption[Effective homonuclear and heteronuclear dipolar couplings for carbons
in the alkyl chain of $8$CB.]{Effective homonuclear $\left(  1/2\pi\right)
\left(  d/\hbar\right)  $ and heteronuclear $\left(  1/2\pi\right)  \left(
b/\hbar\right)  $ dipolar couplings for carbons in the alkyl chain, obtained
from the LG and standard CP experiments performed at $311.5{}\operatorname{K}$
(nematic mesophase).}%
\label{Fig-param2}%
\end{center}
\end{figure}

Table \ref{Tab_dip_coupling_nematic} displays the values of the homonuclear
and heteronuclear couplings for the aliphatic carbons obtained from the
standard CP experiment, and combining this with the LG-CP performed at the
same temperature. Fig \ref{Fig-param2} allows for the comparison of the
$^{13}$C-$^{1}$H and $^{1}$H-$^{1}$H couplings obtained from the two methods.
As expected, an excellent agreement can be observed for the values of $b$. The
novel methodology to estimate $d$ values seems to yield good results within an
error of around $20\%$, which could be easily improved by taking longer time
data. However, there are no many experimental estimations of these homonuclear
couplings and those obtained directly from the $^{1}$H spectra are in good
agreement with the values obtained here.%

\begin{figure}
[tbh]
\begin{center}
\includegraphics[
height=4.0542in,
width=4.6311in
]%
{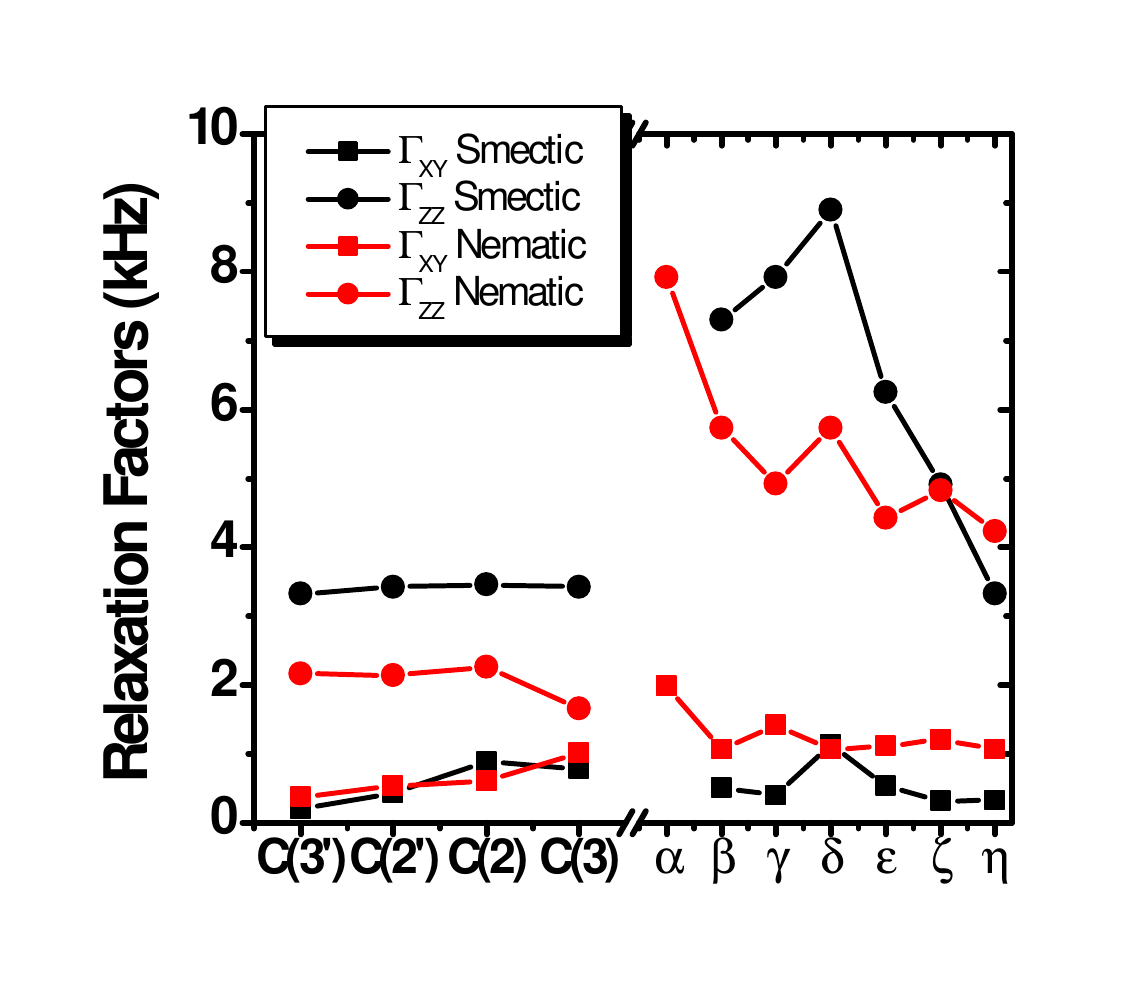}%
\caption[Relaxation factors $\Gamma_{\mathrm{ZZ}}$ and $\Gamma_{\mathrm{XY}}$
obtained by fitting the standard CP data of $8$CB to the anisotropic
model.]{Relaxation factors $\Gamma_{\mathrm{ZZ}}$ and $\Gamma_{\mathrm{XY}}$
obtained by fitting the standard CP data to the anisotropic model [eq.
(\ref{MSaniso})].}%
\label{Fig-param3}%
\end{center}
\end{figure}

With regard to the relaxation of the system in the CP experiments, we can
observe an anisotropic behavior for both phases, quantified by the ratios
$\Gamma_{\mathrm{ZZ}}/\Gamma_{\mathrm{XY}}>1$ (see fig. \ref{Fig-param3}).
Besides, we note that there is an important difference between the behaviors
of the nematic and smectic phases. In the nematic phase, the anisotropy can be
explained with a purely dipolar model. The average anisotropy factor for all
the carbons in the molecule is $(4\pm1).$ In contrast, in the smectic phase
the factor $\Gamma_{\mathrm{ZZ}}/\Gamma_{\mathrm{XY}}>4$ reveals a highly
anisotropic behavior for most of the carbons. This can be appreciated in fig.
\ref{Fog-CP8CB}, where the amplitude of the first maximum is higher than the
following ones, specially for the aliphatic carbons. This fact is not observed
in nematics, giving support to the purely dipolar anisotropy. Different
factors can produce this high anisotropy, one could be $H_{1}$ inhomogeneity,
another could be that our main system is actually larger than $SI_{2}$.
Although both factors would effectively increase the anisotropy, the effect
should be comparable in both phases. Moreover, under the same $H_{1}$
inhomogeneity, we have not observed such anisotropy ratios in molecular
crystals. Alternatively, the large anisotropy observed in the smectic phase
can be originated on the lack of fast fluctuations in this more rigid phase,
which would prevent the application of the extreme narrowing approximation. If
we release this approximation and assume that the spectral densities
$\mathcal{J}(0)$ and $\mathcal{J}(\Sigma/2)$, although different between them,
are approximately constant in an interval of width $2\omega_{0}$, our
calculations indicate that $\Gamma_{\mathrm{ZZ}}$ is proportional to
$\mathcal{J}(0)$, while $\Gamma_{\mathrm{XY}}\propto\mathcal{J}(\Sigma/2).$ As
usually $\mathcal{J}(0)>\mathcal{J}(\Sigma/2),$ this could explain the higher
anisotropy in the smectic phase, where motion is more hindered than in the
nematic phase.

\section{Summary}

We solved the spin dynamics of many-spin systems. First, we started solving
the dynamics of a two-spin system interacting with a fast fluctuating
spin-bath. We extended \cite{Alvarez07a} the solution of MKBE \cite{MKBE74}
starting from the system-environment Hamiltonian to obtain the relaxation
superoperator. We solved the spin dynamics for an anisotropic SE interaction
\cite{Alvarez07a}, where the anisotropy is given by the ratio between the
Ising and XY contributions. M\"{u}ller, \emph{et al.} used a phenomenological
isotropic relaxation for the SE interaction. Here, we emphasize the different
sources of the anisotropy in the decoherence and relaxation processes. The
main difference is that while the XY interaction takes the systems to the
total system equilibrium, the Ising SE interaction takes it to an internal
quasi-equilibrium \cite{JCP03,Alvarez07a}. In the particular case of an
isotropic SE interaction, we reproduce the MKBE solution. These solutions were
obtained neglecting the non-secular terms of the SE interaction with respect
to the system Hamiltonian, i.e. under the condition $b\gg\Gamma_{\mathrm{SE}%
}.$ We extended the solution including the non-secular terms \cite{Alvarez07a}
as will be fully discussed in chapter \ref{Marker_Spin_within_keldysh}.

In section \S \ \ref{M_Q_dyn_in_8CB}, we further extended the MKBE model to
solve the spin dynamics of a three-spin system interacting with a fast
fluctuating spin-bath \cite{JCP03}. Our calculations were performed for
three-spin configurations directly applicable to the $8$CB molecule. This
molecule has two possible configurations for the heteronuclear couplings that
led us to realize that in each space of $M=\pm1/2$ there are only two of the
three eigenstates involved in the dipolar transitions that give rise to the
oscillations. This is a consequence of the symmetry of the system, i.e. the
flip-flop can only occur between the carbon and one combination of the proton
states (the symmetric or the antisymmetric) depending on the relative signs of
the heteronuclear couplings ($b_{1}=b_{2}$ or $b_{1}=-b_{2}$) \cite{JCP03}. We
showed that this affects the oscillation frequency that one measures
experimentally.

In section \S \ \ref{Mark_3spin_experimental_8CB}, we introduced $^{13}%
$C-$^{1}$H cross-polarization experiments which complemented with the detailed
spin dynamics calculations allowed us to obtain separately the homonuclear and
heteronuclear dipolar couplings in CH$_{2}$ systems \cite{JCP03}. The
reliability of the obtained values was tested with a direct determination of
the heteronuclear couplings using CP\ under Lee-Goldburg conditions. Comparing
the standard CP experiments with the LG-CP one, we concluded that the last one
is better to determine directly the $^{13}$C-$^{1}$H couplings. However, the
standard CP allows one to obtain homonuclear couplings in addition to the
heteronuclear couplings. This experiment is also important to get further
information of the system as for example the relaxation phenomena.

From the theoretical analysis of the CH$_{2}$ dynamics, we recognized two
different time scales for the decoherence behavior given by $\Gamma
_{\mathrm{ZZ}}$ and $\Gamma_{\mathrm{XY}}.$ Besides, the CP data showed that
the rate of attenuation of the oscillations is much faster than the rate of
polarization transfer from the bath. This anisotropy could be explained in the
nematic phase by assuming a dipolar interaction Hamiltonian between the
three-spin system and the bath within the extreme narrowing approximation. In
the smectic phase, however, the anisotropy is much more pronounced and it
seems that the short time fluctuations of the bath approximation is not
appropriate. Consideration of a slow motion regime leads to a better agreement
with the experimental observations without resorting to other mechanisms which
operate in both phases \cite{JCP03}.

\chapter{Spin dynamics within another perspective: The Keldysh
formalism\label{Marker_Spin_within_keldysh}}

As we discuss in previous chapters, the characterization and control of spin
dynamics in open and close spin systems of intermediate size remains a problem
of great interest \cite{Loss03}.
However, the quantum interferences of these systems become damped by the lack
of isolation from the environment and one visualizes this phenomenon as
decoherence. Indeed, the inclusion of the degrees of freedom of the
environment may easily become an unsolvable problem and requires
approximations not fully quantified. This motivates a revival of interest on
previous studies in various fields such as Nuclear Magnetic Resonance
\cite{MKBE74}, quantum transport \cite{DAmato-Pastawski} and the
quantum-classical correspondence problem \cite{Paz99,Zurek03} with a view on
their application to emergent fields like the quantum computation
\cite{Lidar02,DeRaedt03,Cory00}
and molecular electronics \cite{Ratner94,Nitzan03,MolElec1,MolElec2}.

The most standard framework adopted to describe the system-environment
interaction is the one that we used in the previous chapter, the generalized
Liouville-von Neumann equation \cite{Abragam,Ernst} in the fast fluctuation
approximation. There, the degrees of freedom of the environment are traced out
to yield a quantum master equation (QME), eq. (\ref{master}). Interactions
with the environment occur at a rate given by the Fermi Golden Rule (FGR)
providing a dissipative mechanism that could induce a non unitary dynamics
into the system. An overall (conservation) balance condition is obtained by
imposing a convergence into the thermal equilibrium state
(\ref{2spin_DM_final_condition}). While sufficient for the most traditional
applications, this approximation leaves aside important memory effects and
interferences in the time domain produced by the coherent interaction between
the system and the bath \cite{Taylor03}.

A less known alternative is provided by the Keldysh formalism \cite{Keldysh64}
in the integral representation proposed by Danielewicz \cite{Danielewicz84}.
On one side, it uses the well known perturbation to infinite order in selected
terms provided by the Feynman diagrams where, under certain conditions that go
beyond the Fermi Golden Rule, the dynamical feedback effects become relevant.
On the other, this integral representation has the advantage of being able to
profit from a Wigner representation for the time-energy domain. This last
representation is particularly meaningful in the fermionic case since it
allows one to define energy states and their occupations simultaneously with
the physical time \cite{GLBE2}. In that case, one can transform the
Danielewicz equation into the generalized Landauer-B\"{u}ttiker equation
(GLBE) \cite{GLBE1,GLBE2} to solve the quantum dynamics of the system. The
Keldysh formalism already inspired original experimental and theoretical
developments in coherent spin dynamics involving quantum interferences in the
time domain. In particular, it was used to develop the notion of polarization
waves leading to mesoscopic echoes \cite{Pastawski95,Pastawski96}, to
establish the influence of chaos on\ time reversal (Loschmidt echoes)
\cite{JCP98,PhysicaA,JalPas01} and to propose a spin projection chromatography
\cite{Danieli04}. A rough account of many-body decoherence enabled the
interpretation of anomalies in \textquotedblleft spin
diffusion\textquotedblright\ experiments as a manifestation of the quantum
Zeno effect \cite{Usaj98}. This technique was applied to a case with an exact
analytical solution \cite{Danieli02} and where more standard approximations
can be obtained
to show the potential of our proposal. Our model represents a single fermion
that can jump between two states while an external fermionic reservoir is
coupled to one of them. This environment provides decoherence due to a through
space Coulomb interaction and can feed the system with an extra particle
through tunneling processes. While the parameters and approximations involved
in this model are especially designed to be mapped to a problem of spin
dynamics (the two-spin system of chapter
\ref{Mark_spin_dynamics_Density_matrix}), it could also be adapted to
represent a double quantum dot charge qubit \cite{Mucciolo-Baranger2005}%
.\textit{ }In that case, a double dot is operated in the gate voltage regime
where there is a single electron which can jump between the two coupled
states. Only one of these states is coupled to an electron reservoir. We
introduce fictitious interactions to obtain a common interaction rate which
leads to a homogeneous non-hermitian effective Hamiltonian. In the specific
model considered, we analyze how different SE interactions, e.g. tunneling to
the leads and through space Coulomb interaction, modify the quantum evolution.
A particular advantage of the fictitious symmetrization is that it leads
naturally to a stroboscopic representation of the SE processes. This leads to
a very efficient numerical algorithm where the quantum dynamics is obtained in
a sequence of time steps. Finally, we resort to the Jordan-Wigner mapping
between fermions and spins to apply the procedure to a spin system. This
allows us to give a first-principle derivation of the self-energies
to explain, in chapter \ref{Sec_QDPT}, the puzzling experimental dynamics
observed in a spin swapping gate \cite{JCP98} (see fig.
\ref{Fig_JCP98_original}).

\section{Two level system dynamics}

\subsection{The system}

Consider a two fermion state system
interacting with the following Hamiltonian $\widehat{\mathcal{H}}_{\mathrm{S}%
}$,%
\begin{equation}
\widehat{\mathcal{H}}_{\mathrm{S}}=E_{0}^{{}}\hat{c}_{0}^{+}\hat{c}_{0}^{{}%
}+E_{1}^{{}}\hat{c}_{1}^{+}\hat{c}_{1}^{{}}-V_{01}\left(  \hat{c}_{0}^{+}%
\hat{c}_{1}^{{}}+\hat{c}_{1}^{+}\hat{c}_{0}^{{}}\right)  , \label{HsTLS}%
\end{equation}
with $\hat{c}_{i}^{+}(\hat{c}_{i}^{{}})$ the standard fermionic creation
(destruction) operators. The $E_{i}$ are the energies of the $i$-th local
states whose spin index is omitted and $V_{12}$ the hopping interaction. In
matrix representation, if we have only one particle in the system, we have a
$2\times2$ matrix for the Hamiltonian
\begin{equation}%
\begin{array}
[c]{ccc}
&
\begin{array}
[c]{cc}%
\left\vert 1,0\right\rangle  & \left\vert 0,1\right\rangle
\end{array}
& \\
\mathcal{H}_{\mathrm{S}}= & \left(
\begin{array}
[c]{cc}%
E_{0} & -V_{01}\\
-V_{01} & E_{1}%
\end{array}
\right)  &
\begin{array}
[c]{c}%
\left\vert 1,0\right\rangle \\
\left\vert 0,1\right\rangle
\end{array}
.
\end{array}
\end{equation}
Here $\left\vert 1,0\right\rangle $ and $\left\vert 0,1\right\rangle $ denote
the states with the particle in the level $0$ and $1$ respectively.

\subsection{System evolution}

\begin{figure}
[tbh]
\begin{center}
\includegraphics[
height=4.3422in,
width=5.3791in
]%
{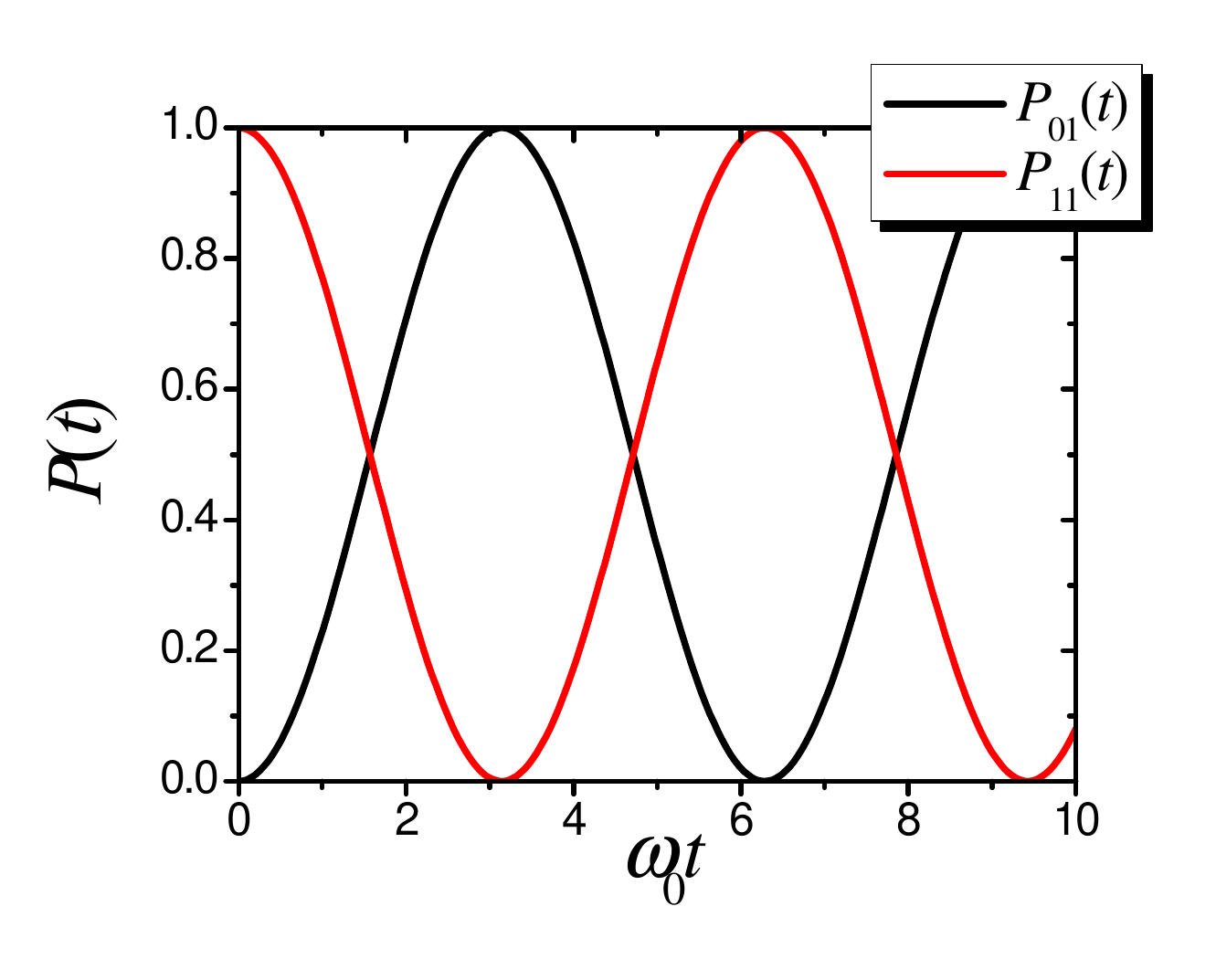}%
\caption[Evolution of the probabilities to find a particle in site $0$ (black
line) and site $1$ (red line) of a two-level system with the initial
conditions at site $1.$]{Evolution of the probabilities to find a particle in
site $0$ (black line) and site $1$ (red line) with the initial conditions at
site $1.$}%
\label{Fig_TLS_probabilities}%
\end{center}
\end{figure}

We are interested in the study of the evolution of an initial local excitation
in the system.
For definiteness, we consider the initial excitation on site $1$ which is
described by the non-equilibrium state $\left\vert \Psi\left(  0\right)
\right\rangle =\hat{c}_{1}^{+}\left\vert 0,0\right\rangle =\left\vert
0,1\right\rangle $ where $\left\vert 0,0\right\rangle $ is the vacuum state.
The probability to find the particle in the state $0$ and $1$ is
\begin{align}
P_{01}\left(  t\right)   &  =\left\vert \left\langle 1,0\right\vert
\Psi\left(  t\right)  \right\vert ^{2},\\
P_{11}\left(  t\right)   &  =\left\vert \left\langle 0,1\right\vert
\Psi\left(  t\right)  \right\vert ^{2},
\end{align}
where
\begin{equation}
\left\vert \Psi\left(  t\right)  \right\rangle =\exp\left\{  -\mathrm{i}%
\widehat{\mathcal{H}}_{\mathrm{S}}~t\right\}  \left\vert \Psi\left(  0\right)
\right\rangle .
\end{equation}
Thus, we obtain if $E_{0}=E_{1}$%
\begin{align}
P_{01}\left(  t\right)   &  =\frac{1}{2}-\frac{1}{2}\cos\left(  \omega
_{0}t\right)  ,\label{P01_isolated_keldysh}\\
P_{11}\left(  t\right)   &  =\frac{1}{2}+\frac{1}{2}\cos\left(  \omega
_{0}t\right)  ,
\end{align}
where $\omega_{0}=2V_{01}/\hbar$ gives the natural oscillation frequency of
the transition between the states $0$ and $1$. The fig.
\ref{Fig_TLS_probabilities}
show the dynamics of these probabilities.

\section{A two level system interacting with a particle reservoir}

\subsection{The system}

\begin{figure}
[ptbh]
\begin{center}
\includegraphics[
height=5.5106in,
width=4.1399in
]%
{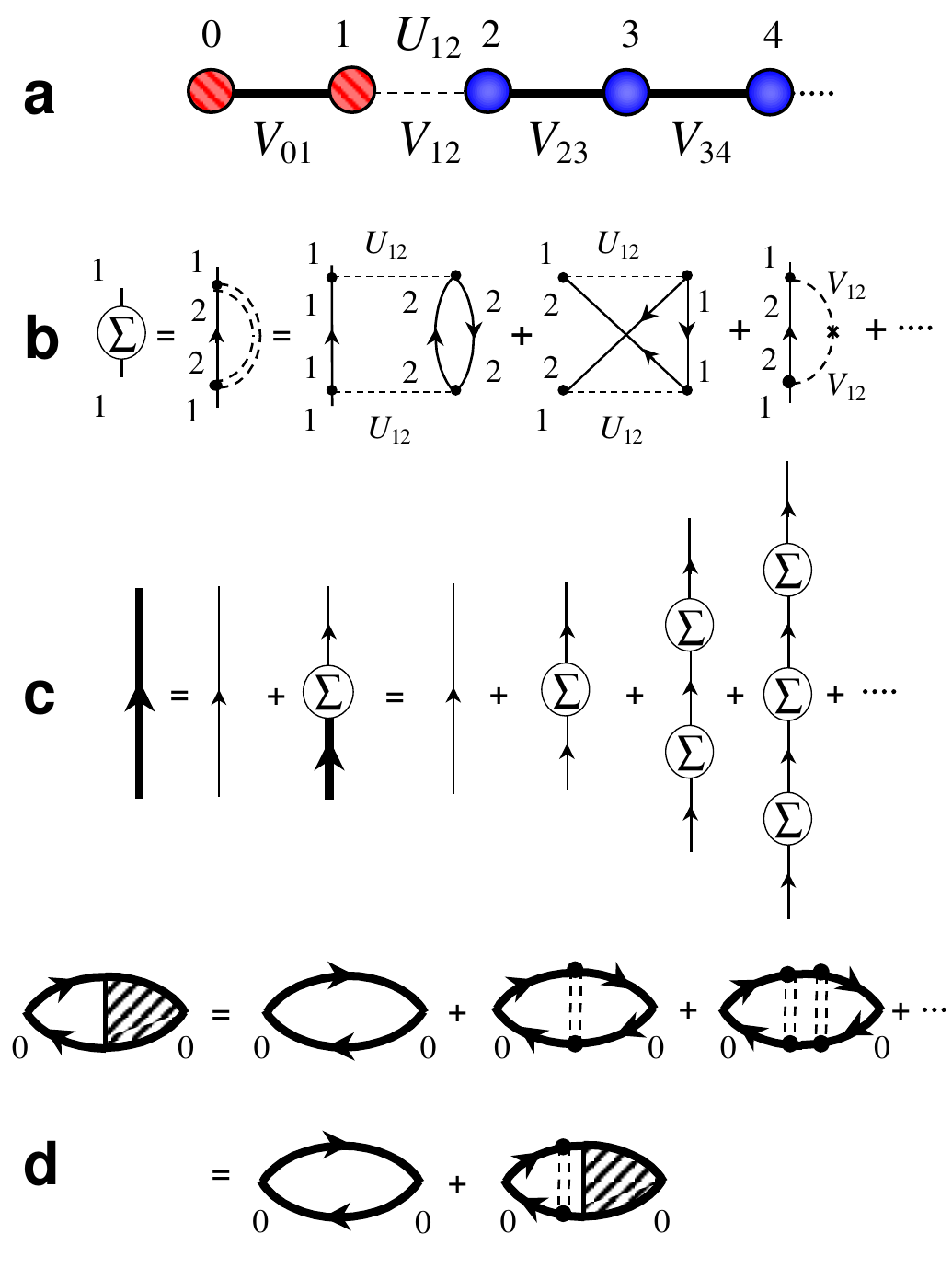}%
\caption[Two-level system interacting with a particle-reservoir scheme and the
Feynman's diagrams of the evolution.]{a) System-environment (SE)
representation. Dashed circles and solid circles represent the system and the
environment states respectively. Dashed lines are hopping interactions while
wiggels lines are through-space Coulomb interactions. b) Self-enegy diagram
summing up the different interactions with the environment in a local basis.
The lines with arrows are exact Green's functions in absence of SE
interactions. The double dashed line represents the effective SE interaction.
c) Retarded Green's function at site $0$. The interaction with the environment
is to infinite order in the self-energy given in b) . d) Particle density
function at site $0$. The double dashed lines represent the effective
interactions local in time and space summed up to infinity order.}%
\label{Fig_system_feynman}%
\end{center}
\end{figure}

Let's consider the electron two-state system of the previous
section\ asymmetrically coupled to an electron reservoir, as shown in fig.
\ref{Fig_system_feynman} a),
with the total Hamiltonian
\begin{equation}
\widehat{\mathcal{H}}=\widehat{\mathcal{H}}_{\mathrm{S}}+\widehat{\mathcal{H}%
}_{\mathrm{E}}+\widehat{\mathcal{H}}_{\mathrm{SE}}.
\end{equation}
\linebreak The system Hamiltonian is the same described in the previous
section (\ref{HsTLS})%
\begin{equation}
\widehat{\mathcal{H}}_{\mathrm{S}}=E_{0}^{{}}\hat{c}_{0}^{+}\hat{c}_{0}^{{}%
}+E_{1}^{{}}\hat{c}_{1}^{+}\hat{c}_{1}^{{}}-V_{01}\left(  \hat{c}_{0}^{+}%
\hat{c}_{1}^{{}}+\hat{c}_{1}^{+}\hat{c}_{0}^{{}}\right)  , \label{Hs}%
\end{equation}
with $\hat{c}_{i}^{+}(\hat{c}_{i}^{{}})$ the standard fermionic creation
(destruction) operators. The hopping interaction $V_{01}$ gives the natural
frequency, $\omega_{0}=2V_{01}/\hbar$, of the transition between the states
$0$ and $1$. The environment has a similar Hamiltonian,%
\begin{equation}
\widehat{\mathcal{H}}_{\mathrm{E}}=\sum_{i=2}^{\infty}E_{i}^{{}}\hat{c}%
_{i}^{+}\hat{c}_{i}^{{}}-\sum_{%
\genfrac{}{}{0pt}{}{i=2}{j>i}%
}^{\infty}V_{ij}^{{}}\left(  \hat{c}_{i}^{+}\hat{c}_{j}^{{}}+\hat{c}_{j}%
^{+}\hat{c}_{i}^{{}}\right)  , \label{He}%
\end{equation}
where the $V_{ij}^{{}}$ determines the topology of the interaction network in
the environment states. The system-environment (SE) interaction is described
by%
\begin{equation}
\widehat{\mathcal{H}}_{\mathrm{SE}}=\sum_{\alpha,\beta=\uparrow,\downarrow
}\left[  U_{12}^{(\mathrm{dir.})}~\hat{c}_{1,\beta}^{+}\hat{c}_{1,\beta}^{{}%
}\hat{c}_{2,\alpha}^{+}\hat{c}_{2,\alpha}^{{}}+U_{12}^{(\mathrm{exch.})}%
~\hat{c}_{1,\alpha}^{+}\hat{c}_{2,\alpha}^{{}}\hat{c}_{2,\alpha}^{+}\hat
{c}_{1,\alpha}^{{}}-V_{12}^{{}}\left(  \hat{c}_{1,\alpha}^{+}\hat{c}%
_{2,\alpha}^{{}}+\hat{c}_{2,\alpha}^{+}\hat{c}_{1,\alpha}^{{}}\right)
\right]  , \label{HSE}%
\end{equation}
The first two terms on the rhs represent the Coulomb interaction of an
electron in site $1$ with an electron in site $2,$ the first site of the
reservoir. $U_{12}^{(\mathrm{dir.})}$ is the standard direct integral and
$U_{12}^{(\mathrm{exch.})}$ is the small exchange integral which we include
for completeness The third term is the hopping interaction between sites $1$
and $2$.

\subsection{System evolution}

\subsubsection{Quantum dynamics within the Keldysh formalism}

Now, we have a many-body system and we consider it in the thermodynamical
many-body equilibrium state $\left\vert \Psi_{\mathrm{eq.}}\right\rangle $. In
the infinity temperature limit, it has all the states equally occupied. We are
interested in the study of the evolution of an initial local excitation in the
system.
Thus, as we consider previously for definiteness, let's consider the initial
excitation with a particle on site $1$ and a hole in site $0$ which is
described by the non-equilibrium state
\begin{equation}
\left\vert \Psi_{\mathrm{n.e.}}\right\rangle =\hat{c}_{1}^{+}\hat{c}_{0}^{{}%
}\left\vert \Psi_{\mathrm{eq.}}\right\rangle .
\end{equation}
The evolution in a complete norm preserving solution is described by the
particle and hole density functions,
\begin{equation}
G_{ij}^{<}\left(  t_{2},t_{1}\right)  =\frac{\mathrm{i}}{\hbar}\left\langle
\Psi_{\mathrm{n.e.}}\right\vert \hat{c}_{j}^{+}\left(  t_{1}\right)  \hat
{c}_{i}^{{}}\left(  t_{2}\right)  \left\vert \Psi_{\mathrm{n.e.}%
}\right\rangle
\end{equation}
and
\begin{equation}
G_{ij}^{>}\left(  t_{2},t_{1}\right)  =-\frac{\mathrm{i}}{\hbar}\left\langle
\Psi_{\mathrm{n.e.}}\right\vert \hat{c}_{i}^{{}}\left(  t_{2}\right)  \hat
{c}_{j}^{+}\left(  t_{1}\right)  \left\vert \Psi_{\mathrm{n.e.}}\right\rangle
,
\end{equation}
that describe spatial and temporal correlations. In these expressions, the
creation and destruction operators are in the Heisenberg representation.
Notice that, in contrast with the equilibrium definitions of $G_{ij}%
^{\lessgtr\,}(t_{2},t_{1}),$ now they have an implicit dependence on the
initial local excitation. The probability amplitude of finding a particle in
site $i$ at time $t_{2}$ when initially was in site $j$ at time $t_{1}$ is
described by the retarded Green's function
of the whole system
\begin{align}
G_{ij}^{\mathrm{R}}\left(  t_{2},t_{1}\right)   &  =\theta\left(  t_{2}%
,t_{1}\right)  \ [G_{ij}^{>}\left(  t_{2},t_{1}\right)  -G_{ij}^{<}\left(
t_{2},t_{1}\right)  ]\nonumber\\
&  =\left[  G_{ji}^{\mathrm{A}}\left(  t_{1},t_{2}\right)  \right]  ^{\ast}.
\end{align}
The reduced density function $\mathbf{G}^{<}\left(  t,t\right)  $, where
matrix indices are restricted to $i,j\in\left\{  0,1\right\}  $, is equivalent
to the single particle $2\times2$ density matrix and $\mathbf{G}^{\mathrm{R}%
}\left(  t_{2},t_{1}\right)  $ is an effective evolution operator\footnote{The
characters in bold are matrix representation of the respective operator.}.
If the system is isolated, the Green's function in its energy representation
is obtained by a Fourier transform (FT) respect to the time interval $\delta
t=t_{2}-t_{1}$
\begin{equation}
\mathbf{G}^{0\mathrm{R}}\left(  \varepsilon,t\right)  =\int\mathbf{G}%
^{0\mathrm{R}}\left(  t+\tfrac{1}{2}\delta t,t-\tfrac{1}{2}\delta t\right)
\exp[\mathrm{i}\varepsilon\delta t/\hbar]\mathrm{d}\delta t,
\end{equation}
where $t=\frac{1}{2}\left(  t_{2}+t_{1}\right)  .$ In a time independent
system:%
\begin{equation}
\mathbf{G}^{0\mathrm{R}}\left(  \varepsilon,t\right)  \equiv\mathbf{G}%
^{0\mathrm{R}}\left(  \varepsilon\right)  =[\varepsilon\mathbf{I}%
-\mathbf{H}_{\mathrm{S}}]^{-1}.
\end{equation}
After including SE interactions, the Green's function defines the reduced
effective Hamiltonian and the self-energy $\mathbf{\Sigma}^{\mathrm{R}%
}(\varepsilon)$ \cite{DAmato90},\emph{ }%
\begin{equation}
\mathbf{H}_{\mathrm{eff.}}(\varepsilon)\equiv\varepsilon\mathbf{I}-\left[
\mathbf{G}^{\mathrm{R}}\left(  \varepsilon\right)  \right]  ^{-1}%
=\mathbf{H}_{\mathrm{S}}+\mathbf{\Sigma}^{\mathrm{R}}(\varepsilon).
\label{Heff}%
\end{equation}
Here, the exact perturbed dynamics is contained in the nonlinear dependence of
the self-energy $\mathbf{\Sigma}^{\mathrm{R}}$ on $\varepsilon$\emph{.} For
infinite reservoirs the evolution with $\mathbf{H}_{\mathrm{eff.}}$ is
non-unitary, hence, the Green's function has poles at the \textquotedblleft
eigenenergies\textquotedblright, $\varepsilon_{\nu}$, that have imaginary
components \cite{DAmato-Pastawski},
\begin{equation}
-2\operatorname{Im}\Sigma^{\mathrm{R}}\left(  \varepsilon_{\nu}\right)
/\hbar=1/\tau_{\mathrm{SE}}^{{}}=2\Gamma_{\mathrm{SE}}/\hbar.
\label{Imaginary_part_sigma_R}%
\end{equation}
These account for the\ \textquotedblleft decay rates\textquotedblright\ into
collective SE eigenstates in agreement with a self-consistent Fermi Golden
Rule (FGR) \cite{Elena06}. Similarly, $\operatorname{Re}\Sigma^{\mathrm{R}%
}\left(  \varepsilon_{\nu}^{{}}\right)  =\operatorname{Re}\varepsilon_{\nu
}-\varepsilon_{\nu}^{0}$ represent the \textquotedblleft
shifts\textquotedblright\ of the system's eigenenergies $\varepsilon_{\nu}%
^{0}.$

The evolution of the density function for the reduced open system is described
using the Keldysh formalism \cite{Keldysh64,Danielewicz84}.
The density function in the Danielewicz form \cite{Danielewicz84} is%
\begin{multline}
\mathbf{G}^{<}\left(  t_{2},t_{1}\right)  =\hbar^{2}\mathbf{G}^{\mathrm{R}%
}\left(  t_{2},0\right)  \mathbf{G}^{<}\left(  0,0\right)  \mathbf{G}%
^{\mathrm{A}}\left(  0,t_{1}\right) \label{Danielewicz_evol}\\
+\int_{0}^{t_{2}}\int_{0}^{t_{1}}\mathrm{d}t_{k}\mathrm{d}t_{l}\mathbf{G}%
^{\mathrm{R}}\left(  t_{2},t_{k}\right)  \mathbf{\Sigma}^{<}\left(
t_{k},t_{l}\right)  \mathbf{G}^{\mathrm{A}}\left(  t_{l},t_{1}\right)  .
\end{multline}
The first term is the \textquotedblleft coherent\textquotedblright\ evolution
while the second term contains \textquotedblleft incoherent
reinjections\textquotedblright\ through the self-energy function,
$\mathbf{\Sigma}^{<}$. This compensates any leak from the coherent evolution
reflected by the imaginary part of $\Sigma^{\mathrm{R}}$ (see \cite{GLBE2}).
The key to solve eq. (\ref{Danielewicz_evol}) is to build up an expression for
the particle (hole) injection and retarded self-energies, $\mathbf{\Sigma
}^{<(>)}\left(  t_{1},t_{2}\right)  $ and
\begin{equation}
\mathbf{\Sigma}^{\mathrm{R}}\left(  t_{1},t_{2}\right)  =\theta\left(
t_{1},t_{2}\right)  [\mathbf{\Sigma}^{>}\left(  t_{2},t_{1}\right)
-\mathbf{\Sigma}^{<}\left(  t_{2},t_{1}\right)  ]. \label{Sigma_R_definition}%
\end{equation}
For this purpose, we use a perturbative expansion on $\widehat{\mathcal{H}%
}_{\mathrm{SE}}$ for the Coulomb interaction \cite{SSC07} and the hopping
interaction \cite{CPL05}.\emph{ }
The first order in the perturbation expansion is the standard Hartree-Fock
energy correction which does not contribute to $\Sigma^{<}$ because it is
real\emph{. }
We focus on the second order terms, with Feynman diagrams
sketched in fig. \ref{Fig_system_feynman} b).

The injection self energy is\emph{ }%
\begin{equation}
\Sigma_{ij}^{\lessgtr}\left(  t_{k},t_{l}\right)  =\left\vert U_{12}^{{}%
}\right\vert ^{2}\hbar_{{}}^{2}G_{22}^{\lessgtr}\left(  t_{k},t_{l}\right)
G_{22}^{\gtrless}\left(  t_{l},t_{k}\right)  G_{11}^{\lessgtr}\left(
t_{k},t_{l}\right)  ~\delta_{i1}^{{}}\delta_{1j}^{{}}+\left\vert V_{12}^{{}%
}\right\vert ^{2}G_{22}^{\lessgtr}\left(  t_{k},t_{l}\right)  ~\delta_{i1}%
^{{}}\delta_{1j}^{{}}, \label{Sigma_Feynman}%
\end{equation}
where $U_{12}=-2U_{12}^{(\mathrm{dir.})}+U_{12}^{(\mathrm{exch.})}$ is the net
Coulomb interaction between an electron in the system and one in the
reservoir. The direct term contributes with a fermion loop and an extra spin
summation which is represented in the $-2$ factor
\cite{Danielewicz84}. The first term in eq. (\ref{Sigma_Feynman}) corresponds
to the direct and exchange self-energy diagrams shown in the last line of fig.
\ref{Fig_system_feynman} b). The first two diagrams schematize the creation of
an electron-hole pair in the environment and its later destruction. The last
term in eq. (\ref{Sigma_Feynman}) and the last diagram of the same figure is
the hopping to site $2$ which allows the electron to perform a full
exploration inside the reservoir. To take into account the different time
scales for the dynamics of excitations in the system and in the reservoir, we
use the time-energy variables \cite{GLBE2}:
the physical time $t_{\mathrm{i}}=\frac{1}{2}\left(  t_{k}+t_{l}\right)  ,$and
the domain of quantum correlations $\delta t_{\mathrm{i}}=t_{k}-t_{l}.$ This
last is related to an energy $\varepsilon$ through a FT \cite{GLBE2}. Thus, in
equilibrium,%
\begin{align}
G_{22}^{<}\left(  \varepsilon,t_{\mathrm{i}}\right)   &  =\mathrm{i}2\pi
~N_{2}\left(  \varepsilon\right)  ~\mathrm{f}_{2}\left(  \varepsilon
,t_{\mathrm{i}}\right)  ,\label{G<(E,t)}\\
G_{22}^{>}\left(  \varepsilon,t_{\mathrm{i}}\right)   &  =-\mathrm{i}%
2\pi~N_{2}\left(  \varepsilon\right)  ~\left[  1-\mathrm{f}_{2}\left(
\varepsilon,t_{\mathrm{i}}\right)  \right]  ,
\end{align}
where $N_{2}\left(  \varepsilon\right)  $ is the local density of states
(LDoS)\ at the surface of the reservoir.
Assuming that the environment stays in the thermodynamical equilibrium and
$k_{\mathrm{B}}T$ is much higher that any energy scale in the bath (high
temperature limit), the occupation factor is
$\ $%
\begin{equation}
\mathrm{f}_{2}\left(  \varepsilon,t_{\mathrm{i}}\right)  =\mathrm{f}_{2}.
\end{equation}
Fourier transforming on $\varepsilon$ one obtains
\begin{equation}
G_{22}^{<}\left(  t_{\mathrm{i}}+\tfrac{\delta t_{\mathrm{i}}}{2}%
,t_{\mathrm{i}}-\tfrac{\delta t_{\mathrm{i}}}{2}\right)  =\mathrm{i}%
2\pi\mathrm{~}g_{2}\left(  \delta t_{\mathrm{i}}\right)  ~\mathrm{f}_{2}%
\end{equation}
and
\begin{equation}
G_{22}^{>}\left(  t_{\mathrm{i}}+\tfrac{\delta t_{\mathrm{i}}}{2}%
,t_{\mathrm{i}}-\tfrac{\delta t_{\mathrm{i}}}{2}\right)  =-\mathrm{i}%
2\pi\mathrm{~}g_{2}\left(  \delta t_{\mathrm{i}}\right)  ~\left(
1-\mathrm{f}_{2}\right)  ,
\end{equation}
where
\begin{equation}
g_{2}\left(  \delta t_{\mathrm{i}}\right)  =\int N_{2}\left(  \varepsilon
\right)  e^{-\mathrm{i}\varepsilon\delta t_{\mathrm{i}}}\frac{\mathrm{d}%
\varepsilon}{2\pi\hbar}.
\end{equation}
Replacing in eq. (\ref{Sigma_Feynman})\emph{ }%
\begin{multline}
\Sigma_{ij}^{\lessgtr}\left(  t_{\mathrm{i}}+\tfrac{\delta t_{\mathrm{i}}}%
{2},t_{\mathrm{i}}-\tfrac{\delta t_{\mathrm{i}}}{2}\right)  =\left\vert
U_{12}^{{}}\right\vert ^{2}\hbar_{{}}^{2}\left(  2\pi\right)  ^{2}\left[
g_{2}\left(  \delta t_{\mathrm{i}}\right)  \right]  ^{2}\mathrm{f}_{2}\left[
1-\mathrm{f}_{2}\right]  G_{11}^{\lessgtr}\left(  t_{\mathrm{i}}+\tfrac{\delta
t_{\mathrm{i}}}{2},t_{\mathrm{i}}-\tfrac{\delta t_{\mathrm{i}}}{2}\right)
\delta_{i1}^{{}}\delta_{1j}^{{}}\label{Sigma_Feynmann_2}\\
\pm\left\vert V_{12}^{{}}\right\vert ^{2}~\mathrm{i}2\pi g_{2}\left(  \delta
t_{\mathrm{i}}\right)  ~\left(
\genfrac{}{}{0pt}{}{\mathrm{f}_{2}}{1-\mathrm{f}_{2}}%
\right)  ~\delta_{i1}^{{}}\delta_{1j}^{{}},
\end{multline}
where the $\left(
\genfrac{}{}{0pt}{}{\mathrm{f}_{2}}{1-\mathrm{f}_{2}}%
\right)  $ associates $\mathrm{f}_{2}$ with $\Sigma^{<}$ and $\left(
1-\mathrm{f}_{2}\right)  $ with $\Sigma^{>}$.

In summary, we are left with the task to evaluate the time dependent self
energies and the integral in eq. (\ref{Danielewicz_evol}). We will focus in
the parametric regime corresponding to the experimental conditions of the spin
swapping gate (cross-polarization) described in section
\S \ \ref{M_2-spin_spin_bath}.

\subsubsection{An environment in the wide band or fast fluctuation regime}

As occurs with the generalized Landauer-B\"{u}ttiker equations for linear
transport, an essential ingredient is the possibility to assign a Markovian
nature to the environment. We are going to see that this appears naturally
from the formalism when the dynamics of excitations within the environment is
faster than the time scales relevant to the system. In order to separate the
different physical time scales involved in the problem, we start changing to
the time-energy variables in eq. (\ref{Danielewicz_evol}). Evaluating in
$t_{2}=t_{1}=t,$ the integrand becomes
\begin{equation}
\int_{0}^{t}\mathrm{d}t_{\mathrm{i}}\int_{-t}^{t}\mathrm{d}\delta
t_{\mathrm{i}}\mathbf{G}^{\mathrm{R}}\left(  t,t_{\mathrm{i}}+\tfrac{\delta
t_{\mathrm{i}}}{2}\right)  \mathbf{\Sigma}^{<}\left(  t_{\mathrm{i}}%
+\tfrac{\delta t_{\mathrm{i}}}{2},t_{\mathrm{i}}-\tfrac{\delta t_{\mathrm{i}}%
}{2}\right)  \mathbf{G}^{\mathrm{A}}\left(  t_{\mathrm{i}}-\tfrac{\delta
t_{\mathrm{i}}}{2},t\right)  .
\end{equation}
The environment unperturbed\ Green's function $g_{2}\left(  \delta
t_{\mathrm{i}}\right)  $ decays within the time scale $\hbar/V_{\mathrm{B}}$
where $V_{\mathrm{B}}$ is the characteristic interaction inside the reservoir.
In the wide band regime ($V_{\mathrm{B}}\gg V_{01}$) $\hbar/V_{\mathrm{B}}$
becomes much shorter than the characteristic evolution time of $G_{11}%
^{\lessgtr}\left(  t_{\mathrm{i}}+\tfrac{\delta t_{\mathrm{i}}}{2}%
,t_{\mathrm{i}}-\tfrac{\delta t_{\mathrm{i}}}{2}\right)  $ given by
$\hbar/V_{01}.$ Then, as the main contribution to the integral on $\delta
t_{\mathrm{i}}$ of eq. (\ref{Danielewicz_evol}) is around the time scale
$\hbar/V_{\mathrm{B}}$ we can replace $G_{11}^{\lessgtr}\left(  t_{\mathrm{i}%
}+\tfrac{\delta t_{\mathrm{i}}}{2},t_{\mathrm{i}}-\tfrac{\delta t_{\mathrm{i}%
}}{2}\right)  $ by $G_{11}^{\lessgtr}\left(  t_{\mathrm{i}},t_{\mathrm{i}%
}\right)  $. Following the same assumption we replace $G^{\mathrm{R}}\left(
t,t_{\mathrm{i}}+\frac{\delta t_{\mathrm{i}}}{2}\right)  $ by $G^{\mathrm{R}%
}\left(  t,t_{\mathrm{i}}\right)  $ and $G^{\mathrm{A}}\left(  t_{\mathrm{i}%
}-\frac{\delta t_{\mathrm{i}}}{2},t\right)  $ by $G^{\mathrm{A}}\left(
t_{\mathrm{i}},t\right)  .$\emph{ }In this fast fluctuation regime, only
$\Sigma_{ij}^{\lessgtr}\left(  t_{\mathrm{i}}+\tfrac{\delta t_{\mathrm{i}}}%
{2},t_{\mathrm{i}}-\tfrac{\delta t_{\mathrm{i}}}{2}\right)  $ depends on
$\delta t_{\mathrm{i}}$ leading to%
\begin{multline}
\Sigma_{ij}^{\lessgtr}\left(  t_{\mathrm{i}}\right)  =\int_{-t}^{t}\Sigma
_{ij}^{\lessgtr}\left(  t_{\mathrm{i}}+\tfrac{\delta t_{\mathrm{i}}}%
{2},t_{\mathrm{i}}-\tfrac{\delta t_{\mathrm{i}}}{2}\right)  \mathrm{d}\delta
t_{\mathrm{i}}\\
=\left\vert U_{12}^{{}}\right\vert ^{2}\hbar_{{}}^{2}\left(  2\pi\right)
^{2}\left[  \int_{-t}^{t}\left[  g_{2}\left(  \delta t_{\mathrm{i}}\right)
\right]  ^{2}\mathrm{d}\delta t_{\mathrm{i}}\right]  \mathrm{f}_{2}\left[
1-\mathrm{f}_{2}\right]  G_{11}^{\lessgtr}\left(  t_{\mathrm{i}}%
,t_{\mathrm{i}}\right)  ~\delta_{i1}^{{}}\delta_{1j}^{{}}\\
\pm\left\vert V_{12}^{{}}\right\vert ^{2}~\mathrm{i}2\pi\left[  \int_{-t}%
^{t}g_{2}\left(  \delta t_{\mathrm{i}}\right)  \mathrm{d}\delta t_{\mathrm{i}%
}\right]  \left(
\genfrac{}{}{0pt}{}{\mathrm{f}_{2}}{1-\mathrm{f}_{2}}%
\right)  \delta_{i1}^{{}}\delta_{1j}^{{}},
\end{multline}
which is local in space and time.\emph{ }This assumption for the time scales
can be seen in fig. \ref{Fig_system_feynman} b) as a collapse of a pair of
black dots, along a vertical line, into a single point. This justifies the
expansion of fig. \ref{Fig_system_feynman} c) and\ the use of the ladder
approximation containing only vertical interaction lines in fig.
\ref{Fig_system_feynman} d).

Assuming $E_{i}=0$ for $i=1,..,\infty$ we obtain for the decay rates
\begin{align}
\frac{1}{\tau_{\mathrm{SE}}}  &  \equiv\tfrac{\mathrm{i}}{\hbar}\left(
\Sigma_{11}^{\mathrm{A}}-\Sigma_{11}^{\mathrm{R}}\right) \nonumber\\
&  =\tfrac{\mathrm{i}}{\hbar}\left(  \Sigma_{11}^{>}-\Sigma_{11}^{<}\right)
\nonumber\\
&  =\left\vert U_{12}^{{}}\right\vert ^{2}~\left(  2\pi\right)  ^{2}\left[
\int_{-t}^{t}\left[  g_{2}\left(  \delta t_{\mathrm{i}}\right)  \right]
^{2}\mathrm{d}\delta t_{\mathrm{i}}\right]  ~\mathrm{f}_{2}\left[
1-\mathrm{f}_{2}\right]  +\tfrac{1}{\hbar}\left\vert V_{12}^{{}}\right\vert
^{2}~2\pi\left[  \int_{-t}^{t}g_{2}\left(  \delta t_{\mathrm{i}}\right)
\mathrm{d}\delta t_{\mathrm{i}}\right] \nonumber\\
&  =\tfrac{2}{\hbar}\left(  \Gamma_{U}+\Gamma_{V}\right)  , \label{decay-rate}%
\end{align}
where we have used $t\gg\hbar/V_{\mathrm{B}}$ to define
\begin{equation}
\Gamma_{U}=\hbar\left\vert U_{12}^{{}}\right\vert ^{2}~2\pi^{2}\left[
\int_{-\infty}^{\infty}\left[  g_{2}\left(  \delta t_{\mathrm{i}}\right)
\right]  ^{2}\mathrm{d}\delta t_{\mathrm{i}}\right]  ~\mathrm{f}_{2}\left[
1-\mathrm{f}_{2}\right]  , \label{Gamma_U-general}%
\end{equation}
the Coulomb decay rate, and
\begin{equation}
\Gamma_{V}=\left\vert V_{12}^{{}}\right\vert ^{2}~\pi\left[  \int_{-\infty
}^{\infty}g_{2}\left(  \delta t_{\mathrm{i}}\right)  \mathrm{d}\delta
t_{\mathrm{i}}\right]  ,
\end{equation}
the hopping decay rate. If one assumes that the environment (\ref{He}) can be
represented by a linear chain with all the hoppings equal to $V_{\mathrm{B}}$
the LDoS is \cite{CPL05}:
\begin{equation}
N_{2}\left(  \varepsilon\right)  =1/\left(  \pi V_{\mathrm{B}}\right)
\sqrt{1-\left(  \frac{\varepsilon}{2V_{\mathrm{B}}}\right)  ^{2}}.
\label{LDoS_Chain}%
\end{equation}
Thus, the Green's function
\begin{equation}
g_{2}\left(  \delta t_{\mathrm{i}}\right)  =\frac{1}{2\pi V_{\mathrm{B}}}%
\frac{J_{1}\left(  \frac{2V_{\mathrm{B}}}{\hbar}\delta t_{i}\right)  }{\delta
t_{i}} \label{Green_function_of_an_infinite_chain}%
\end{equation}
is proportional to the first order Bessel function and decays within a
characteristic time $\hbar/V_{\mathrm{B}}.$ Assuming that $\mathrm{f}_{2}=1/2$
and the integration limits in the $\Gamma$'s expressions are taken to infinity
because $t\sim\hbar/V_{01}\gg\hbar/V_{\mathrm{B}}$ (wide band approximation),
one obtains
\begin{equation}
\Gamma_{U}=\tfrac{2\pi}{\hbar}\left\vert U_{12}^{{}}\right\vert ^{2}\dfrac
{2}{3\pi^{2}V_{\mathrm{B}}} \label{gammaU}%
\end{equation}
and
\begin{equation}
\Gamma_{V}=\tfrac{2\pi}{\hbar}\left\vert V_{12}^{{}}\right\vert ^{2}\frac
{1}{\pi V_{\mathrm{B}}}. \label{gammaV}%
\end{equation}
It is important to remark that the wide band limit can be relaxed because the
FGR holds when time $t$ is in the range \cite{Elena06} $t_{\mathrm{B}%
}<t<t_{\mathrm{R}}\simeq\alpha\hbar/\Gamma_{\mathrm{SE}}\ln(V_{\mathrm{B}%
}/\Gamma_{\mathrm{SE}}),$ where $\alpha$ depends on the van Hove singularities
of the spectral density $\mathcal{J}_{\mathrm{2}}(\varepsilon)=\int
N_{\mathrm{2}}(\varepsilon-\varepsilon^{\prime})N_{\mathrm{2}}(\varepsilon
^{\prime})\mathrm{d}\varepsilon^{\prime}$. Here, $t_{\mathrm{B}}%
=\hbar\mathcal{J}_{\mathrm{2}}(0)\simeq\hbar/V_{\mathrm{B}}$ is the survival
time of an electron-hole excitation at the surface site and $t_{\mathrm{R}}$
characterizes the transition to a power law decay arising from memory effects.
Hence, as long as $\Gamma_{\mathrm{SE}},V_{01}\ll V_{\mathrm{B}},$ the FGR is
valid for times much longer than $\hbar/\Gamma_{\mathrm{SE}}$.

Since the interaction is local in time, the reduced density results:%
\begin{equation}
\mathbf{G}^{<}\left(  t,t\right)  =\hbar^{2}\mathbf{G}^{\mathrm{R}}\left(
t,0\right)  \mathbf{G}^{<}\left(  0,0\right)  \mathbf{G}^{\mathrm{A}}\left(
0,t\right)  +\int_{0}^{t}\mathrm{d}t_{\mathrm{i}}\mathbf{G}^{\mathrm{R}%
}\left(  t,t_{\mathrm{i}}\right)  \mathbf{\Sigma}_{{}}^{<}\left(
t_{\mathrm{i}}\right)  \mathbf{G}^{\mathrm{A}}\left(  t_{\mathrm{i}},t\right)
, \label{GLBE-inhomogeneous}%
\end{equation}
which\ is complemented with%
\begin{equation}
\mathbf{\Sigma}_{{}}^{<}\left(  t_{\mathrm{i}}\right)  =\left(
\begin{array}
[c]{cc}%
0 & 0\\
0 & 2\Gamma_{U}^{{}}\hbar G_{11}^{<}\left(  t_{\mathrm{i}},t_{\mathrm{i}%
}\right)  +2\Gamma_{V}^{{}}\hbar\left(  \frac{\mathrm{i}}{\hbar}\mathrm{f}%
_{2}^{{}}\right)
\end{array}
\right)  .
\end{equation}
Here, the propagators $\mathbf{G}^{\mathrm{R}}\left(  t,0\right)  $ and
$\mathbf{G}^{\mathrm{A}}\left(  0,t\right)  $ that enter in both terms are
obtained from the effective Hamiltonian of the reduced system,%
\begin{equation}
\mathbf{H}_{\mathrm{eff.}}=\left(
\begin{array}
[c]{cc}%
0 & -V_{01}\\
-V_{01} & -\mathrm{i}\Gamma_{\mathrm{SE}}%
\end{array}
\right)  ,
\end{equation}
where $\Gamma_{\mathrm{SE}}$ is energy independent. This results in an
equation of the form of the GLBE. However, the Hamiltonian is asymmetric in
the SE interaction complicating the form of the associated propagator. The
apparent complexity to solve this equation contrasts with the homogeneous case
where the evolution of the GLBE was obtained \cite{GLBE1} through a Laplace
transform. Our strategy will be to induce such form of symmetry.

\subsubsection{A fictitious homogeneous decay}

The main difficulty with the eq. (\ref{GLBE-inhomogeneous}) is that it
involves multiple exponentials. In order to create propagators with an
homogeneous decay, i.e. a single exponential factor, we introduce
\emph{fictitious interactions,} $\mathbf{\Sigma}_{\mathrm{fic.}}^{\mathrm{R}%
},~$with the environment. The symmetric Hamiltonian becomes%
\begin{align}
\mathbf{H}_{\mathrm{sym.}}  &  =\mathbf{H}_{\mathrm{eff.}}+\mathbf{\Sigma
}_{\mathrm{fic.}}^{\mathrm{R}}\nonumber\\
&  =\left(
\begin{array}
[c]{cc}%
0 & -V_{01}\\
-V_{01} & -\mathrm{i}\Gamma_{\mathrm{SE}}%
\end{array}
\right)  +\left(
\begin{array}
[c]{cc}%
-\mathrm{i}\frac{1}{2}\Gamma_{\mathrm{SE}} & 0\\
0 & \mathrm{i}\frac{1}{2}\Gamma_{\mathrm{SE}}%
\end{array}
\right) \nonumber\\
&  =\left(
\begin{array}
[c]{cc}%
-\mathrm{i}\frac{1}{2}\Gamma_{\mathrm{SE}} & -V_{12}\\
-V_{12} & -\mathrm{i}\frac{1}{2}\Gamma_{\mathrm{SE}}%
\end{array}
\right)  .
\end{align}
Here $\mathbf{\Sigma}_{\mathrm{fic.}}^{\mathrm{R}}$ includes the fictitious
interactions which, in the present case, produce a \emph{leak of probability}
in site $0$ at a rate $\Gamma_{\mathrm{SE}}/\hbar$ while in site $1$
\emph{inject probability} at the same rate. Both states of $\mathbf{H}%
_{\mathrm{sym.}}$ interact with the environment independently with the same
characteristic rate $\Gamma_{\mathrm{SE}}/\hbar.$ Note that this rate is half
the real value. The propagators of eq. (\ref{Danielewicz_evol}) have now a
simple dependence on $t$ as
\begin{equation}
\mathbf{G}^{\mathrm{R}}\left(  t,0\right)  =\mathbf{G}^{0\mathrm{R}}\left(
t,0\right)  e^{-\frac{\Gamma_{\mathrm{SE}}}{2}t/\hbar},
\end{equation}
\emph{ }where
\begin{equation}
G_{00}^{0\mathrm{R}}(t,0)=G_{11}^{0\mathrm{R}}(t,0)=\frac{\mathrm{i}}{\hbar
}\cos\left(  \frac{\omega_{0}}{2}t\right)
\end{equation}
and
\begin{equation}
G_{01}^{0\mathrm{R}}(t,0)=G_{10}^{0\mathrm{R}}(t,0)^{\ast}=\frac{\mathrm{i}%
}{\hbar}\sin\left(  \frac{\omega_{0}}{2}t\right)
\end{equation}
are the isolated system propagators. The reduced density evolution is now,%
\begin{multline}
\mathbf{G}^{<}\left(  t,t\right)  =\hbar^{2}\mathbf{G}^{0\mathrm{R}}\left(
t,0\right)  \mathbf{G}^{<}\left(  0,0\right)  \mathbf{G}^{0\mathrm{A}}\left(
0,t\right)  e^{-t/2\tau_{\mathrm{SE}}}\label{Danielewicz_GLBE}\\
+\int_{0}^{t}\mathrm{d}t_{\mathrm{i}}\mathbf{G}^{0\mathrm{R}}\left(
t,t_{\mathrm{i}}\right)  \mathbf{\Sigma}_{\mathrm{sym.}}^{<}\left(
t_{\mathrm{i}}\right)  \mathbf{G}^{0\mathrm{A}}\left(  t_{\mathrm{i}%
},t\right)  e^{-\left(  t-t_{\mathrm{i}}\right)  /2\tau_{\mathrm{SE}}},
\end{multline}
which is similar to the GLBE \cite{GLBE1,GLBE2}. It is easy to see that the
introduction of negative/positive imaginary parts in the diagonal energies of
the effective Hamiltonian produces a decay/growth rates of the elements of the
density function which, being fictitious, must be compensated by a fictitious
injection self-energy
\begin{equation}
\Sigma_{\mathrm{fic.},ij}^{<}\left(  t_{\mathrm{i}}\right)  =-\hbar
\operatorname{Im}\left(  \Sigma_{\mathrm{fic.},ii}^{\mathrm{R}}+\Sigma
_{\mathrm{fic.},jj}^{\mathrm{R}}\right)  G_{ij}^{<}\left(  t_{\mathrm{i}%
},t_{\mathrm{i}}\right)  .
\end{equation}
In our case, this results in an injection that includes the compensation
effects for the symmetrized interaction,%
\begin{multline}
\mathbf{\Sigma}_{\mathrm{sym.}}^{<}\left(  t_{\mathrm{i}}\right)
=\mathbf{\Sigma}_{{}}^{<}\left(  t_{\mathrm{i}}\right)  +\mathbf{\Sigma
}_{\mathrm{fic.}}^{<}\left(  t_{\mathrm{i}}\right) \\
=\left(
\begin{array}
[c]{cc}%
0 & 0\\
0 & 2\Gamma_{U}^{{}}\hbar G_{11}^{<}\left(  t_{\mathrm{i}},t_{\mathrm{i}%
}\right)  +2\Gamma_{V}^{{}}\hbar\left(  \frac{\mathrm{i}}{\hbar}\mathrm{f}%
_{2}^{{}}\right)
\end{array}
\right)  +\left(
\begin{array}
[c]{cc}%
\Gamma_{\mathrm{SE}}\hbar G_{00}^{<}\left(  t_{\mathrm{i}},t_{\mathrm{i}%
}\right)  & 0\\
0 & -\Gamma_{\mathrm{SE}}\hbar G_{11}^{<}\left(  t_{\mathrm{i}},t_{\mathrm{i}%
}\right)
\end{array}
\right)  .
\end{multline}
Here, the second term is proportional to the local density functions
$G_{ii}^{<}\left(  t_{\mathrm{i}},t_{\mathrm{i}}\right)  $ injecting and
extracting density on sites $0$ and $1$ respectively to restore the real
occupation. We can rewrite the last expression to separate the processes that
involve density relaxation (through injection and escape processes) and pure
decoherence (through local energy fluctuations):
\begin{align}
\mathbf{\Sigma}_{\mathrm{sym.}}^{<}\left(  t_{\mathrm{i}}\right)   &
=\Sigma_{\mathrm{i}}^{<}\left(  t_{\mathrm{i}}\right)  +\Sigma_{\mathrm{m}%
}^{<}\left(  t_{\mathrm{i}}\right) \nonumber\\
&  =\mathrm{i}\Gamma_{\mathrm{SE}}\left[  2p_{V}\left(
\begin{array}
[c]{cc}%
0 & 0\\
0 & \left(  \mathrm{f}_{2}-\frac{\hbar}{\mathrm{i}}G_{11}^{<}\left(
t_{\mathrm{i}},t_{\mathrm{i}}\right)  \right)
\end{array}
\right)  +\left(
\begin{array}
[c]{cc}%
\frac{\hbar}{\mathrm{i}}G_{00}^{<}\left(  t_{\mathrm{i}},t_{\mathrm{i}}\right)
& 0\\
0 & \frac{\hbar}{\mathrm{i}}G_{11}^{<}\left(  t_{\mathrm{i}},t_{\mathrm{i}%
}\right)
\end{array}
\right)  \right]  . \label{sigma_stroboscopio}%
\end{align}
Here
\begin{equation}
\frac{\hbar}{\mathrm{i}}G_{11}^{<}\left(  t_{\mathrm{i}},t_{\mathrm{i}%
}\right)  \equiv\frac{\hbar}{\mathrm{i}}\int G_{11}^{<}\left(  \varepsilon
,t_{\mathrm{i}}\right)  \frac{\mathrm{d}\varepsilon}{2\pi\hbar}=\mathrm{f}%
_{1}\left(  t_{\mathrm{i}}\right)
\end{equation}
and
\begin{equation}
\frac{\hbar}{\mathrm{i}}G_{00}^{<}\left(  t_{\mathrm{i}},t_{\mathrm{i}%
}\right)  =\mathrm{f}_{0}\left(  t_{\mathrm{i}}\right)  ,
\end{equation}
while
\begin{equation}
p_{V}=\Gamma_{V}/\Gamma_{\mathrm{SE}}%
\end{equation}
is the weight of the tunneling rate relative to the total SE interaction rate.
As the initial state has the site $2$ occupied we have that
\begin{equation}
\frac{\hbar}{\mathrm{i}}G_{ij}^{<}\left(  0,0\right)  =\delta_{i1}\delta_{1j}.
\end{equation}
Introducing eq. (\ref{sigma_stroboscopio}) into eq. (\ref{Danielewicz_GLBE})
and using
\begin{equation}
\frac{1}{\tau_{\mathrm{SE}}}\equiv\tfrac{2}{\hbar}\Gamma_{\mathrm{SE}}%
\end{equation}
we get two coupled equations for $G_{00}^{<}$ and $G_{11}^{<}$%
\begin{multline}
\tfrac{\hbar}{\mathrm{i}}G_{00}^{<}\left(  t,t\right)  =\left\vert \hbar
G_{01}^{0\mathrm{R}}\left(  t,0\right)  \right\vert ^{2}e^{-t/\left(
2\tau_{\mathrm{SE}}\right)  }+\label{GLBE_probability1}\\
\int\left\vert \hbar G_{01}^{0\mathrm{R}}\left(  t,t_{\mathrm{i}}\right)
\right\vert ^{2}e^{-\left(  t-t_{\mathrm{i}}\right)  /2\tau_{\mathrm{SE}}%
}~2p_{V}~\frac{\mathrm{d}t_{\mathrm{i}}}{2\tau_{\mathrm{SE}}}\left[
\mathrm{f}_{2}-\tfrac{\hbar}{\mathrm{i}}G_{11}^{<}\left(  t_{\mathrm{i}%
},t_{\mathrm{i}}\right)  \right] \\
+\int\left\vert \hbar G_{00}^{0\mathrm{R}}\left(  t,t_{\mathrm{i}}\right)
\right\vert ^{2}e^{-(t-t_{\mathrm{i}})/\left(  2\tau_{\mathrm{SE}}\right)
}\frac{\mathrm{d}t_{\mathrm{i}}}{2\tau_{\mathrm{SE}}}\left[  \tfrac{\hbar
}{\mathrm{i}}G_{00}^{<}\left(  t_{\mathrm{i}},t_{\mathrm{i}}\right)  \right]
\\
+\int\left\vert \hbar G_{01}^{0\mathrm{R}}\left(  t,t_{\mathrm{i}}\right)
\right\vert ^{2}e^{-(t-t_{\mathrm{i}})/\left(  2\tau_{\mathrm{SE}}\right)
}\frac{\mathrm{d}t_{\mathrm{i}}}{2\tau_{\mathrm{SE}}}\left[  \tfrac{\hbar
}{\mathrm{i}}G_{11}^{<}\left(  t_{\mathrm{i}},t_{\mathrm{i}}\right)  \right]
.
\end{multline}%
\begin{multline}
\tfrac{\hbar}{\mathrm{i}}G_{11}^{<}\left(  t,t\right)  =\left\vert \hbar
G_{11}^{0\mathrm{R}}\left(  t,0\right)  \right\vert ^{2}e^{-t/\left(
2\tau_{\mathrm{SE}}\right)  }+\label{GLBE_probability2}\\
\int\left\vert \hbar G_{11}^{0\mathrm{R}}\left(  t,t_{\mathrm{i}}\right)
\right\vert ^{2}e^{-\left(  t-t_{\mathrm{i}}\right)  /2\tau_{\mathrm{SE}}%
}~2p_{V}~\frac{\mathrm{d}t_{\mathrm{i}}}{2\tau_{\mathrm{SE}}}\left[
\mathrm{f}_{2}-\tfrac{\hbar}{\mathrm{i}}G_{11}^{<}\left(  t_{\mathrm{i}%
},t_{\mathrm{i}}\,\right)  \right] \\
+\int\left\vert \hbar G_{10}^{0\mathrm{R}}\left(  t,t_{\mathrm{i}}\right)
\right\vert ^{2}e^{-(t-t_{\mathrm{i}})/\left(  2\tau_{\mathrm{SE}}\right)
}\frac{\mathrm{d}t_{\mathrm{i}}}{2\tau_{\mathrm{SE}}}\left[  \tfrac{\hbar
}{\mathrm{i}}G_{00}^{<}\left(  t_{\mathrm{i}},t_{\mathrm{i}}\right)  \right]
\\
+\int\left\vert \hbar G_{11}^{0\mathrm{R}}\left(  t,t_{\mathrm{i}}\right)
\right\vert ^{2}e^{-(t-t_{\mathrm{i}})/\left(  2\tau_{\mathrm{SE}}\right)
}\frac{\mathrm{d}t_{\mathrm{i}}}{2\tau_{\mathrm{SE}}}\left[  \tfrac{\hbar
}{\mathrm{i}}G_{11}^{<}\left(  t_{\mathrm{i}},t_{\mathrm{i}}\right)  \right]
.
\end{multline}
In each equation, the first term is the probability that a particle initially
at site $1$ be found in site $0$ (or $1$) at time $t$ having survived the
interactions with the environment with a probability $e^{-t/\left(
2\tau_{\mathrm{SE}}\right)  }$. The second term describes the process of
injection/escape of particles enabled by the hopping from/towards the
reservoir, where the last of such processes occurred in the time range
($t_{\mathrm{i}},t_{\mathrm{i}}+\mathrm{d}t_{i}$) with a probability
$2p_{V}\frac{\mathrm{d}t_{\mathrm{i}}}{2\tau_{\mathrm{SE}}}.$ The
injection/escape is produced on site $1$ and fill/empty the site to level it
to the occupation factor $\mathrm{f}_{2}$. The third and fourth terms take
into account the last process of measurement at time $t_{\mathrm{i}}$ due to
the SE interaction with a probability $\tfrac{\mathrm{d}t_{\mathrm{i}}}%
{2\tau_{\mathrm{SE}}}$. This confirms our interpretation that in eq.
(\ref{sigma_stroboscopio}) the dissipation processes are in $\mathbf{\Sigma
}_{\mathrm{i}}^{<}\left(  t\right)  $ while $\mathbf{\Sigma}_{\mathrm{m}}%
^{<}\left(  t\right)  $ involves pure decoherence. It is clear that by
iterating this formula, one gets a series in the form represented in fig.
\ref{Fig_system_feynman} d)\emph{.}

\subsubsection{The dynamics of a swapping gate}

The solution of the coupled eqs. (\ref{GLBE_probability1}) and
(\ref{GLBE_probability2}) involves a Laplace transform.
We consider a parameter range compatible with the spin problem where
$\mathrm{f}_{2}\lesssim1$ while we allow the tunneling relative weight $p_{V}$
in the range $\left[  0,1\right]  $.$\ $In a compact notation, the density
function results:
\begin{equation}
P_{01}\left(  t\right)  =\tfrac{\hbar}{\mathrm{i}}G_{00}^{<}\left(
t,t\right)  =1-a_{0}e^{-R_{0}t}-a_{1}\cos\left[  \left(  \omega+\mathrm{i}%
\eta\right)  t+\varphi_{0}\right]  e^{-R_{1}t}. \label{G11}%
\end{equation}
Here, the decay rates $R_{0},~R_{1}$ and $\eta$, and the oscillation frequency
$\omega$ are real numbers associated with poles of the Laplace transform.
The amplitude $a_{0}$ is also real while, when $\omega=0,$ the amplitude
$a_{1}$ and the initial phase $\varphi_{0}$ acquire an imaginary component
that warrants a real density.
These observables have expressions in terms of adimensional functions of the
fundamental parameters in the model. Denoting
\begin{equation}
x=\omega_{0}\tau_{\mathrm{SE}}%
\end{equation}
and remembering that
\begin{equation}
p_{V}=\Gamma_{V}/\Gamma_{\mathrm{SE}},
\end{equation}
we define
\begin{equation}
\phi\left(  p_{V},x\right)  =\frac{1}{3}\left(  x^{2}-p_{V}^{2}-\frac{1}%
{3}\left(  1-p_{V}\right)  ^{2}\right)  ,
\end{equation}
and
\begin{multline}
\chi\left(  p_{V},x\right)  =\left\{  4\left(  1-p_{V}\right)  \left(
9x^{2}-2\left(  1-p_{V}\right)  ^{2}+18p_{V}^{2}\right)  \right. \\
\left.  +12\left[  3\left(  4x^{6}-\left(  \left(  1-p_{V}\right)
^{2}+12p_{V}^{2}\right)  x^{4}\right.  \right.  \right. \\
\left.  \left.  \left.  +4p_{V}^{2}\left(  5\left(  1-p_{V}\right)
^{2}+3p_{V}^{2}\right)  x^{2}-4p_{V}^{2}\left(  \left(  1-p_{V}\right)
^{2}-p_{V}^{2}\right)  ^{2}\right)  \right]  ^{\frac{1}{2}}\right\}
^{\frac{1}{3}}.
\end{multline}
The observable \textquotedblleft frequency\textquotedblright,
\begin{equation}
\omega+\mathrm{i}\eta=\frac{\sqrt{3}}{2x}\left(  \frac{1}{6}\chi\left(
p_{V},x\right)  +6\frac{\phi\left(  p_{V},x\right)  }{\chi\left(
p_{V},x\right)  }\right)  \omega_{0}, \label{w}%
\end{equation}
is purely real or imaginary, i.e. $\omega\eta\equiv0$. Also,
\begin{align}
R_{0}  &  =\left(  6\frac{\phi\left(  p_{V},x\right)  }{\chi\left(
p_{V},x\right)  }-\frac{1}{6}\chi\left(  p_{V},x\right)  +p_{V}+\frac{1}%
{3}\left(  1-p_{V}\right)  \right)  \frac{1}{\tau_{\mathrm{SE}}},\\
R_{1}  &  =\frac{3}{2}\left(  p_{V}+\frac{1}{3}\left(  1-p_{V}\right)
\right)  \frac{1}{\tau_{\mathrm{SE}}}-\frac{R_{0}}{2},
\end{align}
and
\begin{align}
a_{0}  &  =\frac{1}{2}\frac{2\left(  \omega_{{}}^{2}-\eta_{{}}^{2}\right)
+2R_{1}^{2}-\omega_{0}^{2}}{\left(  \omega_{{}}^{2}-\eta_{{}}^{2}\right)
+\left(  R_{0}^{{}}-R_{1}^{{}}\right)  ^{2}},\\
a_{2}  &  =\dfrac{1}{2\left(  \omega+\mathrm{i}\eta\right)  }\frac{\left(
2R_{0}^{{}}R_{1}^{{}}-\omega_{0}^{2}\right)  \left(  R_{0}-R_{1}\right)
+2\left(  \omega_{{}}^{2}-\eta_{{}}^{2}\right)  R_{0}^{{}}}{\left(  \omega
_{{}}^{2}-\eta_{{}}^{2}\right)  +\left(  R_{0}^{{}}-R_{1}^{{}}\right)  ^{2}%
},\\
a_{3}  &  =\frac{1}{2}\frac{\omega_{0}^{2}+2R_{0}^{2}-4R_{0}^{{}}R_{1}^{{}}%
}{\left(  \omega_{{}}^{2}-\eta_{{}}^{2}\right)  +\left(  R_{0}^{{}}-R_{1}^{{}%
}\right)  ^{2}},\\
a_{1}^{2}  &  =a_{2}^{2}+a_{3}^{2},\;\;\;\;\;\tan\left(  \phi_{0}\right)
=-\frac{a_{2}}{a_{3}}.
\end{align}
The oscillation frequency $\omega$ in eq. (\ref{w}) has a critical point
$x_{\mathrm{c}}$ at a finite value of $x$ showing, as we called, a quantum
dynamical phase transition \cite{JCP06,SSC07} for which $\omega$ and $\eta$ in
eq. (\ref{G11}) exchange their roles as being zero and having a finite value
respectively. A full discussion of this issue for a spin system will be
presented in chapter \ref{Sec_QDPT}. Here, the dynamical behavior changes from
a swapping phase to an overdamped phase. This last regime can be associated
with the Quantum Zeno effect \cite{Misra77} where\ frequent projective
measurements prevent the quantum evolution. Here, this is a dynamical effect
\cite{Usaj98, Pascazio94} produced by interactions with the environment that
freeze the system oscillation.

\begin{figure}
[tbh]
\begin{center}
\includegraphics[
height=4.2376in,
width=5.3558in
]%
{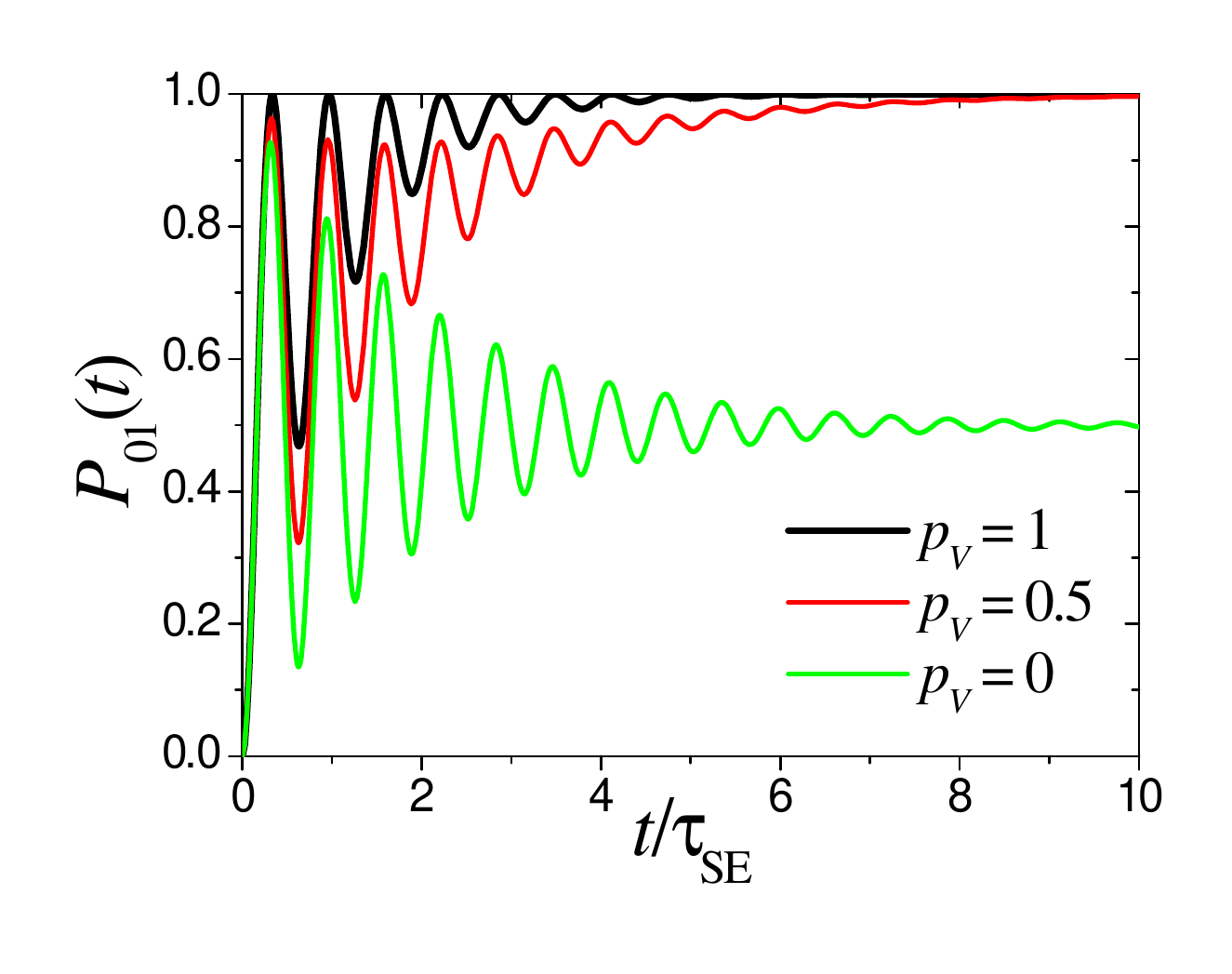}%
\caption[Occupation probability, $P_{01}\left(  t\right)  =$\textrm{i}%
$G_{00}^{<}\left(  t\right)  /\hbar,$ to find at site $0$ a particle when
initially was at site $1$ for a two-level system coupled with a
particle-reservoir.]{Occupation probability, $P_{01}\left(  t\right)
=$\textrm{i}$G_{00}^{<}\left(  t\right)  /\hbar,$ to find a particle at site
$0$ when initially was at site $1.$ Each line corresponds to different kinds,
$p_{V},$ of SE interactions. The plots correspond to $x=V_{01}\tau
_{\mathrm{SE}}/\hbar=10$ belonging to the swapping phase and \textrm{f}%
$_{2}=1$.}%
\label{Fig_G11}%
\end{center}
\end{figure}

Fig. \ref{Fig_G11} shows typical curves of $\tfrac{\hbar}{\mathrm{i}}%
G_{00}^{<}\left(  t,t\right)  $ in the swapping phase. The different colors
correspond to different SE interactions rates, $p_{V}=0$, $0.5$ and $1,$ which
are Coulomb $\left(  \Gamma_{V}=0\right)  $, isotropic $\left(  \Gamma
_{V}=\Gamma_{U}\right)  $ and pure tunneling $\left(  \Gamma_{U}=0\right)  $
interactions rates. The hopping interaction does not conserve the net energy
in the system inducing a dissipation which is manifested through the non
conservation of the number of particles in the system. This is the case of
$p_{V}\neq0$ where the final state of the system has the occupation
probability of the sites equilibrated with the bath occupation (\textrm{f}%
$_{2}$). In fig. \ref{Fig_G11}, this is manifested as the asymptotic
normalized density (occupation probability) of $1$.
However, if $p_{V}=0,$ tunneling is forbidden and the system goes to an
internal quasi-equilibrium as we described in the eq.
(\ref{quasi_equilibrium_2spin}),
i.e. the local excitation is spread inside the system. In this case the
asymptotic occupation probability of site $0$ is $1/2.$
\bigskip

\section{Stroboscopic process \label{Marker_Stroboscopic_process}}

Eq. (\ref{GLBE-inhomogeneous}) has two main difficulties for a numerical
implementation: The first is the evaluation of the system non-unitary
propagators under inhomogeneous perturbations. The second is to keep track of
all previous states of the system to enable the integration over previous
times. We will show that the decay homogenization enables the implementation
of an efficient numerical algorithm. First of all, we identify in expression
(\ref{Danielewicz_GLBE}) that $e^{-t/\left(  2\tau_{\mathrm{SE}}\right)
}=s\left(  t\right)  $ is the system's survival probability to the environment
interruption, i.e., the probability that the system remains coherent, and
$\mathrm{d}t_{\mathrm{i}}/(2\tau_{\mathrm{SE}})=q\left(  t_{\mathrm{i}%
}\right)  \mathrm{d}t_{\mathrm{i}}$ is the \textquotedblleft
interruption\textquotedblright\ probability in a differential time around
$t_{\mathrm{i}}$. The interaction of the environment is discretized in
intervals $\tau_{\mathrm{str.}}$ where it acts instantaneously. This
stroboscopic interaction leads to:%
\begin{align}
s\left(  t\right)   &  =\left(  1-p\right)  ^{n\left(  t\right)
},\label{survivep}\\
q\left(  t\right)   &  =\sum_{m=1}^{\infty}p~\delta\left(  t-m\tau
_{\mathrm{str.}}\right)  \label{interruptionp}%
\end{align}
where
\begin{equation}
n\left(  t\right)  =\mathrm{int}\left(  t/\tau_{\mathrm{str.}}\right)  .
\end{equation}
Here, the stroboscopic interruptions may occur at the discrete times
$m\tau_{\mathrm{str.}}$ with a probability $p$. At time $t$ there were
$n\left(  t\right)  $ possible interruptions. In the joint limit
$\tau_{\mathrm{str.}}\rightarrow0$ and $p\rightarrow0$ such that
\begin{equation}
p/\tau_{\mathrm{str.}}=1/\left(  2\tau_{\mathrm{SE}}\right)  ,
\end{equation}
we can recover the continuous expression. To do this, notice that if $n\left(
t\right)  =n,$ one can write eq. (\ref{survivep}) as%
\begin{equation}
s\left(  t\right)  =\left(  1-p\right)  ^{\frac{\left(  n\tau_{\mathrm{str.}%
}\right)  }{\tau_{\mathrm{str.}}}}=\left(  1-\frac{\tau_{\mathrm{str.}}}%
{2\tau_{\mathrm{SE}}}\right)  ^{\left(  n\tau_{\mathrm{str.}}\right)
/\tau_{\mathrm{str.}}}.
\end{equation}
If $t=n\tau_{\mathrm{str.}}$ then%
\begin{equation}
s\left(  t\right)  =\left(  1-\frac{\tau_{\mathrm{str.}}}{2\tau_{\mathrm{SE}}%
}\right)  ^{t/\tau_{\mathrm{str.}}}.
\end{equation}
By taking the limit $\tau_{\mathrm{str.}}\rightarrow0$ the variable $t$
becomes continuous yielding%
\begin{align}
s\left(  t\right)   &  =\lim_{\tau_{\mathrm{str.}}\rightarrow0}\left(
1-\frac{\tau_{\mathrm{str.}}}{2\tau_{\mathrm{SE}}}\right)  ^{t/\tau
_{\mathrm{str.}}}\nonumber\\
&  =\exp\left[  -t/\left(  2\tau_{\mathrm{SE}}\right)  \right]  ,
\end{align}
which is the continuous expression for $s\left(  t\right)  .$

Then, by substituting $p=\tau_{\mathrm{str.}}/(2\tau_{\mathrm{SE}})$ in eq.
(\ref{interruptionp}) we have%
\begin{equation}
q(t)=\frac{1}{2\tau_{\mathrm{SE}}}\sum_{m=1}^{\infty}\tau_{\mathrm{str.}%
}\delta(t-m\tau_{\mathrm{str.}}).
\end{equation}
In the limit $\tau_{\mathrm{str.}}\rightarrow0,$ $t_{m}=m\tau_{\mathrm{str.}}$
becomes a continuous variable and we can convert the sum into an integral,
leading to continuous expression of the GLBE (\ref{Danielewicz_GLBE}):
\begin{equation}
q(t)=\frac{1}{2\tau_{\mathrm{SE}}}\int_{0}^{\infty}\tau_{\mathrm{str.}}%
\delta(t-t_{m})\frac{\mathrm{d}t_{m}}{\tau_{\mathrm{str.}}}=\frac{1}%
{2\tau_{\mathrm{SE}}}.
\end{equation}

Coming back to the discrete version, we introduce eqs. (\ref{survivep}) and
(\ref{interruptionp}) into the reduced density expression
(\ref{Danielewicz_GLBE}) to obtain%
\begin{multline}
\mathbf{G}^{<}\left(  t,t\right)  =\hbar^{2}\mathbf{G}^{0\mathrm{R}}\left(
t,0\right)  \mathbf{G}^{<}\left(  0,0\right)  \mathbf{G}^{0\mathrm{A}}\left(
0,t\right)  \left(  1-p\right)  ^{n(t)}\\
+\int_{0}^{t}\mathrm{d}t_{\mathrm{i}}\tau_{\mathrm{SE}}\sum_{m=1}^{\infty
}\delta\left(  t_{\mathrm{i}}-t_{m}\right)  \mathbf{G}^{0\mathrm{R}}\left(
t,t_{\mathrm{i}}\right)  \mathbf{\Sigma}_{\mathrm{sym.}}^{<}\left(
t_{\mathrm{i}}\right)  \mathbf{G}^{0\mathrm{A}}\left(  t_{\mathrm{i}%
},t\right)  p\left(  1-p\right)  ^{n(t-t_{\mathrm{i}})}.
\end{multline}
After integration, we obtain%
\begin{multline}
\mathbf{G}^{<}\left(  t,t\right)  =\hbar^{2}\mathbf{G}^{0\mathrm{R}}\left(
t,0\right)  \mathbf{G}^{<}\left(  0,0\right)  \mathbf{G}^{0\mathrm{A}}\left(
0,t\right)  \left(  1-p\right)  ^{n}\label{GLBE_stroboscopic}\\
+\hbar^{2}\sum_{m=1}^{n}\mathbf{G}^{0\mathrm{R}}\left(  t,t_{m}\right)
\delta\mathbf{G}_{\mathrm{inj.}}^{<}\left(  t_{m},t_{m}\right)  \mathbf{G}%
^{0\mathrm{A}}\left(  t,t_{m}\right)  p\left(  1-p\right)  ^{n-m},
\end{multline}
where $n=n(t),$ $t_{m}=m\tau_{\mathrm{str.}}$ and
\begin{equation}
\delta\mathbf{G}_{\mathrm{inj.}}^{<}\left(  t,t\right)  =\frac{2\tau
_{\mathrm{SE}}}{\hbar^{2}}\mathbf{\Sigma}_{\mathrm{sym.}}^{<}\left(  t\right)
.
\end{equation}
In this picture, the evolution between interruptions is governed by the
system's\ propagators
\begin{equation}
\mathbf{G}^{0\mathrm{R}}\left(  t,0\right)  =-\frac{\mathrm{i}}{\hbar}%
\exp[-\mathrm{i}\mathbf{H}_{\mathrm{S}}t/\hbar]
\end{equation}
and
\begin{equation}
\mathbf{G}^{0\mathrm{A}}\left(  0,t\right)  =\mathbf{G}^{0\mathrm{R}}\left(
t,0\right)  ^{\dag}.
\end{equation}
The spin-bath stroboscopically interrupts the system evolution producing the
decay of the coherent beam. This decay is compensated through the reinjection
of probability (or eventually of coherences) expressed in the
\emph{instantaneous interruption function}\textit{,} $\delta\mathbf{G}%
_{\mathrm{inj.}}^{<}\left(  t,t\right)  $, which also contains actual
injection/decay from/to the bath.

\begin{figure}
[ptbh]
\begin{center}
\includegraphics[
height=5.3376in,
width=4.9934in
]%
{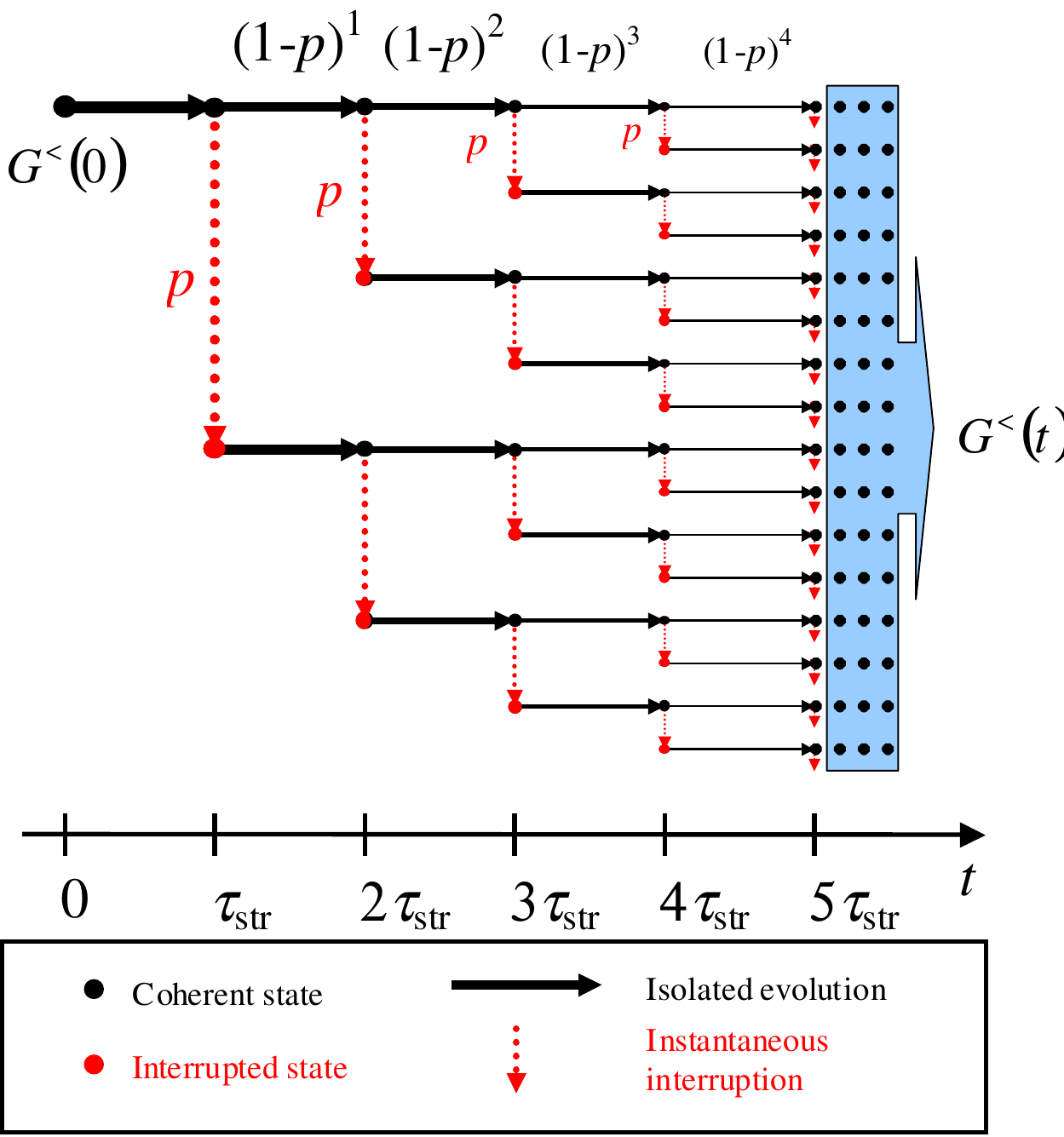}%
\caption[Quantum branching sequence for the stroboscopic evolution.]{Quantum
branching sequence for the stroboscopic evolution. Red dots represent states
with interrupted (incoherent) evolution while the black dots are coherent with
their predecessor. The horizontal continuous arrows represent the isolated
evolution and the vertical dashed lines are the instantaneous interruptions.
Notice the self-similar structure.}%
\label{Fig_stroboscopy}%
\end{center}
\end{figure}

The first term in the rhs of eq. (\ref{GLBE_stroboscopic}) is the coherent
system evolution weighted by its survival probability $\left(  1-p\right)
^{n}.$ This is the upper branch in fig. \ref{Fig_stroboscopy}. The second term
is the incoherent evolution involving all the decoherent branches. The $m$-th
term in the sum represents the evolution that had its \emph{last} interruption
at $m\tau_{\mathrm{str.}}$ and since then survived coherently until
$n\tau_{\mathrm{str.}}$. Each of these terms is represented in fig.
\ref{Fig_stroboscopy} by all the branches with an interrupted state (gray dot,
red online) at the hierarchy level $m$ after which they survive without
further interruptions until $n\tau_{\mathrm{str.}}$. This representation has
an immediate resemblance to that introduced by Pascazio and Namiki to justify
the dynamical Zeno effect \cite{Pascazio94}.%
\emph{ }

As mentioned above, the solutions of eqs. (\ref{GLBE_stroboscopic})
and\ (\ref{Danielewicz_GLBE}) are both computationally demanding since they
involve the storage of all the previous steps and reiterated summations. Thus,
taking advantage of the self-similarity of the hierarchy levels in the
interaction with the environment, we rearrange expression
(\ref{GLBE_stroboscopic}) into a form optimized for numerical computation,%
\begin{multline}
\tfrac{1}{\hbar^{2}}\mathbf{G}^{<}\left(  t_{n+1},t_{n+1}\right)
=\mathbf{G}^{0\mathrm{R}}\left(  t_{n+1},t_{n}\right)  \mathbf{G}^{<}\left(
t_{n},t_{n}\right)  \mathbf{G}^{0\mathrm{A}}\left(  t_{n},t_{n+1}\right)
\left(  1-p\right) \label{G_1step}\\
+\mathbf{G}^{0\mathrm{R}}\left(  t_{n+1},t_{n}\right)  \delta\mathbf{G}%
_{\mathrm{inj.}}^{<}\left(  t_{n},t_{n}\right)  \mathbf{G}^{0\mathrm{A}%
}\left(  t_{n},t_{n+1}\right)  p.
\end{multline}
This equation provides a new computational procedure that only requires the
storage of the density function at a single previous step. Besides, it avoids
random averages required in models that include decoherence through stochastic
or kicked-like perturbations \cite{Teklemarian03,Molmer92}.

\subsection{A nice physical interpretation: The environment as a measurement
apparatus\label{Mark_A_nice_physics_interpretation}}

\begin{figure}
[th]
\begin{center}
\includegraphics[
height=3.9185in,
width=3.0995in
]%
{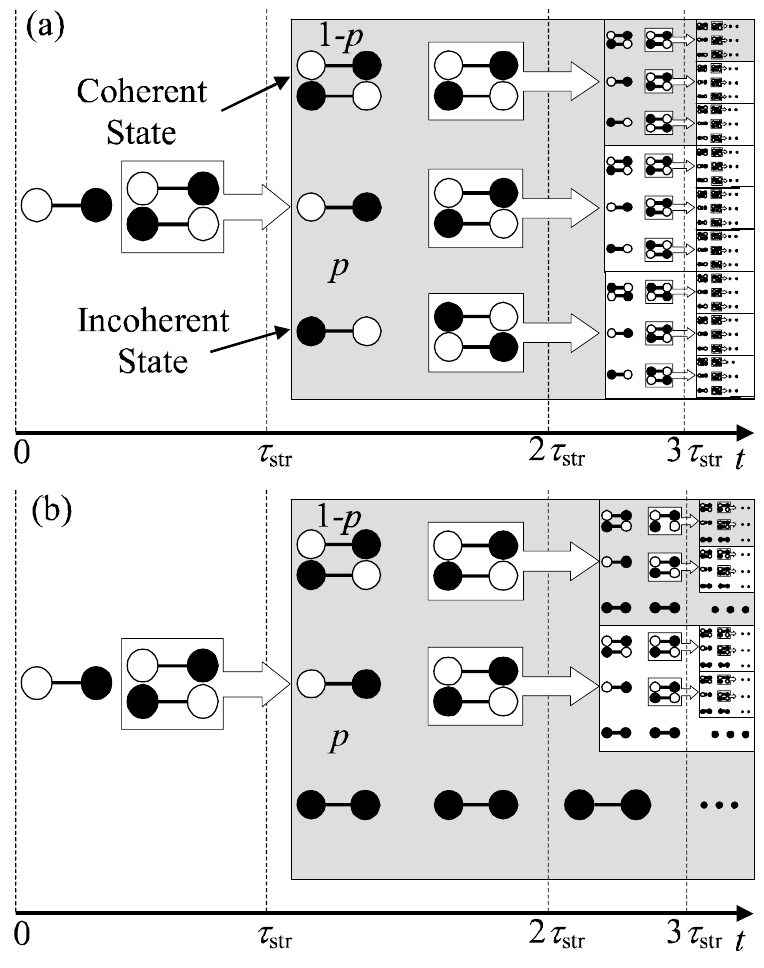}%
\caption[{}Quantum branching sequence for the swapping dynamics.]{{}Quantum
branching sequence for the swapping dynamics. Panel a) stands for a Coulomb
system-environment interaction and b) a pure hopping one. Single states
represent states with interrupted evolution (incoherent) while pairs of states
are coherent superpositions. Notice the self-similar structure.}%
\label{figstrobosc}%
\end{center}
\end{figure}

We introduce our computational procedure operationally for an Coulomb
interaction form ($V_{12}/U_{12}=0$) of $\widehat{\mathcal{H}}_{\mathrm{SE}}$.
The initial state of the isolated two-level system evolves with $\widehat
{\mathcal{H}}_{\mathrm{S}}$. At $\tau_{\mathrm{str.}}$, the particle reservoir
interacts instantaneously with the \textquotedblleft system\textquotedblright%
\ interrupting it with a probability $p$. The actual physical time for the SE
interaction is then obtained as $\tau_{\mathrm{SE}}=\tau_{\mathrm{str.}}/p$.
Considering that the dynamical time scale of the bath ($\tau_{\mathrm{B}%
}\simeq\hbar/V_{\mathrm{B}}$) is much faster than that of the system (fast
fluctuation approximation), the dynamics of site $2$ produces an energy
fluctuation on site $1$ that destroys the coherence of the two-level
\textquotedblleft system\textquotedblright.\ This represents the
\textquotedblleft measurement\textquotedblright\ process that collapses the
\textquotedblleft system\textquotedblright\ state. At time\ $\tau
_{\mathrm{str.}}$, the \textquotedblleft system\textquotedblright\ evolution
splits into three alternatives: with probability $1-p$ the state survives the
interruption and continues its undisturbed evolution, while with probability
$p$ the system is effectively interrupted and its evolution starts again from
each of the \emph{two }eigenstates of $\hat{c}_{1}^{+}\hat{c}_{1}^{{}}$. These
three possible states at $\tau_{\mathrm{str.}}$ evolve freely until the system
is monitored again at time $2\tau_{\mathrm{str.}}$ and a new branching of
alternatives is produced as represented in the scheme of fig.
\ref{figstrobosc} a).%

A similar reasoning holds when $V_{12}\neq0.$ The sequence for isotropic
interaction rates ($\Gamma_{V}=\Gamma_{U}$) is shown in fig. \ref{figstrobosc}
b). The hopping part of $\widehat{\mathcal{H}}_{\mathrm{SE}}$ can inject a
particle. When an interruption occurs, the bath \textquotedblleft
measures\textquotedblright\ at site $2$ and, if found empty, it injects a
particle. The inverse process is not considered because we are assuming that
the occupation factor in the reservoir is over $1/2$ and we are normalizing
the occupation probability to this value. In the figure, this can be
interpreted as a \textquotedblleft pruning\textquotedblright\ of some
incoherent branches increasing the global coherence. In the next section, we
will map the fermion system to a spin system to describe spin dynamics. Thus,
this \textquotedblleft pruning\textquotedblright\ explains why the rate
of\emph{ \textit{decoherence is greater when }}the\emph{\textit{ Ising }%
}(Coulomb) part of $\widehat{\mathcal{H}}_{\mathrm{SE}}$\emph{ dominates over
the }XY\emph{ }(hopping)\emph{ }part\emph{. }This occurs with the dipolar
interaction and, in less degree, with the isotropic one. This contrasts with a
pure XY interaction where the survival of spin coherences is manifested by a
\textquotedblleft spin wave\textquotedblright\ behavior.\cite{Madi97}\emph{
}The empty site of the bath is refilled and de-correlates in a time much
shorter than the swapping between sites $0$ and $1$ (fast fluctuation
approximation). Consequently, the injection can only occur from the bath
toward the system.

\section{Application to spin systems through the Jordan-Wigner
transformation\label{M_Keldysh_applied_to_spin_systems}}

We apply this procedure to the spin system described in section
\S \ \ref{M_2-spin_spin_bath} providing a first principle derivation of the
phenomenological equations employed there. We consider a system with $M=2$
spins $1/2$ coupled to a spin environment with the following Hamiltonian
\begin{equation}
\widehat{\mathcal{H}}=\widehat{\mathcal{H}}_{\mathrm{S}}+\widehat{\mathcal{H}%
}_{\mathrm{E}}+\widehat{\mathcal{H}}_{\mathrm{SE}},
\end{equation}
where the system Hamiltonian, $\widehat{\mathcal{H}}_{\mathrm{S}},$ is
described by eq. (\ref{H_system})%
\begin{align}
\widehat{\mathcal{H}}_{\mathrm{S}}  &  =\hbar\Omega_{\mathrm{Z}}^{{}}\left(
\hat{S}_{{}}^{z}+\hat{I}_{1}^{z}\right)  +\tfrac{1}{2}b\left(  \hat{S}_{{}%
}^{+}\hat{I}_{1}^{-}+\hat{S}_{{}}^{-}\hat{I}_{1}^{+}\right) \nonumber\\
&  =\hbar\Omega_{\mathrm{Z}}^{{}}\left(  \hat{S}_{{}}^{z}+\hat{I}_{1}%
^{z}\right)  +b\left(  \hat{S}_{{}}^{x}\hat{I}_{1}^{x}+\hat{S}_{{}}^{y}\hat
{I}_{1}^{y}\right)  .
\end{align}
The environment Hamiltonian is described by%
\begin{equation}
\widehat{\mathcal{H}}_{\mathrm{E}}=\sum_{i\geq2}\hbar\Omega_{\mathrm{Z}}^{{}%
}\hat{I}_{i}^{z}+%
{\textstyle\sum\limits_{\genfrac{}{}{0pt}{1}{i\geq2}{j>i}}}
\tfrac{1}{2}b_{ij}^{{}}\left(  \hat{I}_{i}^{+}\hat{I}_{j}^{-}+\hat{I}_{i}%
^{-}\hat{I}_{j}^{+}\right)  .
\end{equation}
This equation, in contrast with eq. (\ref{H_E_2_spin_spin_bath_DM}), does not
contain the Ising term. Thus, we have here an XY interaction inside the
spin-bath instead of the dipolar interaction of section
\S \ \ref{M_2-spin_spin_bath}. The other difference is that the SE interaction
described by
\begin{equation}
\widehat{\mathcal{H}}_{\mathrm{SE}}=a_{12}^{{}}\hat{I}_{1}^{z}\hat{I}_{2}%
^{z}+\tfrac{1}{2}b_{12}^{{}}\left(  \hat{I}_{1}^{+}\hat{I}_{2}^{-}+\hat{I}%
_{1}^{-}\hat{I}_{2}^{+}\right)  ,
\end{equation}
has only one connection between the environment ($\hat{I}_{2}$) and the system
($\hat{I}_{1}$) while the given by eq. (\ref{H_SE_2_spin_spin_bath_DM}) has
many. The purpose of the last simplification is only for a simplified
description, but it could be generalized. Remember that, this spin-spin
interaction is Ising if $b_{12}/a_{12}=0,$ and $XY$, isotropic (Heisenberg) or
the truncated dipolar (secular) if $a_{12}/b_{12}=0,1,-2,$ respectively.

In the previous sections of this chapter we have observed the similarity of
the results with the obtained ones in section \S \ \ref{M_2-spin_spin_bath}.
Thus, to use the results obtained within Keldysh formalism we map the spin
system into a fermionic system using the Jordan-Wigner Transformation (JWT)
\cite{Lieb61},
\begin{equation}
\hat{I}_{i}^{+}=\hat{c}_{i}^{+}\exp\left\{  \mathrm{i}\pi\sum_{j=1}^{i-1}%
\hat{c}_{j}^{+}\hat{c}_{j}^{{}}\right\}  .
\end{equation}
The previous Hamiltonians become\footnote{Note that the Jordan-Wigner
transformation maps a linear many-body $XY$ spin Hamiltonian into a system of
non-interacting fermions. This leads us to solve a one-body problem, reducing
the dimension of the Hilbert space from $2^{N}$ to $N$ states that represent
local excitations \cite{Lieb61}.
}%
\begin{align}
\widehat{\mathcal{H}}_{\mathrm{S}}  &  =\hbar\Omega_{\mathrm{Z}}^{{}}\left(
\hat{c}_{0}^{+}\hat{c}_{0}^{{}}+\hat{c}_{1}^{+}\hat{c}_{1}^{{}}-\hat
{1}\right)  +\tfrac{1}{2}b\left(  \hat{c}_{0}^{+}\hat{c}_{1}^{{}}+\hat{c}%
_{1}^{+}\hat{c}_{0}^{{}}\right)  ,\label{HS_spins}\\
\widehat{\mathcal{H}}_{\mathrm{E}}  &  =\sum_{i\geq2}\hbar\Omega_{\mathrm{Z}%
}^{{}}\left(  \hat{c}_{i}^{+}\hat{c}_{i}^{{}}-\tfrac{1}{2}\hat{1}\right)  +%
{\textstyle\sum\limits_{\genfrac{}{}{0pt}{1}{i\geq2}{j>i}}}
\tfrac{1}{2}b_{ij}^{{}}\left(  \hat{c}_{i}^{+}\hat{c}_{j}^{{}}+\hat{c}_{j}%
^{+}\hat{c}_{i}^{{}}\right)  ,\label{HE_spins}\\
\widehat{\mathcal{H}}_{\mathrm{SE}}  &  =a_{12}^{{}}\left(  \hat{c}_{1}%
^{+}\hat{c}_{1}^{{}}-\tfrac{1}{2}\hat{1}\right)  \left(  \hat{c}_{2}^{+}%
\hat{c}_{2}^{{}}-\tfrac{1}{2}\hat{1}\right)  +\tfrac{1}{2}b_{12}^{{}}\left(
\hat{c}_{1}^{+}\hat{c}_{2}^{{}}+\hat{c}_{2}^{+}\hat{c}_{1}^{{}}\right)  ,
\label{HSE_spins}%
\end{align}
where the index $0$ represents the spin $S.$ Here, the system interacts with
the environment through the site $2$ (the surface site of the bath). In the
last Hamiltonians, the terms proportional to the identity do not contribute to
the dynamics because they only change the total energy by a constant number.
This Hamiltonian, as we described in section \S \ \ref{M_2-spin_spin_bath}, is
a standard cross-polarization experiment (swapping gate) in NMR \cite{MKBE74}%
.
In this experiment, the site $S$ is a $^{13}$C and the site $I_{1}$ a $^{1}$H
while the environment is a $^{1}$H spin bath. The typical experimental
Hartmann-Hahn condition \cite{Hartmann62} equals the values of the\ effective
energies at the $^{13}$C and the $^{1}$H sites to optimize the polarization
transfer. The SE interaction has terms linear in the number operators $\hat
{c}_{1}^{+}\hat{c}_{1}^{{}}$ and $\hat{c}_{2}^{+}\hat{c}_{2}^{{}},$ that only
change the energy of the sites $1$ and $2$ respectively. Thus, the
Hartmann-Hahn implementation, compensates the change of energy produced by the
environment through these linear terms. Finally, we have Hamiltonians
equivalent to those in eqs. (\ref{Hs},\ref{He},\ref{HSE}) where the site
energies are equal, and $V_{01}=-\frac{b}{2},~V_{ij}=-\tfrac{b_{ij}}{2}$,
$U_{12}^{\mathrm{(dir.)}}=a_{12}^{{}}$ and $U_{12}^{\mathrm{(ex.)}}=0.$

The spin dynamics of the system is described by the spin correlation function
\cite{Danieli04,DanieliThesis}:
\begin{equation}
P_{i1}(t)=\frac{\left\langle \Psi_{\mathrm{eq.}}\right\vert \hat{I}_{i}%
^{z}(t)\hat{I}_{1}^{z}(0)\left\vert \Psi_{\mathrm{eq.}}\right\rangle
}{\left\langle \Psi_{\mathrm{eq.}}\right\vert \hat{I}_{1}^{z}(0)\hat{I}%
_{1}^{z}(0)\left\vert \Psi_{\mathrm{eq.}}\right\rangle },
\label{spin_correlations}%
\end{equation}
which gives the local polarization at time $t$ on the $i$-th spin with an
initial local excitation on the $1$-th spin at time $t=0.$ Here, $\left\vert
\Psi_{\mathrm{eq.}}\right\rangle $ is the thermodynamical many-body
equilibrium state and
\begin{equation}
\hat{I}_{i}^{z}(t)=e^{\mathrm{i}\widehat{\mathcal{H}}t/\hbar}\hat{I}_{i}%
^{z}e^{-\mathrm{i}\widehat{\mathcal{H}}t/\hbar}%
\end{equation}
are the spin operators in the Heisenberg representation. After the JWT, the
initial local excitation on site $1$ is described by the non-equilibrium
state
\begin{equation}
\left\vert \Psi_{\mathrm{n.e.}}\right\rangle =\hat{c}_{1}^{+}\left\vert
\Psi_{\mathrm{eq.}}\right\rangle .
\end{equation}
In the experimental high temperature regime, $k_{\mathrm{B}}T$ much\ larger
than any energy scale of the system, the spin correlation function becomes
\cite{DanieliThesis}
\begin{equation}
P_{i1}(t)=\tfrac{2\hbar}{\mathrm{i}}G_{ii}^{<\,}(t,t)-1. \label{Pol_spin}%
\end{equation}
Notice that $G_{ii}^{<\,}(t,t)$ implicitly depends on the initial local
excitation at site $1$. Here, $G_{ii}^{<\,}(t,t)$ is the reduced density
function of sites $0$ and $1$ and can be split into the contributions
$G_{ii}^{<\,_{(N)}}(t_{2},t_{1})$ from each subspace with $N$ particles (or
equivalently $N$ spins up) in the following way,
\begin{equation}
G_{ii}^{<\,}(t,t)=\sum_{N=1}^{M}\dfrac{\left(
\genfrac{}{}{0pt}{1}{M-1}{N-1}%
\right)  }{2^{M-1}}G_{ii}^{<\,_{(N)}}(t,t),
\end{equation}
and analogous for the hole density function. The initial condition in this
picture is described by
\begin{equation}
G_{ij}^{<_{(N)}}(0,0)=\tfrac{\mathrm{i}}{\hbar}\left(  \tfrac{N-1}{M-1}%
\delta_{ij}+\tfrac{M-N}{M-1}\delta_{i1}\delta_{1j}\right)  ,
\label{Initial_condition_spins}%
\end{equation}
where the first term is the equilibrium density (identical occupation for all
the sites) and the second term is the non-equilibrium contribution where only
site $1$ is excited. Thus, we have an expression like (\ref{Danielewicz_evol})
for each $N$-th subspace \cite{CPL05}.
For this two-spin system
the $-1$ term of eq. (\ref{Pol_spin}) is canceled out by the background
evolution, i.e. the evolution of the first term of eq.
(\ref{Initial_condition_spins}) plus the evolution of the second term of eq.
(\ref{Danielewicz_evol}) for the $N=2$ subspace. As a consequence, the
observable dynamics only depends on the initial local excitation at site $1,$%
\begin{equation}
G_{ij}^{<_{\left(  1\right)  }}(0,0)=\tfrac{\mathrm{i}}{\hbar}\delta
_{i1}\delta_{1j},
\end{equation}
and evolves in the $1$-th particle subspace,
\begin{equation}
P_{i1}(t)=\tfrac{\hbar}{\mathrm{i}}G_{ii}^{<\,_{(1)}}(t,t).
\label{Pi1_one_particle_without_background}%
\end{equation}
Finally, the solution of the polarization $P_{01}(t),$ with $\Gamma
_{\mathrm{XY}}\leftrightarrow\Gamma_{V}$ and $\Gamma_{\mathrm{ZZ}%
}\leftrightarrow\Gamma_{U},$ is the same that was obtained in eq.
(\ref{G11}).

\subsection{Keldysh formalism versus the generalized quantum master
equation\label{Mark_Keldysh_vs_DM}}

Typical solutions of the quantum master equation for a spin swapping
\cite{JCP03,Alvarez07a}, described in chapter
\ref{Mark_spin_dynamics_Density_matrix},
were obtained following that of M\"{u}ller\emph{ et al.} \cite{MKBE74}. They
considered an \emph{isotropic interaction} with the spin environment,
represented\ by a phenomenological relaxation rate $R=\Gamma_{\mathrm{XY}%
}=\Gamma_{\mathrm{ZZ}}=1/\left(  4\tau_{\mathrm{SE}}\right)  .$ Within the
fast fluctuation approximation and neglecting non-secular terms, this leads
to
\begin{equation}
P^{^{\mathrm{MKBE}}}(t)=1-\frac{1}{2}\exp\left[  -Rt\right]  -\frac{1}{2}%
\cos(\omega t)\exp\left[  -\frac{3}{2}Rt\right]  , \label{MKBE_expression}%
\end{equation}
that is obtained from eq. (\ref{Msz_2spin_MKBEg}) in the isotropic case. This
expression is used in most of the experimental fittings \cite{Pratima96,JCP98}%
.
Our eq. (\ref{G11}) reproduces this result with $1/\tau_{\phi}\equiv
R\simeq1/\left(  4\tau_{\mathrm{SE}}\right)  $ by considering an isotropic
relation between $\Gamma_{\mathrm{ZZ}}$ and $\Gamma_{\mathrm{XY}},$ i.e.
$p_{\mathrm{XY}}=1/2$ under the condition $1/\left(  4\tau_{\mathrm{SE}%
}\right)  \ll b/\hbar.$ Note that our microscopic derivation, for an XY chain
as environment, does not imply that $\Gamma_{\mathrm{ZZ}}=\Gamma_{\mathrm{XY}%
}$ comes from an isotropic SE interaction. This conclusion arrives from the
expressions (\ref{gammaU}) and (\ref{gammaV}).

Returning to the comparison between expressions (\ref{MKBE_expression}) and
(\ref{G11}) at short times $t\ll\tau_{\mathrm{SE}},$ one can see that the MKBE
swapping probability growths exponentially with a rate $1/\left(
4\tau_{\mathrm{SE}}\right)  $. In contrast, our solution manifests that the
polarization growths quadratically on time, $\left(  \frac{1}{2}%
b/\hbar\right)  ^{2}t^{2},$ as we anticipated in
\S \ \ref{Mark_non_secular_2spin}. In eq. (\ref{G11}) this is made possible by
the phase $\phi_{0}$ in the cosine. In the parametric region, $b\tau
_{\mathrm{SE}}\ll\hbar$, \ where MKBE is not valid, our model enables the
manifestation of the quantum Zeno effect \cite{Misra77,Usaj98,Pascazio02}.
This means that the bath interrupts the system through measurements too
frequently, freezing its evolution. Here, this is a dynamical effect
\cite{Usaj98, Pascazio94} produced by interactions with the environment that
freezes the system oscillation. At longer times, $t\gg\tau_{\mathrm{SE}}$, one
gets
\begin{equation}
1-P_{01}\left(  t\right)  \sim\left(  1+2\left(  b/\hbar\right)  ^{2}%
\tau_{\mathrm{SE}}^{2}\right)  \exp\left[  -(b/\hbar)^{2}\tau_{\mathrm{SE}%
}^{{}}t\right]  ,
\end{equation}
and the quantum Zeno effect is manifested in the reduction of the decay rate
$1/\tau_{\phi}\propto\left(  b/\hbar\right)  ^{2}\tau_{\mathrm{SE}}$ as
$\tau_{\mathrm{SE}}$ gets smaller than $\hbar/b$.
This surprising dependence deserves some interpretation. First, we notice that
a strong interaction with the bath makes the $^{1}$H spin to fluctuate,
according to the Fermi golden rule, at a rate $1/\tau_{\mathrm{SE}}^{{}}$. The
effect on the $^{13}$C is again estimated in a fast fluctuation approximation
as
\begin{equation}
1/\tau_{\phi}\propto\left(  b/\hbar\right)  ^{2}\tau_{\mathrm{SE}}%
~\propto\left(  b/\hbar\right)  ^{2}\left[  \left(  a_{12}^{2}/\hbar
^{2}+b_{12}^{2}/\hbar^{2}\right)  \tau_{\mathrm{B}}\right]  ^{-1}.
\label{Tau_phi_quadratic_Zeno}%
\end{equation}
This \textquotedblleft nesting\textquotedblright\ of two Fermi golden rule
rates is formally obtained from a continuous fraction evaluation of the
self-energies \cite{DAmato90,CPL05}
involving an infinite order perturbation theory. Another relevant result is
that the frequency depends not only on $b$, but also on $\tau_{\mathrm{SE}}$.
A remarkable difference between the quantum master equation and our
formulation concerns the final state. In the quantum master equation
$\sigma_{\infty}$ must be hinted beforehand, while here it is reached
dynamically from the interaction with the spin-bath. Here, the reduced
density, whose trace gives the system polarization, can fluctuate until it
reaches its equilibrium value.

We mentioned in section \S \ \ref{Mark_non_secular_2spin} that by including
the non-secular terms of the SE interaction in the generalized QME, we obtain
the expression (\ref{G11}) for the polarization evolution. However, here we
obtained the solution from a microscopic model using an XY linear chain as the
spin environment, without forcing the final state. The Keldysh formalism gives
us another perspective to discuss about the physics of quantum evolutions and
the interpretation of the approximations made.

It is important to remark that within the fast fluctuation approximation, the
temporal and parametric behavior of expression (\ref{G11}) does not depend on
the environment Hamiltonian. This only appears as an intensity in the values
of $\Gamma_{\mathrm{ZZ}}$ and $\Gamma_{\mathrm{XY}}$ and the relation between
them (the anisotropy).

\section{Memory effects of the spin-bath\label{Marker_memory_effects_keldysh}}

\begin{figure}
[tbh]
\begin{center}
\includegraphics[
height=4.0058in,
width=5.4916in
]%
{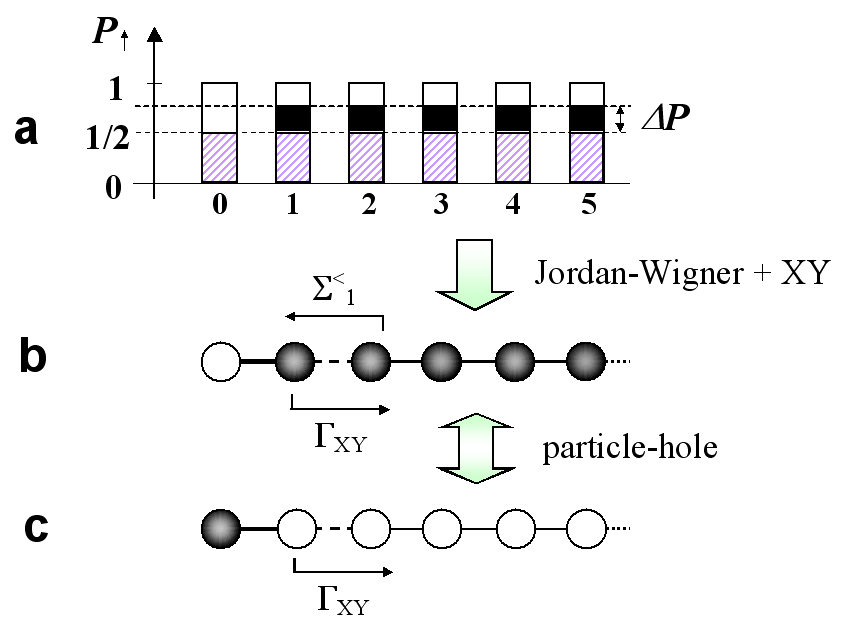}%
\caption[Schematic representations of the spin system at time $t=0$ and its
Jordan-Wigner transformation.]{a) Schematic representation of the spin system
at time $t=0.$ The shaded region stands for the thermodynamic equilibrium
state at high temperature and establish a background probability level. The
black filling represents the excess of probability over the equilibrium state
which is responsible for the observed dynamics. In b) the same system as in a)
after the JWT, that is, under the particle point of view. Note that in this
situation the background contribution is removed and the dynamics is described
by the excess of probability $\Delta P$. In c) we represent the complementary
problem of the case b). Here the black filling stands for the hole that
represent the excitation. In this representation it is easier to calculate the
memory effects in the bath.}%
\label{fig--System}%
\end{center}
\end{figure}

The $^{13}$C polarization, $P_{01}(t)$, in the Keldysh formalism arises from
the coherent evolution of the initial particle density, for which the
environment is a \textquotedblleft sink\textquotedblright, and an incoherent
contribution where the bath acts as a particle \textquotedblleft
source\textquotedblright. This can be compared with the complementary
framework. Instead of dealing with a \textquotedblleft
particle\textquotedblright\ problem we consider it as a \textquotedblleft
hole\textquotedblright\ problem [fig. \ref{fig--System} b) and c)
respectively].
On these grounds, at $t=0,$ all the sites are occupied except for the
\textquotedblleft hole\textquotedblright\ excitation at the $0$-th site. See
fig. \ref{fig--System} c) where the black color stands for the hole
excitation. At later times this excitation evolves in the system and also
propagates through the reservoir. The \textquotedblleft
environment\textquotedblright\ does not have holes to inject back into the
\textquotedblleft system\textquotedblright\ but those evolved coherently from
the initial hole (i.e. $\Sigma^{<}\equiv0$). Here the environment is a perfect
\textquotedblleft sink\textquotedblright. Thus all the dynamics would be
coherent, in the sense previously explained. If we add the result obtained in
this case with that of eq. (\ref{G11}) we obtain a \emph{one} for all times
consequence of the particle-hole symmetry. This is a particularly good test of
the consistency of the formalism because in each result the \textquotedblleft
environment\textquotedblright\ is set in a different framework. It also shows
that the background polarization does not contribute to the dynamics as we
obtained in (\ref{Pi1_one_particle_without_background}).

This \textquotedblleft hole\textquotedblright\ picture can help us to get a
very interesting insight on the dynamics in a case where the memory on the
environment becomes relevant. Consider, for example, the case $V_{01}%
=V_{12}=V_{ij}=V$ and $E_{0}=E_{1}$. The finite version of this effective
Hamiltonian applies to the actual experiments reported in ref. \cite{Madi97}.
In this case, the simplifying approximations of the fast fluctuations regime
are not justified. However, the exact dynamics of the system can be
analytically obtained if one considers an infinite chain. This enables the use
of eq. (\ref{LDoS_Chain}) to evaluate $\Sigma^{\mathrm{R}}$, eq.
(\ref{Sigma_R_definition}). Then we can calculate the propagator through eq.
(\ref{Heff}) to solve eq. (\ref{Danielewicz_evol}). The integration gives the
first Bessel function, hence:
\begin{equation}
P_{01}(t)=1-\left\vert \frac{\hbar}{tV}J_{1}(2tV/\hbar)\right\vert ^{2}.
\label{eq--Bessel-1}%
\end{equation}
A first observation is that the frequency above is roughly increased by a
factor of two as compared with that in eq. (\ref{P01_isolated_keldysh}).
Since the maxima of $P_{01}$ are zeroth of the Bessel function it is clear
that the frequency increases slightly with time. These are \emph{memory
effects of the environment}\textit{ }that are dependent on the interplay
between the spectral density of the bath and that of the system.

We notice that the memory effect can also appear in other condition for the
bath. For example, if the proton nuclei have random polarizations and the
density excitation is at site $0$, i.e. in fig. \ref{fig--System} a)
$\mathrm{f}_{n}(\varepsilon)=\frac{1}{2}$ for $n=1,..$ representing the $^{1}%
$H sites filled up to the shaded region; and the $^{13}$C site with an
occupation $\frac{1}{2}+\Delta P$. In this case the excitation propagates over
a background level (shaded region) that does not contribute to the dynamics.
The schematic view of this initial condition is equivalent to that of fig.
\ref{fig--System} c) where now the black filling represents a particle
excitation. The solution of the polarization is the first Bessel function,
\begin{equation}
P_{00}(t)=\left\vert \frac{\hbar}{tV}J_{1}(2tV/\hbar)\right\vert ^{2}%
\end{equation}
as we obtain in (\ref{Green_function_of_an_infinite_chain}). Apart from the
finite size mesoscopic effect, this is precisely the situation observed\ in
ref. \cite{Madi97}, although without enough resolution for a quantitative
comparison. The effect of a progressive modification of the swapping frequency
is often observed in many experimental situations such as CP experiments.
Depending on the particular system, the swapping frequency can accelerate or
slow down. Reported examples are fig. $5$ on ref. \cite{JCP98} and fig. $4$ on
ref. \cite{Chattah04}. We can observe the same effect in the figure
\ref{Fog-CP8CB} for a three-spin system.
This simple example solved so far shows that environmental correlations have
fundamental importance in the dynamics and deserve further attention.

\section{Summary}

We have solved the Schr\"{o}dinger equation within the Keldysh formalism for
an open system within the wide band regime (fast fluctuation approximation) in
the environment \cite{CPL05,SSC07}. We have shown a method
\cite{SSC07,Alvarez07b} that involves the transformation of the density
function expressed in the Danielewicz integral form into a Generalized
Landauer B\"{u}ttiker Equation. This was possible by resorting to Wigner
time-energy variables to perform the fast fluctuation approximation for the
environment which leads to interactions local in time. This results in an
injection of quantum waves without definite phase relation with the initial
state. The model proposed allowed us to consider the effect of the environment
over the system via the decay of the initial state followed with an incoherent
injection. Further on, we effectively symmetrized the system-environment
interactions transforming them into a spatially homogeneous process
\cite{Alvarez07b}. This has an uniform system-environment interaction rate
leading to a simple non-hermitian propagator. The original multi-exponential
decay processes are recovered by an injection density function. Moreover,
through discretization of the GLBE, we built a stroboscopic process which is
the basis for an optimal numerical algorithm where the quantum dynamics is
calculated in discrete time steps \cite{JCP06,Alvarez07b}.

We applied \cite{JCP06,Alvarez07b} these techniques to a two-spin system
coupled to a spin-bath improving the result obtained through the application
of the secular approximation \cite{MKBE74} in the standard density matrix
calculation given in section \S \ \ref{M_2-spin_spin_bath}. One improvement is
the manifestation of the quantum Zeno effect that leads to novel
interpretation of previous experiments \cite{JCP98} which will be discussed in
chapter \ref{Sec_QDPT}. The arising of the Zeno effect showed in the
decoherence time, $1/\tau_{\phi}^{{}}\propto\left(  b/\hbar\right)  _{{}}%
^{2}\tau_{\mathrm{SE}}^{{}}$, can be interpreted as a \textquotedblleft
nested\textquotedblright\ Fermi golden rule rate emphasizing the
non-perturbative nature of the result \cite{JCP06}. We observed that the
solution of the Keldysh formalism is also obtained within the generalized QME
if we include the non-secular terms of the SE interaction \cite{Alvarez07a}.
However, within the Keldysh formalism we derived it from a microscopic model
for the entire system (system plus environment) and the final state must not
be hinted beforehand. The Keldysh formalism gives us another perspective to
discuss about the physics of the quantum evolution and the interpretation of
the approximations made. Moreover, of particular interest, the Keldysh
formalism allows us to include temporal correlations within the spin-bath in a
model which has exact solution. On one side, it enabled us to show a novel
result: memory effects can produce a progressive change of the swapping
frequency \cite{CPL05}. On the other side, these results will serve to test
approximate methods developed to deal with complex correlations.

In general, our analytical results based in the spin-particle mapping allow a
deeper understanding of the polarization dynamics. They may constitute a
starting point for the study of other problems.

\chapter{Environmentally induced Quantum Dynamical Phase
Transition\label{Sec_QDPT}}

Experiments on quantum information processing involve atoms in optical traps
\cite{Myatt00}, superconducting circuits \cite{Vion02} and nuclear spins
\cite{cory03,Awschalom2003} among others. As we discussed in previous
chapters, the system to be manipulated interacts with an environment
\cite{Myatt00,Gurvitz03,Zurek03,Barret04} that perturbs it, smoothly degrading
its quantum dynamics with a decoherence rate, $1/\tau_{\phi}$, proportional to
the system-environment (SE) interaction $\hbar/\tau_{\mathrm{SE}}$.
Strikingly,\ there are conditions where the decoherence rate\ can become
perturbation independent \cite{Usaj-Mol98,PhysicaA}. This phenomenon is
interpreted \cite{JalPas01,Beenakker01,Cucchietti04} as the onset of a
Lyapunov phase, where $1/\tau_{\phi}=\min\left[  1/\tau_{\mathrm{SE}}%
,\lambda\right]  $ is controlled by the system's own complexity $\lambda$.
Describing such a transition, requires expressing the observables (outputs) in
terms of the controlled parameters and interactions (inputs) beyond the
perturbation theory. We are going to show that this is also the case of the
two-spin system treated in sections \S \ \ref{M_2-spin_spin_bath} and
\S \ \ \ref{M_Keldysh_applied_to_spin_systems}, a simple swapping gate that is
an essential building block for quantum information processing. While the
swapping operation was recently addressed in the field of NMR in liquids
\cite{Madi98, Freeman99} with a focus on quantum computation, in the
introduction we showed that the pioneer experiments were performed in solid
state NMR by M\"{u}ller \emph{et al.} \cite{MKBE74}. They obtained a swapping
frequency $\omega$ determined by a two-spin dipolar interaction $b$, and a
decoherence rate $1/\tau_{\phi}\equiv R$ that, in their model, was fixed by
interactions with the environment $1/\left(  4\tau_{\mathrm{SE}}\right)  $.
This dynamical description was obtained by solving a generalized Liouville-von
Neumann equation as in section \S \ \ref{M_2-spin_spin_bath}.
As we anticipate previously, more recent experiments which span the internal
interaction strength \cite{JCP98} (see fig. \ref{Fig_JCP98_original}) hinted
that there is a critical value of this interaction when a drastic change in
the behavior of the swapping frequency and relaxation rates occurs. Since this
is not predicted by the standard approximations in the quantum master equation
\cite{MKBE74}, this motivated us to deepen into the physics of the phenomenon
leading to develop the new theoretical result of chapter
\ref{Marker_Spin_within_keldysh}.

In the first part of this chapter, we present a set of $^{13}$C-$^{1}$H
cross-polarization NMR data, swept over a wide range of a control parameter
(the ratio between internal interactions and SE interaction strengths).
These results clearly show that the transition\textit{ }between the two
expected dynamical regimes for the $^{13}$C polarization, an oscillating
regime and an over-damped regime, is not a smooth cross-over. Indeed, it has
the characteristics of critical phenomena where a divergence of the
oscillation period at a given critical strength of the control parameter is
indicative of the nonanalyticity of this observable \cite{Horsthemke-Lefever,
Sachdev}. The data are interpreted by solving the swapping dynamics between
two coupled spins (qubits) interacting with a spin bath.
With this purpose we use the microscopic model proposed in section
\S \ \ref{M_Keldysh_applied_to_spin_systems}
to describe the cross-polarization (swapping operation) using the Keldysh
formalism. Within this picture, the overdamped regime arises because of the
quantum Zeno effect \cite{Misra77,Usaj98,Pascazio02}, i.e. the environment
\textquotedblleft measures\textquotedblright\ the system so frequently that
prevents its evolution. Such quantum freeze can arise as a pure dynamical
process governed by strictly unitary evolutions \cite{Pascazio94,Usaj98}.The
analytical solution confirms that there is a critical value of the control
parameter where a bifurcation occurs. This is associated with the switch among
dynamical regimes: the \emph{swapping phase}\textit{ }and the \emph{Zeno
phase}. Consequently, we call this phenomenon a \emph{Quantum Dynamical Phase
Transition}.

A major challenge in quantum control is the decoupling between the system and
the environment. Many techniques \cite{Taylor05,Petta05,Ardavan06} are
developed to avoid the loss of information induced by decoherence. We will
show here how the criticality described above is useful to \textquotedblleft
isolate\textquotedblright\ a spin-pair. For this purpose, we will extend the
model to a three interacting spin system where only one is coupled to the
environment. We show that beyond a critical interaction, the two spins not
directly coupled to the environment oscillate with their bare frequency and
relax more slowly. In a two-spin system there is always a critical point that
depends on the anisotropy relation of the SE interaction quantified as the
ratio between the Ising and XY terms. However, in the three-spin system, the
decoherence rate has a\ smooth cross-over from proportional to inversely
proportional to the SE interaction. This cross-over approaches a critical
transition as the anisotropy of the SE interaction goes from a purely XY to an
Ising form.

\section{Experimental evidence}

\begin{figure}
[tbh]
\begin{center}
\includegraphics[
height=4.7357in,
width=4.2609in
]%
{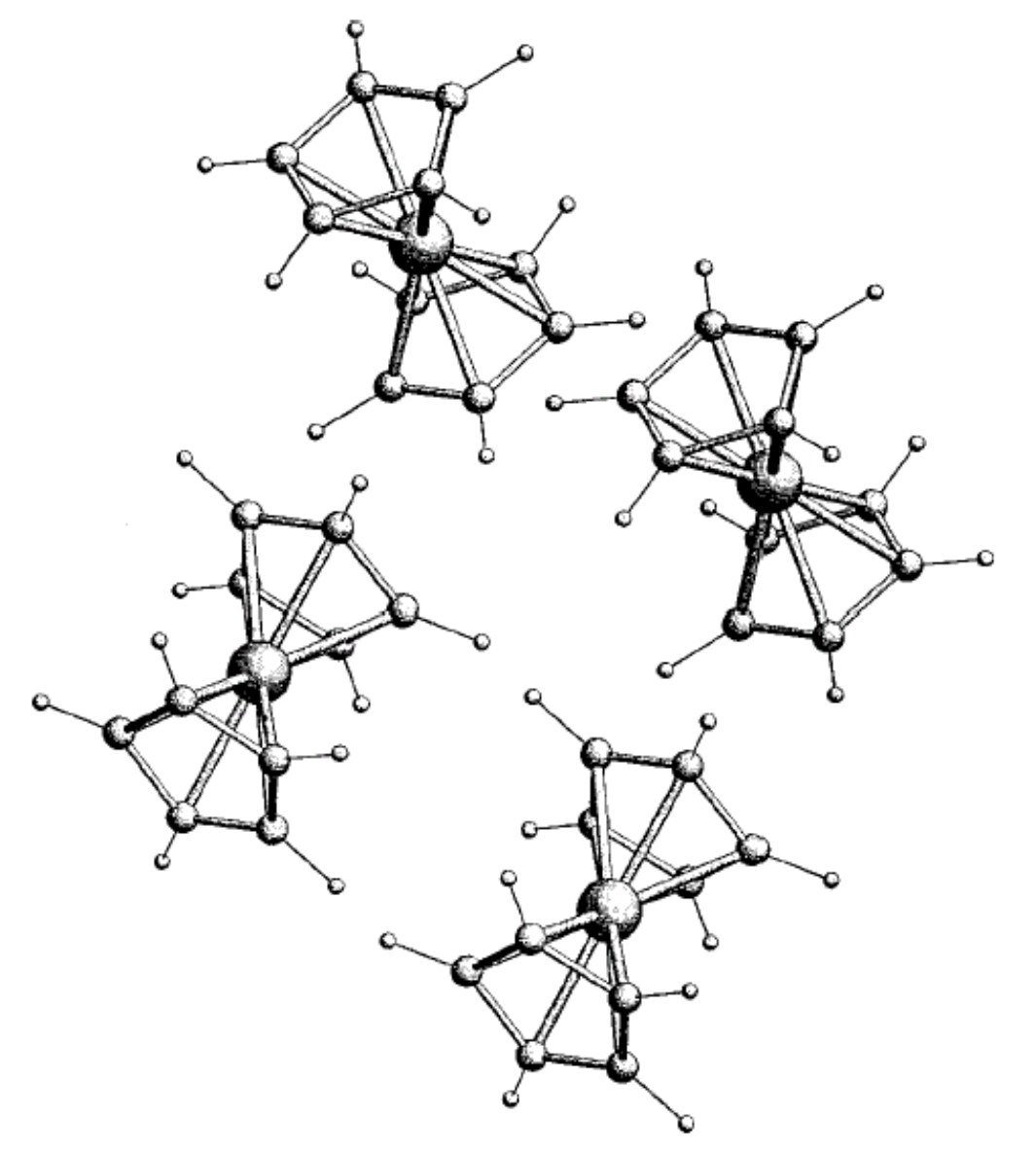}%
\caption[Crystalline structure of ferrocene, Fe(C$_{5}$H$_{5}$)$_{2}$%
.]{Crystalline structure of ferrocene, Fe(C$_{5}$H$_{5}$)$_{2}$ in its (room
temperature) monociclic form with space group P2$_{1}/a$. Two unit cells are
included in the graph.}%
\label{Fig_Ferroceno_molecule}%
\end{center}
\end{figure}

\begin{figure}
[ptb]
\begin{center}
\includegraphics[
height=5.246in,
width=5.1024in
]%
{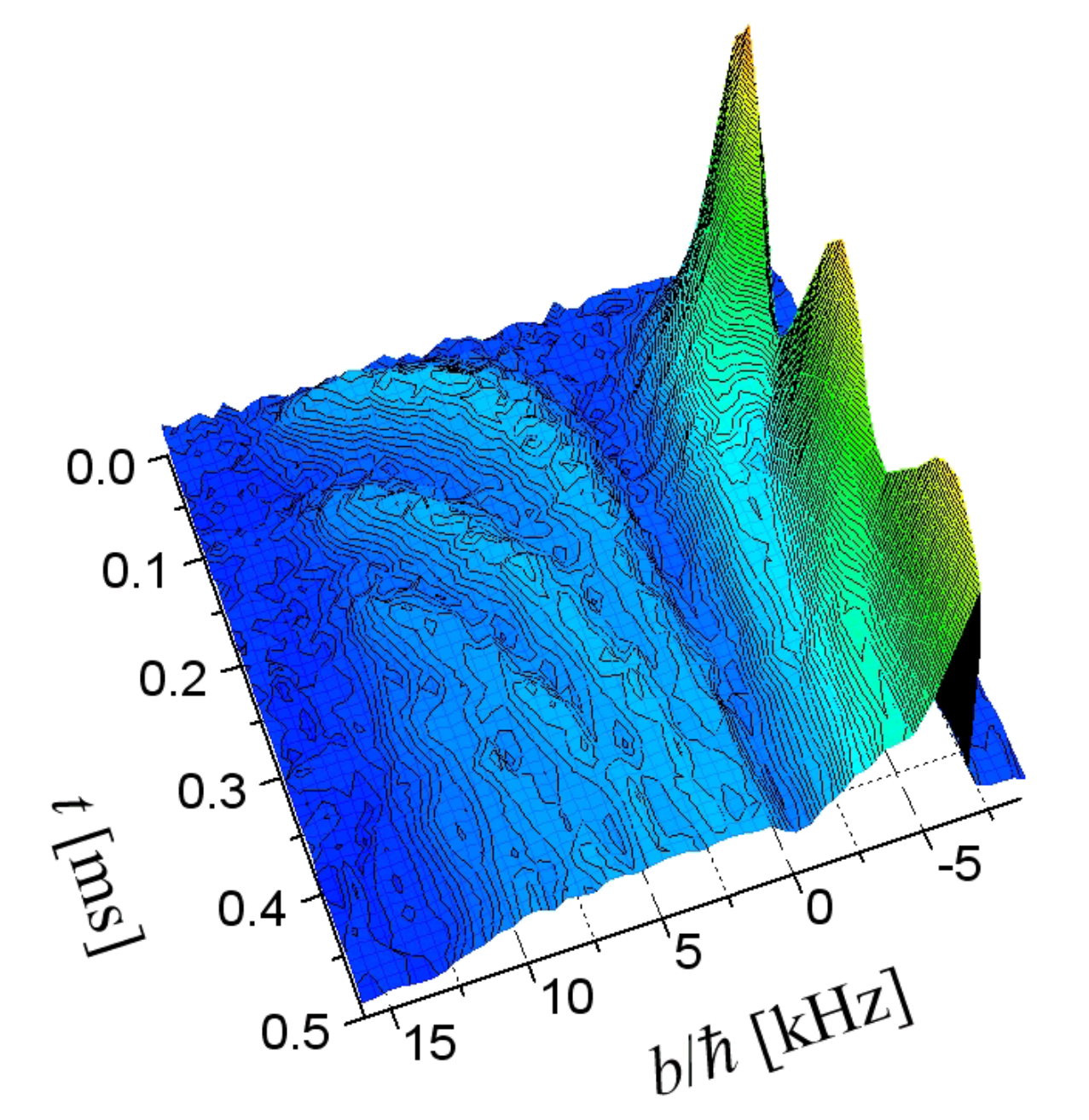}%
\caption{Raw experimental data of $^{13}$C polarization as a function of the
contact time and spin-spin coupling $b(\theta)$ for a spin swapping dynamics
in a $^{13}$C-$^{1}$H system.}%
\label{Fig3D-Ferroceno}%
\end{center}
\end{figure}

\begin{figure}
[ptbh]
\begin{center}
\includegraphics[
height=4.6631in,
width=7.1174in
]%
{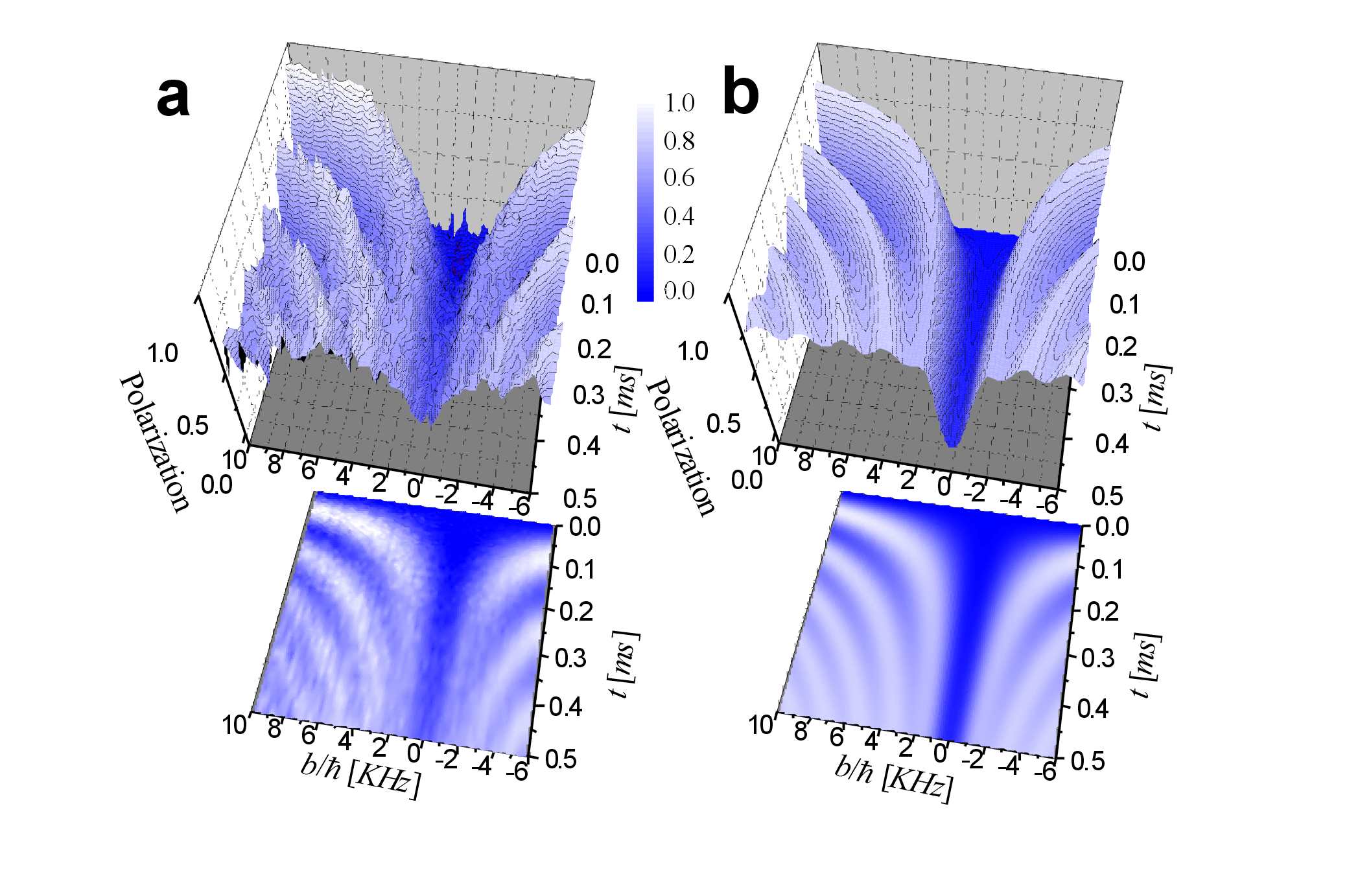}%
\caption[Experimental and theoretical spin swapping dynamics in $^{13}$%
C-$^{1}$H evidencing a Quantum Dynamical Phase Transition.]{Spin swapping
dynamics in $^{13}$C-$^{1}$H. a): Experimental $^{13}$C polarization in
Fe(C$_{5}$H$_{5}$)$_{2}$ as a function of the contact time $t_{\mathrm{c}}$
and spin-spin coupling $b(\theta)$. b): Numerical simulations of the $^{13}$C
polarization obtained from eq. (\ref{PolCST}) for different values of $b$, a
dipolar system-environment interaction $\left(  \left\vert a_{12}%
/b_{12}\right\vert =2\right)  $ and a constant value for $\tau_{\mathrm{SE}}$
($\tau_{\mathrm{SE}}=0.275\operatorname{ms}$) obtained by fitting the
experimental data in the regime where the MKBE expression is valid. Projection
plots in the $b-t$ plane show a canyon where the oscillation period diverges
indicating a Quantum Dynamical Phase Transition.}%
\label{Figteoexp3D}%
\end{center}
\end{figure}

The cross-polarization experiments exploit the fact that in polycrystalline
ferrocene Fe(C$_{5}$H$_{5}$)$_{2}$ (see fig. \ref{Fig_Ferroceno_molecule}),
%
one can select a pair of interacting spins, i.e. a $^{13}$C and its directly
bonded $^{1}$H, arising on a molecule with a particular orientation. This is
because the cyclopentadienyl rings perform fast thermal rotations ($\approx%
\operatorname{ps}%
$) around the five-fold symmetry axis, leading to a time averaged $^{13}%
$C-$^{1}$H interaction. The new dipolar constant depends \cite{Slichter} only
on the angle $\theta$ between the molecular axis and the external magnetic
field $H_{0}$ and the angles between the internuclear vectors and the rotating
axis, which in this case are $90^{\circ}$. Thus, the effective coupling
constant is
%

\begin{equation}
b=\frac{1}{2}\frac{\mu_{0}\ \gamma_{\mathrm{H}}\,\gamma_{\mathrm{C}}%
\,\hbar^{2}}{4\pi r_{\mathrm{HC}}^{3}}\frac{\left\langle 3\cos^{2}%
\theta-1\right\rangle }{2}, \label{b}%
\end{equation}
where $\gamma$'s are the gyromagnetic factors and $r_{\mathrm{HC}}^{{}}$ the
internuclear distance. Notice that $b(\theta)$ cancels out at the magic angle
$\theta_{\mathrm{m}}\simeq54.74^{\circ}.$ As the chemical shift anisotropy of
$^{13}$C is also averaged by the rotation and also depends on $\theta$ as
$\left\langle 3\cos^{2}\theta-1\right\rangle ,$ it is straightforward to
assign each frequency in the $^{13}$C spectrum to a dipolar coupling $b$.
Thus, all possible $b$ values are present in a single polycrystalline
spectrum.
The swapping induced by $b$ is turned on during the \textquotedblleft contact
time\textquotedblright\ $t_{\mathrm{c}}$, when the sample is irradiated with
two radio frequencies fulfilling the Hartmann-Hahn condition
\cite{Hartmann62,Slichter}.
At $t=0,$ there is no polarization at $^{13}$C while the $^{1}$H system is
polarized. The polarization is transferred forth and back in the $^{13}%
$C-$^{1}$H pairs while the other protons inject polarization into these pairs.
We show the raw experimental data of $^{13}$C polarization as a function of
the contact time and $b(\theta)$ in fig. \ref{Fig3D-Ferroceno} a).
In fig. \ref{Figteoexp3D} a), the polarizations have been normalized to their
respective values at the maximum contact time ($3%
\operatorname{ms}%
$) for each $\theta$ when it saturates. It can be appreciated in the figure
that the oscillation frequency is roughly proportional to $\left\vert
b\right\vert ,$ showing that this is the dominant interaction in the dynamics.
This is consistent with the fact that\ the next $^{13}$C-$^{1}$H coupling
strength with a non-directly bonded proton is roughly $b/8$ and, as all the
intramolecular interactions, also scales with the angular factor $\left\langle
3\cos^{2}\theta-1\right\rangle $.%

A noticeable feature in these experimental data is the\ presence of a
\textquotedblleft canyon\textquotedblright,\ in the region $\left\vert
b\right\vert <2%
\operatorname{kHz}%
$, where oscillations (swapping) disappear. The white hyperbolic stripes in
the contour plot at the bottom evidence a swapping period $2\pi/\omega$ that
diverges for a non-zero critical interaction. This divergence is the signature
of a critical behavior.

The standard procedure to characterize the cross-polarization experiment in
ferrocene and similar compounds was described in section
\S \ \ref{M_2-spin_spin_bath} that is derived from the MKBE model
\cite{MKBE74}. There the $^{13}$C polarization exchanges with that of its
directly bonded $^{1}$H, which, in turn, interacts isotropically with other
protons that constitute the environment. Their solution is
(\ref{MKBE_expression})
\begin{equation}
P^{^{\mathrm{MKBE}}}(t)=1-\frac{1}{2}\exp\left[  -\frac{t}{\tau_{\phi}%
}\right]  -\frac{1}{2}\cos(\omega t)\exp\left[  -\frac{3}{2}\frac{t}%
{\tau_{\phi}}\right]  ), \label{eq-MKBE}%
\end{equation}
where the decoherence rate becomes determined by the rate of interaction with
the environment $\frac{1}{2}\Gamma_{\mathrm{SE}}/\hbar=1/\left(
4\tau_{\mathrm{SE}}\right)  \rightarrow1/\tau_{\phi}\equiv R,$
while the swapping frequency is given by the two-spin dipolar interaction,
$b/\hbar\rightarrow\omega$. A dependence of the inputs $b$ and $\tau
_{\mathrm{SE}}$ on $\theta$ should manifest in the observables $\omega$ and
$\tau_{\phi}.$
However, working on a polycrystal, each $\tau_{\mathrm{SE}}(\theta)$ value
involves a cone of orientations of neighboring molecules and a rough
description with single average value for the SE interaction rate is suitable.

\begin{figure}
[tbh]
\begin{center}
\includegraphics[
height=3.9349in,
width=4.1252in
]%
{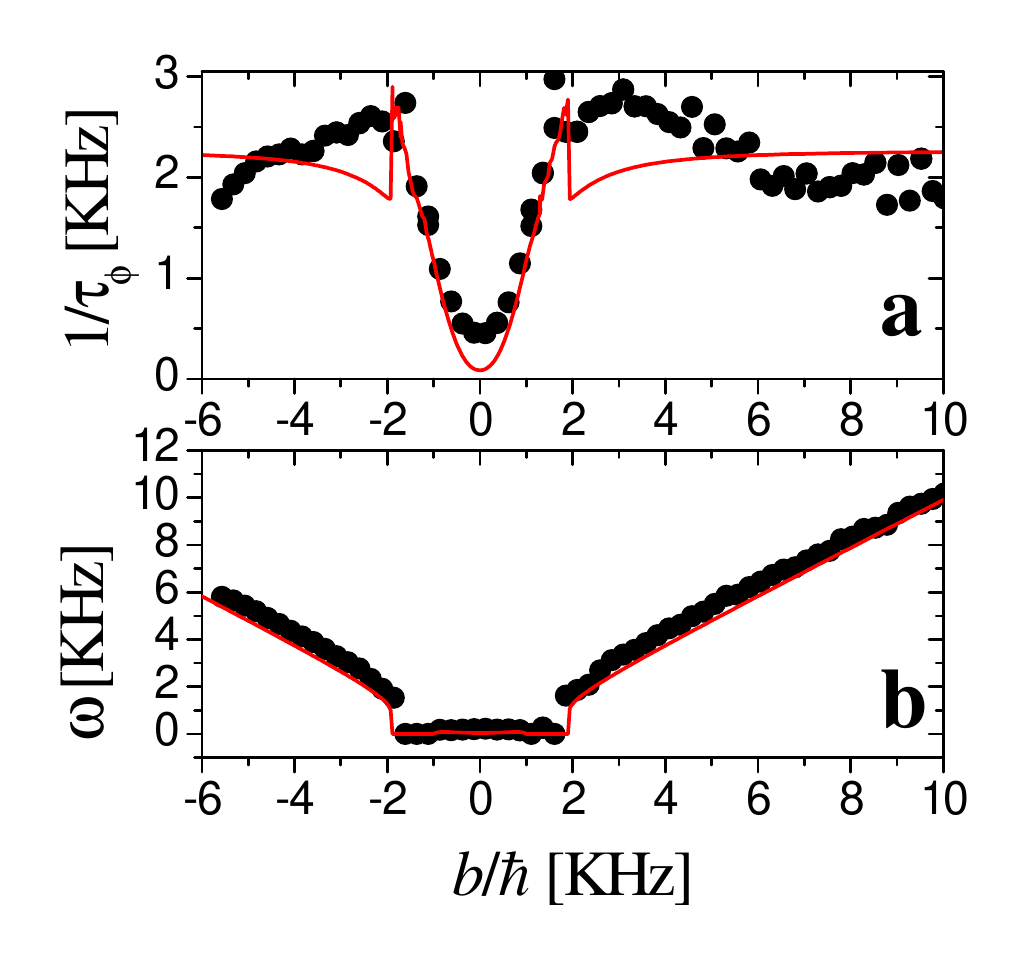}%
\caption[Experimental and theoretical decoherence rate $1/\tau_{\phi}$ and
frequency $\omega$ in the spin swapping of a $^{13}$C-$^{1}$H system.]%
{Decoherence rate $1/\tau_{\phi}$ and frequency $\omega$ in the spin swapping
of a $^{13}$C-$^{1}$H system. Data points are obtained by fitting cross
polarization experiments to the expression $P^{^{\mathrm{MKBE}}}(t)$. The zero
plateau in the frequency and the parabolic behavior of $1/\tau_{\phi}$ in the
region $b\tau_{\mathrm{SE}}\ll\hbar$ are indicative of an over-damped Zeno
phase. Solid lines are the prediction of our model assuming a constant
$\tau_{\mathrm{SE}}=0.275\operatorname{ms}$.}%
\label{Figexp}%
\end{center}
\end{figure}

We have performed non-linear least square fittings of the experimental points
to the equation $P^{^{\mathrm{MKBE}}}(t)$ for the whole $^{13}$C spectra of
ferrocene in steps of $\approx80%
\operatorname{Hz}%
$ and contact times ranging from $2%
\operatorname{\mu s}%
$ to $3%
\operatorname{ms}%
$. The $1/\tau_{\phi}$ and $\omega$ parameters obtained from these fits are
shown as dots in fig. \ref{Figexp}. The proportionality of the frequency with
$b$ for orientations that are far from the magic angle is verified. In this
region a weak variation of $1/\tau_{\phi}$ around $2.2%
\operatorname{kHz}%
$ reflects the\ fact that $1/\left(  2\tau_{\mathrm{SE}}\right)  $ does not
depend on $\theta$. A drawback of this simple characterization is that it
tends to overestimate the width of the canyon because of limitations of the
fitting procedure when eq. (\ref{eq-MKBE}) is used around the magic angle.

In spite of the MKBE theoretical prediction, one observes that the frequency
becomes zero abruptly and the relaxation rate suddenly drops with a quadratic
behavior when $b_{\mathrm{c}}\simeq2%
\operatorname{kHz}%
$. The minimum of the parabola occurs at the magic angle, when $b=0.$ Then,
all the polarization reaching the $^{13}$C at this orientation originates from
protons outside the molecule. Then, the rate of $0.5%
\operatorname{kHz}%
$ obtained at this minimum constitutes an experimental estimation of this
mechanism. This has to be compared with the almost constant value of
$1/(4\tau_{\mathrm{SE}})=1/\tau_{\phi}\simeq2.2%
\operatorname{kHz}%
$ observed outside the magic angle neighborhood. This justifies neglecting the
$J$-coupling and the direct relaxation of the $^{13}$C polarization through
the dipolar interaction with protons outside the molecule. In the following we
describe our stroboscopic model that accounts for the \textquotedblleft
anomalous\textquotedblright\ experimental behavior.

\section{Theoretical description}

\subsection{The system}

Let us consider $M$ coupled $1/2$ spins with a Hamiltonian:
\begin{equation}
\widehat{\mathcal{H}}=\widehat{\mathcal{H}}_{\mathrm{Z}}+%
{\textstyle\sum\limits_{i<j}}
\left[  a_{ij}\hat{I}_{i}^{z}\hat{I}_{j}^{z}+b_{ij}\left(  \hat{I}_{i}^{+}%
\hat{I}_{j}^{-}+\hat{I}_{i}^{-}\hat{I}_{j}^{+}\right)  /2\right]  ,
\label{SpinHamiltonian}%
\end{equation}
where
\begin{equation}
\mathcal{H}_{\mathrm{Z}}=\sum_{i=1}^{M}\hbar\left(  \Omega_{\mathrm{Z}}%
+\delta\Omega_{i}\right)  I_{i}^{z}%
\end{equation}
is the Zeeman energy, with a mean Larmor frequency $\Omega_{\mathrm{Z}}$. As
we mention in previous chapters, the second term is the spin-spin interaction:
$b_{ij}/a_{ij}=0$ is Ising, and $a_{ij}/b_{ij}=0,1,-2$ gives an $XY$, an
isotropic (Heisenberg) or the truncated dipolar (secular), respectively.

In order to describe the experimental system, we use the model described in
sections \S \ \ref{M_2-spin_spin_bath} and
\S \ \ref{M_Keldysh_applied_to_spin_systems}. Hence, let us take the first
$N=2$ spins, $I_{0}\equiv S$ (a $^{13}$C) and $I_{1}$ (its directly bonded
$^{1}$H), as the \textquotedblleft system\textquotedblright\ where the
swapping $\left\vert \downarrow,\uparrow\right\rangle \rightleftarrows
\left\vert \uparrow,\downarrow\right\rangle $ occurs under the action of
$b_{01}$.\ The other $M-N$ spins (all the other $^{1}$H), with $M\rightarrow
\infty$, are the spin-bath or \textquotedblleft environment\textquotedblright.
This limit enables the application of the Fermi Golden Rule or a more
sophisticated procedure to obtain a meanlife $\tau_{\mathrm{SE}}$ for the
system levels. We will not need much detail for the parameters of the
spin-bath in eq. (\ref{SpinHamiltonian}) except for stating that it is
characterized by an energy scale $d_{\mathrm{B}}$ which leads to a very short
correlation time $\tau_{\mathrm{B}}\simeq\hbar/d_{\mathrm{B}}$. Thus, this
spin system can be mapped into a fermion particle system using the
Jordan-Wigner transformation \cite{Lieb61}
as we described in section \S \ \ref{M_Keldysh_applied_to_spin_systems}. Under
the experimental conditions, $\delta\Omega_{i}=0,$ $a_{01}=0$ and $b_{01}=b,$
the system Hamiltonian becomes (\ref{HS_spins})%
\begin{equation}
\widehat{\mathcal{H}}_{\mathrm{S}}=\hbar\Omega_{\mathrm{Z}}\left(  \hat{c}%
_{0}^{+}\hat{c}_{0}^{{}}+\hat{c}_{1}^{+}\hat{c}_{1}^{{}}-\hat{1}\right)
+\tfrac{1}{2}b\left(  \hat{c}_{0}^{+}\hat{c}_{1}^{{}}+\hat{c}_{1}^{+}\hat
{c}_{0}^{{}}\right)  .
\end{equation}
The Jordan-Wigner transformation maps a linear many-body $XY$ spin Hamiltonian
into a system of non-interacting fermions. Spins $I_{i}$ with $2\leq i\leq
M-1$ are interacting among them. To simplify the presentation, and without
loss of generality, we consider a \emph{single} connection between the system
and the spin-bath $a_{1j}=b_{1j}=0,$ $j=3\ldots\infty$ and $a_{0j}=b_{0j}=0,$
$j=2\ldots\infty$ to use the SE interaction described by eq. (\ref{HSE_spins}%
)
\begin{equation}
\widehat{\mathcal{H}}_{\mathrm{SE}}=a_{12}\left(  \hat{c}_{1}^{+}\hat{c}%
_{1}^{{}}-\tfrac{1}{2}\right)  \left(  \hat{c}_{2}^{+}\hat{c}_{2}^{{}}%
-\tfrac{1}{2}\right)  +\tfrac{1}{2}b_{12}\left(  \hat{c}_{1}^{+}\hat{c}%
_{2}^{{}}+\hat{c}_{2}^{+}\hat{c}_{1}^{{}}\right)  .
\end{equation}
In the experimental initial condition, all spins are polarized with the
exception of $S$ \cite{MKBE74} (see section \S \ \ref{M_2-spin_spin_bath}).
Thus in the high temperature limit ($\hbar\Omega_{\mathrm{Z}}/k_{\mathrm{B}%
}T\equiv s\ll1$),
the reduced density operator is
\begin{equation}
\hat{\sigma}\left(  0\right)  =\frac{\hbar}{\mathrm{i}}\hat{G}^{<}\left(
0\right)  =\left(  \hat{1}+s\hat{I}_{1}^{z}\right)  /\operatorname{Tr}\left\{
\hat{1}\right\}
\end{equation}
which under the Jordan-Wigner transformation becomes
\begin{equation}
\frac{\left(  1-s/2\right)  }{\operatorname{Tr}\left\{  \hat{1}\right\}  }%
\hat{1}+\frac{s}{\operatorname{Tr}\left\{  \hat{1}\right\}  }\hat{c}_{1}%
^{+}\hat{c}_{1}^{{}}.
\end{equation}
Since the first term does not contribute to the dynamics
(\S \ \ref{M_Keldysh_applied_to_spin_systems}), we retain only the second term
and normalize\ it to the occupation factor. This means that site $1$ is empty
while site $2$ and sites at the particle reservoir are \textquotedblleft
full\textquotedblright. This describes the tiny excess $\Delta p$ above the
mean occupation $1/2$.
To find the dynamics of the reduced density matrix of the \textquotedblleft
system\textquotedblright\
\begin{equation}
\hat{\sigma}\left(  t\right)  =\frac{\hbar}{\mathrm{i}}\hat{G}^{<}\left(
t\right)  ,
\end{equation}
we will take advantage of the particle representation and use the
\emph{integral} form \cite{Keldysh64,Danielewicz84} of the Keldysh formalism,
eq. (\ref{Danielewicz_evol}), instead of the standard Liouville-von Neumann
differential equation, eq. (\ref{master}), as we described in section
\S \ \ref{M_Keldysh_applied_to_spin_systems}. There, any perturbation term is
accounted to infinite order ensuring the proper density normalization.
The interaction with the bath is \emph{local} and\textit{, }because of the
fast dynamics in the bath, it can be taken as \emph{instantaneous
}\cite{SSC07,Alvarez07b}\textit{.} Hence, the evolution is further simplified
into an integral form of the Generalized Landauer B\"{u}ttiker Equation (GLBE)
\cite{GLBE1,GLBE2} for the particle density \cite{SSC07,Alvarez07b}. There,
the environment plays the role of a local measurement apparatus
\S \ \ref{Mark_A_nice_physics_interpretation}.

\subsection{Analytical solution}

After all these assumption, we have the spin system used in section
\S \ \ref{M_Keldysh_applied_to_spin_systems} where we apply the spin-fermion
mapping to use the Keldysh formalism.
In that section, we obtain that the observable dynamics is consequence of a
initial local excitation on site $1,$
\begin{equation}
G_{ij}^{<\left(  1\right)  }(0,0)=\tfrac{\mathrm{i}}{\hbar}\delta_{i1}%
\delta_{1j},
\end{equation}
and evolve in the $1$-st particle subspace,
\begin{equation}
P_{i1}(t)=\tfrac{\hbar}{\mathrm{i}}G_{ii}^{<\,_{(1)}}(t,t).
\end{equation}
After that, the solution of the polarization $P_{01}(t)$ (experimental $^{13}%
$C polarization) is the same that was obtained in (\ref{G11}),%
\begin{equation}
P_{01}(t)=1-a_{0}e^{-R_{0}t}-a_{1}\cos\left[  \left(  \omega+\mathrm{i}%
\eta\right)  t+\phi_{0}\right]  e^{-R_{1}t}, \label{PolCST}%
\end{equation}
where $V_{01}=-\frac{b}{2},~V_{ij}=-\tfrac{b_{ij}}{2}$, $U_{12}%
^{\mathrm{(dir.)}}=a_{12}^{{}}$ and $U_{12}^{\mathrm{(ex.)}}=0$
and we substitute in the present solution $\Gamma_{\mathrm{XY}}\leftrightarrow
\Gamma_{V},$ $\Gamma_{\mathrm{ZZ}}\leftrightarrow\Gamma_{U}$ and
$p_{\mathrm{XY}}\leftrightarrow p_{V}.$
In spite of appearance, the last equation has a \emph{single fitting
parameter}. This is because the real functions $\omega,$ $\eta,$ $R_{0}$ and
$R_{1}$ as well as $a_{0}$, $a_{1}$ and $\phi_{0}$ depend exclusively on $b,$
$\tau_{\mathrm{SE}}$ and $p_{\mathrm{XY}}$. Besides, $b$ and $p_{\mathrm{XY}}$
are determined from crystallography and the anisotropy of the magnetic
interaction ($p_{\mathrm{XY}}=1/5$ for dipolar) respectively. The phase
transition is ensured by the condition $\omega\eta\equiv0$. The anisotropy
ratio $\Gamma_{\mathrm{XY}}/\Gamma_{\mathrm{SE}}\rightarrow p_{\mathrm{XY}}$
accounts for the observed competition (see section
\S \ \ref{M_2-spin_spin_bath} and \S \ \ref{M_SI2_bath_secular}) between the
Ising and XY terms of $\widehat{\mathcal{H}}_{\mathrm{SE}}$. The Ising
interaction drives the \textquotedblleft system\textquotedblright\ to the
internal quasi-equilibrium (\ref{quasi_equilibrium_2spin}). In contrast, the
XY term allows the thermal equilibration with the bath \cite{JCP03}.

\subsection{Comparison with the experiments}

In order to see how well our model reproduces the experimental behavior we
plot the $^{13}$C polarization with realistic parameters. Since the system is
dominated by the dipolar SE interaction \cite{JCP03}, as we described in
chapter \ref{Mark_spin_dynamics_Density_matrix}
we take $\left\vert a_{12}/b_{12}\right\vert =2.$ We introduce $b$ with its
angular dependence according to eq. (\ref{b}) and we select a constant value
for $\tau_{\mathrm{SE}}=0.275%
\operatorname{ms}%
$ representative of the $b\gg\hbar/\tau_{\mathrm{SE}}$ regime. Since here, we
are only interested in the qualitative aspects of the critical behavior of the
dynamics, there is no need to introduce $\tau_{\mathrm{SE}}(\theta)$ as a
fitting parameter. These evolutions are normalized at the maximum contact time
($3%
\operatorname{ms}%
$) experimentally acquired. They are shown in fig. \ref{Figteoexp3D} b) where
the qualitative agreement with the experimental observation of a canyon is
evident. Notice that the experimental canyon is less deep than the theoretical
one. This is due to intermolecular $^{13}$C-$^{1}$H couplings neglected in the
model.
We will show that the analytical expression of eq. (\ref{PolCST}) allows one
to determine the edges of the canyon which are the critical points of what we
will call a \emph{Quantum Dynamical Phase Transition }(QDPT).

\section{Quantum Dynamical Phase Transition}

Our \emph{quantum} observable (the local spin polarization) is a binary random
variable. The dynamics of its ensemble average (swapping probability), as
described by\ eq. (\ref{PolCST}), depends parametrically on the
\textquotedblleft noisy\textquotedblright\ fluctuations of the environment
through $\tau_{\mathrm{SE}}$. Thus, following\ Horsthemke and Lefever
\cite{Horsthemke-Lefever},
one can identify the precise value for $\tau_{\mathrm{SE}}$ where a
qualitative change in\ the functional form of this probability occurs as the
\emph{critical point of a phase transition}\textit{.} This is evidenced by the
functional change (nonanalyticity) of the dependence of the observables (e.g.
the swapping frequency $\omega$) on the control parameter $b\tau_{\mathrm{SE}%
}/\hbar$. Since\ the control parameter switches among \emph{dynamical regimes}
we call this phenomenon a \emph{Quantum Dynamical Phase Transition}\textit{. }

It should be remarked that the effect of other spins on the two spin system
introduces non-commuting perturbing operators (symmetry breaking
perturbations) which produce non-linear dependences of the observables. While
this could account for cross-over among the limiting dynamical regimes, it
does not ensure a phase transition. A true phase transition needs a
\emph{non-analyticity} in these functions which is only enabled by taking the
thermodynamic limit of an infinite number of spins \cite{Sachdev}. In our
formalism, this is incorporated through the imaginary part of the energy,
$\hbar/\tau_{\mathrm{SE}}$, evaluated from the Fermi Golden Rule
(\ref{Imaginary_part_sigma_R}).

\begin{figure}
[ptbh]
\begin{center}
\includegraphics[
height=4.9908in,
width=3.8917in
]%
{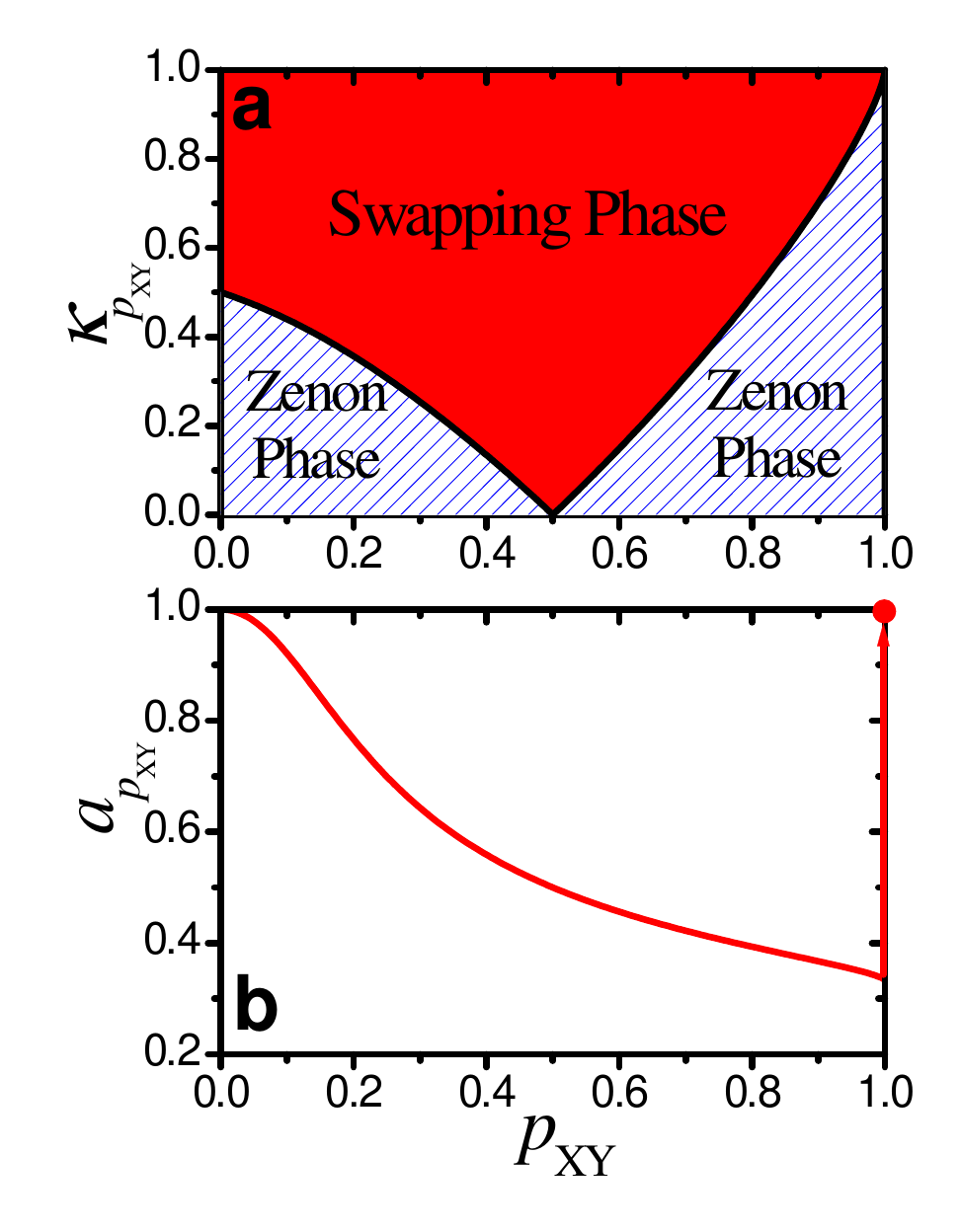}%
\caption[Critical value of the swapping frequency and $a_{p_{\mathrm{XY}}}$ as
a function of $p_{\mathrm{XY}}$ (anisotropy of the SE interaction).]{a)
Critical value of the swapping frequency as a function of $p_{\mathrm{XY}}$
(anisotropy of the SE interaction). b) The parameter $a_{p_{\mathrm{XY}}}$ of
the swapping frequency as a function of $p_{\mathrm{XY}}.$}%
\label{Fig_apxy_kpxy_QDPT}%
\end{center}
\end{figure}

When the SE interaction rates are \emph{anisotropic} ($\Gamma_{\mathrm{ZZ}%
}\neq\Gamma_{\mathrm{XY}}$),
there is a functional dependence of $\omega\ $on $\tau_{\mathrm{SE}}$ and $b$
yielding a critical value for their product, $b\tau_{\mathrm{SE}}%
/\hbar=k_{p_{\mathrm{XY}}},$ where the dynamical regime switches. One
identifies two parametric regimes: 1- The \emph{swapping phase}\textit{,}
which is a form of sub-damped dynamics, when $b\tau_{\mathrm{SE}%
}/k_{p_{\mathrm{XY}}}>\hbar$ ($\eta=0$ in eq. (\ref{PolCST})). 2- A \emph{Zeno
phase}, with an over-damped dynamics for $b\tau_{\mathrm{SE}}%
/k_{p_{\mathrm{XY}}}<\hbar$ as a consequence of the strong coupling with the
environment\ (\emph{zero frequency}, i.e. $\omega=0$, in eq. (\ref{PolCST})).
In the neighborhood of the critical point the swapping frequency takes the
form:
\begin{equation}
\omega=\left\{
\begin{array}
[c]{cc}%
a_{p_{\mathrm{XY}}}^{{}}\sqrt{\left(  b/\hbar\right)  _{{}}^{2}%
-k_{p_{\mathrm{XY}}}^{2}/\tau_{\mathrm{SE}}^{2}} & b\tau_{\mathrm{SE}%
}/k_{p_{\mathrm{XY}}}>\hbar\\
0 & b\tau_{\mathrm{SE}}/k_{p_{\mathrm{XY}}}\leq\hbar
\end{array}
\right.  . \label{freq-crit}%
\end{equation}
and $\eta$ becomes%
\begin{equation}
\eta=\left\{
\begin{array}
[c]{cc}%
0 & b\tau_{\mathrm{SE}}/k_{p_{\mathrm{XY}}}>\hbar\\
a_{p_{\mathrm{XY}}}^{{}}\sqrt{k_{p_{\mathrm{XY}}}^{2}/\tau_{\mathrm{SE}_{{}}%
}^{2}-\left(  b/\hbar\right)  _{{}}^{2}} & b\tau_{\mathrm{SE}}%
/k_{p_{\mathrm{XY}}}\leq\hbar
\end{array}
\right.  ,
\end{equation}
where%
\begin{equation}
k_{p_{\mathrm{XY}}}^{2}=\frac{1}{12}\left\{  \left[  \left(  p_{\mathrm{XY}%
}-1\right)  ^{2}\chi\left(  p_{\mathrm{XY}}\right)  \right]  ^{\frac{1}{3}%
}+\zeta\left(  p_{\mathrm{XY}}\right)  +19p_{\mathrm{XY}}^{2}+\frac{\left(
p_{\mathrm{XY}}-1\right)  ^{\frac{4}{3}}\zeta\left(  p_{\mathrm{XY}}\right)
}{\left[  \chi\left(  p_{\mathrm{XY}}\right)  \right]  ^{\frac{1}{3}}%
}\right\}  ,
\end{equation}%
\begin{equation}
\chi\left(  p_{\mathrm{XY}}\right)  =-5291p_{\mathrm{XY}}^{4}%
-1084p_{\mathrm{XY}}^{3}+546p_{\mathrm{XY}}^{2}-4p_{\mathrm{XY}}+1+24\sqrt
{3}p_{\mathrm{XY}}\sqrt{\left(  28p_{\mathrm{XY}}^{2}-2p_{\mathrm{XY}%
}+1\right)  ^{3}},
\end{equation}%
\begin{equation}
\zeta\left(  p_{\mathrm{XY}}\right)  =-215p_{\mathrm{XY}}^{2}-2p_{\mathrm{XY}%
}+1,
\end{equation}
and%
\begin{equation}
a_{p_{\mathrm{XY}}}^{2}=\frac{1}{2}\frac{\left(  f_{1}^{~~2/3}+36f_{2}\right)
\left(  -f_{3}f_{1}^{\ \ 2/3}+36f_{2}f_{3}+f_{1}f_{4}\right)  }{f_{1}%
^{~~5/3}f_{4}}, \label{orden1}%
\end{equation}%
\begin{equation}
f_{1}=36k_{p_{\mathrm{XY}}}^{2}-8+24p_{\mathrm{XY}}+48p_{\mathrm{XY}}%
^{2}-36k_{p_{\mathrm{XY}}}^{2}p_{\mathrm{XY}}-64p_{\mathrm{XY}}^{3}+12f_{4},
\end{equation}%
\begin{equation}
f_{2}=1/3k_{p_{\mathrm{XY}}}^{2}-1/3p_{\mathrm{XY}}^{2}-1/9\left(
1-p_{\mathrm{XY}}\right)  ^{2},
\end{equation}%
\begin{equation}
f_{3}=-f_{4}+p_{\mathrm{XY}}f_{4}-6k_{p_{\mathrm{XY}}}^{2}-16p_{\mathrm{XY}%
}^{4}+20p_{\mathrm{XY}}^{3}-10p_{\mathrm{XY}}^{2}+13p_{\mathrm{XY}}%
^{2}k_{p_{\mathrm{XY}}}^{2}-2k_{p_{\mathrm{XY}}}^{2}p_{\mathrm{XY}%
}+k_{p_{\mathrm{XY}}}^{2},
\end{equation}%
\begin{multline}
f_{4}=\left(  12k_{p_{\mathrm{XY}}}^{2}+96p_{\mathrm{XY}}^{4}k_{p_{\mathrm{XY}%
}}^{2}-120p_{\mathrm{XY}}^{3}k_{p_{\mathrm{XY}}}^{2}+60p_{\mathrm{XY}}%
^{2}k_{p_{\mathrm{XY}}}^{2}-39k_{p_{\mathrm{XY}}}^{2}p_{\mathrm{XY}}%
^{2}\right. \nonumber\\
\left.  +6k_{p_{\mathrm{XY}}}^{2}p_{\mathrm{XY}}-3k_{p_{\mathrm{XY}}}%
^{2}-48p_{\mathrm{XY}}^{4}+48p_{\mathrm{XY}}^{3}-12p_{\mathrm{XY}}^{2}\right)
^{\frac{1}{2}}.
\end{multline}
In equation (\ref{orden1}) the functions $f_{1},$ $f_{2},$ $f_{3}$ and $f_{4}$
only depend of $p_{\mathrm{XY}}.$ Thus, the parameters $a_{p_{\mathrm{XY}}}$
and $k_{p_{\mathrm{XY}}}$ only depend on $p_{\mathrm{XY}}$ which is determined
by the origin of the interaction Hamiltonian. For typical interaction
Hamiltonians the values of these parameters $\left(  p_{\mathrm{XY}%
},k_{p_{\mathrm{XY}}},a_{p_{\mathrm{XY}}}\right)  $ are: Ising $\left(
0,\frac{1}{2},1\right)  $, dipolar $\left(  \frac{1}{5},0.3564,0.8755\right)
$, isotropic $\left(  \frac{1}{2},0,\frac{1}{\sqrt{2}}\right)  $ and $XY$
$\left(  1,1,1\right)  $.
Fig. \ref{Fig_apxy_kpxy_QDPT}
shows the parameters $a_{p_{\mathrm{XY}}}$ and $\kappa_{p_{\mathrm{XY}}}$ as a
function of $p_{\mathrm{XY}}$. It is important to note that when the SE
interaction rates are isotropic ($\Gamma_{\mathrm{XY}}=\Gamma_{\mathrm{ZZ}}$)
the critical point is zero, thus, the oscillation frequency is finite for all
values of the SE interaction rate $\Gamma_{\mathrm{SE}}/\hbar$. While the
oscillation could be very attenuated it is ensured a swapping with a frequency
proportional to $b.$ The swapping period is
\begin{equation}
T\simeq\frac{T_{0\mathrm{c}}^{3/2}}{\sqrt{2}a_{p_{\mathrm{XY}}}^{{}}}\ \left(
T_{0\mathrm{c}}-T_{0}\right)  ^{-1/2},
\end{equation}
where
\begin{equation}
T_{0}=\frac{2\pi\hbar}{b}%
\end{equation}
is the isolated two spin period and its critical value
\begin{equation}
T_{0\mathrm{c}}=\frac{2\pi\tau_{\mathrm{SE}}}{k_{p_{\mathrm{XY}}}},
\end{equation}
determines the region where the period $T$ diverges as is observed in fig.
\ref{Figteoexp3D}. The estimated value of $\tau_{\mathrm{SE}}=0.275%
\operatorname{ms}%
$ and dipolar SE interactions yield a critical value for the $^{13}$C-$^{1}$H
coupling of $b_{\mathrm{c}}/\hbar=2\pi/T_{0\mathrm{c}}=1.3%
\operatorname{kHz}%
\mathrm{.}$

\begin{figure}
[ptbh]
\begin{center}
\includegraphics[
height=5.0436in,
width=5.2607in
]%
{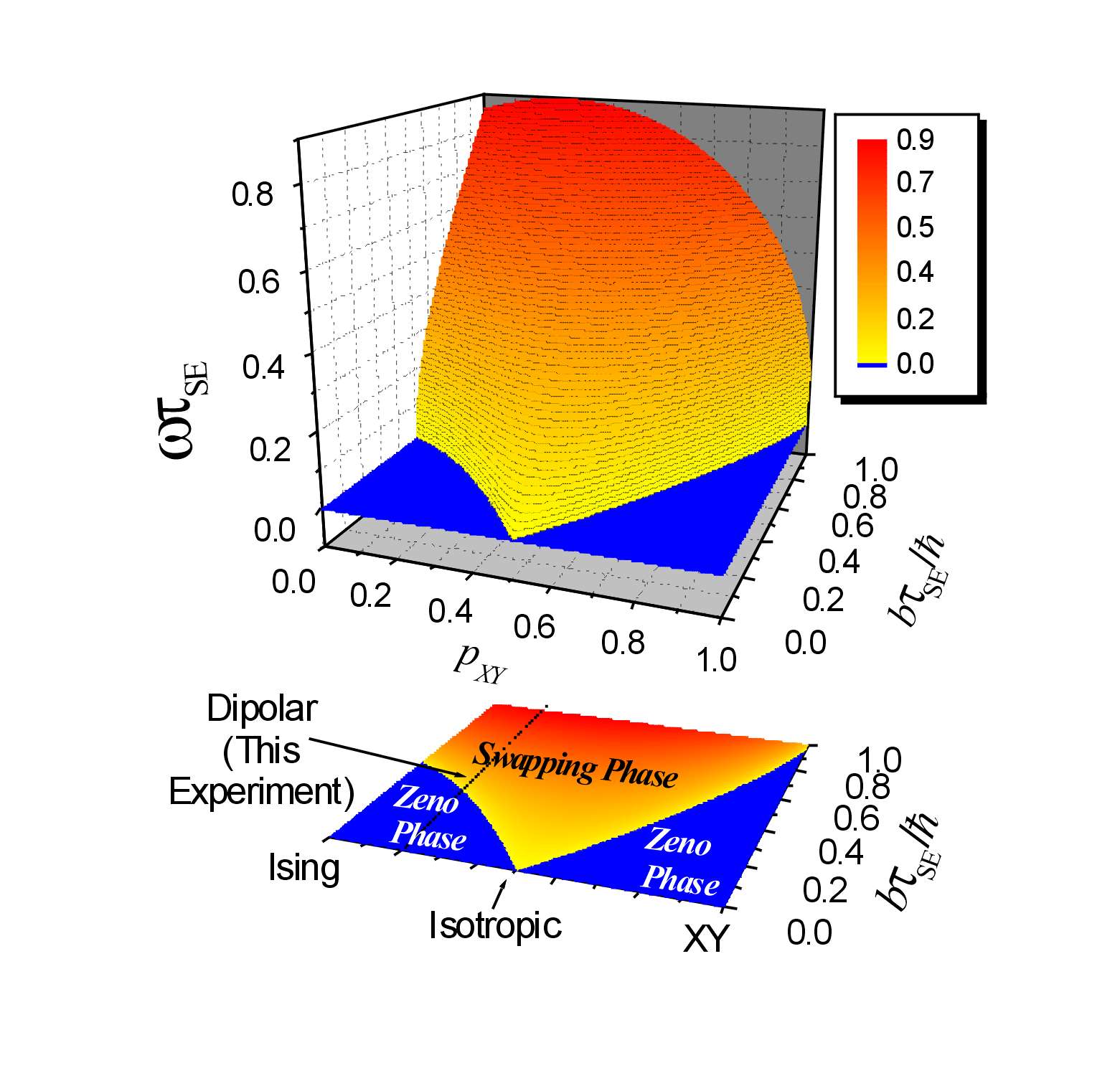}%
\caption[Quantum dynamical phase diagram for the spin swapping operation.]%
{Quantum dynamical phase diagram for the spin swapping operation. The figure
shows the frequency dependence on system-environment (SE) interaction
anisotropy $p_{\mathrm{XY}}$ and the ratio among the internal and the SE
interaction $b\tau_{\mathrm{SE}}/\hbar$. The projection over the
$b\tau_{\mathrm{SE}}/\hbar-p_{\mathrm{XY}}$ plane determines the phase diagram
where the transition between the swapping phase into the Zeno phase
($\omega=0)$ is manifested. Values of $p_{\mathrm{XY}}$ for typical SE
interaction Hamiltonian are indicated in the contour plot.}%
\label{phasediag}%
\end{center}
\end{figure}

The complete phase diagram that accounts for the anisotropy of the SE
interactions is shown in fig. \ref{phasediag}. There, the frequency dependence
on $p_{\mathrm{XY}}$ and $b\tau_{\mathrm{SE}}/\hbar$ is displayed. At the
critical line, $k_{p_{\mathrm{XY}}}^{2}$ as a function of $p_{\mathrm{XY}},$
the frequency becomes zero setting the limits between both dynamical phases.%

The two dynamical phases can now be identified in the NMR experiments which up
to date defied explanation. The experimental setup provides a full scan of the
parameter $b\tau_{\mathrm{SE}}$ through the phase transition that is
manifested when the frequency goes suddenly to zero (fig. \ref{Figexp} b) and
the relaxation rate [fig. \ref{Figexp} a)] changes its behavior decaying
abruptly. The fact that $1/\tau_{\phi}$ tends to zero when $b\ll\hbar
/\tau_{\mathrm{SE}}$ confirms the Zeno phase predicted by our model. In this
regime, $1/\tau_{\phi}$ is quadratic on $b$ as prescribed by eq.
(\ref{Tau_phi_quadratic_Zeno}),%
\begin{equation}
1/\tau_{\phi}\propto\left(  b/\hbar\right)  ^{2}\tau_{\mathrm{SE}}.
\end{equation}
To make the comparison between the two panels of fig. \ref{Figteoexp3D}%
\ quantitative, we fit the predicted dynamics of fig. \ref{Figteoexp3D} b)
with $P^{^{\mathrm{MKBE}}}\left(  t\right)  ,$ following the same procedure
used to fit the experimental data. The solid line in fig. \ref{Figexp} show
the fitting parameters $1/\tau_{\phi}$ and $\omega$ in excellent agreement
with the experimental ones.

We point out that eq. (\ref{eq-MKBE}) is used to fit both the experiments and
the theoretical prediction of eq. (\ref{PolCST}) because it constitutes a
simple, thought imperfect, way to extract the \textquotedblleft
outputs\textquotedblright\ (oscillation frequency\ $\omega$ and a decoherence
time~$\tau_{\phi}$). While the systematic errors shift the actual critical
value of the control parameter,\ $b/\hbar$, from $1.3%
\operatorname{kHz}%
$ to $2%
\operatorname{kHz}%
$, eq. (\ref{eq-MKBE}) yields a simplified way to \textquotedblleft
observe\textquotedblright\ the transition.

\section{Signatures of a QDPT in a three-spin system coupled to a spin-bath}

\begin{figure}
[ptbh]
\begin{center}
\includegraphics[
height=5.1768in,
width=4.0439in
]%
{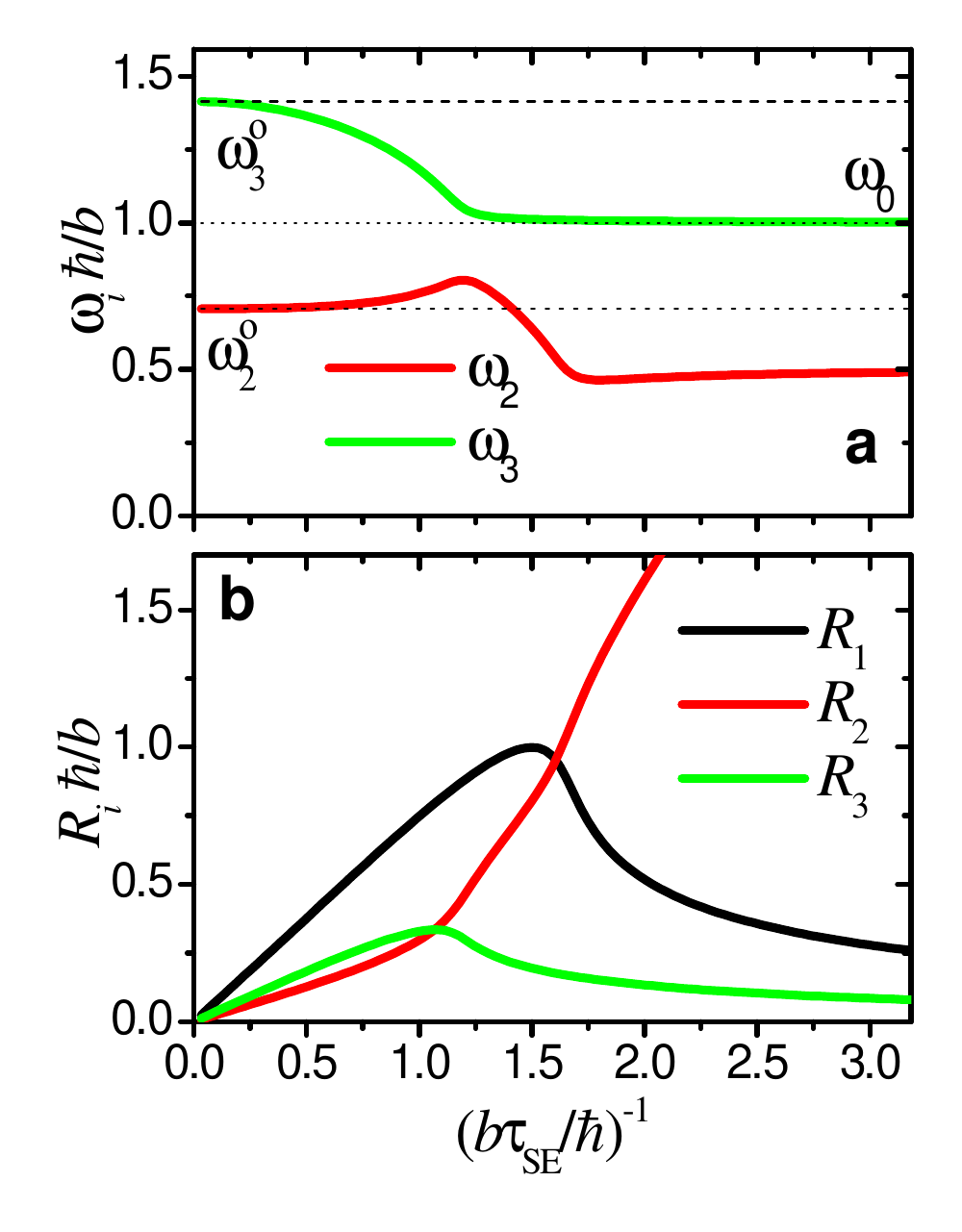}%
\caption[Frequencies involved in the temporal evolution of the polarization in
the $3$-spin system in presence of a spin-bath as a function of $\left(
b\tau_{\mathrm{SE}}/\hbar\right)  ^{-1}$ and the different relaxation rates of
the polarization.]{a) Frequencies involved in the temporal evolution of the
polarization in the $3$-spin system as a function of $\left(  b\tau
_{\mathrm{SE}}/\hbar\right)  ^{-1}$. Dashed lines represent the isolated
system. Dot line correspond to two spins decoupled from the environment. b)
Different relaxation rates of the polarization.}%
\label{Fif_w_R}%
\end{center}
\end{figure}

\begin{figure}
[ptbh]
\begin{center}
\includegraphics[
height=6.1973in,
width=4.4053in
]%
{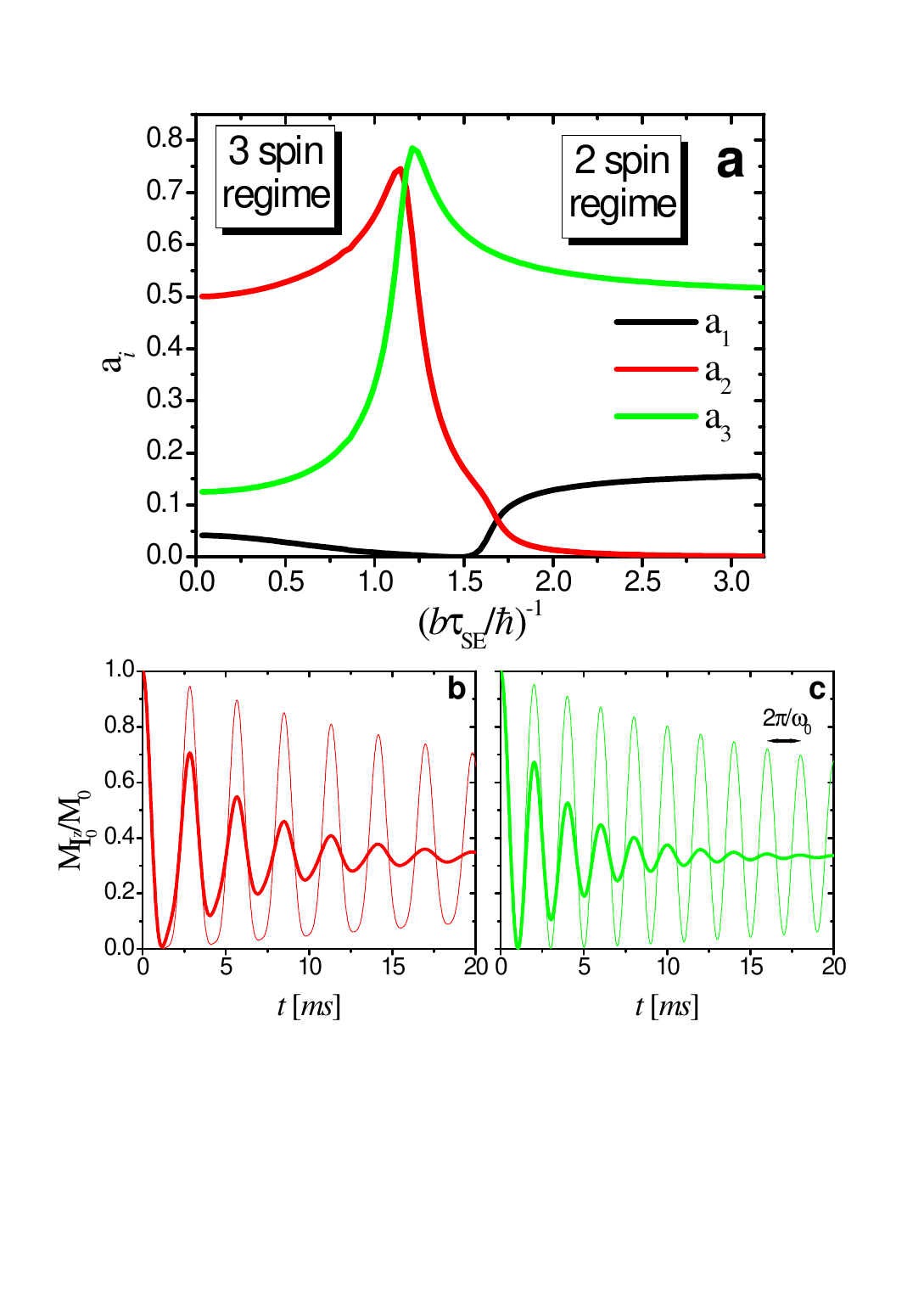}%
\caption[Coefficients (weights) of the different terms of polarization
equation of a $3$-spin system in presence of a spin-bath. At the critical
region there is a switch between \ the $2$-spin and the $3$-spin regime.
Temporal evolutions of the polarization in the $2$-spin and $3$-spin regimes
respectively for different $\tau_{\mathrm{SE}}$.]{a) Coefficients (weights) of
the different terms of eq. (\ref{M3spins}). At the critical region there is a
switch between \ the $2$-spin and the $3$-spin regime. b) and c) Temporal
evolutions of the polarization in the $2$-spin and $3$-spin regimes
respectively for different $\tau_{\mathrm{SE}}.$ In b) $b/\hbar=2\pi
\times1\operatorname{kHz}$ and $\tau_{\mathrm{SE}}=1.43\operatorname{ms}$ for
the thick line and $\tau_{\mathrm{SE}}=10\operatorname{ms}$ for the thin line.
In c) $b/\hbar=2\pi\times1\operatorname{kHz}$ and $\tau_{\mathrm{SE}%
}=0.1\operatorname{ms}$ for the thick line and $\tau_{\mathrm{SE}%
}=0.01\operatorname{ms}$ for the thin line.}%
\label{Fig_as_pol}%
\end{center}
\end{figure}

We consider a three-spin system with the following Hamiltonian
\begin{equation}
\widehat{\mathcal{H}}_{\mathrm{S}}=\hbar\Omega_{\mathrm{Z}}\left(  \hat
{I}_{-1}^{z}+\hat{I}_{0}^{z}+\hat{I}_{1}^{z}\right)  +\tfrac{1}{2}b\left(
\hat{I}_{-1}^{+}\hat{I}_{0}^{-}+\hat{I}_{-1}^{-}\hat{I}_{0}^{+}\right)
+\tfrac{1}{2}b\left(  \hat{I}_{0}^{+}\hat{I}_{1}^{-}+\hat{I}_{0}^{-}\hat
{I}_{1}^{+}\right)  ,
\end{equation}
where we add one spin to the system. However, in this case there is not a
interaction between the spin $I_{-1}$ and the $I_{1}$ as there was in section
\S \ \ref{M_Q_dyn_in_8CB}. The environment Hamiltonian and the SE interaction
Hamiltonians remain as in the previous section,%
\begin{align}
\widehat{\mathcal{H}}_{\mathrm{E}}  &  =\sum_{i\geq2}\hbar\Omega_{\mathrm{Z}%
}^{{}}\hat{I}_{i}^{z}+%
{\textstyle\sum\limits_{\genfrac{}{}{0pt}{1}{i\geq2}{j>i}}}
\tfrac{1}{2}b_{ij}^{{}}\left(  \hat{I}_{i}^{+}\hat{I}_{j}^{-}+\hat{I}_{i}%
^{-}\hat{I}_{j}^{+}\right)  ,\\
\widehat{\mathcal{H}}_{\mathrm{SE}}  &  =a_{12}^{{}}\hat{I}_{1}^{z}\hat{I}%
_{2}^{z}+\tfrac{1}{2}b_{12}^{{}}\left(  \hat{I}_{1}^{+}\hat{I}_{2}^{-}+\hat
{I}_{1}^{-}\hat{I}_{2}^{+}\right)  .
\end{align}
Also, the environment is coupled only to one spin of the system. We solve the
generalized quantum master equation (\ref{master}), using the same procedure
as in section \S \ \ref{Mark_non_secular_2spin},
without neglecting non-secular terms of the relaxation superoperator.
Considering now the initial condition
\begin{equation}
\hat{\sigma}\left(  0\right)  =\frac{\left(  \hat{1}+\beta_{\mathrm{B}}%
\hbar\Omega_{0,I}\hat{I}_{-1}^{z}\right)  }{\mathrm{Tr}\left\{  \hat
{1}\right\}  }%
\end{equation}
and the spin-bath polarized,
we obtain for the magnetization of site $-1$%
\begin{align}
M_{I_{-1}^{z}}\left(  t\right)   &  =\mathrm{Tr}\left\{  \hat{I}_{-1}^{z}%
\hat{\sigma}\left(  t\right)  \right\} \label{M3spins}\\
&  =M_{0}\left(  1-a_{0}e^{-R_{0}t}-a_{1}e^{-R_{1}t}+a_{2}\sin\left(
\omega_{2}t+\phi_{2}\right)  e^{-R_{2}t}+a_{3}\sin\left(  \omega_{3}t+\phi
_{3}\right)  e^{-R_{3}t}\right)  .
\end{align}
The coefficients $a_{i},$ $R_{i}$, $\omega_{i}$ and $\phi_{i}$ are real and
they are functions of $b,$ $1/\tau_{\mathrm{SE}}\ $and $p_{\mathrm{XY}}.$ If
$p_{\mathrm{XY}}\neq0$ the final state has all the spins polarized because a
net transfer of magnetization from the spin-bath is possible. However, for an
Ising SE interaction, $p_{\mathrm{XY}}=0,$ we obtain that $R_{0}=0$ and
$1-a_{0}=1/3$ (the asymptotic polarization) because the final state is the
quasi-equilibrium of the $3$-spin system as we described in section
\S \ \ref{M_SI2_bath_secular}, eq. (\ref{quasi-equilibrium_3spin_DM}). In fig.
\ref{Fif_w_R}
we show the frequencies $\omega_{2}$ and $\omega_{3}$ and the different
relaxation rates as a function of $\left(  b\tau_{\mathrm{SE}}/\hbar\right)
^{-1}$
when the SE interaction is Ising $(p_{\mathrm{XY}}=0)$.
Two changes, each resembling the critical behavior shown by two spin systems
are observed (see fig. \ref{Figexp}). The same phenomenon occurs in figure
\ref{Fig_as_pol} a)
where the coefficients $a_{i}$ are shown. The polarization evolution of an
isolated $3$-spin system is
\begin{equation}
M_{I_{-1}^{z}}\left(  t\right)  =\frac{M_{0}}{8}\left(  3+4\cos\left(
\omega_{2}^{o}t\right)  +\cos\left(  \omega_{3}^{o}t\right)  \right)
\end{equation}
where
\begin{equation}
\omega_{2}^{o}=\frac{\sqrt{2}}{4}\frac{b}{\hbar}%
\end{equation}
and
\begin{equation}
\omega_{3}^{o}=\frac{\sqrt{2}}{2}\frac{b}{\hbar}%
\end{equation}
are the natural frequencies. When $\left(  b\tau_{\mathrm{SE}}/\hbar\right)
^{-1}<<1,$ we observe that $\omega_{2}\rightarrow\omega_{2}^{o},$ $\omega
_{3}\rightarrow\omega_{3}^{o},$ $a_{1}\rightarrow1/3-3/8=-1/24,$
$a_{2}\rightarrow1/2$ and $a_{3}\rightarrow1/8$ as expected for an isolated
$3$-spin dynamics. The dependence of $\omega_{3}$ as a function of $\left(
b\tau_{\mathrm{SE}}/\hbar\right)  ^{-1}$ is similar to that of the swapping
frequency of fig. \ref{Figexp}. However, instead of becoming zero when the SE
interaction increase, it suddenly stabilizes at $\omega_{0}=b/\hbar$, the bare
$2$-spin Rabi frequency. At the same point $\omega_{2}$, $R_{2}$ and $R_{3}$
also have a sudden change. While $R_{2}$ and $R_{3}$ initially grew as
$\left(  b\tau_{\mathrm{SE}}/\hbar\right)  ^{-1}$, there $R_{2}$ increases the
growing speed while $R_{3}$ begins to decay as in the Zeno phase (see fig.
\ref{Figexp}). Moreover, looking at the behavior of the coefficients $a_{2}$
and $a_{3}$, we observe a form of switch between them, $a_{2}$ suddenly goes
down and $a_{3}$ goes up. These coefficients are the weight of the different
frequency contributions in the time evolution. Both changes together, on the
decoherence rates and on the weight coefficients of the different oscillatory
terms, beyond the critical interaction (region) leads the system to oscillate
with the bare Rabi frequency of the two spins decoupled from the environment.
If we continue increasing the control parameter $\left(  b\tau_{\mathrm{SE}%
}/\hbar\right)  ^{-1},$ this effect is enhanced by the next transition. After
the second transition, $R_{1}$ begins to decrease as in the Zeno phase
behavior of fig. \ref{Figexp}. As the term of eq. \ref{M3spins} that relax
with $R_{1}$ leads the system to the $3$-spin quasi-equlibrium,
when it goes down this final state is approached more slowly. This effect
tries to avoid the interaction between the $2$-spins not coupled to the
environment and the $I_{2}$ spin. After the second transition, the coefficient
$a_{1}$ goes to zero abruptly leading to a more pronounced \textquotedblleft
isolation\textquotedblright\ of the $2$-spins. After that, we can characterize
two dynamical regimes: One which is characterized by the $3$-spin dynamics
when $\left(  b\tau_{\mathrm{SE}}/\hbar\right)  ^{-1}\lesssim1$ and the second
one, when $\left(  b\tau_{\mathrm{SE}}/\hbar\right)  ^{-1}\gtrsim1,$ have the
$2$-spin behavior.

Fig. \ref{Fig_as_pol} b) and c) show the temporal evolution of the
magnetization of eq. (\ref{M3spins}) on the $3$-spins and the $2$-spin regimes
respectively.
While in fig. \ref{Fig_as_pol} b) the two frequency contributions are evident,
in fig. \ref{Fig_as_pol} c) only the bare Rabi frequency is manifested. In
each graph we show two curves with different SE interactions. In fig.
\ref{Fig_as_pol} b), we show that increasing the SE interaction the
decoherence rates increase. However, in the $2$-spin regime [fig.
\ref{Fig_as_pol} c)] when it is increased the decoherence rate decrease
leading to a better \textquotedblleft isolation\textquotedblright. It is
important to take into account that while the relaxation rates goes to zero
smoothly the swapping frequency acquire the bare value near the critical
point. Another fact to remark is that this effect is more pronounced when the
anisotropy of the SE interaction is close to a pure Ising SE interaction while
an increase in the XY nature leads a further smoothing of the transition.
The reason is that, when $p_{\mathrm{XY}}\neq0$,
there is a net transfer of magnetization to the system which is redistributed
between the three spins, this redistribution begins to be slower at the second
transition when $R_{3}$ goes down. In contrast, for a purely Ising
interaction, there is no net polarization transfer and a purely decoherent
process at site $3$ freezes its dynamics but its fast energy fluctuations
prevent the interaction with the other spins.

\section{Summary}

We found experimental evidence that environmental interactions can drive a
swapping gate through a \emph{Quantum Dynamical Phase Transition} towards an
over-damped or Zeno phase \cite{JCP06}. The NMR implementation of a spin
swapping in a $^{13}$C-$^{1}$H system enables the identification and
characterization of this phase transition as a function of the ratio
$b\tau_{\mathrm{SE}}/\hbar$ between the internal and SE interaction. In
chapter \ref{Marker_Spin_within_keldysh}, we developed a microscopic model
\cite{SSC07,Alvarez07b} that describes both phases and the critical region
with great detail, showing that it depends only on the nature of the
interaction \cite{JCP06,Alvarez07b}. In particular, the phase transition does
not occur if the SE interaction gives isotropic interaction rates,
$\Gamma_{\mathrm{ZZ}}=\Gamma_{\mathrm{XY}}$. The phase transition is
manifested not only in the observable swapping frequency but also in the
decoherence rate $1/\tau_{\phi}$. While a perturbative estimation through the
standard Fermi golden rule would tend to identify this rate with the SE
interaction, i.e. $1/\tau_{\phi}\cong1/\tau_{\mathrm{SE}}^{{}}$ $\simeq\left(
d_{23}^{{}}/\hbar\right)  ^{2}\tau_{\mathrm{B}},$ as it occurs well inside the
swapping phase, both rates differ substantially\ as the system enters in the
Zeno phase ($b\tau_{\mathrm{SE}}^{{}}\leq k_{p_{\mathrm{XY}}}\hbar$). Here,
the decoherence rate switches to the behavior $1/\tau_{\phi}^{{}}%
\propto\left(  b/\hbar\right)  _{{}}^{2}\tau_{\mathrm{SE}}^{{}}$. In the Zeno
phase, the system's free evolution decays very fast with a rate $\tau
_{\mathrm{SE}}^{-1}$. In spite of this, one can see that the initial state as
a whole has a slow decay (its dynamics becomes almost frozen) because it is
continuously fed by the environment. Since the $\tau_{\mathrm{SE}}^{{}}$ has
become the correlation time for the spin directly coupled to the environment,
$1/\tau_{\phi}^{{}}$ provided by our calculation can be interpreted as a
\textquotedblleft nested\textquotedblright\ Fermi golden rule rate emphasizing
the non-perturbative nature of the result \cite{JCP06}. Based on the wealth of
this simple swapping dynamics, we can foresee applications that range from
tailoring the environments for a reduction of their decoherence on a given
process to using the observed critical transition in frequency and decoherence
rate as a tracer of the environment's nature. This led us to extend the model
to a $3$-spin system and to show that beyond a critical region the two spins
become almost decoupled from the environment oscillating with the bare Rabi
frequency and relaxing more slowly \cite{Alvarez07a}. While in the two spin
swapping gate the dynamical transition is critical, in the $3$-spin system the
criticality is smoothed out. However, enough abruptness remains to give the
possibility to use it to \textquotedblleft isolate\textquotedblright\ a
two-spin system with a finite system-environment interaction. Thus, these
applications open new opportunities for both, the field of quantum information
processing and the general physics and chemistry of open quantum systems
\cite{SSC07}.

\chapter{Polarization transfer enhancement by selective pruning in NMR}

Inspired in the stroboscopic model discussed in section
\S \ \ref{Marker_Stroboscopic_process}, we propose a new NMR pulse sequence to
improve the transfer of polarization through a specific pathway in a system of
many interacting spins. The sequence effectively prunes branches of spins,
where no polarization is required, during the polarization transfer procedure.
In this way, a local polarization excitation is transferred more efficiently
from a chosen \textquotedblleft source spin\textquotedblright\ to a
\textquotedblleft target spin\textquotedblright. Simulations of the spin
dynamics with real values of chemical shifts and $J$-couplings in the $^{13}$C
backbone of leucine for a couple of sources and targets are performed.
Enhancements of a factor of up to $300\%$ in the target polarization are
observed without sacrificing the relevant times. Possible applications and
potential fundamental contributions to engineered decoherence are discussed.

\section{The pruner sequence\label{Marker_pruner_sequence}}

One standard procedure to transfer polarization within molecules in the liquid
state, called TOCSY (Total Correlation Spectroscopy)
\cite{TocsyErnst,TocsyBax85}
is based on the isotropic interaction (or $J$-interaction). In general, one
has a mixing Hamiltonian which makes possible the spin-spin interaction.
The mixing Hamiltonian in the rotating frame for an isotropic interaction is%
\begin{equation}
\widehat{\mathcal{H}}_{\mathrm{mix}}^{\mathrm{iso}}=\sum_{i\neq j}J_{ij}^{{}%
}\left(  \hat{I}_{i}^{z}\hat{I}_{j}^{z}+\hat{I}_{i}^{y}\hat{I}_{j}^{y}+\hat
{I}_{i}^{x}\hat{I}_{j}^{x}\right)  ,
\end{equation}
where $J_{ij}$ is the scalar coupling and $\hat{I}_{i}^{u}$ $\left(
u=x,y,z\right)  $ are the spin operators. As this Hamiltonian connects all the
possible pairs of spins, an initial local excitation would spread over all of
them. The idea behind our new sequence is to effectively \textquotedblleft
disconnect\textquotedblright\ those spins which do not belong to the
\textquotedblleft selected transfer pathway\textquotedblright. In order to
achieve this goal, one makes use of the \emph{differences in chemical shifts}
in the spin system.

\begin{figure}
[tbh]
\begin{center}
\includegraphics[
height=1.6855in,
width=4.3517in
]%
{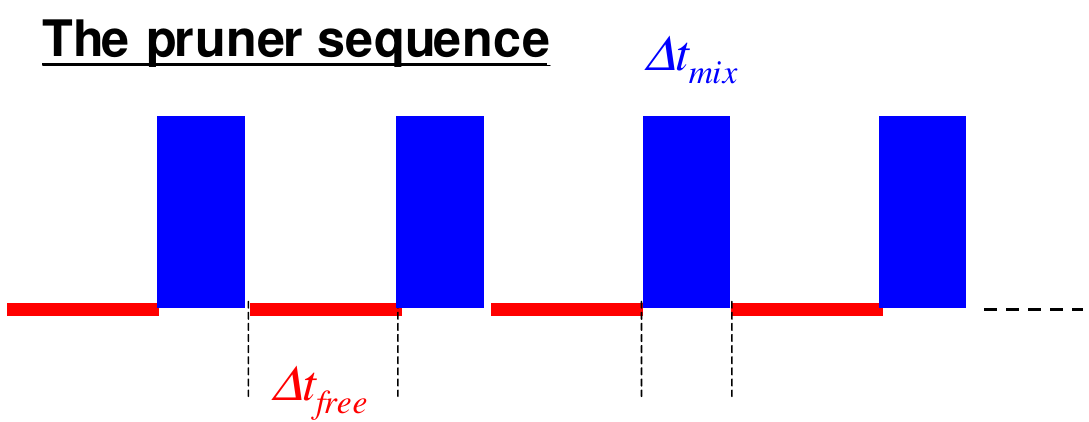}%
\caption{Schematic representation of the pruner sequence.}%
\label{Fig_the_pruner_sequence}%
\end{center}
\end{figure}

\textquotedblleft The pruner\textquotedblright\ sequence stroboscopically
interrupts the mixing evolution performed under $\widehat{\mathcal{H}%
}_{\mathrm{mix}}^{\mathrm{iso}}$ with free evolutions, that in the rotating
frame are given by%
\begin{equation}
\widehat{\mathcal{H}}_{\mathrm{free}}=\sum_{i}\hbar\Delta\Omega_{i}^{{}}%
\hat{I}_{i}^{z}+\sum_{i\neq j}J_{ij}^{{}}\hat{I}_{i}^{z}\hat{I}_{j}^{z},
\end{equation}
where $\Delta\Omega_{i}$ are the off-resonance shifts as defined in
\ref{offresonance} and $\Delta_{ij}=\hbar(\Delta\Omega_{i}-\Delta\Omega
_{j})\gg J_{ij}$. Thus, there is a time interval, $\Delta t_{\mathrm{mix}},$
with the mixing evolution under the Hamiltonian $\widehat{\mathcal{H}%
}_{\mathrm{mix}}^{\mathrm{iso}}$ followed by a free evolution time step
$\Delta t_{\mathrm{free}}$. This sequence of evolutions are repeated
successively as schematized in fig. \ref{Fig_the_pruner_sequence}.
The adjustment of the free evolution time, $\Delta t_{\mathrm{free}}$, must be
a least common multiple of the inverses of the chemical shift differences
\begin{equation}
\Delta_{ij}=\hbar\left(  \Delta\Omega_{i}-\Delta\Omega_{j}\right)
\end{equation}
between the spins connected in the pathway, i.e.
\begin{equation}
\Delta t_{\mathrm{free}}=2\pi n_{ij}\hbar/\left\vert \Delta_{ij}\right\vert
\end{equation}
where $n_{ij}$ is a natural number. Thus, all the spins in the pathway
\emph{are in phase} because their relative phase is proportional to $2\pi$,
while the others are dephased in each free evolution.

\section{Numerical simulation on the L-leucine molecule}

\begin{figure}
[tbh]
\begin{center}
\includegraphics[
height=4.2272in,
width=5.4025in
]%
{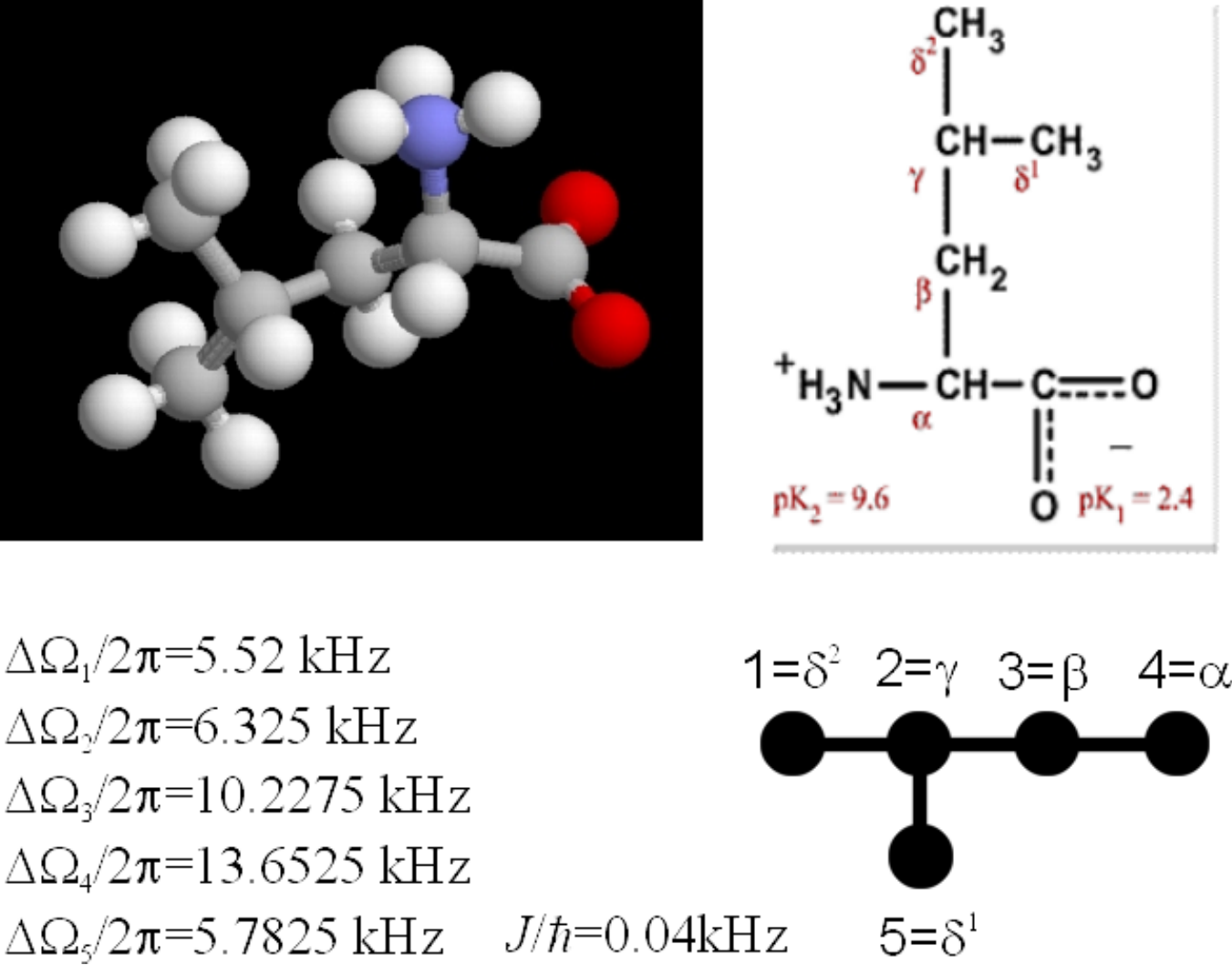}%
\caption[Molecule of L-leucine.]{Molecule of L-leucine showing the numbering
of the carbons used in the spin dynamics calculation. The carboxylate carbon
is not considered in the simulations because as it resonates at quite
different frequency is easily decoupled from the others. The $^{13}$C chemical
shifts in a $23\operatorname{T}$ \ magnet at pH $6.89$ are $\Delta\Omega
_{1}/2\pi=5.52\operatorname{kHz}\mathrm{,}$ $\Delta\Omega_{2}%
=6.325\operatorname{kHz}$, $\Delta\Omega_{3}=10.2275\operatorname{kHz}%
\mathrm{,}$ $\Delta\Omega_{4}=13.6525\operatorname{kHz}$ and $\Delta\Omega
_{5}=5.7825\operatorname{kHz}\mathrm{,}$ and $J/\hbar=0.04\operatorname{kHz}%
\mathrm{\ }$for all the $^{13}$C-$^{13}$C nearest neighbor couplings.}%
\label{Fig_L-leucine}%
\end{center}
\end{figure}

\begin{figure}
[tbh]
\begin{center}
\includegraphics[
height=4.3033in,
width=6.1635in
]%
{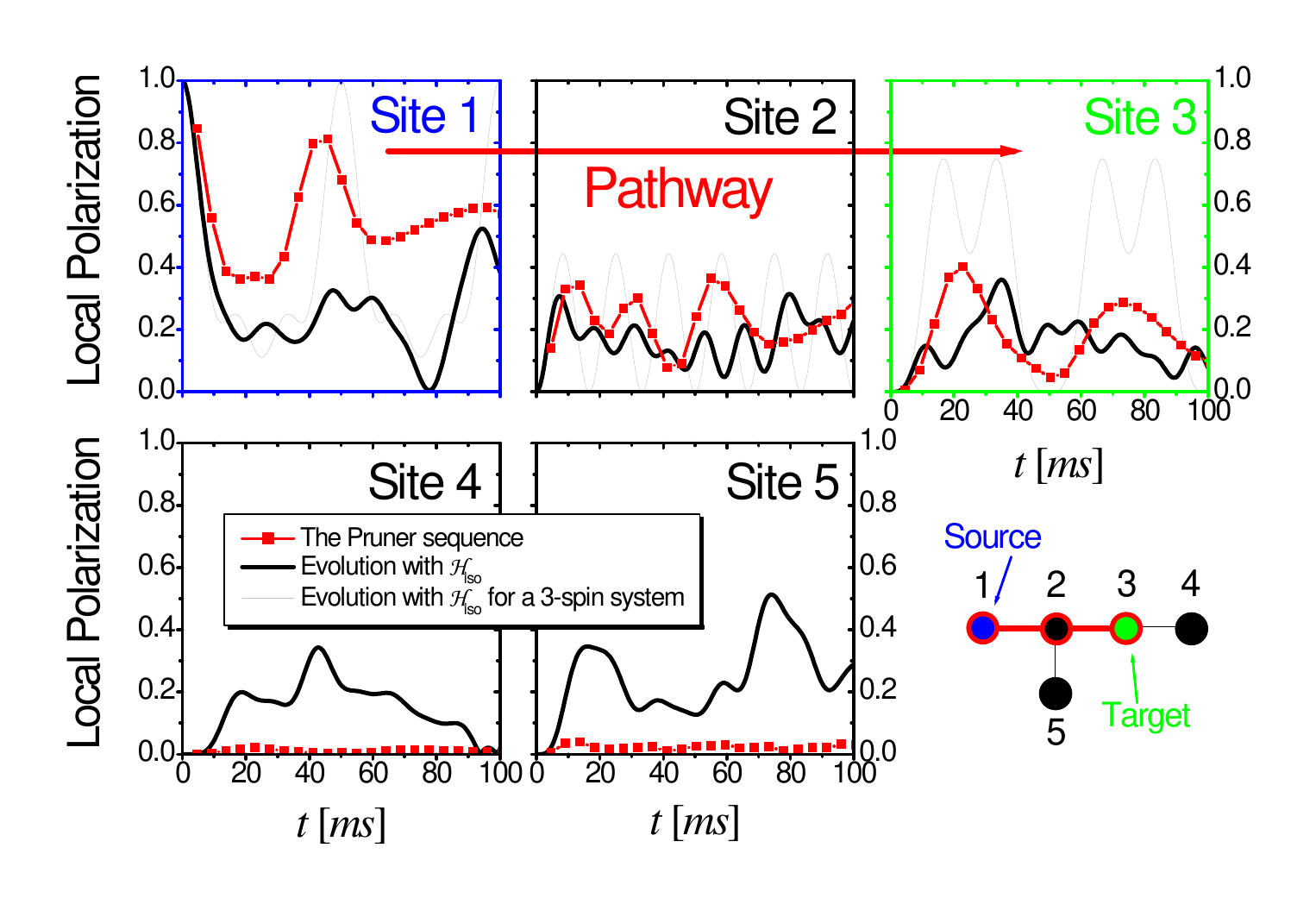}%
\caption[Local polarization evolution under the pruner sequence at different
$^{13}$C sites in an L-lecucine molecule. Selected pathway ($1-2-3$).]{Local
polarization evolution under the pruner sequence at different $^{13}$C sites
in L-leucine. The initial local excitation at site $1$ (the source) is
transfered through the selected pathway ($1-2-3$), to site $3.$ The free
evolution time is $\Delta t_{\mathrm{free}}=1.2744\operatorname{ms}%
=6\times\hbar/\Delta_{13}\approx\hbar/\Delta_{12}\approx5\times\hbar
/\Delta_{23}$ and $\Delta t_{\mathrm{iso}}=3.3\operatorname{ms}\approx\frac
{1}{8}\frac{\hbar}{J}.$ A gain of approx. $270\%$ relative to the first
maximum of the isotropic evolution is observed at site $3.$}%
\label{Fig_L-leucine_1-2-3_path}%
\end{center}
\end{figure}

\begin{figure}
[tbh]
\begin{center}
\includegraphics[
height=4.3033in,
width=6.1678in
]%
{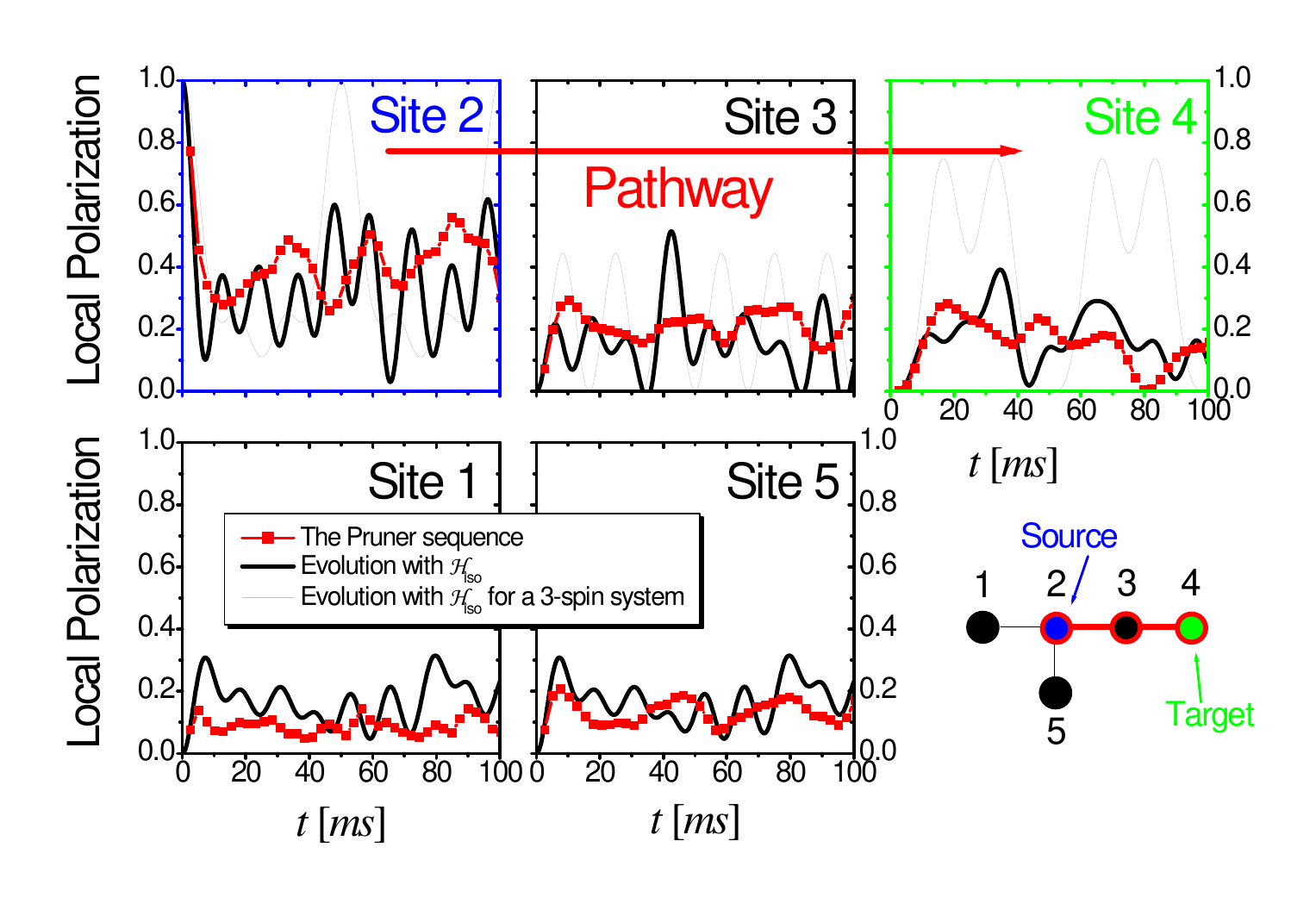}%
\caption[Numerical evolution of the local polarization under the pruner
sequence at different $^{13}$C sites in an L-leucine molecule. Selected
pathway ($2-3-4$).]{Numerical evolution of the local polarization under the
pruner sequence at different $^{13}$C sites in an L-leucine molecule. The
$^{13}$C chemical shifts and nearest neighbor $J$-couplings are given in fig.
\ref{Fig_L-leucine}. The initial local excitation at site $2$ (the source) is
transfered through the selected pathway ($2-3-4$), to site $4$ (the target).
The free evolution time is $\Delta t_{\mathrm{free}}=0.27292\operatorname{ms}%
=2\times\hbar/\Delta_{24}\approx\hbar/\Delta_{23}\approx2\times\hbar
/\Delta_{34}$ and $\Delta t_{\mathrm{iso}}=2.3\operatorname{ms}\approx\frac
{1}{11}\frac{\hbar}{J}.$ A gain of $152\%$ relative to the first maximum of
the isotropic evolution is observed at site $4.$}%
\label{Fig_L-leucine_2-3-4_path}%
\end{center}
\end{figure}

Simulations of a local excitation dynamics under \textquotedblleft the
pruner\textquotedblright\ sequence were performed for different topologies and
initial conditions. The topologies, chemical shifts and $J$-couplings to
exemplify the operation of the sequence were extracted from amino acid
molecules\footnote{Particular values for leucine in D$_{2}$O at pH $6.89$ were
taken from the National Institute of Advanced Industrial Science and
Technology (AIST) of Japan database (SDBS-$^{13}$C NMRSDBS No.
1142CDS-00-770).}. Here, we present the simulations that were done in the
L-leucine molecule (see fig. \ref{Fig_L-leucine}).
%
We find that the pruning is more effective for
\begin{equation}
\Delta t_{\mathrm{mix}}\ll\hbar/J_{ij}^{\mathrm{path}}%
\end{equation}
where $J_{ij}^{\mathrm{path}}$ are the $J$-couplings in the pathway. However,
polarization transfer becomes much slower than in a regular isotropic mixing
sequence.
Thus, after numerical explorations, we found that for a practical experimental
situation, a good compromise for an enhancement of the signal is found if the
time interval for evolution under $\widehat{\mathcal{H}}_{\mathrm{mix}%
}^{\mathrm{iso}}$ is
\begin{equation}
\Delta t_{\mathrm{mix}}\sim\frac{1}{10}\left(  \hbar/J_{ij}^{\mathrm{path}%
}\right)  .
\end{equation}
This allows one to observe a polarization transfer that is larger than the one
obtained by the isotropic mixing. In general, for arbitrary chemical shift
differences $\Delta_{ij},$ it is difficult to obtain a short time $\Delta
t_{\mathrm{free}}$. Then, one must find an approximate solution to avoid
excessive long times. Therefore, the particular enhancement factor depends on
the number of spins necessary to reach the target: the more spin steps the
lower efficiency of the overall sequence. In pathways with three spins, the
enhancement factor in the polarization transfer is approximately three. In the
case of a completely enriched carbon-13 backbone of an L-leucine residue (fig.
\ref{Fig_L-leucine_1-2-3_path})
with the initial excitation at site $1$, one can get a polarization at site
$3$ which is $270\%$ larger than that obtained with $\widehat{\mathcal{H}%
}_{\mathrm{mix}}^{\mathrm{iso}}$ alone. The enhancement factor is calculated
by comparing the first maximum in each evolution. The black lines in fig.
\ref{Fig_L-leucine_1-2-3_path} show the polarization evolution on each site of
the molecule under an isotropic mixing. The red lines show the evolution under
the pruner sequence. It is remarkable the effectiveness of the pruning
manifested in the non observable polarization at sites not belonging to the
selected pathway, i.e. sites $4$ and $5.$ The gray lines show for comparison
the polarization evolution under an isotropic mixing for a three-spin system
composed exclusively by carbons $1,$ $2$ and $3.$ This is the optimal result
that one can expect. In the same molecule, by choosing the source at site $2$
and site $4$ as the target (fig. \ref{Fig_L-leucine_2-3-4_path})
one obtains a gain of approximately $152\%$. There, we can observe that the
pruning is not working so good as in the previous example. This seems to be a
consequence of the fact that carbons $1$ and $5$ do not differ much in
chemical shift from carbon $2,$ requiring more cycles of the sequence to
dephase appreciably from the selected carbons.

\subsection{Alternative mixing Hamiltonians}

\begin{figure}
[tbh]
\begin{center}
\includegraphics[
height=4.3033in,
width=6.1635in
]%
{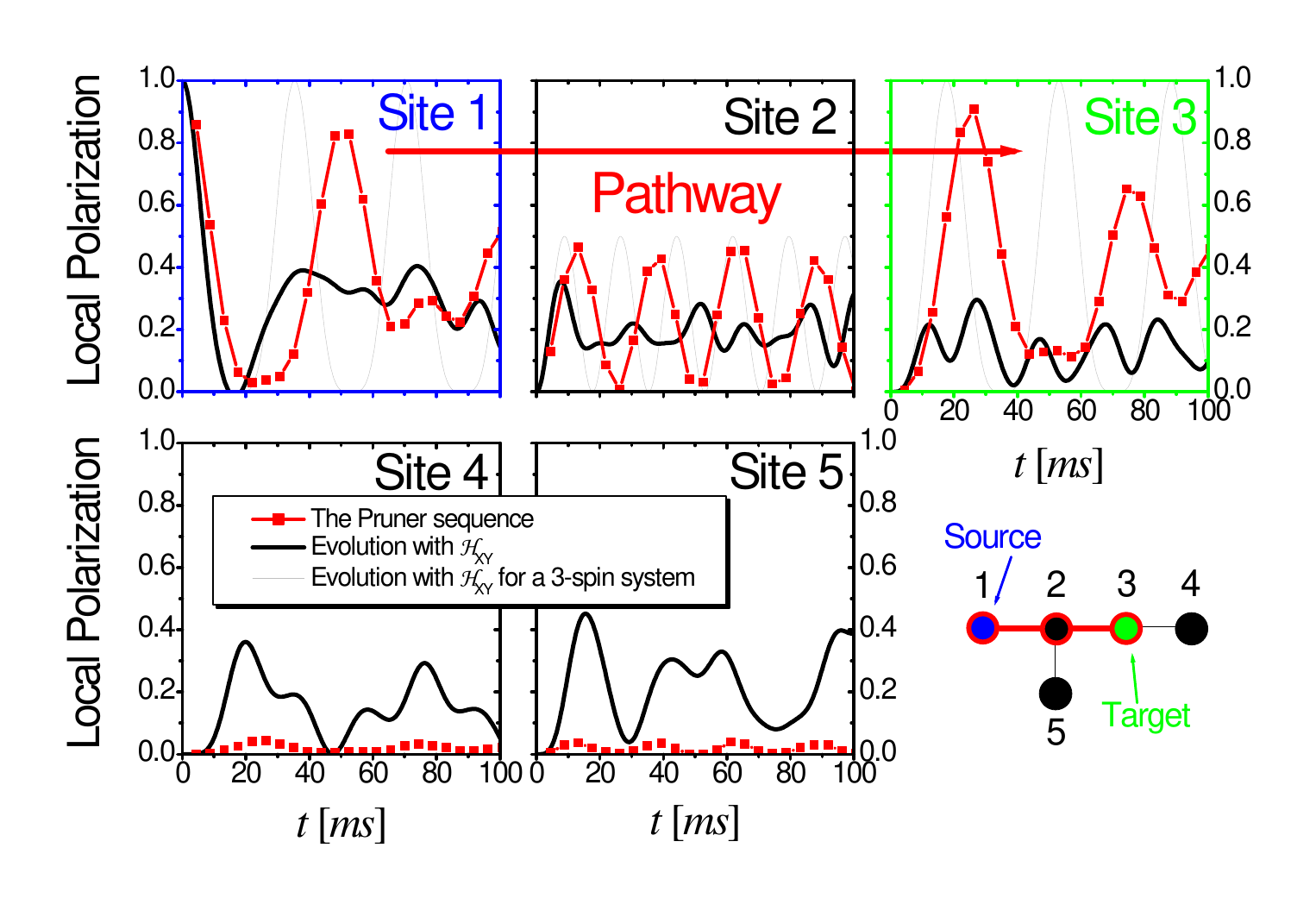}%
\caption[Local polarization evolution with the pruner sequence at different
$^{13}$C sites in an L-leucine molecule under an XY mixing Hamiltonian.]{Local
polarization evolution with the pruner sequence at different $^{13}$C sites in
an L-leucine molecule under an XY mixing Hamiltonian. The initial local
excitation at site $1$ (the source) is transfered through the selected pathway
($1-2-3$), to site $3.$ The free evolution time is $\Delta t_{\mathrm{free}%
}=1.2744\operatorname{ms}=6\times\hbar/\Delta_{13}\approx\hbar/\Delta
_{12}\approx5\times\hbar/\Delta_{23}$ and $\Delta t_{\mathrm{iso}%
}=3.1\operatorname{ms}\approx\frac{1}{8}\frac{\hbar}{J}.$ A gain of approx.
$420\%$ relative to the first maximum of the isotropic evolution is observed
at site $3.$}%
\label{Fig_L-leucine_1-2-3_path_XY}%
\end{center}
\end{figure}

\begin{figure}
[tbh]
\begin{center}
\includegraphics[
height=4.3033in,
width=6.1635in
]%
{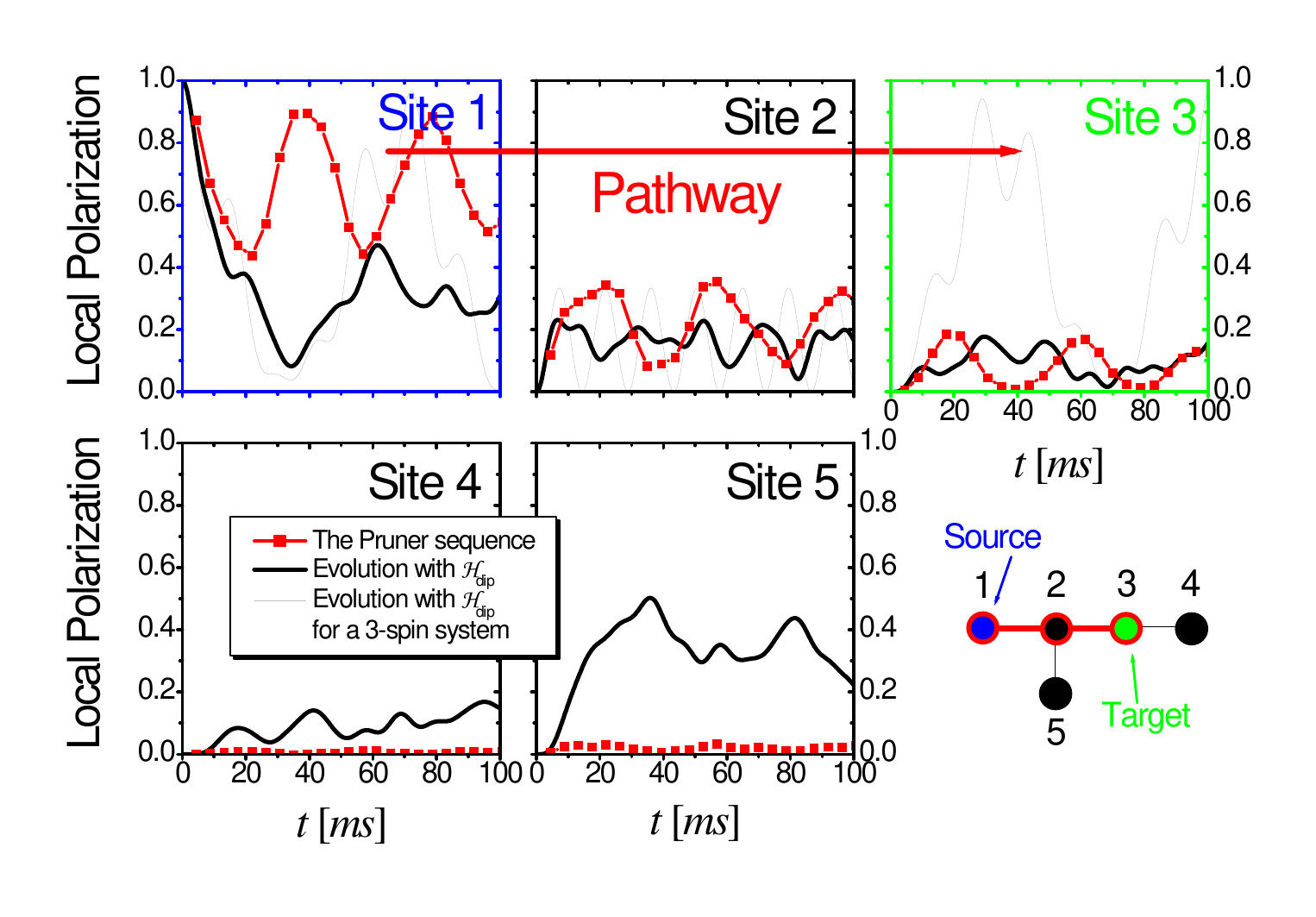}%
\caption[Local polarization evolution with the pruner sequence at different
$^{13}$C sites in an L-leucine molecule under a truncated dipolar mixing
Hamiltonian.]{Local polarization evolution with the pruner sequence at
different $^{13}$C sites in an L-leucine molecule under a truncated dipolar
mixing Hamiltonian. The initial local excitation at site $1$ (the source) is
transfered through the selected pathway ($1-2-3$), to site $3.$ The free
evolution time is $\Delta t_{\mathrm{free}}=1.2744\operatorname{ms}%
=6\times\hbar/\Delta_{13}\approx\hbar/\Delta_{12}\approx5\times\hbar
/\Delta_{23}$ and $\Delta t_{\mathrm{iso}}=3.1\operatorname{ms}\approx\frac
{1}{8}\frac{\hbar}{J}.$ A gain of approx. $234\%$ relative to the first
maximum of the isotropic evolution is observed at site $3.$}%
\label{Fig_L-leucine_1-2-3_path_Dip}%
\end{center}
\end{figure}

As we mention in section \S \ \ref{Marker_pruner_sequence}, the pruner
sequence stroboscopically interrupts the spin-spin interaction to
\textquotedblleft disconnect\textquotedblright\ those spins which do not
belong to the \textquotedblleft selected transfer pathway\textquotedblright.
Here, we will show what happens if we change the mixing Hamiltonian. We
consider the following mixing Hamiltonians%
\begin{align}
\widehat{\mathcal{H}}_{\mathrm{mix}}^{\mathrm{XY}}  &  =\sum_{i\neq j}%
J_{ij}^{{}}\left(  \hat{I}_{i}^{y}\hat{I}_{j}^{y}+\hat{I}_{i}^{x}\hat{I}%
_{j}^{x}\right)  ,\\
\widehat{\mathcal{H}}_{\mathrm{mix}}^{\mathrm{dip}}  &  =\sum_{i\neq j}%
J_{ij}^{{}}\left(  2\hat{I}_{i}^{z}\hat{I}_{j}^{z}-\hat{I}_{i}^{y}\hat{I}%
_{j}^{y}-\hat{I}_{i}^{x}\hat{I}_{j}^{x}\right)  ,
\end{align}
where the first is an XY (planar) Hamiltonian and the second one is a
truncated dipolar. We mention in previous chapters that the XY Hamiltonian is
less \textquotedblleft diffusive\textquotedblright, i.e., the intensities of
the coherences are less attenuated. Moreover, an XY dynamics can be solved
exactly in some systems \cite{Feldman98,Feldman99,Danieli02} as we have done
in section \S \ \ref{Marker_memory_effects_keldysh} to test new theoretical
methods. This Hamiltonian is experimentally achievable in liquid \cite{Madi97}
and solid-state \cite{FeldmanXY,Cory06} NMR while the truncated dipolar
Hamiltonian is the natural interaction in solid-state NMR
\cite{Abragam,Ernst,Slichter}. We apply these mixing Hamiltonians to the
L-leucine molecule to show the main differences in the pruner sequence
performance. Note that with the purpose of comparing the effect of the
different anisotropies in the mixing Hamiltonian, we have kept the chemical
shift and coupling values of the isotropic interaction in the liquid-state. A
real solid-state $\widehat{\mathcal{H}}_{\mathrm{mix}}^{\mathrm{dip}}$ would
require knowledge of the $^{13}$C-$^{13}$C dipolar couplings and orientation
with respect to the external magnetic field. Figures
\ref{Fig_L-leucine_1-2-3_path_XY} and \ref{Fig_L-leucine_1-2-3_path_Dip} show
the polarization dynamics for a source at site $1$ for an XY and a dipolar
Hamiltonian respectively.
The pruner sequence (red lines) is optimized to transfer polarization to the
site $3$ (the target) through site $2.$ It is noticeable the performance of
our sequence under the XY Hamiltonian. It almost reproduces the isolated
three-spin system dynamics (gray lines). The Ising term of the mixing
Hamiltonian originates the decrease of the performance in the other
Hamiltonians because it produces an effective energy difference between the
sites of the pathway. This detuning does not allow for very high polarization
transfers. This can be seen in the polarization dynamics of site $1,$ where
for the XY mixing (fig. \ref{Fig_L-leucine_1-2-3_path_XY}) the polarization
transfer is almost complete, whereas in the dipolar (fig.
\ref{Fig_L-leucine_1-2-3_path_Dip}) and isotropic (fig.
\ref{Fig_L-leucine_1-2-3_path}) cases, only around half of the polarization is
transferred. In spite of this, if the goal is the efficiency of the pruning
and not the amount of polarization transfer, the pruner sequence works quite
well even for isotropic or dipolar interactions.

\subsection{Step by step pruning of the branches}

\begin{figure}
[pth]
\begin{center}
\includegraphics[
height=4.3803in,
width=5.8954in
]%
{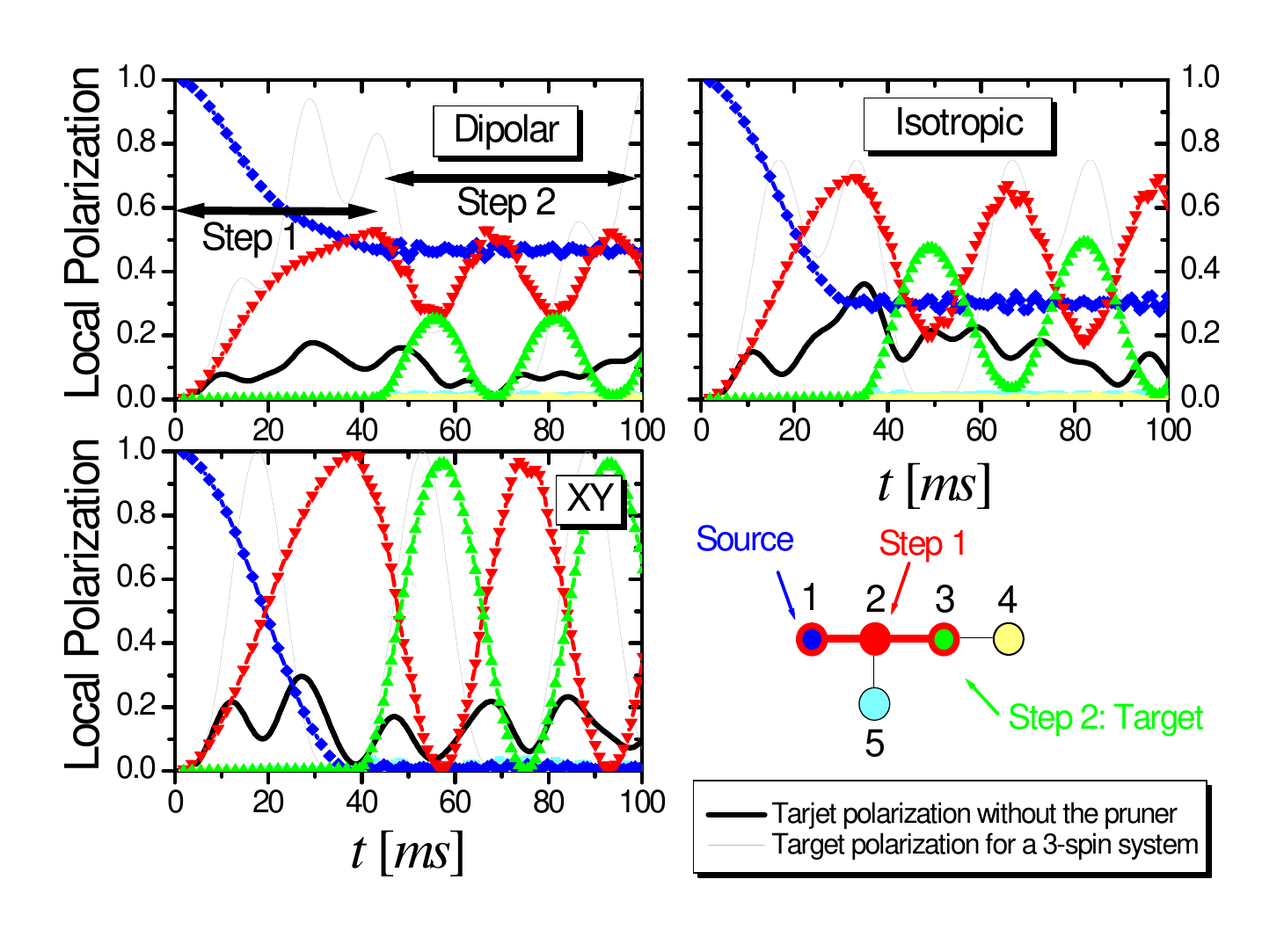}%
\caption[Local polarization evolution at different $^{13}$C sites in an
L-leucine molecule under a step by step version of the pruner sequence for XY,
isotropic and dipolar mixing Hamiltonians.]{Local polarization evolution at
different $^{13}$C sites in an L-leucine molecule under a step by step version
of the pruner sequence for XY, isotropic and dipolar mixing Hamiltonians. The
scattered lines show the polarization evolution with this pruner sequence,
whereas the solid lines show the evolution under the mixing Hamiltonian for
the five spins (black) and a three-spin chain (gray). The initial local
excitation at site $1$ (the source) is transfered to site $2$ in the first
step of the sequence. Within this step the free evolution time is $\Delta
t_{\mathrm{free}}=1.2422\operatorname{ms}=\hbar/\Delta_{12}$ and $\Delta
t_{\mathrm{iso}}=0.6\operatorname{ms}\ll\frac{\hbar}{J}.$ Within the second
part of the sequence, the free evolution time $\Delta t_{\mathrm{free}%
}=0.2562\operatorname{ms}=\hbar/\Delta_{23}$ optimizes the polarization
transfer between sites $2$ and $3$ once the polarization arrived at site $2.$}%
\label{Fig_L-leucine_step_by_step-1-2-3_path}%
\end{center}
\end{figure}

We have shown how to \textquotedblleft isolate\textquotedblright\ a selected
group of spins. Now, we will show the suitability of this technique to choose
a first group of spins during a time interval and then to select another group
to transfer the polarization from the previous one. In order to do this,
firstly, we optimize the pruner sequence to transfer the polarization from
site $1$ to site $2$. Then, once the polarization arrived at site $2$, we
optimize the sequence to transfer the magnetization to site $3.$ Fig.
\ref{Fig_L-leucine_step_by_step-1-2-3_path} shows this step by step pruning
for different mixing Hamiltonians.
We observe only the interaction between site $1$ (blue line) and site $2$ (red
line) during the the first step of the sequence where $\Delta t_{\mathrm{free}%
}^{(1\text{\textrm{-step}})}=2\pi\hbar/\left\vert \Delta_{12}\right\vert .$ It
is very similar to the isolated two-spin polarization evolution until these
spins are disconnected in the second step of the sequence by keeping only the
interaction between sites $2$ and $3$. Within this step, we use $\Delta
t_{\mathrm{free}}^{(2\text{\textrm{-step}})}=2\pi\hbar/\left\vert \Delta
_{23}\right\vert $ to optimize the transfer between sites $2$ and $3.$ Thus,
in the figures, one can observe the two-spin dynamics between the sites $2$
and $3$ while the others remain frozen. The figures show essentially a pure
two-spin polarization evolution. This resembles the quantum Zeno phase
predicted in chapter \ref{Sec_QDPT} for the swapping operation
\cite{JCP06,SSC07} which arises when the system-environment interaction is
much stronger than the system interaction. There, the Ising system-environment
interaction, in the stroboscopic version \cite{JCP06,Alvarez07b}, acted as a
measurement process freezing the quantum oscillations. Moreover, the
manifestation of this phase in a three-spin system leads to \textquotedblleft
isolate\textquotedblright\ the two spins not directly connected to the
environment \cite{Alvarez07a}. Here, the pruner sequence has stroboscopic free
evolutions that act as an environment producing an effective Ising
system-environment interaction over the \textquotedblleft
non-selected\textquotedblright\ spins, isolating the selected ones. This
resembles the applications of engineered reservoir techniques for fundamental
studies of decoherence \cite{Myatt00,Paz01}.

\section{Summary}

We developed a new NMR pulse sequence, \textquotedblleft the
pruner\textquotedblright, to improve the transfer of polarization through a
specific pathway in a system of many interacting spins \cite{Alvarez07d}. The
sequence effectively prunes branches of spins, where no polarization is
required, during the polarization transfer procedure. We obtained a remarkable
enhancement, higher than $200\%$, of the polarization transfer with respect to
the standard methods. Moreover, by changing the mixing Hamiltonian to an XY
interaction, we obtain a gain of about $400\%$. Therefore, it is a very
practical tool for NMR applications where a signal gain is of great
importance. However, from a fundamental point of view, it is the starting
point of an engineered application of the stroboscopic model
\cite{JCP06,Alvarez07b} described in section
\S \ \ref{Marker_Stroboscopic_process} which seems to manifest a spin dynamics
within the quantum Zeno phase \cite{JCP06,SSC07}. Moreover, it can help to go
deeper in the understanding of decoherence processes and consequently of the
environmentally induced quantum dynamical phase transition \cite{JCP06,SSC07}.

\chapter{Entanglement as a tool for ensemble spin dynamics
calculations\label{Mark_entanglement_vs_ensemble}}

The quantum time evolution of a qubit cluster of intermediate size is of great
interest because its potential applications to quantum information processing
(QIP)\ \cite{Kane98,BD2000}, and structural characterization through Nuclear
Magnetic Resonance \cite{Pastawski95,Madi97,JCP03,Ramanathan05}. Experimental
realizations and control of a pure-state dynamics is still one of the major
challenges in nowadays quantum physics \cite{QCRoadmap04}.
Therefore, one generally has to deal with ensemble evolutions, as done in
previous chapters for the study of NMR spin dynamics and in related works
\cite{Pastawski95,Madi97,Ramanathan05}. There, the dynamics is from an initial
mixed-state that, as discussed in the Introduction, gives rise to the
development of the ensemble quantum computation \cite{Cirac04,Suter05}. Hence,
it is necessary to find a way to calculate or simulate the ensemble evolution
efficiently. Theoretical solutions of the ensemble dynamics arise from the
integration of the Liouville von-Neumann equation (see chapter
\ref{Mark_spin_dynamics_Density_matrix}) or alternative methods like the
Keldysh formalism, discussed in chapter \ref{Marker_Spin_within_keldysh}
which, for some special cases, yield exact analytical results
\cite{FeldLacelle02,CPL05}. However, in the most general situations, one has
to resort to numerical solutions. There are different numerical methods to
solve the time dependent evolution, such as exact diagonalization, the
Suzuki-Trotter Product-Formula (STPF) and the Chevyshev Polynomial among
others \cite{Raedt04}. These methods are limited by the exponential growth of
the Hilbert space dimension with the system size \cite{Raedt04}. The STPF
allows one to treat systems of greater dimension than the exact
diagonalization method which is presently limited to around $20$ qubits. The
use of the STPF to obtain the dynamics of a pure initial state avoids the
storage of the Hamiltonian matrix to be diagonalized. The computation of an
ensemble dynamics with an initial mixed-state, requires the calculation of the
\emph{individual} evolutions of each of its components. These are of the order
of the Hilbert space dimension demanding long computational times. Here, in
order to overcome this limitation, we take profit of the quantum parallelism
\cite{Loss02parallelism} to calculate the ensemble dynamics through a pure
entangled state evolution using the STPF. Apparently, the underlying physical
mechanism that makes possible these efficient simulations is the rapid
intrinsic decoherence of highly correlated many-qubit systems
\cite{Suter04,Suter06,Cory06,Sanchez07}.

We consider a system of $M$ spins $1/2$ to calculate the $M$-qubits dynamics.
We compare the time evolution of a local polarization (experimentally observed
by NMR) \cite{ZME92,JCP98},
between an initial ensemble of local excitations and an initial pure entangled
state. This entangled state is built in as a superposition of each of the
ensemble components where the complex coefficients of the linear combination
are the thermal weights with random phases. We show that the contribution of
the coherences of the pure initial state to the dynamics can be neglected due
to a self averaging property arising on the destructive interference of the
random phases. To show the relevance of the number of independent phases,
which are of the order of the Hilbert space dimension, we also calculate the
same evolution with an initial pure product state where the number of
independent phases are of the order of the system size. We compare the
entangled state and ensemble calculations in two contrasting topological
configurations: one with relevant mesoscopic effects and other where the
disorder makes these effects negligible.

\section{Ensemble vs. pure entangled state evolution}

\begin{figure}
[tbh]
\begin{center}
\includegraphics[
height=5.5089in,
width=4.5325in
]%
{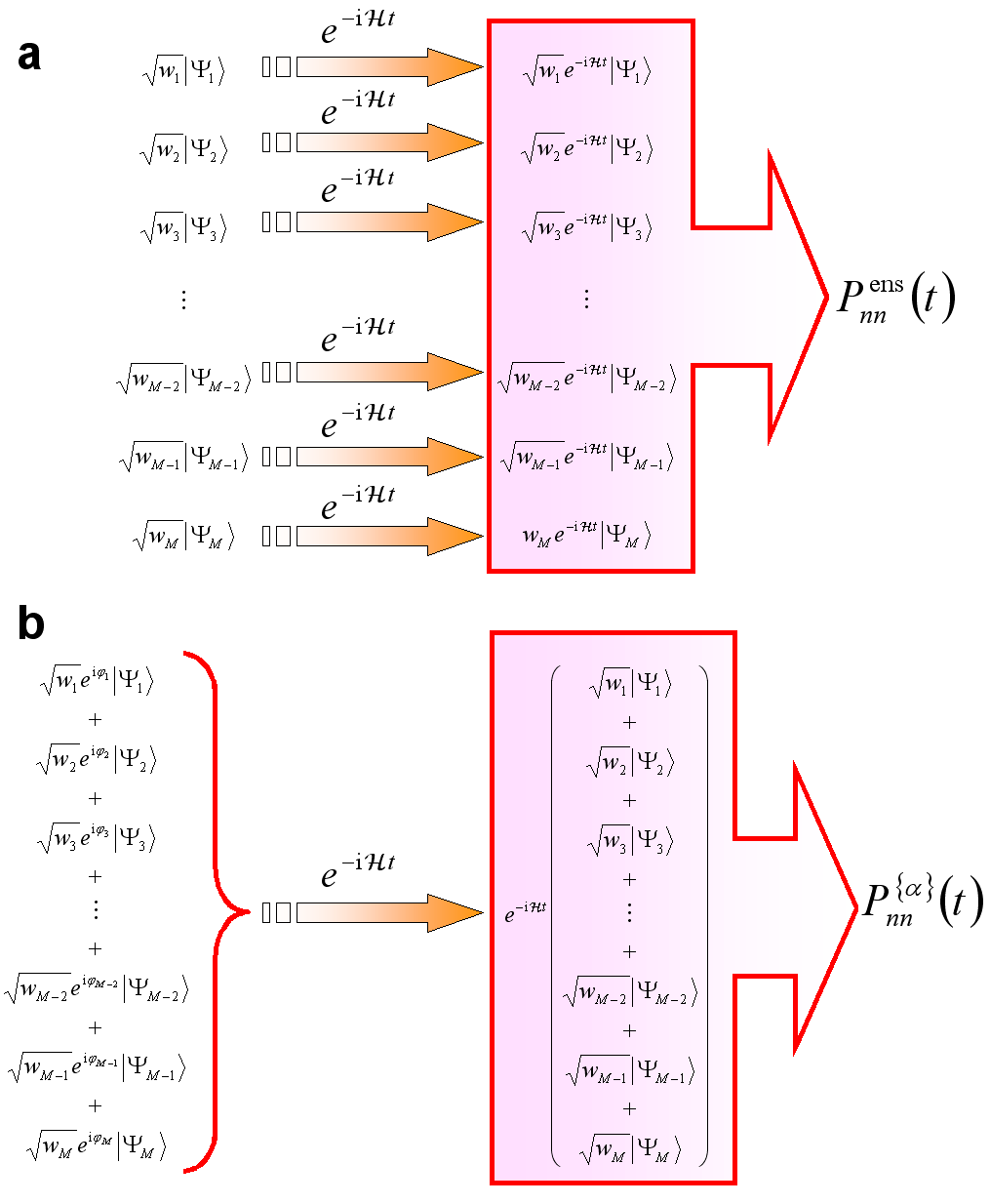}%
\caption[Quantum evolution schemes of an ensemble and an entangled
pure-state.]{Quantum evolution schemes of an ensemble [panel a)] and an
entangled pure-state [panel b)]. Each $\left\vert \Psi_{i}\right\rangle $
represents a simple tensor product state.}%
\label{Fig_ensemble_vs_entaglement_scheme}%
\end{center}
\end{figure}

Let us start considering a system of $M$ spins $1/2$. As we are interested in
the evolution of a local excitation,
we take the ensemble of all states $\left\vert \Psi_{i}\right\rangle $
representing the many-spin states in the product base along the quantization
axis ($z$), with the $n$-th site polarized (spin \emph{up}).
Each of these states has a thermal weight $w_{i}$. The local spin dynamics
through this ensemble can be obtained by calculating a generalization of the
auto-correlation function $P_{n,n}^{\mathrm{ens}}(t)$ \cite{Pastawski95}%
\begin{equation}
P_{nn}^{\mathrm{ens}}(t)=2\left[
{\displaystyle\sum\limits_{i=1}^{2^{M-1}}}
{\displaystyle\sum\limits_{f=1}^{2^{M-1}}}
w_{i}\left\vert \left\langle \Psi_{f}\right\vert e^{-\mathrm{i}\widehat
{\mathcal{H}}t/\hbar}\left\vert \Psi_{i}\right\rangle \right\vert ^{2}%
-\frac{1}{2}\right]  . \label{eq--PCFunc}%
\end{equation}
The quantity $\left\vert \left\langle \Psi_{f}\right\vert e^{-\mathrm{i}%
\widehat{\mathcal{H}}t}\left\vert \Psi_{i}\right\rangle \right\vert ^{2}%
$represents the probability of finding the $n$-th site polarized in the state
$\left\vert \Psi_{f}\right\rangle $ at time $t$ provided that the same site
was polarized at $t=0$ in the state $\left\vert \Psi_{i}\right\rangle $. The
sum over all possible initial and final states $\left\vert \Psi_{i}%
\right\rangle $ and $\left\vert \Psi_{f}\right\rangle $, gives the amount of
the $z$ component of the local polarization at time $t$ on\ the $n$-th site.
As the value of the polarization runs from $-1$ to $1$, the eq.
(\ref{eq--PCFunc}) is properly renormalized. It represents the standard
thermal mixture calculation obtained with the density matrix \cite{Madi97}
(see section \S \ \ref{Mark_spin_dynamics_Density_matrix})
\begin{equation}
P_{nn}^{\mathrm{ens,DM}}\left(  t\right)  =\frac{\mathrm{Tr}\left\{  \hat
{I}_{n}^{z}\hat{\sigma}\left(  t\right)  \right\}  }{\mathrm{Tr}\left\{
\hat{I}_{n}^{z}\hat{\sigma}\left(  0\right)  \right\}  },
\end{equation}
where
\begin{align}
\hat{\sigma}(t)  &  =\hat{U}(t)\hat{\sigma}(0)\hat{U}^{-1}(t),\\
\hat{\sigma}\left(  0\right)   &  =\frac{\hat{1}+\beta_{\mathrm{B}}\hbar
\Omega_{0,I}\hat{I}_{n}^{z}}{\mathrm{Tr}\left\{  \hat{1}\right\}  }%
\end{align}
with
\begin{equation}
\hat{U}\left(  t\right)  =e^{-\mathrm{i}\widehat{\mathcal{H}}t/\hbar}.
\end{equation}
Within the Keldysh formalism, using equations (\ref{spin_correlations}) and
(\ref{Pol_spin}), we obtain
\begin{equation}
P_{nn}^{\mathrm{ens,K}}(t)=\frac{\left\langle \Psi_{\mathrm{eq.}}\right\vert
\hat{I}_{n}^{z}(t)\hat{I}_{n}^{z}(0)\left\vert \Psi_{\mathrm{eq.}%
}\right\rangle }{\left\langle \Psi_{\mathrm{eq.}}\right\vert \hat{I}_{n}%
^{z}(0)\hat{I}_{n}^{z}(0)\left\vert \Psi_{\mathrm{eq.}}\right\rangle }%
=\tfrac{2\hbar}{\mathrm{i}}G_{nn}^{<\,}(t,t)-1.
\end{equation}
The expression (\ref{eq--PCFunc}) involves $2^{M-1}$ different dynamics for
each of the initial states, see fig. \ref{Fig_ensemble_vs_entaglement_scheme}
a).
This number is directly related to the dimension of the Hilbert space, and
thus the number of required evolutions increases exponentially with $M$,
leading to large computational times. Our goal is to extract essentially the
same information hidden in this ensemble calculation but at a lower cost. To
do this we exploit the parallel behavior of quantum systems
\cite{Loss02parallelism}.
This is achieved by replacing the ensemble by only \emph{one}
\emph{pure-state}, see fig. \ref{Fig_ensemble_vs_entaglement_scheme} b),
obtained as a linear combination of the components of the ensemble, that is,
\begin{align}
P_{nn}^{\{\alpha\}}(t)  &  =2\left[
{\displaystyle\sum\limits_{f=1}^{2^{M-1}}}
\left\vert \left\langle \Psi_{f}\right\vert e^{-\mathrm{i}\widehat
{\mathcal{H}}t/\hbar}\left\vert \Psi_{\mathrm{pure}}^{\{\alpha\}}\right\rangle
\right\vert ^{2}-\frac{1}{2}\right] \label{eq--PCFuncLC}\\
&  =2\left[
{\displaystyle\sum\limits_{f=1}^{2^{M-1}}}
\left\vert \left\langle \Psi_{f}\right\vert e^{-\mathrm{i}\widehat
{\mathcal{H}}t/\hbar}%
{\displaystyle\sum\limits_{i=1}^{2^{M-1}}}
\alpha_{i}\left\vert \Psi_{i}\right\rangle \right\vert ^{2}-\frac{1}%
{2}\right]  .
\end{align}
Here $\{\alpha\}$ denotes the set of all the $\alpha_{i}=\sqrt{w_{i}%
}e^{\mathrm{i}\varphi_{i}}$, with $\varphi_{i}$ an arbitrary phase, involved
in the initial pure state. Note that the substantial difference between eq.
(\ref{eq--PCFunc}) and eq. (\ref{eq--PCFuncLC}) is that the sum in the former
is outside the square modulus while in the latter is inside. \ Rewriting eq.
(\ref{eq--PCFuncLC}), it becomes%
\begin{multline}
P_{nn}^{\{\alpha\}}(t)=2\left[
{\displaystyle\sum\limits_{f,i=1}^{2^{M-1}}}
w_{i}\left\vert \left\langle \Psi_{f}\right\vert e^{-\mathrm{i}\widehat
{\mathcal{H}}t/\hbar}\left\vert \Psi_{i}\right\rangle \right\vert ^{2}%
-\frac{1}{2}\right] \label{eq--PCFuncLC_exp}\\
+2%
{\displaystyle\sum\limits_{\genfrac{}{}{0pt}{}{f,i,j=1}{i\neq j}}^{2^{M-1}}}
\alpha_{i}\alpha_{j}^{\ast}\left\langle \Psi_{f}\right\vert e^{-\mathrm{i}%
\widehat{\mathcal{H}}t/\hbar}\left\vert \Psi_{i}\right\rangle \left\langle
\Psi_{j}\right\vert e^{\mathrm{i}\widehat{\mathcal{H}}t/\hbar}\left\vert
\Psi_{f}\right\rangle \\
=P_{nn}^{\mathrm{ens}}(t)+2%
{\displaystyle\sum\limits_{\genfrac{}{}{0pt}{}{f,i,j=1}{i\neq j}}^{2^{M-1}}}
\alpha_{i}\alpha_{j}^{\ast}\left\langle \Psi_{f}\right\vert e^{-\mathrm{i}%
\widehat{\mathcal{H}}t/\hbar}\left\vert \Psi_{i}\right\rangle \left\langle
\Psi_{j}\right\vert e^{\mathrm{i}\widehat{\mathcal{H}}t/\hbar}\left\vert
\Psi_{f}\right\rangle ,
\end{multline}
where we can see that the \emph{cross terms }make the difference between
$P_{nn}^{\mathrm{ens}}(t)$ and $P_{nn}^{\{\alpha\}}(t).$ It is evident that
the choice of the phases $\varphi_{i}$ has a relevant role in the equivalence
between eq. (\ref{eq--PCFuncLC}) and eq. (\ref{eq--PCFunc}). For
$M\rightarrow\infty$ and choosing the $\varphi_{i}$ randomly distributed,
where $\left\vert \Psi_{\mathrm{pure}}^{\{\alpha\}}\right\rangle $ is an
\emph{entangled state}\footnote{If the $\varphi_{i}$ are randomly chosen, the
pure state $\left\vert \Psi_{\mathrm{pure}}^{\{\alpha\}}\right\rangle $ has a
high probability to be an entangled state (non-factorable).},
we observe that the \emph{cross terms} in eq. (\ref{eq--PCFuncLC_exp}) average
to zero and
\begin{equation}
P_{nn}^{\mathrm{ens}}(t)=P_{nn}^{\{\alpha\}}(t).
\end{equation}
This \emph{self averaging} property of a randomly correlated pure-state,
assisted by the large dimension of the Hilbert space, suggests an
\emph{\textquotedblleft intrinsic decoherence\textquotedblright}
characteristic of very complex systems. As we are considering the whole
system, one does not expect to observe decoherence as defined in chapter
\ref{Mark_introduction}. Nevertheless, the observable involves a single spin
of the system and consequently we cannot distinguish between a mixed-state and
a pure entangled one, which effectively reflects a loss of information.
However, for finite size systems, this self averaging property depends on
their particular characteristics as shown below. When the self averaging
property is not satisfied, an extra average over initial states is mandatory
to force the equality between $P_{nn}^{\mathrm{ens}}(t)$ and $P_{nn}%
^{\{\alpha\}}(t)$. Then,
\begin{equation}
P_{nn}^{N_{\alpha}}(t)=\frac{1}{N_{\alpha}}%
{\displaystyle\sum\limits_{\{\alpha\}}^{N_{\alpha}}}
P_{nn}^{\{\alpha\}}(t), \label{eq--PCFuncLC_aver}%
\end{equation}
where $N_{\alpha}$ denotes the number of different realizations of the sets
$\{\alpha\}$. One has to perform $N_{\alpha}$ evolutions in the last
expression, while for an ensemble dynamics one needs $2^{M-1}$. Thus, the
calculation time is reduced by a factor of $N_{\alpha}/2^{M-1}$ where, for an
appropriate choice of the $\{\alpha\}$ coefficients, $N_{\alpha}$ $\ll2^{M-1}$.

\section{Application to spin-systems with different coupling networks}

\subsection{The systems}

In typical situations of high-field solid-state NMR \cite{Abragam,Ernst}, the
spin interaction Hamiltonian $\widehat{\mathcal{H}}$ can be expressed by eq.
(\ref{SpinHamiltonian}). One can effectively eliminate the Zeeman contribution
by working on-resonance in the rotating frame \cite{Slichter}. Then, we can
focus in the spin-spin interaction,%
\[
\widehat{\mathcal{H}}=%
{\textstyle\sum\limits_{i<j}^{M}}
\left[  a_{ij}\hat{I}_{i}^{z}\hat{I}_{j}^{z}+\tfrac{1}{2}b_{ij}\left(  \hat
{I}_{i}^{+}\hat{I}_{j}^{-}+\hat{I}_{i}^{-}\hat{I}_{j}^{+}\right)  \right]  ,
\]
where, $b_{ij}/a_{ij}=0$ represents an Ising-like coupling, $a_{ij}/b_{ij}=0$
an $XY$ Hamiltonian, $a_{ij}/b_{ij}=1$ the isotropic one, and $a_{ij}%
/b_{ij}=-2$ a dipolar (secular) Hamiltonian truncated with respect to a Zeeman
field along the $z$ axis.

In order to show the potential of the proposal summarized in eq.
(\ref{eq--PCFuncLC_aver}), we will apply it to two different spin systems
which have well differentiated kinds of dynamics:

a) A \emph{ladder} of spins interacting through an XY Hamiltonian, as shown in
fig. \ref{Fig_Ent_vs_Ens_ladder} a). There, $a_{ij}=0,$ $b_{i,i+1}%
=b_{i+M/2,i+M/2+1}=b_{x}$ and $b_{i,i+M/2}=b_{y}$. Here, the dynamics presents
long lived recurrences, showed in the black line of fig.
\ref{Fig_Ent_vs_Ens_ladder}, due to the ordered topology
\cite{Pastawski95,Madi97}.

b)\emph{ }A \emph{star }system in which all the spins are interacting with
each other through a dipolar coupling, $a_{ij}/b_{ij}=-2,$ with a Gaussian
random distribution with zero mean and $\sigma^{2}$ variance for the
intensities of the couplings [see fig. \ref{Fig_Ent_vs_Ens_ladder} b)].
In this case, the local polarization decays with a rate proportional to the
square root of the \emph{local }second moment of the Hamiltonian and no
recurrences are observed. The black line of fig. \ref{Fig_Ent_vs_Ens_star}
shows the local polarization of this system.
\begin{figure}
[tbh]
\begin{center}
\includegraphics[
height=4.2774in,
width=3.154in
]%
{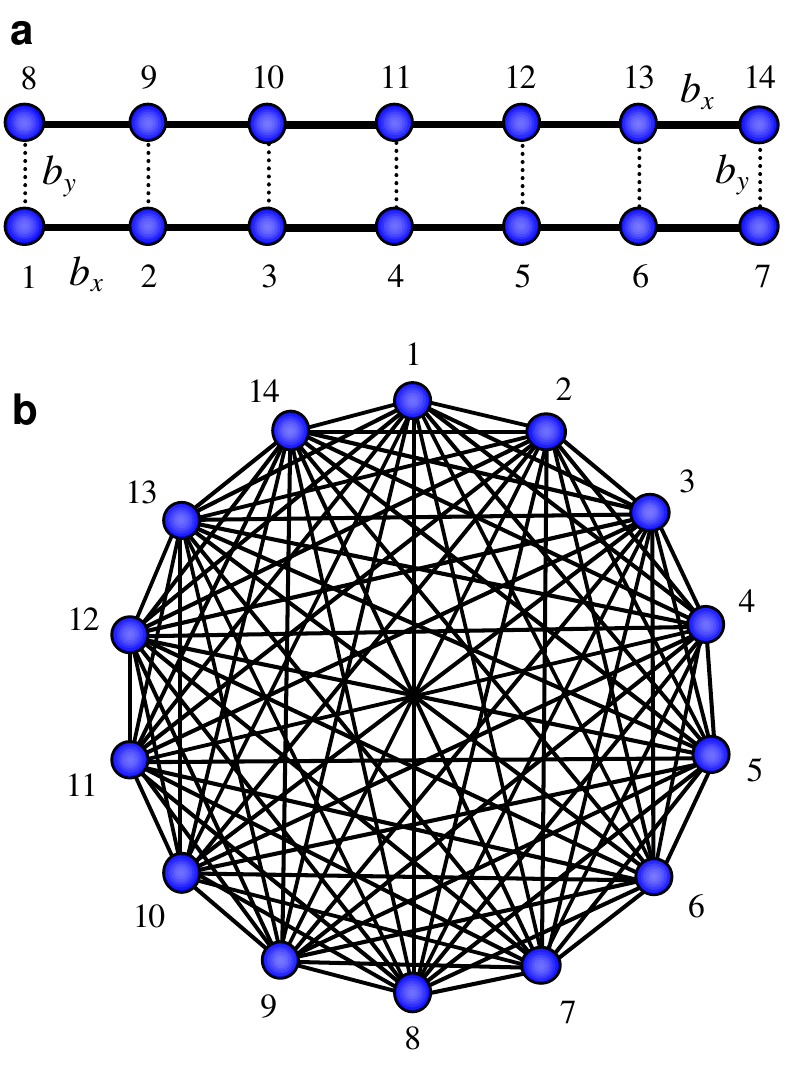}%
\caption[Many-spin systems: ladder of spins and star systems.]{Panel a) shows
a ladder of spins interacting through an $XY$ Hamiltonian where the horizontal
couplings are given by $b_{x}$ and the vertical by $b_{y}.$ Panel b) contains
a star system in which all the spins interact with each other through a
dipolar coupling, $a_{ij}/b_{ij}=-2,$ with gaussian random distribution with
zero mean and $\sigma^{2}$ variance.}%
\label{Fig_ladder_star_diagram}%
\end{center}
\end{figure}

In order to compare eqs. (\ref{eq--PCFunc}) and (\ref{eq--PCFuncLC_aver}), and
the dependence of eq. (\ref{eq--PCFuncLC_aver}) on the choice of the phases
$\varphi_{i}$, we calculate the evolution for two types of initial states in
the infinite temperature limit, i.e. $w_{i}=1/2^{M-1}$. This limit corresponds
to the NMR experimental condition \cite{Abragam}. The particular initial
conditions are states with a local excitation on the $n$-th site over a
background level which is determined by the zero magnetization of the others
$M-1$ spins. These states are used in several NMR experiments
\cite{ZME92,Pastawski95,Madi97}.

The pure entangled state has the phases $\varphi_{i}$ randomly chosen and
$\left\vert \Psi_{\mathrm{pure}}^{\{\alpha\}}\right\rangle $ becomes%
\begin{equation}
\left\vert \Psi_{\mathrm{ent}}^{\{\alpha\}}\right\rangle =%
{\textstyle\sum_{i=1}^{2^{M-1}}}
\tfrac{1}{\sqrt{2^{M-1}}}e^{-\mathrm{i}\varphi_{i}}\left\vert \Psi
_{i}\right\rangle . \label{eq--entstate}%
\end{equation}
The correlation function, eq. (\ref{eq--PCFuncLC_aver}), calculated with this
state will be called $P_{nn}^{\mathrm{ent},N_{\alpha}}(t).$

The second case is a product (unentangled) state, built with the $n$-th spin
\emph{up} and all the others in a linear combination of spins \emph{up} and
\emph{down,} with equal probability and randomly correlated. Assuming $n=1$,
we have%
\begin{equation}
\left\vert \Psi_{\mathrm{prod}}^{\{\alpha\}}\right\rangle =\left\vert
\uparrow\right\rangle _{1}%
{\textstyle\bigotimes}
\prod\limits_{m=2}^{M}\left\vert \rightarrow\right\rangle _{m},
\label{eq--alestate}%
\end{equation}
where
\[
\left\vert \rightarrow\right\rangle _{m}=\tfrac{1}{\sqrt{2}}\left(  \left\vert
\downarrow\right\rangle +\left\vert \uparrow\right\rangle e^{-\mathrm{i}%
\varphi_{m}}\right)  .
\]
Note that this state is a particular case of (\ref{eq--entstate}) with a
special correlation on the phases $\varphi_{i}$. Here, the correlation
function (\ref{eq--PCFuncLC_aver}) will be identified as $P_{nn}%
^{\mathrm{prod},N_{\alpha}}(t)$.

\subsection{Quantum evolution}

\begin{figure}
[ptbh]
\begin{center}
\includegraphics[
height=5.5106in,
width=4.3716in
]%
{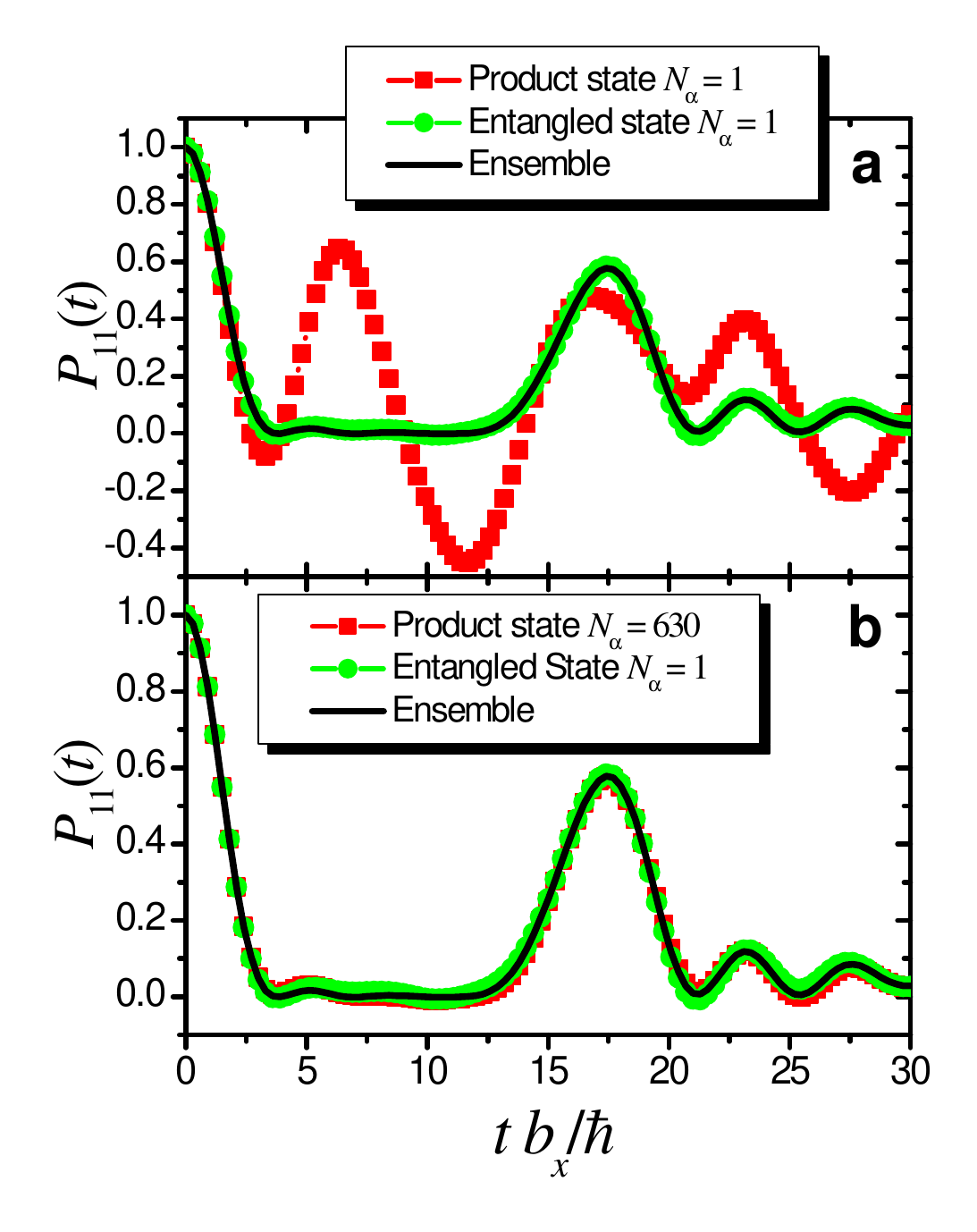}%
\caption[Local polarization evolutions of a $14$-spin ladder system.]%
{Polarization evolution of the local spin dynamics in a $14$-spin ladder
system of an ensemble compared with an entangled pure-state dynamics. The
ratio between the $x$ and $y$ coupling in the ladder is given by $b_{y}%
/b_{x}=1/10.$ The black line shows the ensemble local polarization,
$P_{11}^{\mathrm{ens}}(t),$ and the red and green lines correspond to the
evolution obtained with $P_{11}^{\mathrm{prod},N_{\alpha}}(t)$ and
$P_{11}^{\mathrm{ent},N_{\alpha}}(t)$ respectively. The upper panel shows the
dynamics of eq. (\ref{eq--PCFuncLC_aver}) for $N_{\alpha}=1$. The lower panel
shows the dynamics for $P_{11}^{\mathrm{prod},N_{\alpha}}(t)$ and
$P_{11}^{\mathrm{ent},N_{\alpha}}(t)$, where $N_{\alpha}$ is the lower value
for which each curve reproduces the ensemble dynamics.}%
\label{Fig_Ent_vs_Ens_ladder}%
\end{center}
\end{figure}

The local polarization, $P_{11}^{\mathrm{ens}}(t),$ obtained with eq.
(\ref{eq--PCFunc}), for the ladder system composed of $14$ spins is shown in
fig. \ref{Fig_Ent_vs_Ens_ladder} with black line.
Red and green lines correspond to the temporal evolution of eq.
(\ref{eq--PCFuncLC_aver}), $P_{11}^{\mathrm{prod},N_{\alpha}}(t)$ and
$P_{11}^{\mathrm{ent},N_{\alpha}}(t)$ respectively. The upper panel shows the
dynamics of eq. (\ref{eq--PCFuncLC_aver}) for $N_{\alpha}=1$. The agreement
between $P_{11}^{\mathrm{ens}}(t)$ and $P_{11}^{\mathrm{ent},1}(t)$ is
excellent, while $P_{11}^{\mathrm{prod},1}(t)$ has a dynamics far away from
that of the ensemble. The difference between the dynamics of the two initial
pure states is due to the different number of independent random phases
(uncorrelated phases) of each state. In the entangled state, there are
$2^{M-1}$ independent random phases that make possible the cancellation of the
second term in the rhs of eq. (\ref{eq--PCFuncLC_exp}). However, the number of
independent phases for the product state is $M-1$. This implies that there are
multiple correlations between the phases in the \emph{cross terms }impeding
their self cancellation for low values of $M$.

The lower panel in fig. \ref{Fig_Ent_vs_Ens_ladder} shows the dynamics for
$P_{11}^{\mathrm{prod},N_{\alpha}}(t)$ and $P_{11}^{\mathrm{ent},N_{\alpha}%
}(t)$, where $N_{\alpha}$ is the lower value for which each curve reproduces
the ensemble dynamics. Note that the relation between $N_{\mathrm{prod}}$ and
$N_{\mathrm{ent}}$ comes from the equivalence in the number of independent
phases
\begin{equation}
N_{\mathrm{ent}}2^{M-1}\simeq N_{\mathrm{prod}}(M-1).
\end{equation}
This equation for the particular case of fig. \ref{Fig_Ent_vs_Ens_ladder} a)
is%
\begin{equation}
8192=N_{\mathrm{ent}}2^{M-1}\simeq N_{\mathrm{prod}}(M-1)=7969.
\end{equation}
The statistical theory of the density matrix \cite{Blum} is based in the
random correlation nature of a real system\ to describe it as an
\emph{ensemble state in the thermodynamic limit}. Here, we observe that even
for small numbers like $14,$ the equivalence between a randomly correlated
pure state with an ensemble state remains as a consequence of the dimension of
the Hilbert space which grows exponentially with $M.$

\begin{figure}
[ptbh]
\begin{center}
\includegraphics[
height=5.546in,
width=4.3206in
]%
{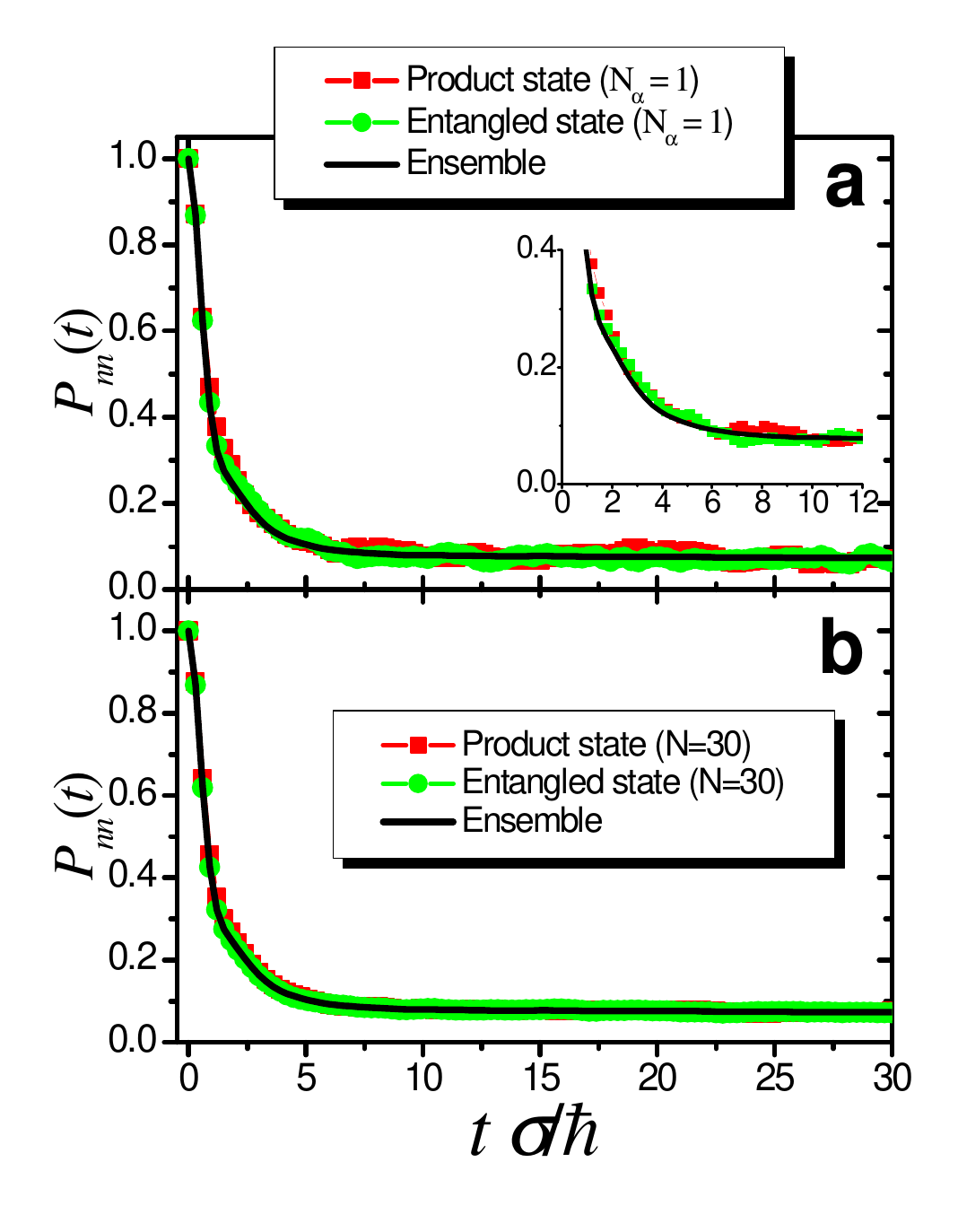}%
\caption[Local polarization evolutions in a star system of $14$ spins.]%
{Polarization evolution of the local spin dynamics in a star system of $14$
spins. The ensemble dynamics compared with an entangled pure-state evolution
is showed. The dipolar interaction $a_{ij}/b_{ij}=-2$ is given by a random
normal distribution with zero mean and $\sigma$ variance. The black line shows
the ensemble local polarization, $P_{nn}^{\mathrm{ens}}(t),$ and the red and
green lines correspond to the evolution obtained with $P_{nn}^{\mathrm{prod}%
,N_{\alpha}}(t)$ and $P_{nn}^{\mathrm{ent},N_{\alpha}}(t)$ respectively. Panel
a) shows the dynamics of eq. (\ref{eq--PCFuncLC_aver}) for $N_{\alpha}=1$ and
panel b) the dynamics for $P_{nn}^{\mathrm{prod},N_{\alpha}}(t)$ and
$P_{nn}^{\mathrm{ent},N_{\alpha}}(t)$, where $N_{\alpha}$ is the lower value
for which each curve reproduces the ensemble dynamics.}%
\label{Fig_Ent_vs_Ens_star}%
\end{center}
\end{figure}

In fig. \ref{Fig_Ent_vs_Ens_star}
it is shown the same calculations observed in fig. \ref{Fig_Ent_vs_Ens_ladder}
but for a star system. This system shows a spin \textquotedblleft
diffusion\textquotedblright\ behavior for the polarization due to the
topological complexity yielding the lack of recurrences for long times. For
$N_{\alpha}=1$, the upper panel of fig. \ref{Fig_Ent_vs_Ens_star} shows that
the agreement between the ensemble dynamics and both $P_{nn}^{\mathrm{prod}%
,1}(t)$ and $P_{nn}^{\mathrm{ent},1}(t)$ curves is good only for values of the
polarization higher than $0.2$. For times where $P_{nn}^{\mathrm{ens}}(t)$ is
close to zero, the \emph{cross terms} in eq. (\ref{eq--PCFuncLC_exp}) become
relevant and both $P_{nn}^{\mathrm{prod},1}(t)$ and $P_{nn}^{\mathrm{ent}%
,1}(t)$ are different from the ensemble curve.
Note that in contrast with the ladder case, in this case, the evolutions
$P_{nn}^{\mathrm{prod},1}(t)$ and $P_{nn}^{\mathrm{ent},1}(t)$ are similar.
For higher values of $N_{\alpha}$ one obtains a better agreement as in the
ladder system. It is important to note that the difference between
$P_{nn}^{\mathrm{ens}}(t)$, $P_{nn}^{\mathrm{prod},N_{\alpha}}(t)$ and
$P_{nn}^{\mathrm{ent},N_{\alpha}}(t)$ would not be appreciable in a real
experiment with a typical signal to noise relation.

The ensemble calculation of eq. (\ref{eq--PCFunc}) needs to project the
evolution of \emph{every} $2^{M-1}$ initial states, $2^{13}=8192$ for this
$14$-spin system, into the same number of possible final states. Instead of
this, if one starts with an entangled state or with a random product state,
the number of initial states gets significantly reduced. Even for the worst
case in which one needs to do $N_{\alpha}\sim630$ averages to obtain a good
agreement with the ensemble curve (low panel of fig.
\ref{Fig_Ent_vs_Ens_ladder}), this number represents a small fraction, lower
than $8\%,$ of the $2^{13}$ initial states of the ensemble. For the case where
only one pure entangled state is enough to mimic the ensemble dynamics, the
number of evolutions is reduced to a number around the $0.01\%$ of the
$2^{13}$ of the ensemble. This shows the potentiality of the method.

\section{Summary}

In summary, in order to overcome the limitations of the numerical calculations
of an ensemble spin dynamics when one increases the number of spins,\textbf{
}we develop a novel numerical method \cite{Alvarez07c}. It exploits the
delicate property of the quantum superpositions and quantum parallelisms
\cite{Loss02parallelism} to reproduce the ensemble dynamics through a pure
entangled state evolution. The method is useful to use the Suzuki-Trotter
product-formula in an ensemble evolution. We showed that the contribution of
the coherences of the pure initial state to the dynamics can be neglected due
to a self averaging property arising on the destructive interferences of the
randomly correlated pure-state. The underlying physical mechanism that makes
possible these efficient simulations seems to be related to the observation
that higher order coherences in highly correlated states decay faster than
those of lower order \cite{Suter04,Suter06,Cory06,Sanchez07}. This suggests
that the \emph{self averaging} property assisted by the large dimension of the
Hilbert space could be involved in what is called \emph{\textquotedblleft
intrinsic decoherence\textquotedblright}. The concept developed here can be
used in two-ways: on one side, it allows for very efficient dynamical
calculations of common experimental situations where big ensembles are
involved. On the other side, it sheds light on how to experimentally prepare
entangled states for specific purposes.

\chapter{Conclusion and final remarks}

In this thesis we had a dive into quantum dynamics focused in the decoherence
phenomenon. Usually, one wants to manipulate a particular system but
inevitably other degrees of freedom interact with it changing its dynamics. In
order to continue using the potentialities of the system of interest, this led
us to study how these external degrees of freedom disturb the system dynamics
While this work was done within the NMR field, the workhorse of quantum
mechanics, our results could be applicable to many fields because they involve
fundamental concepts of quantum mechanics.

In the beginning, we studied the cross-polarization technique for molecular
characterization purposes. Through the well known methodology of the
generalized Liouville-von Neumann quantum master equation, typically used in
the NMR field, we incorporated the degrees of freedom of the environment. This
mechanism was described in chapter \ref{Mark_spin_dynamics_Density_matrix}
where we solved the spin dynamics of many-spin systems. Interested in the
$8$CB characterization, we interpreted cross-polarization experiments over
this liquid crystal in several mesophases \cite{JCP03}. The $8$CB molecule, as
we described in section \S \ \ref{M_Q_dyn_in_8CB}, could be represented by a
three-spin system coupled with a spin-environment. Thus, we began with a
simple system to understand the cross-polarization dynamics. Based on the
M\"{u}ller, \emph{et al.} model \cite{MKBE74}, we solved the two-spin dynamics
interacting with a fast fluctuating spin-bath. We reobtained their solution
and then we extended the model to a three-spin system interacting with a
spin-bath. As this molecule has two possible configurations for the
heteronuclear coupling, this led us to note that in each space of $M=\pm1/2$
there are only two of the three eigenstates that are involved in the dipolar
transitions that give rise to the oscillations. This was explained as a
consequence of the symmetry of the system, i.e. the flip-flop can occur only
between the carbon and one (the symmetric or the antisymmetric) combination of
the proton states depending on the relative signs of the heteronuclear
couplings ($b_{1}=b_{2}$ or $b_{1}=-b_{2}$) \cite{JCP03}. Hence, the frequency
of the oscillation depends of the different configurations. The experimental
data were well fitted to the analytical polarization expression concerning the
oscillation frequency. However, the relaxation process was not well described
using a direct extension of the MKBE model with an isotropic
system-environment interaction. This was manifested by the fact that the
experimental cross-polarization data of the $8$CB molecule showed that the
rate of attenuation of the oscillations is much faster than the rate of
polarization transfer from the bath. Consequently, we extended the MKBE model
to obtain a relaxation superoperator that takes into account this phenomenon
arising from an anisotropic system-environment interaction
\cite{JCP03,Alvarez07a}. The anisotropy is given by the ratio between the
interaction rates of the Ising and XY term. We emphasized the different roles
of the anisotropy of the system-environment interaction on decoherence and
relaxation processes. The main difference is that while the XY interaction
takes the system to the total system equilibrium, the Ising system-environment
interaction takes it to an internal quasi-equilibrium \cite{JCP03,Alvarez07a}.
The introduced anisotropy could be explained in the nematic phase by assuming
a dipolar system-environment interaction Hamiltonian within the extreme
narrowing approximation. Nevertheless, in the smectic phase the anisotropy is
much more pronounced. Hence, we extended the model outside the fast
fluctuation approximation to take into account slower motions. In this way, a
better agreement with the experimental observations was obtained without
resorting to other mechanisms which operate in both phases \cite{JCP03}. This
detailed spin dynamics calculations allowed us to obtain separately the
homonuclear and heteronuclear dipolar couplings in CH$_{2}$ systems which
constitute the $8$CB molecule \cite{JCP03}. We tested the reliability of the
results with an experimental direct determination of the heteronuclear
couplings using cross-polarization under Lee-Goldburg conditions.

Within a fundamental point of view, we observed that the solutions based on
the MKBE model did not describe the quantum quadratic short time behavior nor
the features displayed in fig. \ref{Fig_JCP98_original} within the $b\ll
\Gamma_{\mathrm{SE}}$ region. In order to overcome these limitations, a
further improvement was done including non-secular terms of the
system-environment interaction to extend the solution. This solution describes
well the experimental short time behavior and led us, together with the spin
dynamics analysis within the Keldysh formalism (chapter
\ref{Marker_Spin_within_keldysh}), to enrich our perspectives.

Inspired in the generalized Landauer-B\"{u}ttiker equation within the Keldysh
formalism, we obtained that the initially phenomenological stroboscopic model
\cite{JCP06}, in the continuous form, led to the non-secular solution within
the generalized quantum master equation. This induced us to obtain analytical
results of the spin dynamics from microscopic derivation and characterization
of the system-environment interaction \cite{CPL05,SSC07,Alvarez07b}. To do
that, we solved the Schr\"{o}dinger equation within the Keldysh formalism for
an open system under the wide band regime in the environment (fast fluctuation
approximation). Within this formalism, assisted by the Jordan-Wigner
transformation that maps a spin system into a fermion one, an exact solution
for an XY linear chain is obtainable. This allowed us not only to include
memory effects within the spin-bath \cite{CPL05}, discussed in section
\S \ \ref{Marker_memory_effects_keldysh}, but it enabled us to test
approximation methods to solve the more complex systems of section
\S \ \ref{M_Keldysh_applied_to_spin_systems}. To describe the spin-bath under
the fast fluctuation approximation that leads to interactions local in time,
we resorted to the Wigner time-energy variables. This allowed us to transform
the density function expressed in the Danielewicz integral form into a
generalized Landauer-B\"{u}ttiker equation. By applying this technique to a
two-spin system coupled to a spin-bath \cite{JCP06,Alvarez07b}, we improved
the results obtained through the secular approximation within the standard
density matrix formalism. Further on, we effectively symmetrized the
system-environment interactions virtually transforming them into a spatially
homogeneous process \cite{Alvarez07b}. This involved a uniform
system-environment interaction rate that leads to a simple non-hermitian
propagator while the original multi-exponential decay processes were recovered
by an injection density function. This method led us to interpret the
phenomenon of decoherence and the system-environment interaction within a
special view: The environment as a measurement apparatus \cite{JCP06}. There,
the decoherence takes another perspective: The loss of information of the
system entity, i.e. how the isolated system dynamics is degraded in a
characteristic time given by the fictitious homogeneous decoherence time. The
homogenization procedure enabled the microscopic derivation of the
stroboscopic model for the system-environment interaction \cite{Alvarez07b}.
While, the stroboscopic process may not be the best description of reality, it
provides an optimal numerical algorithm to calculate quantum dynamics in
discrete time steps \cite{JCP06,Alvarez07b}.

While the dynamics obtained through the Keldysh formalism are reproduced by
the non-secular solution within the generalized quantum master equation, they
derive from a microscopic model of the entire system (system plus environment)
and the final state must not be hinted beforehand. The Keldysh formalism gives
us another perspective to discuss about the physics of the quantum time
evolution and the interpretations of the approximations made. For example, the
arising of the quantum Zeno effect shown in the decoherence time,
$1/\tau_{\phi}^{{}}\propto\left(  b/\hbar\right)  _{{}}^{2}\tau_{\mathrm{SE}%
}^{{}}$, could be interpreted as a \textquotedblleft nested\textquotedblright%
\ Fermi golden rule rate emphasizing the non-perturbative nature of the result
\cite{JCP06}.

The manifestation of the quantum Zeno effect led us to a novel interpretation
of previous experiments \cite{JCP98} shown in fig. \ref{Fig_JCP98_original}.
In chapter \ref{Sec_QDPT}, we found experimental evidence that environmental
interactions can drive a swapping gate (two-spin system) through a
\emph{Quantum Dynamical Phase Transition} towards an over-damped or Zeno phase
\cite{JCP06}. The NMR spin swapping experiments in a $^{13}$C-$^{1}$H system
enable the identification and characterization of this phase transition as a
function of the ratio $b\tau_{\mathrm{SE}}/\hbar$ between the internal and
system-environment interaction. The developed microscopic model
\cite{SSC07,Alvarez07b} for the swapping operation describes both phases and
the critical region with great detail, showing that it depends only on the
nature of the interaction \cite{JCP06,Alvarez07b}. In particular, it shows
that the phase transition does not occur if the system-environment interaction
gives isotropic interaction rates, $\Gamma_{\mathrm{ZZ}}=\Gamma_{\mathrm{XY}}%
$. Within the standard approximations typically used to solve the generalized
quantum master equation, one tends to think that $\Gamma_{\mathrm{ZZ}}%
=\Gamma_{\mathrm{XY}}$ corresponds to an isotropic system-environment
interaction. However, a careful microscopic derivation within the Keldysh
formalism showed us that this is not necessarily correct. It is also important
to mention that the spin-bath occupation factor modifies the $\Gamma
_{\mathrm{ZZ}}$ rate. For the description of the quantum dynamical phase
transition, it is crucial to distinguish between inputs and output parameters.
One can visualize applications that range from tailoring the environments for
a reduction of their decoherence on a given process to using the observed
critical transition in frequency and decoherence rate as a tracer of the
environment's nature. For example, we extended the model to a $3$-spin system
to show that beyond a critical region two spins become almost decoupled from
the environment oscillating with the bare Rabi frequency and relaxing more
slowly \cite{Alvarez07a}.

Inspired in the stroboscopic model and the arising of the environmentally
induced quantum Zeno phase, we developed a new NMR pulse sequence.
\textquotedblleft The pruner\textquotedblright\ stroboscopically interrupts
the system evolution to improve the transfer of polarization through a
specific pathway in a system of many interacting spins \cite{Alvarez07d}. The
sequence effectively prunes branches of spins, where no polarization is
required, during the polarization transfer procedure. We obtained a remarkable
enhancement of the polarization transfer with respect to the standard methods.
Therefore, it is a very practical tool for NMR applications where a signal
gain is of great importance. However, from a fundamental point of view, it is
the starting point of an engineered application of the stroboscopic model
\cite{JCP06,Alvarez07b} described in section
\S \ \ref{Marker_Stroboscopic_process} which seems to manifest a spin dynamics
within the quantum Zeno phase \cite{JCP06,SSC07}. Moreover, it can help to go
deeper in the understanding of decoherence processes and consequently of the
environmentally induced quantum dynamical phase transition \cite{JCP06,SSC07}.

Finally, in order to study the spin dynamics of larger systems, we developed a
novel numerical method that exploits quantum parallelism \cite{Alvarez07c}.
This provides a tool to overcome the limitations of standard numerical
calculations of ensemble spin dynamics for high number of spins. Hence, this
numerical method constitutes the starting point to extend the study of the
quantum dynamical phase transition to larger systems and how it is involved in
the irreversibility phenomena. Moreover, the observation that the contribution
of many coherences of an entangled randomly correlated state to the dynamics
can be neglected due to a self averaging property,
opens a fundamental question: Is this \emph{self averaging} property involved
in what is called \emph{\textquotedblleft intrinsic
decoherence\textquotedblright}?

\backmatter

\bibliographystyle{amsalpha}
\bibliography{biblio}

\end{document}